# Solar System Physics for Exoplanet Research

Horner, J.[1][*], Kane, S. R.[2], Marshall, J. P.[3], Dalba, P. A.[2][†], Holt, T. R.[1,4], Wood, J.[1,5], Maynard-Casely, H. E.[6], Wittenmyer, R.[1], Lykawka, P. S.[7], Hill, M.[2], Salmeron, R.[8,9,1], Bailey, J.[10], Löhne, T.[11], Agnew, M.[12], Carter, B. D.[1], Tylor, C. C. E.[1]

[1] Centre for Astrophysics, University of Southern Queensland, Toowoomba, QLD 4350, Australia
[2] Department of Earth and Planetary Sciences, University of California, Riverside, CA 92521, USA
[3] Academia Sinica, Institute of Astronomy and Astrophysics, 11F Astronomy-Mathematics Building, NTU/AS campus, No. 1, Section 4, Roosevelt Rd., Taipei 10617, Taiwan
[4] Southwest Research Institute, Department of Space Studies, Boulder, CO. USA
[5] Hazard Community and Technical College, 1 Community College Drive, Hazard, Ky, USA 41701
[6] Australian Nuclear Science and Technology Organisation, Locked Bag 2001, Kirrawee DC, NSW, Australia
[7] Kindai University, 3–4–1 Kowakae, Higashiosaka City, Osaka, 577–8502, Japan
[8] Research School of Astronomy and Astrophysics, Australian National University, Canberra, ACT 2611, Australia
[9] Mathematical Sciences Institute, Australian National University, Canberra, ACT 2601, Australia
[10] School of Physics, University of New South Wales, Sydney, New South Wales, Australia
[11] Astrophysikalisches Institut, Friedrich-Schiller-Universität Jena, Schillergäßchen 2–3, 07745 Jena, Germany
[12] Centre for Astrophysics and Supercomputing, Swinburne University of Technology, Hawthorn, Victoria 3122, Australia

[*] **e-mail:** jonathan.horner@usq.edu.au
[†]: National Science Foundation Astronomy & Astrophysics Postdoctoral Fellow

**ABSTRACT**

Over the past three decades, we have witnessed one of the great revolutions in our understanding of the cosmos - the dawn of the Exoplanet Era. Where once we knew of just one planetary system (the Solar system), we now know of thousands, with new systems being announced on a weekly basis. Of the thousands of planetary systems we have found to date, however, there is only one that we can study up-close and personal - the Solar system.

In this review, we describe our current understanding of the Solar system for the exoplanetary science community - with a focus on the processes thought to have shaped the system we see today. In section one, we introduce the Solar system as a single well studied example of the many planetary systems now observed. In section two, we describe the Solar system's small body populations as we know them today - from the two hundred and five known planetary satellites to the various populations of small bodies that serve as a reminder of the system's formation and early evolution. In section three, we consider our current knowledge of the Solar system's planets, as physical bodies. In section four we discuss the research that has been carried out into the Solar system's formation and evolution, with a focus on the information gleaned as a result of detailed studies of the system's small body populations. In section five, we discuss our current knowledge of planetary systems beyond our own - both in terms of the planets they host, and in terms of the debris that we observe orbiting their host stars.

As we learn ever more about the diversity and ubiquity of other planetary systems, our Solar system will remain the key touchstone that facilitates our understanding and modelling of those newly found systems, and we finish section five with a discussion of the future surveys that will further expand that knowledge.

# 1. INTRODUCTION

Prior to the discovery of the first planets around other stars (Gamma Cephei Ab and HD 114762b, initially thought to be brown dwarfs (Campbell et al., 1988; Latham et al., 1989); the planets orbiting pulsars PSR 1257+12 (Wolszczan & Frail, 1992) and PSR B1620-26 (Thorsett, Arzoumanian & Taylor, 1993); and 51 Pegasi, the first planet around a Sun-like star (Mayor & Queloz, 1995)), our ideas on how planetary systems formed and evolved were based purely on observations of the Solar system. A wide range of scenarios had been proposed, of varying complexity, all of which attempted to explain the minutiae of our Solar system.

For many years, it was believed that the Solar system was formed as a result of an encounter between the Sun and a passing star, which was thought to have pulled a tongue of material either from our star, or that which it encountered, that subsequently fragmented to form the planets (e.g. Lyttleton, 1936; Woolfson, 1964). Such a scenario would suggest that the formation of planetary systems should be an incredibly rare and violent process, since the likelihood of passing stars experiencing such close encounters (nearly colliding with one another) is almost vanishingly small.

The theory which held sway at the start of the epoch of exoplanetary discovery argued instead that the planets had formed from a primordial circumstellar disc around the Sun, and that their formation was essentially a gentle, leisurely process (e.g. Swedenborg, 1734; Laplace, 1796; Edgeworth, 1949; Safronov, 1969[1]; Lissauer, 1993). The expectation was that, if we were we to discover planets around other stars, those planetary systems would strongly resemble our own. The fact that those first detected systems (and, indeed, the vast majority of those discovered since) were dramatically different to our own led to a thorough re-examination of how planetary systems could form.

Since the discovery of the first planets orbiting other stars, the fields of exoplanetary and Solar system science have remained largely separate – one dealing with a small amount of information about a large number of systems, the other dealing with our almost overwhelmingly detailed knowledge of our own Solar system. In an attempt to help remedy the disconnect between the two fields, in this work, we present a review of our current knowledge and understanding of the Solar system.

In the sections that follow, we detail the various populations of objects known in the Solar system and describe the theories that attempt to explain how they arrived in their current locations. We include some of the more speculative ideas within Solar system astronomy, in order to ensure that this review is as clear and comprehensive as possible. Once all these concepts have been discussed, we move on to a discussion of our burgeoning knowledge of exoplanetary systems, with a particular focus on highlighting the similarities and differences between those systems and our own. Since this is by nature a lengthy review, spanning a wide variety of topics, we here present a short bulleted summary of the various topics described in this work, together with pointers to the relevant section, and indications of any exoplanetary systems to which those ideas may apply (where appropriate).

- **In section 2, we describe the Solar system in its entirety, as known today.**

- **In section 3, we introduce the Solar system's planets, describing them as physical objects – from their atmospheres to our current understanding of their interiors and compositions.**

- **In section 4, we give a brief overview of the key processes thought to have occurred in the Solar system's formation. That section comprises the following subsections:**

    - **In section 4-1, we describe the pivotal role played by giant impacts in the formation of the Solar system we observe today.**

---

[1] For an English language translation of Safronov (1969), we direct the interested reader to Safronov (1972).

- o **In section 4-2, we discuss the impact rate within our Solar system in more general terms.**

- o **In section 4-3, we consider the migration, both inward and outward, of the giant planets of the Solar system.**

- o **In section 4-4, we discuss the ongoing debate over the origin of water on the Earth.**

- o **In section 4-5, we discuss theories that invoke the presence of planet-mass objects moving sufficiently far from the Sun that they have, to date, escaped detection.**

- o **In section 4-6, we consider the formation and evolution of the Oort cloud, which might provide evidence on the type of cluster environment in which the Sun formed.**

- o **In section 4-7, we detail the formation and distribution of dust in the Solar system.**

- o **In section 4-8, we discuss the recent discoveries of rings orbiting some of the Solar system's small bodies**

- o **In section 4-9, we describe the various non-gravitational effects that can significantly affect the orbits of both dust and kilometre-sized bodies, before moving on to detail the various mechanisms by which material is lost from the Solar system's small body reservoirs.**

- **In section 5, we turn our attention to exoplanetary science, highlighting how our knowledge of the Solar system can (and does) inform our understanding of other planetary systems. That section comprises the following subsections:**

  - o **In section 5.1, we describe the known demographics of exoplanets.**

  - o **In section 5.2, we discuss how our Solar system would look as an exoplanetary system.**

  - o **In section 5.3, we discuss debris and dust in exoplanetary systems.**

  - o **In section 5.4, we discuss the comparative analysis of planetary systems.**

  - o **In section 5.5, we discuss super-Earths and sub-Neptune - two classes of planet that are not found within the Solar system.**

  - o **In section 5.6, we discuss the atmospheres of exoplanets, and their relation to those in the Solar system.**

  - o **In section 5.7, we describe research into the dynamics of exoplanetary systems.**

  - o **In section 5.8, we describe current and future exoplanet surveys and missions.**

- **Finally, in section 6, we draw together our conclusions.**

## 2. OUR SOLAR SYSTEM – A PLETHORA OF SMALL BODIES

The Solar system contains a wide variety of objects, many of which have unusual or unexpected features that hold information on their formation, and the formation of the Solar system itself (de Pater & Lissauer, 2015). Aside from the Sun, the most massive objects in the Solar system are the planets[2] – the gas giants, Jupiter and Saturn; the ice giants, Uranus and Neptune; and the terrestrial planets[3], Mercury, Venus, Earth and Mars. Six of these eight planets host one or more *natural satellites* – with over two hundred distributed between them (the great majority of which orbit either Jupiter or Saturn). These satellites in turn range in size from the tiny (kilometre-sized, or smaller) to those that are almost planet-like (the Moon, the Galilean satellites of Jupiter, Saturn's Titan, all of which are comparable in diameter to Mercury, albeit significantly less massive). The satellites of the outer planets are often broken down into two distinct classes (e.g. Kuiper, 1951a, though he counted our Moon as a third class by itself).

The *regular satellites* orbit close to their host planet, moving on prograde orbits that have both low inclinations (with respect to the planet's equatorial plane) and low eccentricities (e.g. Mosqueira & Estrada, 2003a, b; Sasaki, Stewart & Ida, 2010). The *irregular satellites*, by contrast, are scattered much more widely around their host planets, and move on orbits that can be both prograde and retrograde (e.g. Nesvorný et al., 2003; Jewitt & Haghighipour, 2007; Holt et al., 2018). These objects often move on highly inclined and eccentric orbits, and are typically (with the exception of Neptune's giant moon Triton) relatively small. Indeed, the largest irregular satellite after Triton is Saturn's Phoebe, which is just over 200 km in diameter (Thomas, 2010).

| **Name** | $a$ (AU) | $e$ | $i$ (º) | $M/M_{Earth}$ | $R/R_{Earth}$ | $N_{Sat}$[4] |
|---|---|---|---|---|---|---|
| Mercury | 0.387 | 0.206 | 7.00 | 0.0553 | 0.383 | 0 |
| Venus | 0.723 | 0.00673 | 3.39 | 0.815 | 0.950 | 0 |
| Earth | 1.00 | 0.0164 | 0.00345 | 1.00 | 1.00 | 1 |
| Mars | 1.52 | 0.0934 | 1.85 | 0.107 | 0.532 | 2 |
| Jupiter | 5.20 | 0.00487 | 1.30 | 318 | 11.0 | 79 (71) |
| Saturn | 9.58 | 0.00508 | 2.49 | 95.2 | 9.14 | 82 (58) |
| Uranus | 19.1 | 0.00474 | 0.772 | 14.5 | 3.98 | 27 (9) |
| Neptune | 30.2 | 0.00869 | 1.78 | 17.1 | 3.86 | 14 (8) |

*Table 1:* *The key physical parameters of the planets, and the number of satellites they possess. The masses (M) and mean radii (R) were taken from the JPL's Solar system dynamics page[5] on 26th July 2019. The osculating orbital elements (semi-major axis, a, eccentricity, e, and inclination, i) were obtained using the JPL HORIZONS web-interface[6], and are valid on 25th March 2019. For ease of reading, all values are given to three significant figures - the actual values are known to a far greater level of precision. The number given in parentheses in the $N_{Sat}$ column details how many of the planet's satellites are classified as irregular.*

---

[2] We provide an overview of the current state of our knowledge of the planets as physical objects in section three, below.

[3] We note that the term 'terrestrial planet' can sometimes be taken as inferring a degree of similarity with the Earth that is not entirely justified for such a diverse group of objects, and urge the reader to remember that 'terrestrial' does not equal 'Earth-like'.

[4] The number of satellites for each planet was taken from the JPL Planetary Satellite Discovery Circumstances page (at http://ssd.jpl.nasa.gov/?sat_discovery), which gives the details of each known satellite's discovery circumstances. The site was accessed on 26th July 2019, at which time it had last been updated on 22nd February 2019.

[5] *http://ssd.jpl.nasa.gov/?planet_phys_par*, accessed on 26th July 2019

[6] *http://ssd.jpl.nasa.gov/horizons.cgi#top*, accessed on 26th July 2019

In addition to the major planets, five objects are currently classified as *dwarf planets*[7] by the International Astronomical Union - namely Ceres, Pluto, Haumea, Makemake and Eris. These five objects reside in two distinct reservoirs (the *Asteroid belt* and the *trans-Neptunian disc*), each of which contains vast numbers of smaller objects. It is likely that, as more observations are made to determine the size and shape of known trans-Neptunian objects, and astronomical surveys delve farther into the outer reaches of our Solar system, the number of known dwarf planets will increase significantly (for an up-to-date tally, we direct the interested reader to the Minor Planet Centre website[8]).

The two reservoirs mentioned above, in addition to the *Oort cloud* (a third reservoir which stretches out to halfway to the nearest star), are dynamically stable on timescales comparable to the age of our Solar system, and as such, contain a vast number of objects (with the Asteroid belt, the least populous, hosting an estimated $10^6$ objects of diameter 1 km or greater; e.g. Tedesco & Desert, 2002). In addition to these three famous reservoirs, two additional regions of stability are thought to each house at least as many objects as are present in the Asteroid belt[9] - the *Jovian and Neptunian Trojans* (e.g. Jewitt et al., 2000; Sheppard & Trujillo, 2006). Trojans are objects moving in essentially the same orbit as their host planet, which librate around the $L_4$ and $L_5$ Lagrange points of that planet's orbit (Figure 1), such that, on average, they remain either 60 degrees ahead, or 60 degrees behind the planet in its orbit.

---

[7] Dwarf planets, as defined by the International Astronomical Union, are objects sufficiently massive that their self-gravity can overcome their material strength. Dwarf planets therefore assume a shape determined by hydrostatic equilibrium – usually nearly spherical, with varying degrees of oblateness dependent on the object's spin rate. However, they are not sufficiently massive to have dynamically cleared the area around their orbit, and are not in orbit around a planet. This definition of "dwarf planet" remains somewhat controversial, with a vocal minority continuing to protest the demotion of Pluto from planet to dwarf. In passing, we note that the timescale for an object to clear the region around its orbit increases with the orbital period of that object, such that if the Earth were placed sufficiently far from the Sun, it would be unable to do so within the age of our Solar system, and hence would also be classified as a dwarf planet. For the full text of the IAU resolution that defined the nature of planets, we direct the interested reader to http://www.iau.org/public_press/news/detail/iau0603/. Details of an alternative metric, based on the timescale required for objects to clear their orbital zone (through the ejection or accretion of other objects on similar orbits), can be found in Margot, 2015.

[8] http://www.minorplanetcenter.net/dwarf_planets

[9] We note that, although the number of objects in the Jovian Trojan population greater than 1 km in diameter likely rivals, or exceeds, those contained in the Asteroid belt, the total mass of the Jovian Trojan population is significantly smaller than that contained in the Asteroid belt.

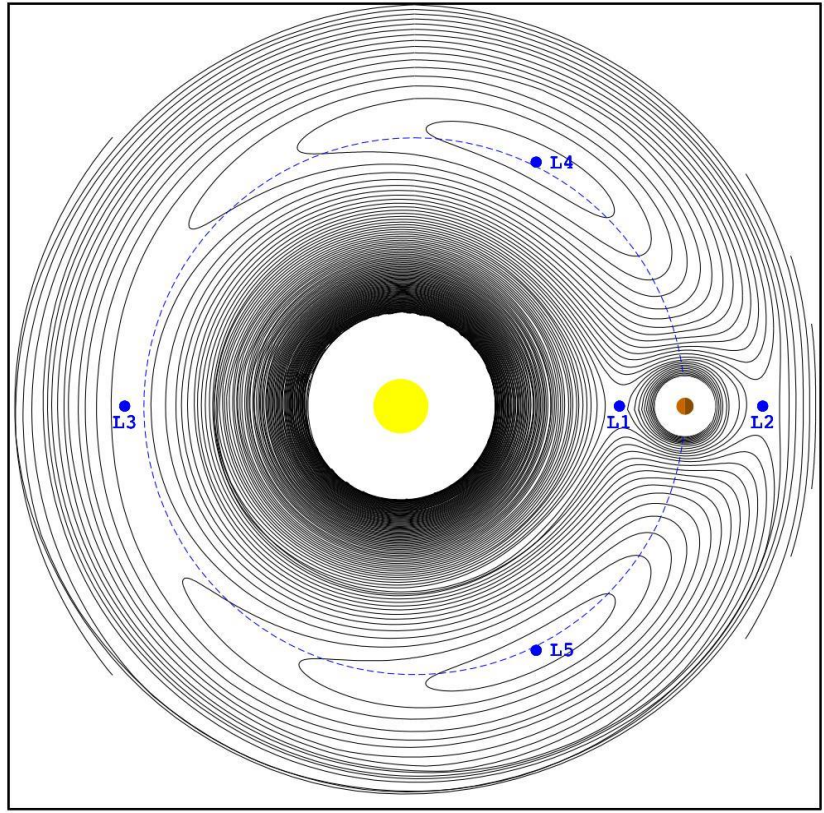

*Figure 1: The locations of the five Lagrange points in the case of the restricted three-body problem, originally plotted for Horner & Lykawka, 2011, and reproduced here with the permission of Oxford University Press. Solid lines connect areas of equal gravitational potential. All of the Lagrange points are equilibrium points – and hence are regions in which objects can move on orbits that are more dynamically stable than might otherwise be expected. $L_1$, $L_2$ and $L_3$ are similar to saddle points – although they are local plateaus in the gravitational potential, a small displacement is enough to carry an object away from the region of relative stability, back to a chaotic orbit. By contrast, $L_4$ and $L_5$ are broad regions in which an object can remain dynamically stable on long timescales. The populations of Jovian and Neptunian Trojans, which are thought to number at least a million objects of diameter 1 km or greater, are concentrated within these tadpole-like regions ahead and behind the location of their host planet in its orbit. The figure shows the potentials for a scenario where the primary:secondary mass ratio matches that between the Sun and Jupiter – i.e. where the primary is approximately 1050 times the mass of the secondary.*

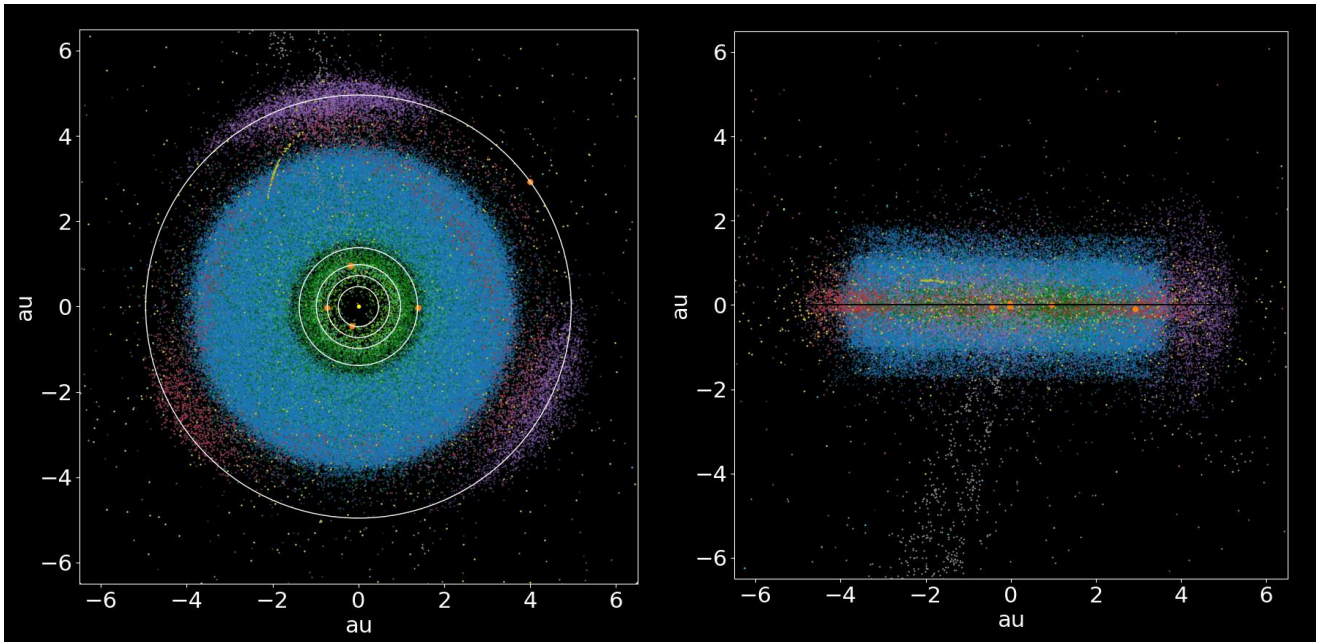

*Figure 2: The distribution of objects in the inner six au of the Solar system, shown in Cartesian coordinates[10]. The right panel shows a 'top-down' view of the system, whilst the left panel shows a 'side-on' view. The data plotted were taken from the JPL HORIZONS database[11], and the positions of the objects are shown at epoch 2000-01-01 00:00:00 UT. The various small body populations are colour coded as follows: the near-Earth asteroids (NEAs) in green (with the sub-populations distinguished by distinct shares, as follows: Atiras in aquamarine, Atens in chartreuse, Apollos in sea green and Amors in dark green); main-belt asteroids in blue, the Hilda asteroids in red, the Jovian Trojans in purple, the Centaurs in brown, long-period comets in grey, Jupiter family comets in olive and Halley-type comets in cyan. The locations of the five innermost planets (Mercury, Venus, Earth, Mars and Jupiter) are marked in orange, with their orbits shown in white. The large number of objects moving on highly eccentric orbits that extend towards the 11 o'clock position (top) and almost vertically down from the plane of the system (bottom plot) are the sungrazing comets (e.g. Marsden, 1967, 1989), thousands of which have been discovered in recent years by the Solar Heliospheric Observatory, SOHO (e.g. Battams & Knight, 2017). We note that individual plots for each of the various small body populations are provided in the appendix, for the interested reader. The 'arc' feature of Jupiter family comets at the outer edge of the asteroid belt (top left of the belt) is the result of the fragmentation of comet 73P/Schwassmann-Wachmann 3 in 1995 (e.g. Crovisier et al., 1996). Almost 70 fragments of that comet are now known, and in the 24 years since the parent comet disintegrated, those fragments have begun to disperse around the parent's orbit[12].*

In Figure 2, we plot the instantaneous locations of all known bodies in the inner Solar system, at the locations they would have occupied at 00:00:00 UT on 1st January, 2000. The two clouds of Jovian Trojans are clearly seen sharing Jupiter's orbit, marked in purple, as are the Hilda asteroids (in red), trapped in 3:2 mean-motion resonance with Jupiter[13], which trace a triangular pattern with the vertices located at the Jovian $L_3$, $L_4$ and $L_5$ Lagrange points. The most prominent feature, however, is the *Asteroid belt*, which can be clearly seen, with its members coloured blue. The belt stretches from just beyond the orbit of Mars to a loosely defined outer edge well within the orbit of Jupiter, and is thought to contain more than a million objects of diameter greater than 1 km (e.g. Tedesco & Desert, 2002). To date, more than 700,000 asteroids have been discovered[14] - the vast majority of which (523,824[15]) have given numerical designations to denote that their orbits are considered to be well constrained on the basis of observations over an extended period of time. The overwhelming majority of those objects lie within the Asteroid belt - as can be seen in Figures 2 and 3.

---

[10] In the appendix, we present a series of figures in black and white showing the same information as that contained in the population figures within this section of the paper. Those figures show the individual populations one at a time, rather than all together, and are suitable for printing out in black-and-white, and for those who might otherwise struggle to differentiate between the variety of colours required to show all populations in a single figure.

[11] The JPL Horizons database is a tool that can be used to generate ephemerides for all known Solar system bodies. The web interface to the database can be found at https://ssd.jpl.nasa.gov/horizons.cgi , though this offers a limited interface to the full dataset. As noted on the website, full access is available through a dedicated telnet interface, at horizons.jpl.nasa.gov, port 6775.

[12] In 2022, the debris from the disintegration of 73P/Schwassmann-Wachmann will come very close to Earth, potentially causing a significant meteor storm from the τ Herculid meteor shower (e.g. Wiegert et al., 2005). A second potential storm could occur in 2049, when the debris once again comes particularly close to the Earth.

When the instantaneous locations of the asteroids are plotted within the Solar system (as can be seen in Figure 2), the bulk appear to be constrained within a puffed-up torus of material that initially appears to be remarkably unstructured. However, when one instead plots the orbital elements of the objects in the inner Solar system (Figure 3), it becomes obvious that the region beyond the orbit of Mars is actually highly structured. The main Asteroid belt is threaded by narrow regions where few asteroids are found - features known as the Kirkwood gaps, which mark the location of mean-motion resonances between the asteroids and Jupiter. Exterior to the main concentration of objects in the Asteroid belt, the Hildas (in red) and Trojans (in purple) are narrow bands of objects, trapped in mean-motion resonance with Jupiter (the 3:2 and 1:1 resonances, respectively).

Figure 4 shows a more detailed view of the main Asteroid belt. By zooming in in this fashion, the complex structure of the belt is more clearly visible. The tendency for the asteroids to move on orbits that avoid certain semi-major axes (the locations of mean-motion resonance with Jupiter) can clearly be seen. Similarly, certain regions of inclination space appear favoured, and some regions contain objects spread over a far wider range of orbital eccentricities than others. This complexity is the result of the interplay between mean-motion resonances, secular resonances, and also the ongoing collisional attrition of the belt, with collisional families of asteroids causing certain regions to be more densely populated than their neighbours (a topic we will revisit in more detail in section 4.1).

---

[13] A mean-motion resonance (hereafter MMR) occurs when the orbital period of one body is approximately in integer ratio with that of another. The Trojans, mentioned earlier, are therefore said to be in 1:1 resonance with their host planet, while Pluto, which completes two orbits of the Sun for every three completed by Neptune, is said to be trapped in the Neptunian 2:3 MMR. By convention, resonances are described by a ratio *n:m*. When *m > n*, the resonance is *exterior* to the major body involved, and when *n > m*, the resonance is interior to the orbit of that object.

[14] We note, here, that some of these 700,000+ objects have diameters less than 1 km – and that the observed population of objects larger than 1 km remains incomplete.

[15] As of 11th April, 2019, based on the information at the JPL Solar System Dynamics page, at https://ssd.jpl.nasa.gov/?sb_elem

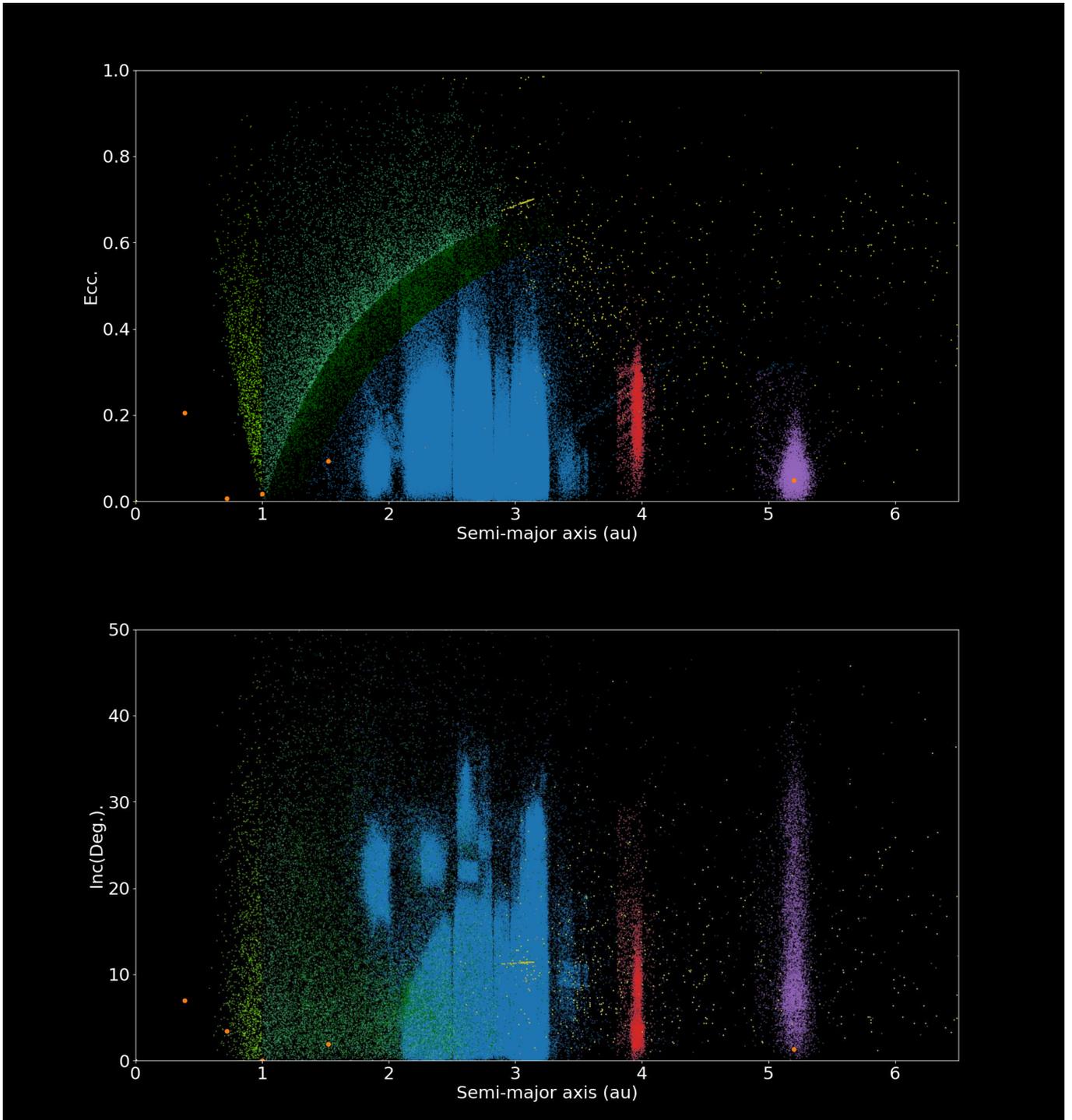

*Figure 3: The observed distribution of the small bodies in the inner Solar system, as a function of their semi-major axis, eccentricity, and inclination. The various small body populations are plotted in different colours, using the same colour scheme as for Figure 2 (near-Earth asteroids in green, with the sub-populations as follows: Atiras in aquamarine; Atens in chartreuse, Apollos in sea green and Amors in dark green. The Main-Belt Asteroids are shown in blue, the Hildas in red, Jovian Trojans in purple, and the Jupiter family comets in olive). The inner planets (Mercury, Venus, Earth, Mars & Jupiter - orange) are also indicated. The important role of mean-motion resonances can be clearly seen in the main belt, in the form of gaps where few objects are found, and beyond, in the form of the Hilda and Jupiter Trojan populations, constrained to narrow bands in semi-major axis. The various individual populations of object plotted here are also plotted, individually, in the appendix, as a resource for the interested reader.*

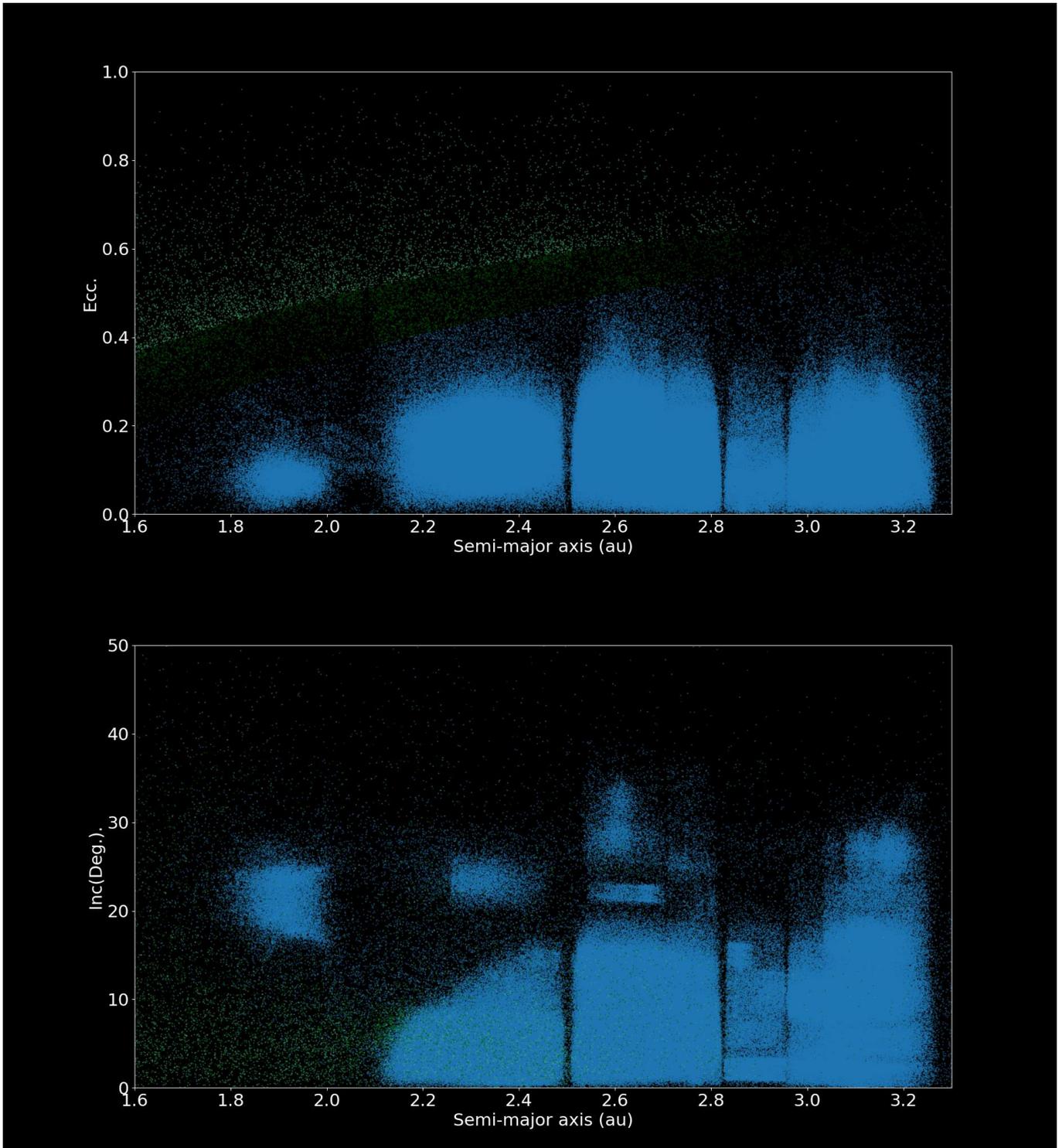

*Figure 4: The observed distribution objects in the Asteroid belt, plotted in semi-major axis-eccentricity space (top) and semi-major axis-inclination space (bottom). The data were taken from the JPL Horizons database, In this plot, The Apollo NEO's are indicated by sea green, and the Amor NEOs by dark green. Main belt asteroids, including Mars crossing asteroids, are indicated in blue. By zooming in on the belt in this manner, a vast amount of fine structure is visible - including the impact of secular and mean-motion resonances throughout the belt, as well as the collisional disruption of asteroids creating asteroid families.*

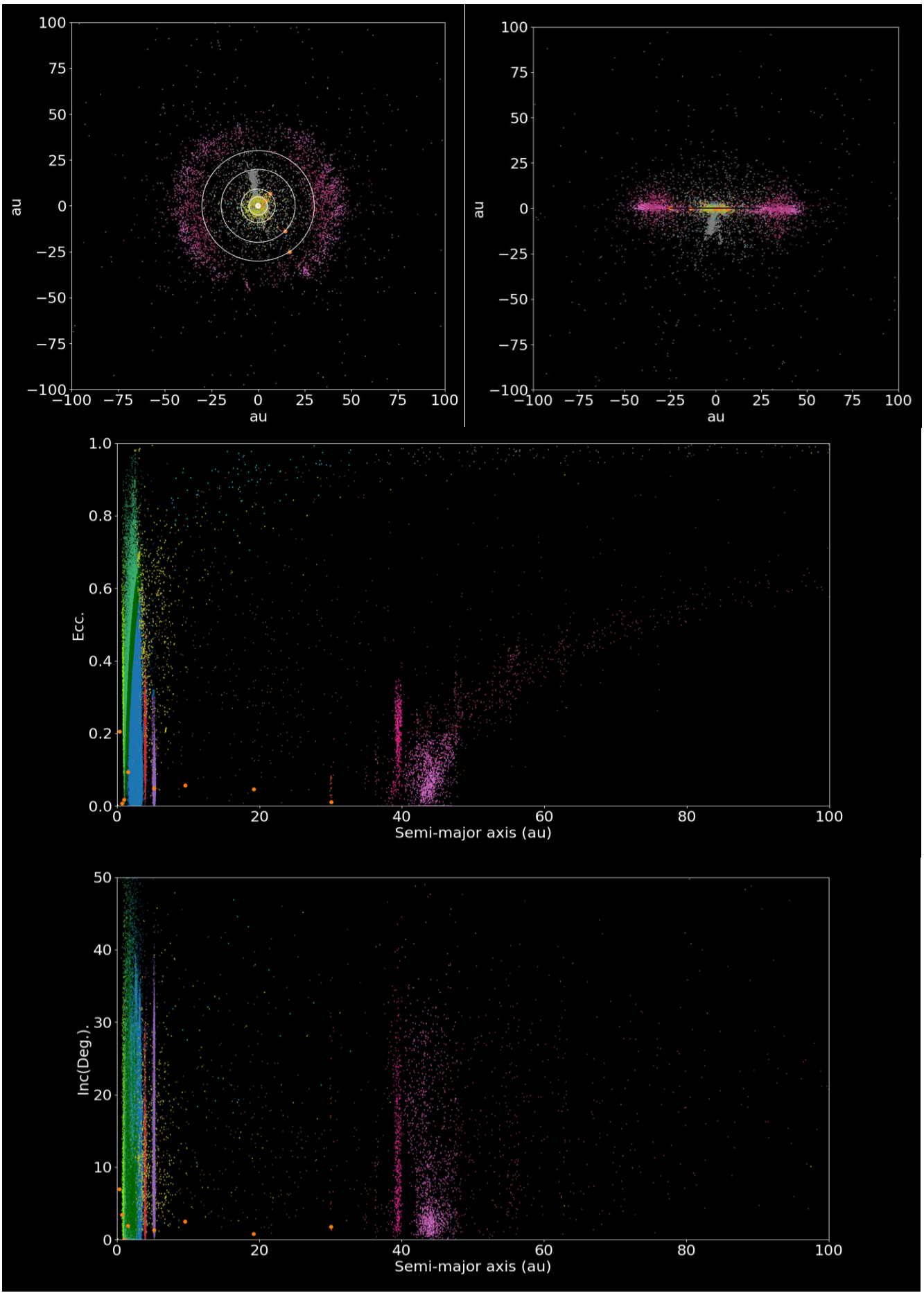

*Figure 5: The distribution of bodies in the Solar system. Top: the Solar system as seen face-on (left) and seen edge-on (right), showing the distribution of objects in Cartesian coordinates. The positions of the objects are those they occupied on 1st January 2000. The middle and lower panels show the orbital elements of the objects in semi-major axis-eccentricity (a-e) and semi-major axis-inclination (a-i) space, respectively. The objects in the inner Solar system are coloured as in Figure 3. Outer Solar system objects are also coloured according to their classification. As was the case in the inner Solar system, populations of resonant objects stand out in the middle and lower plots as concentrations of objects centred on a specific semi-major axis. Centaurs (brown) are shown between Jupiter and Neptune. The Neptune Trojans (orange-red) can be seen at 30 au, and the Plutinos (deep pink) at 39.5 au, just interior to the objects that make up the classical Edgeworth-Kuiper belt (orchid, between ~40 and 48 au). To higher eccentricities, the Scattered Disc objects (maroon) can be seen spreading outward in a curved population in the middle plot - objects whose perihelia fall between ~30 and 40 au that move on eccentric, chaotic orbits. Two cometary populations are shown, the Jupiter family comets (olive) and the Halley type comets (cyan).*

The *trans-Neptunian region* contains a vast number of objects, ranging in size up to the dwarf planets (such as Pluto and Eris), which are shown (along with those objects whose objects cross those of the outer planets) in Figure 5. The stable populations of objects beyond Neptune's orbit are typically broken down into a number of sub-categories. The *Edgeworth-Kuiper belt* (named for two of the scientists to discuss its potential existence – see Edgeworth, 1943, 1949; Kuiper, 1951b), or *Classical belt*, stretches between the approximate location of the Neptunian 2:3 mean-motion resonance and its 1:2 resonance (located at ~39.4 and ~47.8 au, respectively). Objects in the Edgeworth-Kuiper belt typically move on orbits with inclinations less than 30º, and are themselves often broken down into two sub-populations, which seem to display somewhat different physical characteristics, and are known as the "hot" and "cold" populations (e.g. Levison & Stern, 2001; Elliot et al., 2005; Petit et al., 2011; Fraser et al., 2014). The bulk of such objects known to date move on orbits characterised by both low eccentricities and low inclinations (typically of order a few degrees), and are observed to be characteristically red in colour. These objects are known as the *cold classical component* of the disc. The "*hot*" component contains objects with significantly greater orbital inclinations, and often somewhat higher eccentricities. These objects are often observed to be somewhat bluer than their "cold" counterparts, which might be evidence that they experience more frequent and energetic collisions (a result of their dynamically more excited state), and are hence less "weathered" (Fraser & Brown, 2012).

In addition to the Edgeworth-Kuiper belt, the trans-Neptunian population contains two further types of object that are dynamically stable on timescales comparable to the age of the Solar system – the *resonant* and *detached* populations. The resonant objects (e.g. Gladman et al., 2012) move on orbits trapped within one of Neptune's MMRs - for example the so-called *Plutinos* (named for the dwarf planet Pluto, the largest member; e.g. Yu & Tremaine, 1999; Chiang & Jordan, 2002) are trapped within Neptune's 2:3 MMR. These resonant objects can move on orbits that are significantly more excited than those of the Classical population, with some displaying particularly large inclinations and eccentricities. Indeed, many of the Plutinos move on orbits that cross that of Neptune. The nature of the resonance in which they are trapped ensures that they never encounter the planet, however - whenever they cross its orbit, it is located well ahead, or well behind the crossing point. The detached population move on orbits with semi-major axes greater than the location of Neptune's 1:2 resonance, and have perihelion distances (closest approach to the Sun) sufficiently far from Neptune that it cannot significantly perturb their orbits over the age of the Solar system. Typically, detached objects pass perihelion at least 40 au from the Sun (Gladman et al., 2008), although there are a few exceptions whose perihelia lie at slightly smaller radii.

The final population of trans-Neptunian objects is known as the *Scattered Disc* (Duncan & Levison, 1997). Scattered Disc objects typically move on dynamically excited orbits, with perihelia closer to Neptune than the location of its 2:3 resonance. As such, they can experience dynamical perturbations from that planet sufficient to result in significant variations in their orbits over the age of the Solar system. Though many of these objects are nominally dynamically unstable on a wide range of timescales, the lifetime of the population is such that the great majority would be expected to remain as Scattered Disc objects for timescales comparable to the lifetime of our Solar system. Trans-Neptunian objects that are highly dynamically unstable due to strong interactions with Uranus or Neptune are often described as *scattering objects*. Gladman et al. (2008) classify such objects as being those that experience such interactions within a 10 Myr window of orbital evolution, whilst the remaining unstable TNOs are classified as *scattered objects*.

The *Oort cloud* (sometimes called the Öpik-Oort cloud, named for the researchers who first proposed its existence; Öpik, 1932; Oort, 1950), the third large reservoir of small bodies in the Solar system, stretches approximately halfway to the nearest star, and is thought to contain up to $10^{11}$ cometary nuclei of diameter 1 km or greater. Objects within the cloud itself are much too far from the Sun to be detected directly, but the existence of the cloud is inferred from the ongoing flux of material sourced from that region to the inner Solar system - a group of objects known as the Oort Cloud Comets. The Oort cloud is sufficiently far from the Sun that its members are only tenuously bound to our star, and are continually tugged and tweaked by the effects of the galactic tide and the gravitational influence of passing stars. These continual perturbations act to nudge new comets towards the inner Solar system, replacing those that are removed from the system by ejection, collision with an object in that region, and fragmentation.

In addition to the dynamically stable reservoirs of debris detailed above, the Solar system is also cluttered with innumerable unstable objects - including the *Jupiter-family*, *Halley-type*, and *long-period comets*, the *Centaurs*, and the *near-Earth asteroids* (e.g. Wiegert & Tremaine, 1999; Bottke et al., 2002; Morbidelli et al., 2002; Horner et al., 2003; Jewitt, 2009). By nature, these populations are highly dynamically unstable, and would typically decay to nothing on timescales far shorter than the age of the Solar system. As such, they must be continually replenished in order to support the supposedly steady-state populations we observe today.

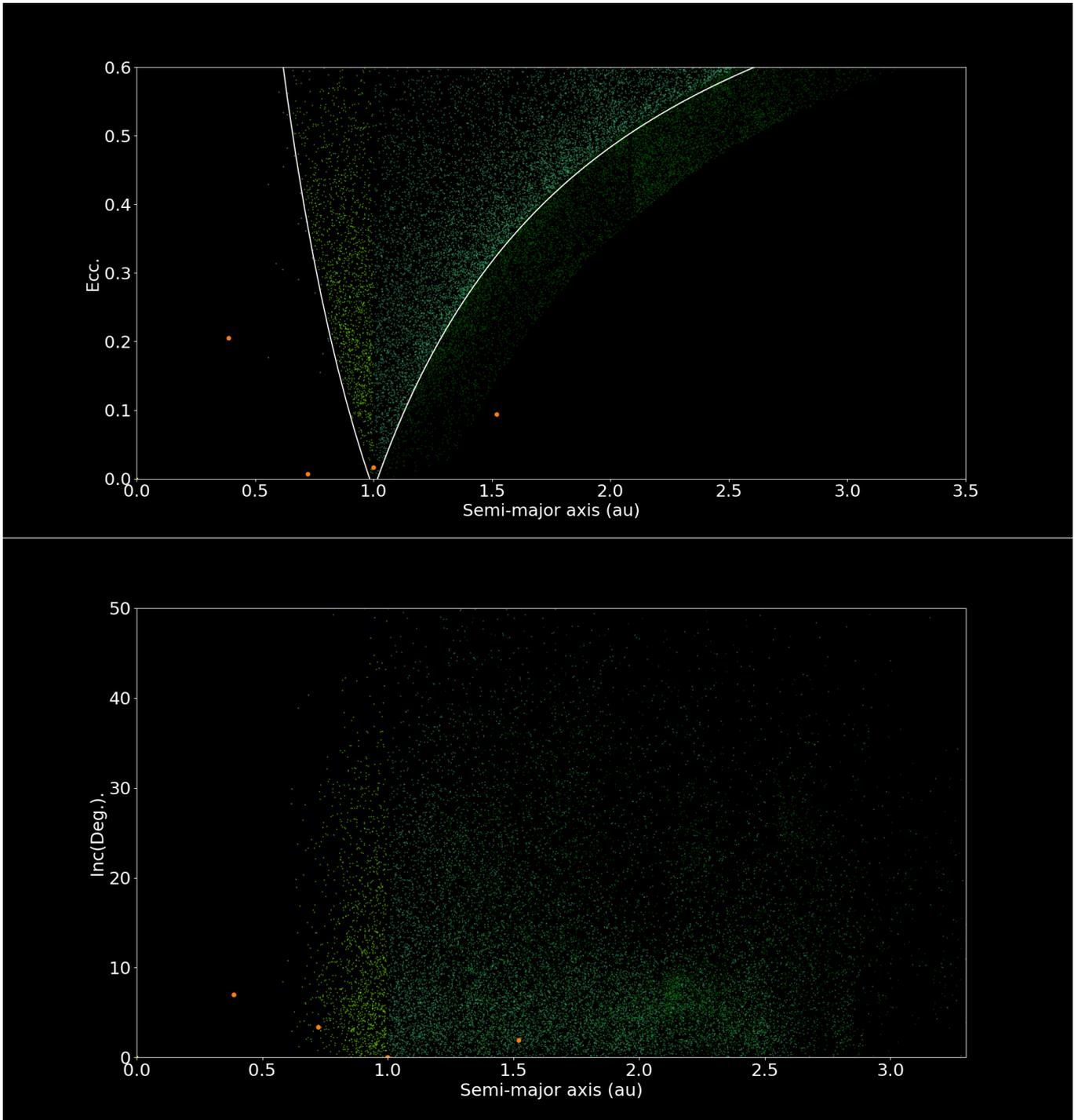

*Figure 6: The orbital element distribution of the known near-Earth asteroids, in semi-major axis-eccentricity space (top) and semi-major axis-inclination space (bottom). The four sub-groups within the near-Earth asteroid population are shown in different shades of green, to match the colour scheme used in Figures 2 - 5. The Atira asteroids are shown in aquamarine, the Atens in chartreuse, Apollos in sea green and Amors in dark green. Plots for the individual sub-populations (and all other populations of Solar system small body) are given in the appendix. The impact of observational bias is clearly seen here, particularly in the upper panel. Objects are easier to detect when closer to Earth than farther away - and the smallest (but most numerous) objects can only be discovered during close approaches to our planet. For that reason, the greatest population in a-e space is bounded by lines of constant perihelion = 0.9833 au (curving outward to the right) and aphelion = 1.0167 au (moving inwards toward higher eccentricities), which we show in white in the top panel. The wedge bounded by these two lines contains those objects that can reach a heliocentric distance at a distance within the bounds set by Earth's perihelion and aphelion distances and can therefore experience very close encounters with our planet.*

The *near-Earth asteroids (NEAs)* are a population of primarily rocky and metallic objects moving on short-period orbits that either cross, or approach, the orbit of the Earth, and their distribution is shown in Figure 6. The aphelia of the orbits of near-Earth asteroids (the point at which they are furthest from the Sun) either lie within the main Asteroid belt, or significantly closer to the Sun. The population of NEAs are often broken into three sub-populations. The *Aten asteroids* are those objects whose semi-major axes are less than 1 au, and so have orbital periods less than one year. A small fraction of these objects (~2%) follow orbits entirely within that of the Earth (e.g. Bottke et al., 2002) – this subset of the Aten population are sometimes referred to as *Apohele asteroids* or *Atiras*, to distinguish them from the other Atens. The *Apollo asteroids* have semi-major axes greater than 1 au, and move on orbits that cross that of the Earth. The *Amors* move on orbits with perihelia greater than that of the Earth, meaning that they do not, currently, cross our planet's orbit. The orbits of NEAs are unstable, and they have typical dynamical lifetimes orders of magnitude shorter than the age of the Solar system. The population is continually replenished by a slow trickle of fresh objects from the main Asteroid belt, typically as a result of the perturbations of distant Jovian MMRs (e.g. Wetherill, 1988, Morbidelli et al., 2002), coupled with the influence of the non-gravitational Yarkovsky effect, which results in the gradual orbital drift of objects in the size range ~1m to ~10km, a size range typical of the fragments generated by collisions between asteroids. We discuss the Yarkovsky effect in more detail in section 4.9.2, below.

The *Jupiter-family comets* are a population of objects that typically move on orbits with aphelia located very close to the orbit of Jupiter. In the same way that the bulk of the near-Earth asteroid population originated in the Asteroid belt, the primary parent population for the Jupiter-family comets are the dynamically unstable *Centaurs* (which move on orbits which cross those of the outer planets, having perihelia located between the orbits of Jupiter and Neptune). The Centaurs, in turn, are thought to have their origins primarily in the trans-Neptunian population, with the main sources thought to be the Scattered Disc (e.g. Duncan & Levison, 1997; Di Sisto & Brunini, 2007; Volk & Malhotra, 2008), the Edgeworth-Kuiper belt (e.g. Levison & Duncan, 1997; Lowry et al., 2008; Volk & Malhotra, 2013), the Plutinos (e.g. di Sisto, Brunini & de Elía, 2010) and the Neptune and Jovian Trojans (e.g. Horner & Lykawka, 2010a; Horner, Lykawka & Müller, 2012; Di Sisto, Ramos & Gallardo, 2019), although it seems likely that there is at least some contribution from objects originating in the inner Oort cloud (Emel'yanenko, Asher & Bailey, 2005; Brasser et al., 2012a; de la Fuente Marcos & de la Fuente Marcos, 2014). Both the Centaurs and Jupiter-family comets are dynamically unstable on timescales much shorter than the age of the Solar system, and so must be continually replenished from those reservoirs in order to maintain the populations we see today[16] (e.g. Duncan et al., 1988; Duncan & Levison, 1997; Levison & Duncan, 1997; Horner et al., 2004a,b; Pál et al., 2015; Grazier et al., 2018; Grazier, Horner & Castillo-Rogez, 2019).

The *Halley-type comets* are cometary bodies that move on orbits that take tens or hundreds of years to complete. In comparison to the Jupiter-family and long-period comets, their orbits are remarkably stable, and they can remain on the same planet-crossing orbit for many revolutions around the Sun. There remains significant debate as to their origin (see e.g. Emel'Yanenko & Bailey, 1998; Wiegert & Tremaine, 1999; Levison, Dones & Duncan, 2001; Levison et al., 2006; Emel'Yanenko, Asher & Bailey, 2007; Wang & Brasser, 2014; Nesvorný et al., 2017). Some studies have suggested that the amount of time it would take for a comet to evolve onto such a stable orbit from the short-period comet population is so great that they should have faded (through loss of volatiles) or disintegrated long before they can attain such orbits. By contrast, if the Halley-types are primarily captured from the long-period, isotropic comet flux, similar studies have suggested that their number should be far greater than we currently observe - implying that most Halley-type comets must fade below detectability remarkably rapidly, either through devolatilisation or disintegration. Compared to the populations of Jupiter-family comets, Centaurs, and long-period comets, very few Halley-types are known, but they remain an enigma that awaits a solution.

---

[16] This is particularly true of those objects that display cometary activity, since their numbers are reduced not only as a result of their dynamical instability, but also as a natural outcome of their ongoing activity, which leads to their eventual devolatilisation. That process leaves some cometary nuclei dormant – and essentially indistinguishable from asteroidal bodies. It can also lead to the total disintegration of the comet in question – as witnessed on several occasions in the past few hundred years (such as was the case for comet 3D/Biela, whose disintegration in the early 1840s gave birth to the Andromedid meteor shower; e.g. Jenniskens & Vaubaillon, 2007; and comet 73P/Schwassmann-Wachmann 3, which fragmented in 1995; Crovisier et al., 1996). We note that comet 73P is a particularly interesting example of such a disintegration process, as it is one whose evolution is still unfolding. Having initially fragmented in 1995, the single parent comet clearly no longer exists – but a large number of fragments of that comet are still being tracked. These fragments are identified with sub-designations (e.g. 73P-A, 73P-B, and so on). Once those fragments themselves disintegrate, or are lost, the eventual fate for comet 73P will be a change in designation, to become 73D/Schwassmann-Wachmann 3.

The long-period comets are cometary (icy) bodies moving on orbits that take many hundreds, or even millions of years to complete. Their aphelia lie far beyond the orbit of Neptune, and many are observed on their first passage through the inner Solar system. Their origin is the Oort cloud - they are a continual trickle of material being removed from that vast reservoir by the galactic tide, passing stars, and even close passages of giant molecular clouds. Many long-period comets are ejected from the Solar system on their first pass through its inner reaches, as a result of distant perturbations by the planets (primarily Jupiter). Those long-period comets that are observed on their first passage through the inner Solar system are often called 'Oort cloud comets' or 'Dynamically New' comets, to distinguish them from those comets that move on shorter, more tightly bound orbits that must be the result of modification through the course of previous perihelion passages.

The *Main Belt Comets* are an unusual population of small bodies that exhibit both cometary and asteroidal properties. Trapped on dynamically stable orbits in the Asteroid belt, they seem likely to be primordial objects that formed in the belt, and have remained there ever since (e.g. Haghighipour, 2009). The first such object to be identified was the asteroid 7968 Elst-Pizarro, which was observed to sport a short tail in September 1996 (e.g. Boehnhardt et al., 1996). Further activity was observed at subsequent perihelion passages (e.g. Hsieh et al., 2010), even though 7968 Elst-Pizzaro's orbit keeps it well within the bounds of the asteroid belt. In recent years, several more Main Belt Comets have been identified, and a variety of different mechanisms have been proposed to explain the observed activity (such as the collision of a small asteroid with the observed Main Belt Comet, liberating dust; the release of dust as a result of the rapid or unstable rotation of the object; the sudden exposure of previously buried volatile material; or even the rotational disruption of a larger parent body; e.g. Prialnik & Rosenberg, 2009; Jewitt et al., 2011, 2013; Hirabayashi et al., 2014). Based on the number of Main Belt Comets observed by the Pan-STARRS1 survey telescope, Hsieh et al. (2015) suggest that, at any given epoch, there should be roughly 140 Main Belt Comets amongst the population of the outer part of the Asteroid belt, which suggests that the next generation of survey telescopes may well reveal significantly more about this fascinating new population of Solar system small bodies.

In much the same way that asteroids are given a numerical designation once their orbits are sufficiently well constrained by ongoing observation, periodic comets that have been seen at multiple apparitions are also 'numbered'. To date, 375 numbers have been given to periodic comets with well determined orbits - split between Encke-type comets, Halley-type, Jupiter-Family Comets, and active Centaurs such as 95P/Chiron[17]. However, several of those comets have fragmented, with the fragments carrying the same name and number as their parent body, but with the addition of a letter at the end of their name to distinguish between them - these fragments bring the total number of cometary bodies with designated numbers to 455, as of 5th February 2019.

The naming convention for numbered comets is to give the number, followed by a letter (typically P, for 'Periodic'), followed by a /, then the name (or names) of those who either discovered the comet, were the first to calculate its orbit, or successfully identified that several cometary apparitions over the years were actually the same object. Several of the numbered comets have disintegrated or been lost - for these comets, the letter 'P' is typically replaced with 'D'. The naming convention for non-periodic comets (or those that have only been seen at a single apparition) takes the form C/1995 O1 Hale-Bopp - the C denotes a cometary body, 1995 is the discovery year, O1 indicates that the comet was the first discovered in the 15th/24th of the year, and Hale-Bopp denotes the discoverers (in this case Alan Hale and Thomas Bopp). If an object once thought to be cometary is later determined to be asteroidal, the 'C' at the beginning of the designation is replaced with an 'A', whilst if no reliable orbit can be determined for an object, the 'C' is replaced with an 'X'. Finally, the discovery of the first interstellar objects passing through the Solar system, in 2017, led to the introduction of the prefix 'I' (for interstellar). That object, now known as 1I/'Oumuamua, is unlikely to be the last interstellar interloper discovered traversing the Solar system, and was remarkably well studied (e.g. Meech et al., 2017; Jewitt et al., 2017; Bannister et al., 2017; Fitzsimmons et al., 2018), despite only remaining visible for a short period of time due to its high velocity. Indeed, whilst this paper was under review, a second interstellar interloper, comet 2I/Borisov, was discovered, passing through perihelion beyond the orbit of Mars on 8$^{th}$ December 2019. Like 1I/'Oumuamua, 2I/Borisov generated a wealth of global interest, and has already been the subject of a number of research studies (e.g. Fitzsimmons et al., 2019; Jewitt & Luu, 2019; Opitom et al., 2019; Guzik et al., 2020; Hallatt & Wiegert, 2020; McKay et al., 2020).

---

[17] Chiron holds the distinction of being the first object to ever be given both an asteroidal and cometary designation. It is most commonly referred to as 2060 Chiron.

Table 2 presents a summary of the number of objects currently known across all the small body populations discussed above, along with the approximate boundaries that delineate membership of those populations. Where they are available in the literature, we have also detailed the estimated number of objects in those populations whose diameter exceeds a certain value (1 km, by default, unless otherwise stated). We also distinguish between populations that are held to be dynamically stable on long (Gyr) timescales and those that are unstable. Finally, for the unstable populations, we also name the accepted source population - as we described above.

| Population | Inner Edge (au) | Outer Edge (au) | $N_{known}$ | $N_{est}$ (> 1km) | Stability | Main Parent Population |
|---|---|---|---|---|---|---|
| NEA - Aten | ... | $a < a_{Earth}$ | 1538 [1] | ... | Unstable | Asteroid Belt |
| NEA - Apollo | $q<Q_{Earth}$ | ... | 10151 [1] | ... | Unstable | Asteroid Belt |
| NEA - Amor | $q>Q_{Earth}$ | ... | 8538 [1] | ... | Unstable | Asteroid Belt |
| NEA - Total | ... | $q \sim 1.3$ | 20227 [1] | 990 [8] | Unstable | Asteroid Belt |
| JFCs | ... | $q \sim 4$, $2 \leq T_J^{18} \leq 3$ | 646 [2] | (>) 300 >2.3km [9] | Unstable | Centaurs |
| Encke-Type + Main Belt Comets | ... | $q \sim 4$ $T_J > 3^{19}$ | 57 [2] | ... | Stable | ... |
| Halley Type | ... | $q \sim 4$ $T_J \leq 2$ $P \sim 200$ yr | 100 [2] | (>) 100 >2.3km [10] | Unstable | Debated |
| Long Period Comets | $P > 200$ yr[20] | $q \sim 4$ | 2720 [2] | | Unstable | Oort Cloud |
| Centaurs | $q \sim 4$ | $q \sim 32$ | 421 [3] + 121 [6] = **542** | ~44300 [11] | Unstable | Debated - TNO |
| Asteroid Belt | $\sim 1.3$ $a > 2$ | $\sim 4$ $a < 4$ | 735534 [4] | $\sim 10^6$ [12] | Stable | ... |
| Earth Trojans | ~1 | ~1 | 1 [5] | ... | Unstable | NEA |
| Mars Trojans | ~1.524 | ~1.524 | 9 [5] | ... | Unstable | Asteroid Belt |
| Jovian Trojans | ~ 4.8 | ~ 5.6 | 7079 [5] | $> 10^6$ [13] | Stable | ... |
| Uranus Trojans | ~19.2 | ~19.2 | 1 [5] | ... | Unstable | Centaurs |
| Neptunian Trojans | ~ 29.5 | ~ 30.5 | 23 [5] | $10^6 - 10^7$ [14] | Stable | ... |
| Scattered Disc | $q \sim 32$ | $q \sim 40$ | 380 [3] + 580 [6] = **960** | $> 10^9$ [15] | ~Stable | Other TNOs |
| Detached Objects | $q \sim 40$ | ... | 57 [3] | ... | Stable | ... |
| Resonant TNOs | various | various | 670 [7] | > 50000 (>100 km) | Stable | ... |

---

[18] $T_J$ is the object's Tisserand Parameter, with respect to Jupiter, given by $T_J = \frac{a_J}{a} + 2 \cos i \sqrt{\frac{a(1-e^2)}{a_J}}$, where $a$ is the object's semi-major axis, $a_J$ the semi-major axis of Jupiter, $i$ the object's orbital inclination, and $e$ its eccentricity. For more information on the Tisserand parameter, we direct the interested reader to page 3 of Horner et al. (2003), and references therein.

[19] Note - Encke-type comets are those whose aphelion is interior to that of Jupiter, sufficiently displaced from that planet's orbit that their Tisserand Parameter with respect to Jupiter is greater than three. Such objects are therefore included here, paired with the Main Belt Comets, rather than as a Jupiter-Family Objects, despite the disparate formation mechanisms proposed for the two classes of object.

[20] Historically, the threshold that divided the long and short-period comets was set at 200 years. However, that definition became somewhat blurred in 2002 by the discovery of comet 153P/Ikeya-Zhang, which follow up observations revealed was identical to a comet observed in 1661, as well as those seen in 877 and 1273 AD (e.g. Hasegawa & Nakano, 2003). With an outbound orbital period of 366.5 years, Ikeya-Zhang holds the record for the longest orbital period of any periodic comet to have been observed at more than one apparition, and blurred the admitted arbitrary dividing line between those comets considered 'periodic' (or short-period) and the single-apparition long-period comets.

| | | | [16] | | |
|---|---|---|---|---|---|
| Classical EK belt | ~ 40 | ~ 50 | 1339 [6] | > $10^9$ [17] | Stable | ... |
| Oort Cloud | ... | ~ 200 000 | 0 | > $10^{11}$ [18] | Stable | ... |

*Table 2: The populations of small bodies within the Solar system. Boundaries given are typically somewhat flexible, and are based primarily on the work of Horner et al., 2003 and Gladman et al., 2008. Some objects fall into multiple categories at once - for example, if an object currently classified as a near-Earth asteroid were observed to be outgassing, it would also be categorised as a cometary body, most likely in the Jupiter family. Another example of such a multiply-categorised object is the Centaur Chiron, which is also counted in many lists of short-period comets, since it displays a coma around perihelion. The distances quoted in the inner and outer edge columns refer to semi-major axes unless stated otherwise. In those columns, and throughout this work, a refers to an object's semi-major axis, in au, P its orbital period, q its perihelion distance, and Q its aphelion distance.*

*[1] The number of objects in these categories was taken from the Minor Planet Center's 'Unusual Minor Planets' page[21], on 26th July 2019.*

*[2] The number of objects in these categories was calculated using the boundaries denoted in the table, applied to an ASCII table of cometary orbital parameters downloaded from the JPL Solar System Dynamics page[22] on 5 February 2019*

*[3] The number of objects in these categories was calculated based on the Minor Planet Center's list of Centaurs and Scattered Disc objects[23], accessed on 4 February 2019*

*[4] The number of Main Belt Asteroids was calculated using ASCII tables of numbered and unnumbered asteroids downloaded from the JPL Solar System Dynamics page on 6 February 2019, and applying the cuts detailed in the table*

*[5] The number of objects in these categories was taken from the Minor Planet Center's 'Trojans' page[24], on 26th July 2019.*

*[6] The number of objects in these categories was calculated using the parameters in this table, applied to the Minor Planet Center's list of 'Trans-Neptunian Objects'[25], on 6 February 2019. We note that more than 700 objects in that list would be classified as Scattered Disc objects or Centaurs, following the classification scheme detailed in the table, giving the alternate number detailed above. It seems that this dichotomy is born of the manner by which the Minor Planet Centre divides information. between the Centaurs and Scattered Disc object list (mentioned in [3]) and the trans-Neptunian object list. The former contains those objects with semi-major axes less than 30 au, or greater than 50 au, whilst the latter contains all those objects with 30 au < a < 50 au.*

*[7] The Minor Planet Centre does not maintain statistics on the number of TNOs that are trapped in mean-motion resonance. However, Dr Wm Robert Johnston maintains an archive of TNOs, with orbital parameters taken from MPC circulars, in "Johnston's Archive"[26]. The number of resonant TNOs given in this work was therefore taken from that list, which, at the time of writing, was last updated on 7 October 2018. Of these 670 objects, 439 are Plutinos (in Neptune's 2:3 resonance).*

*[8] Harris & D'Abramo, 2008*

*[9] Brasser & Wang, 2015; this estimate considers cometary nuclei with diameter > 2.3 km, and perihelion distance <2.5 au, and hence represents a lower limit in terms of the criteria described in this table.*

---

[21] https://www.minorplanetcenter.net/iau/lists/Unusual.html

[22] https://ssd.jpl.nasa.gov/?sb_elem

[23] https://www.minorplanetcenter.net/iau/lists/Centaurs.html

[24] https://www.minorplanetcenter.net/iau/lists/Trojans.html

[25] https://www.minorplanetcenter.net/iau/lists/t_tnos.html

[26] http://www.johnstonsarchive.net/astro/tnoslist.html

*[10] Wang & Brasser, 2014; this estimate considers cometary nuclei with diameter > 2.3km, and perihelion distance <1.8 au, and hence is a lower limit in terms of the criteria in this table.*
*[11] Horner et al., 2004a*
*[12] Tedesco & Desert, 2002*
*[13] Yoshida & Nakamura, 2005*
*[14] Sheppard & Trujillo, 2006*
*[15] Gomes et al., 2008*
*[16] Gladman et al., 2012; note, the value given here is for objects with D > 100 km*
*[17] Durda & Stern, 2000*
*[18] Emel'yanenko, Asher & Bailey, 2007*

# 3 THE PLANETS

Even though the Solar system's six innermost planets have been known since time immemorial, aside from the Earth, the detailed study of the planets as physical objects had to wait for the dawn of the Space Age. In the past fifty years, we have sent robot envoys to all of the planets – although the ice giants (Uranus and Neptune) have, to date, only seen a single fleeting visit – the fly-by of the Voyager 2 spacecraft (in 1986 and 1989, respectively; e.g. Smith et al., 1986, 1989). In this section, we provide a brief overview of the current state of our knowledge of the planets as physical objects – from their atmospheres to their bulk composition and interiors.

*Mercury*

Mercury, the closest planet to the Sun, is a shattered husk of a world. In terms of its uncompressed bulk density, it is the densest of the planets[27] – the result of a composition that is dominated by metals, rather than silicates or volatiles. From the Earth, Mercury's surface features remain elusive – a result of the planet's small size, and the challenges involved in observing it (since it always remains with some 28 degrees of the Sun in the sky). It was long thought to be trapped in 1:1 mean-motion resonance with the Sun – spinning once on its axis for every orbit – a theory finally that was finally disproved as a result of radar observations of the planet, carried out in 1965 (e.g. Pettengill & Dyce, 1965; McGovern, 1965). Rather than spinning once per orbit, those observations revealed that Mercury instead spins on its axis once every 59 days, whilst it orbits the Sun in just under 88 days. As a result, the planet is trapped in a 3:2 spin-orbit resonance – meaning that it completes three full rotations in the time it takes to orbit the Sun twice (e.g. Liu & O'Keefe, 1965). No other object is known to be trapped in such a resonance – and it seems likely that the resonance nature of the planet's rotation is maintained, in part, as a result of the planet's eccentric orbit. Were Mercury's orbit circular, it seems likely that the tidal interaction between it and the Sun would continue to slow its rotation, driving it towards an eventual 1:1 spin-orbit resonance. However, as Mercury's orbit is eccentric, the strength of the tides raised upon Mercury by the Sun vary markedly through the course of each orbit – strongest at perihelion, and weakest at aphelion.

The result of Mercury being trapped in this spin orbit resonance is that the Solar day on Mercury's surface (i.e. the time between two consecutive sunrises for most locations on the planet) is twice the length of Mercury's year. More peculiar still, however, is the combination of Mercury's spin and its eccentric orbit. As the velocity of Mercury in its orbit around the Sun reaches its peak, at perihelion, the easterly motion of the Sun across the sky that results from the planet's motion exceeds the westerly motion that results from Mercury's constant rotation. As a result, the motion of the Sun across Mercury's sky appears to slow as the planet approaches perihelion, then reverses through the perihelion passage (with the Sun moving from west to east). Finally, as Mercury recedes from perihelion, the situation reverses, and the Sun resumes its normal east-to-west motion across the sky. As a result, from some locations on Mercury's surface, the Sun would appear to rise in the east in the morning, slow down, reverse direction, and set once again, before rising a second time and moving across the sky as normal! At other locations, the same kind of behaviour would be seen at sunset – the Sun would sink below the western horizon, before rising again, moving west-to-east. It would then slow, turn, and resume its motion, setting once again and ushering in a lengthy 88 day night!

Another result of this peculiar rotation is that there are two 'heat poles' on the surface of Mercury – locations where, every second orbit, the Sun reverses direction whilst close to the zenith. Not only is the Sun overhead (or close to) at perihelion – hence maximising the instantaneous flux falling on the surface – but it remains there for an extended period of time, thanks to the retrograde motion induced by the speed of the planet's perihelion passage. A huge impact feature, one of the largest impact basins in the

---

[27] Mercury's uncompressed mean density is approximately 5.3 g/cm$^3$ (e.g. Cameron et al., 1988), which is markedly higher than the uncompressed mean density of the Earth (at 4.45 g/cm$^3$; Lewis, 1972), which is the second densest of the planets, when measured in this manner. It should be noted that the Earth's bulk density (5.514 g/cm$^3$) is higher than that of Mercury (5.427 g/cm$^3$) – the result of the Earth's significantly higher mass and the resultant compression of our planet under its own gravity. The uncompressed mean density reflects the chemical makeup of a planet, ignoring the effect of gravity – and reveals Mercury to be particularly metal-rich compared to the other terrestrial planets.

Solar system, lies close to one of these two 'heat poles', with its name (Caloris Planitia, or the Caloris basin; e.g. Gault et al., 1977) intended to invoke the fact it is likely to be one of the hottest locations on the planet, at local noon. The basin shows evidence of ancient volcanism that dates from some time after the basin's formation (e.g. Thomas et al., 2014), indicating that Mercury remained volcanically active until less than a billion years ago.

For many years, it was believed Mercury had no atmosphere – but the first spacecraft to visit Mercury, *Mariner 10*, revealed that the planet has an extremely tenuous exosphere (e.g. Broadfoot et al., 1974, 1976) – a result confirmed by the more recent *Messenger* spacecraft (e.g. Solomon et al., 2008; Zurbuchen et al., 2008). That exosphere is not considered bound to Mercury – rather, it is continually generated and denuded as a result of the interaction between the Solar wind and the planet's surface, which leads to a continual erosion and ejection of ionised material from that surface. As such, Mercury's tenuous exosphere consists primarily of a mix of Solar helium, and native sodium, potassium and oxygen ions, with other ionised species forming smaller components – as detailed in Zurbuchen et al., 2008. Interestingly, there is evidence that the impact of meteoroids upon Mercury's surface plays an important role in the generation of its exosphere (e.g. Killen & Hahn, 2015, who suggest that passages of the planet through the Taurid meteor stream could lead to annual enhancements in the amount of Calcium found in Mercury's exosphere, and Pokorný, Saranotos & Janches, 2017, who invoke the seasonal variation in the impact flux at Mercury as a potential origin for the observed dawn-dusk asymmetry in the planet's exosphere).

The *Mariner 10* spacecraft gave us our first close-up look at the Solar system's innermost planet. In addition to revealing Mercury's tenuous exosphere, it also revealed that Mercury has a significant intrinsic magnetic field, enough to generate a magnetosphere and bow shock (e.g. Ness et al., 1974, 1975), which might indicate the presence of an internal dynamo, and hence that Mercury's core is still in the process of freezing (e.g. Stevenson et al., 1983). The structure of that magnetic field was revealed in more detail by the *MESSENGER* spacecraft, which made three flybys of the innermost planet in 2008 and 2009, before moving into orbit around the planet in March 2011 (e.g. Solomon et al., 2007, 2008). The spacecraft revealed that Mercury's magnetic field is tilted by ~5 degrees from the planet's rotation axis (e.g. Anderson et al., 2008, 2010), is approximately 1% the strength of the Earth's magnetic field (at the planet's surface) and is centred on a point that is displaced by just over 400 km to the north of the planet's centre (Alexeev et al., 2010). For a detailed review of our knowledge of Mercury's magnetic field following the *MESSENGER* mission, we direct the interested reader to Anderson et al., 2010.

*Mariner 10* also provided our first images of Mercury's surface, and first measurement of the planet's mass and density. Through those flybys, the spacecraft imaged around 45% of the planet's surface, revealing a scarred, cratered world, similar in appearance to the Moon (e.g. Murray et al., 1974). In addition to the abundant impact features of all sizes, more recent observations carried out by *MESSENGER* have revealed that some ~27% of Mercury's surface is covered by smooth planes, most of which are thought to be volcanic in origin (e.g. Denevi et al., 2013). In the region surrounding the Caloris basin, *MESSENGER* revealed a region of hummocky plains, covering approximately 2% of the planet's surface, which might be the result of eject from the impact that formed the basin (Denevi et al., 2013). The imagery returned of Mercury also revealed unusual thrust fault features, known as 'lobate scarps', whose origins likely reflect the contraction of the planet as it cooled, following its formation (e.g. Solomon, 1977; Watters et al., 1998), with a potential contribution coming from changes in the planet's oblateness as its spin slowed to the current ~59 day period (e.g. Melosh, 1977).

Based on the observations made by *Mariner 10*, researchers were able to determine that Mercury is differentiated, with a large iron-nickel core that is markedly larger and more massive (as a fraction of the planet's mass) than the cores of the other terrestrial planets (e.g. Murray et al., 1974; Siegfried & Solomon, 1974; Gault et al., 1977). These results were refined considerably as a result of the *MESSENGER* mission – which enabled the construction of a detailed model of Mercury's gravity field (Smith et al., 2012). Those observations revealed that the planet's crust varies in thickness as a function of latitude – being thickest near the equator, and thinnest near the poles. The distribution of mass in Mercury's interior was found to be consistent with the planet having an iron-rich liquid outer core – a result that ties in nicely with the idea that Mercury's magnetic field is generated by an interior dynamo, in much the same way as that of the Earth. As we discuss later (in section 4.1.1), Mercury's anomalously massive core, and high bulk density, are key pieces of evidence that suggest the planet was once victim to a celestial hit-and-run collision, which shattered the proto-Mercury, and stripped it of much of its silicaceous mantle material (e.g. Cameron et al., 1988, Benz et al., 2007).

For a detailed review of our knowledge of Mercury prior to the *MESSENGER* mission, we direct the interested reader to the book 'Mercury', published as part of the Space Sciences Series of ISSI (Balogh et al., 2008), whilst Rothery (2015) provides a detailed overview of our understanding of the planet, following the new data delivered by *MESSENGER*. In the coming years, our knowledge and understanding of the Solar system's smallest planet will likely be revolutionised once again, with the arrival of the *BepiColombo* mission (e.g. Benkhoff et al., 2010). Launched in October 2018, *BepiColombo* is following a convoluted path to Mercury – and is scheduled to move into orbit around the iron planet in December 2025. Before this, the spacecraft will use a mixture of gravitational assists (through flybys of the Earth, Venus, and Mercury itself) and Solar-electrical propulsion to modify its orbit, and gradually reduce its encounter velocity, with respect to Mercury, to one sufficiently slow as to effect the orbit insertion. The spacecraft will then split into two components – the Mercury Planetary Orbiter, and the Mercury Magnetospheric Orbiter (*Mio*), which will observe Mercury until at least mid-2027. For more details on the goals of the *BepiColombo* mission, we direct the interested reader to Benkhoff et al. (2010).

*Venus*
Venus is the second planet from the Sun, and one of the Solar system planets known since antiquity. It is typically the third brightest object visible from Earth, after the Sun and Moon, and is frequently visible soon after sunset or before sunrise as the "evening star" or "morning star". The apparent brightness of Venus largely stems from the combination of its proximity to the Sun and the Earth, the reflecting/scattering properties of its atmosphere, and its size.

Due to its similarity in size to the Earth (0.95 Earth radii), Venus is frequently referred to as our planet's "twin" or "sibling". Furthermore, the mass of Venus is 82% of that of the Earth, which results in the planet having a similar bulk density to our own (5.204 g cm$^{-3}$ compared to Earth's 5.514 g cm$^{-3}$). It is therefore considered reasonable to assume that Venus has both a comparable bulk composition, and underwent a similar formation process to the Earth (e.g. Zharkov, 1983). Up until as recently as the early 1960s, the Venusian surface was theorised to have a range of temperate surface environments, including steamy jungle landscapes. Ground-based microwave observations (Mayer et al. 1958) combined with data from the NASA *Mariner* (e.g. Barath et al., 1964) and Soviet *Venera* space programs (e.g. Avduevskij et al., 1971; Keldysh, 1977) finally revealed the extent to which the surface of our sister planet diverged from a temperate Earth-like environment (e.g. Sagan, 1962; Ronca & Green, 1970; Marov, 1978).

The first suggestions that Venus might not be a true twin to the Earth came in the early part of the 20$^{th}$ Century. Pioneering spectroscopic observations by Slipher (1903) strongly suggested that the planet's rotation must be far slower than that of the Earth. Indeed, his observations found no evidence of any rotation – leading Slipher to conclude "*A glance at the table will show that the errors of observation were small and that there is no evidence of a short rotation period for the planet. A rotation period of twenty-four hours would incline the planetary lines one-third of a degree, a quantity quite large in comparison with the errors of observation.*". Confirmation of Venus' incredibly slow rotation came with the advent of radar observations, in the 1960s (Goldstein & Carpenter, 1963), which revealed both that the planet rotates incredibly slowly (with a period of slightly more than 243 days), and that that rotation is retrograde in nature (i.e. occurs in the opposite direction to its orbital motion)[28]. As a result, Venus is the planet with the slowest rotation (both in terms of rotational velocity at the planet's equator, and rotation period) and the greatest axial tilt (with a value of around 177°). As we discuss in section 4.1.1, a number of explanations have been put forward to explain Venus' peculiar rotation – ranging from torques exerted on the planet by its thick atmosphere (e.g. Dobrovolskis & Ingersoll, 1980) to the chaotic evolution of the spin under the gravitational influence of the other planets (Correia et al., 2001a, b), or even the effects of a giant collision, early in the planet's youth (Dormand & McCue, 1987; Dormand & McCue 1993). Regardless of the true origin of the planet's slow spin, it places the planet in an unusual situation – unlike the Earth and Mars, whose atmospheres rotate along with their host planets, the atmosphere of Venus instead super-rotates (aside from the layers adjacent to the planet's surface) – with the upper layers of the atmosphere rotating with a period of approximately four days (e.g. Schubert et al., 1980).

The atmosphere of Venus is dominated by $CO_2$ (96.5%), with the remainder consisting mostly of $N_2$ (3.5%). Other trace atmospheric constituents include $SO_2$ and clouds of $H_2SO_4$ that extend from 25km to 50km altitudes above the surface. The thickness of the atmosphere creates enormous pressure at the planet's surface, with an average surface pressure of 92 bars, equivalent to the pressure at an ocean depth of almost 1km. The scattering properties of the thick atmosphere produce a relatively high Bond albedo of 0.76 (Haus et al., 2016). Venus receives 1.91 times the incident flux received at Earth, but less than 3% of that incident flux penetrates to the Venusian surface. However, the Venusian climate is locked in a post-runaway greenhouse state, with an average surface temperature of 735K (Walker, 1975).

The present state of the Venusian atmosphere is presumed to have been caused through the planet's passage through a threshold whereby the insolation exceeded the outgoing thermal radiation limit for a moist atmosphere (Komabayashi, 1967; Ingersoll, 1969; Nakajima et al., 1992; Goldblatt & Watson, 2012; Goldblatt et al., 2013). Moreover, it is possible that Venus may have retained substantial surface liquid water until as recently as a billion years ago (Way et al., 2016), although it has also been argued that the planet's water may never have condensed after its accretion (Hamano et al., 2013). The evidence for a previously habitable Venus is critically important both for the general evolution of terrestrial planets and the studies of Venus that are applicable to exoplanets (Kane et al., 2018, 2019b).

---

[28] Thanks to the retrograde rotation of Venus, the time between consecutive sunrises on the planet (the Solar day) is approximately 117 days. From the surface of Venus (if it were possible to see through the planet's thick clouds), the Sun would appear to rise in the west, and set in the east.

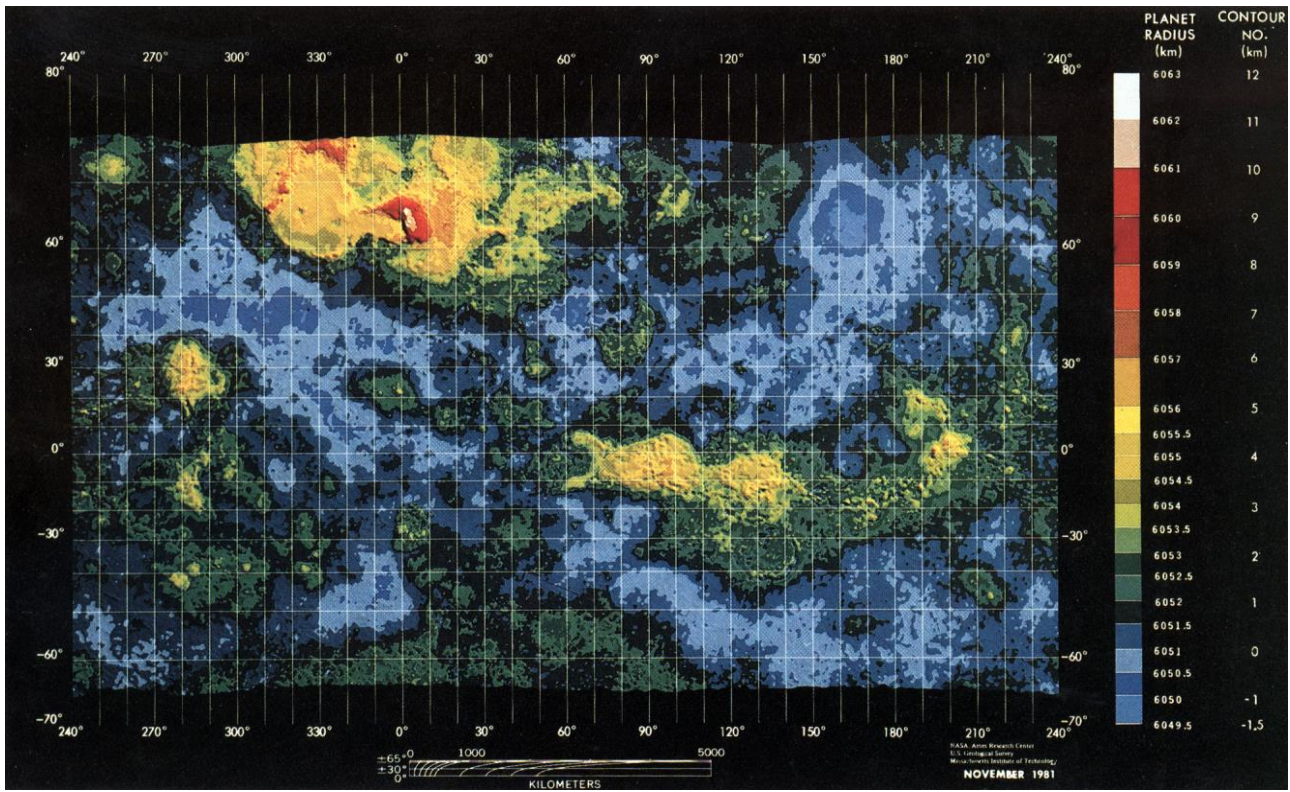

*Figure 7:* *Topographical map of the Venus surface, based on Pioneer Venus orbiter observations. Credit: NASA Ames Research Center, US Geological Survey, Massachusetts Institute of Technology, and the NASA Space Science Data Coordinated Archive; image is public domain.*

The interior and geology of Venus retain numerous outstanding questions, including the size and state of the core, refractory elemental abundances, seismic activity and density variations with depth, and convection within the mantle with relation to current surface geology. *Pioneer Venus* produced one of the first complete topographical maps of Venus via radar mapping (Pettengill et al., 1980), shown in Figure 7. The map reveals the complexity of the surface, including the dominant highlands of Ishtar Terra in the northern hemisphere and Aphrodite Terra along the equator (Ivanov & Head, 2011).

Images from the JAXA *Akatsuki* spacecraft show strong evidence for stationary waves in the upper atmosphere, shown in Figure 8, where the wave features are centred above the Aphrodite Terra highland (Fukuhara et al., 2017b). Such atmospheric phenomena are explained by stationary gravity waves caused by deep atmosphere wounds over the highland regions, and emphasize the importance of interactions between planetary atmospheres and surface topography. Indeed, topographical maps of the Venusian surface yield the appearance of analogues to continents and ocean basins. However, unlike Earth, the surface of Venus does not represent a broad distribution of ages (Head, 2014).

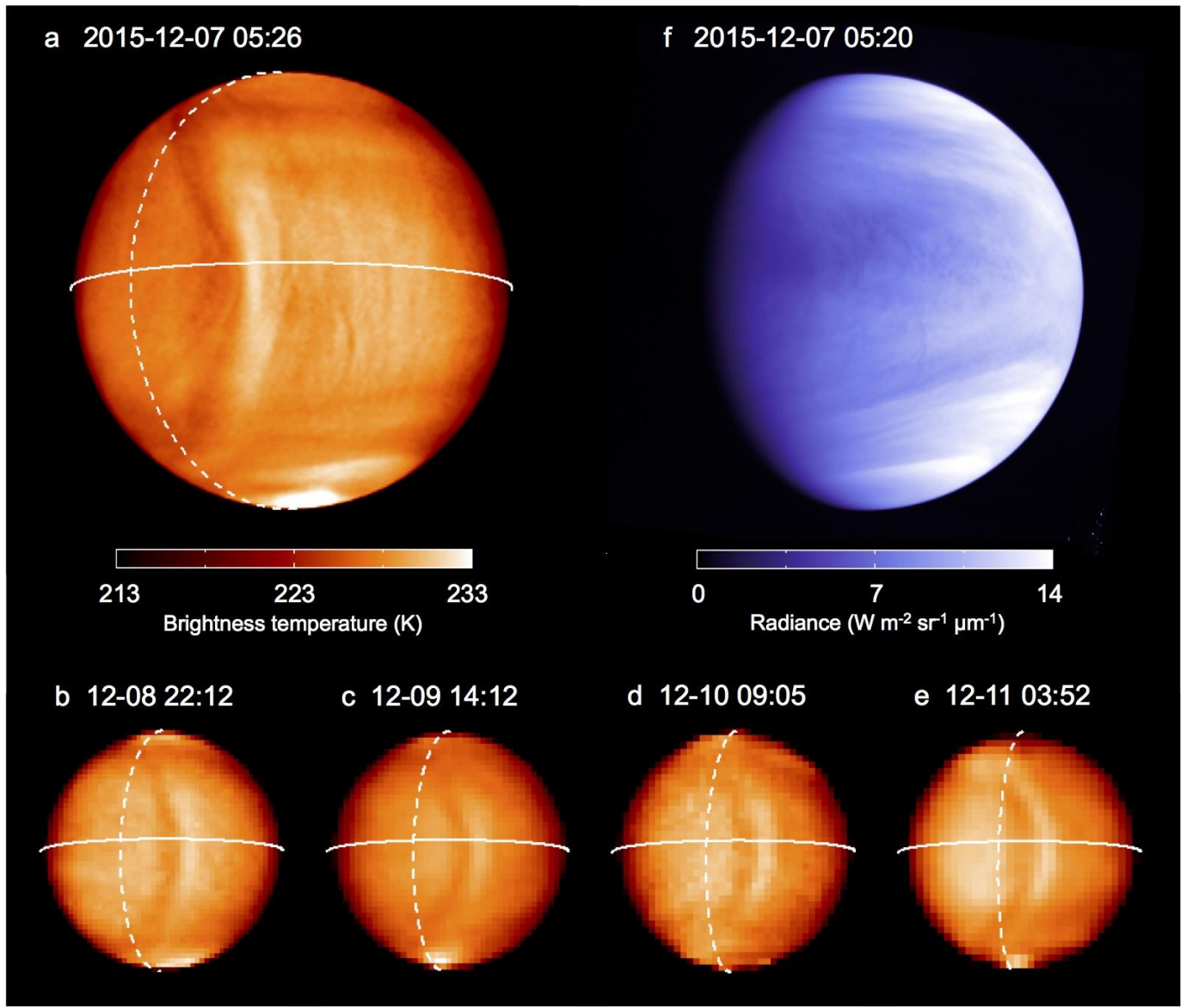

*Figure 8: Stationary wave above the Aphrodite Terra highland, based on Akatsuki spacecraft observations. Credit: JAXA. Image from Fukuhara et al., 2017a; and is reproduced here with permission from the publisher, Springer Nature. images a – e were obtained using the LIR instrument, which observes at 10 µm, whilst image f was taken by the UVI instrument at a wavelength of 283 nm.*

Analyses of surface crater counts have shown that the average age of the surface is ~750 Ma (Schaber et al., 1992; McKinnon et al., 1997), which is less than 20% of the total age of the planet. The precise mechanisms through which resurfacing scenarios could have occurred is still a matter of some debate (Bougher et al., 1997) and include catastrophic overturn (Parmentier & Hess, 1992) and mantle convection mechanism changes (Herrick, 1994) as possible explanations. Answering such questions regarding the interior and atmosphere will shed new light on the evolution of habitable terrestrial planets.

*Earth*
The Earth is, of course, by far the best studied of the planets, and, as a result, is the one whose nature is the best understood. As the cradle of life, our planet remains the only known inhabited world – a fact that helps to drive the global community's fascination with the search for another 'Earth-like' world – a phrase which means different things to different people. Whilst it is beyond the scope of this review to describe all of the contemporary research into our terrestrial home, we here provide a general overview of our planet's primary physical characteristics, and describe those that mark the Earth as distinct from its neighbours in the Solar system.

The Earth is the largest and most massive of the Solar system's four terrestrial planets. As our home, it is by far the best studied of all of the Solar system's planets – although thanks to our ongoing exploration using robotic spacecraft, it is fair to say that we have better imagery of the surfaces of the Moon or Mars than we do of the bottom of Earth's oceans.

The Earth is accompanied by a single natural satellite – the Moon – whose unusually large size and peculiar composition hint at a violent origin (as we discuss in more detail in section 4.1.1). The Moon is gradually receding from the Earth, as a result of their mutual tidal interaction – a process that is also slowing the Earth's spin, as angular momentum is transferred between the two bodies. As a result, in the past the Earth-Moon distance was markedly smaller, and our planet's spin faster than that we observe today. In a recent study, de Winter et al. (2020) used core samples taken through a fossilised bivalve (*torreites sanchezi*) that lived 70 million years ago to show that the year, at that time, was 372 days long. Since it is unreasonable to assume that the Earth's orbital semi-major axis was markedly larger at that time, those results suggest that the length of the Solar day on Earth at the time was 23 hours 31 minutes – some 29 minutes shorter than that we observe today. As well as offering a fascinating insight into life in the late Cretaceous, that work offers a promising new technique that could allow researchers to accurately track the evolution of the Earth's spin over time. Such work is particularly important, since it seems likely that, shortly after the formation of the Moon, the Earth may have spun with a period of just a few hours (e.g. Canup, 2012; Ćuk & Stewart, 2012; Wisdom & Tian, 2015).

Unlike Venus and Mars, the Earth has an appreciable magnetic field, one that is approximately one hundred times stronger than that of Mercury. Earth's strong magnetic field is thought to be produced by a deep internal dynamo generated within the liquid outer core of our planet (e.g. Buffett, 2000). This magnetic field produces a large magnetosphere, which acts to deflect the majority of the Solar wind. Those charged particles that succeed in penetrating the Earth's magnetic field follow the field lines down to hit Earth's atmosphere in its polar regions, driving continual aurora at high magnetic latitudes. At times of particularly energetic Solar activity, those aurorae are driven equator-wards, with the most energetic events causing aurorae that can be seen close to the equator. It has been argued that the presence of Earth's strong magnetic field has played an important role in ensuring the ongoing habitability of our planet – with the atmospheric degradation experienced by Mars usually held up as an example of what can happen to the atmosphere of a terrestrial planet without the twin protections of a strong magnetic shield, and outgassing and recycling of the atmosphere driven by plate tectonics. For an overview of the impact of plate tectonics and the Earth's magnetic field on our planet's habitability (along with the various other factors that may have combined to make our planet a suitable host for life), we direct the interested reader to Horner & Jones (2010c), and references therein.

The orientation of Earth's magnetic field is far from fixed – with the location of the magnetic north pole on Earth's surface having drifted markedly over the past three centuries. In the mid-1800s, the northern magnetic pole was located more than twenty degrees from the geographical north pole. In recent years, the orientation of Earth's magnetic axis has been shifting rapidly, with the spin and magnetic poles currently separated by less than four degrees (Chulliat et al., 2019). On longer timescales, there is evidence that Earth's magnetic field can vary markedly in strength, and its orientation can reverse. Such reversals have been demonstrated in numerical models of the Earth's dynamo (Kuang and Bloxham, 1997) and, indeed, the periodic reversals in Earth's magnetic field provide an elegant piece of evidence for the ongoing process of plate tectonics on Earth. When iron-bearing rock forms, it freezes in place a record of the orientation of Earth's magnetic field. It was observed that changes in the remnant magnetism in rocks across regions signified that not only had the Earth's magnetism changed during geological time, but also gave large clues as to how the continents had moved relative to each other (Runcorn, 1956).

The dynamics of the Earth's surface reflects its interior structure, the deepest layers of which having been probed by seismic investigations. Indeed, the Earth is almost unique amongst the planets in that it has been subject to a detailed seismic study of its interior, although the NASA *InSight* mission, currently on the surface of Mars, is aiming to remedy this, with the goal of using seismic observations to investigate the red planet's interior, as well as the flux of impacts it experiences (e.g. Banerdt et al., 2012; Daubar et al., 2018). In the case of the Earth, our ability to continually monitor and study seismic activity has allowed the interior of our planet to be studied in remarkable detail. By monitoring the propagation of seismic waves generated by earthquakes through our planet we have been able to discern that the Earth has two-component core, a liquid outer core (Oldham, 1906) and a solid inner core (Lehmann, 1936). Surrounding this is a plastically deforming mantle that comprises most of the volume of the planet, and itself can be divided into several layers that are characterised by changes in mineralogy (e.g. Tackley et al., 1994). It is the heat flow within the mantle, through convection, that is the main driving force for movement of Earth's crust, in the form of plate tectonics – a process that is notably absent on the surfaces of the other terrestrial planets. Interestingly, it has been argued that the Earth can only support plate tectonics as a result of abundant water trapped within the mantle. Indeed, the lack of global plate tectonics on Venus, whose size, mass, and composition are similar to those of the Earth, has often been attributed to the lack of water on the planet's scorching surface (e.g. Nimmo & McKenzie, 1998; O'Neill, Jellinek & Lenardic, 2007). Even water, however, might not have been enough on its own to allow Earth to begin and sustain plate tectonics – with a recent study suggesting that modern tectonic processes might well have been initially triggered by the impacts that bombarded the Earth in the billion years after its formation (e.g. O'Neill et al., 2019).

The Earth's atmosphere is deep and complex. By volume, when dry, it comprises approximately 78% nitrogen, 21% oxygen, and 0.93% argon, 0.04% carbon dioxide[29], and traces of a wide variety of other elements and compounds. In the lower atmosphere, water vapour provides an additional component, though the amount of $H_2O$ can vary quite dramatically, from almost none in deserts and at times of particularly low humidity, to being super-saturated. The Earth's atmosphere is noticeably stratified, and is typically broken down into several major layers. Closest to the ground is the troposphere, within which the great bulk of the planet's weather and clouds can be found. Within the troposphere, in general, the temperature decreases with increasing altitude. The second layer of Earth's atmosphere, the stratosphere, sees this trend reverse, with temperature once again increasing as one rises farther above Earth's surface. This temperature inversion may well have played a major role in helping to ensure that the Earth has not become dehydrated over the aeons since it formed, by serving as a 'cold trap'. Water vapour which rises too high in the atmosphere will condense out, freezing, effectively forcing it to fall back to lower altitudes. As a result, almost all of our planet's water vapour remains trapped beneath this layer – well below the altitudes at which solar ultraviolet radiation (most of which is blocked by the ozone layer) would dissociate that water vapour, allowing its constituent hydrogen to escape from our planet. Above the stratosphere, the Earth's atmosphere begins to cool once again – a region known as the mesosphere, which extends to an altitude of approximately 80 km above the ground. Then, once again, the temperature of the atmosphere begins to climb, rapidly – a region known as the thermosphere. Beyond the thermosphere lies the exosphere – the region where the Earth's atmosphere gradually thins until it becomes indistinguishable from the interplanetary medium.

Perhaps the most striking feature of the Earth is the presence of liquid water across our planet's surface. Indeed, oceans, rivers, and lakes cover something like 70% of the Earth's surface – a stark contrast to the surfaces of the other terrestrial planets. The origin of that water remains heavily debated – a topic we discuss in more detail in section 4.4. Whilst it has been suggested that both Mars and Venus were once warm and wet, like the Earth, our planet is the only one of the three whose climate has remained suited to the presence of widespread surface liquid water. Over the course of the Earth's history, our planet's climate has remained remarkably stable – a fact made even more remarkable by the fact that, over the same time period, the Sun's luminosity has increased by an estimated ~30%. The fact that the Earth's climate was clement at a time when the Sun was just 70% as luminous as it is today led to a conundrum known as the 'faint young Sun problem'. For many years, the Earth's early warm climate appeared to be at odds with the fact that the young Sun was so faint. For a detailed overview of the 'faint young Sun problem', we direct the interested reader to Feulner (2012), and references therein.

---

[29] The current level of $CO_2$ in Earth's atmosphere is slightly over 0.04% (or 400 parts per million) and is climbing as a result of human activity. At the time of writing (25th March, 2020), the latest $CO_2$ level measured at Mauna Loa Observatory in Hawaii was 415.34 ppm (0.041534%; as per https://www.co2.earth/daily-co2). Prior to the industrial revolution, the level of atmospheric $CO_2$ was lower, at around 280 ppm (0.028%).

Despite its apparent stability on billion years timescales, the climate of the Earth does vary to some degree on shorter timescales. The most famous of these periodic oscillations are known as the Milankovitch cycles – and are the result of the continual perturbation of the Earth by the gravitational influence of the other planets in the Solar system. For a more detailed discussion of the Milankovitch cycles and the history of the Earth's climate variability, we direct the interested reader to Horner et al. (2020), and references therein.

*Mars*
Mars, the outermost of the terrestrial planets, is probably the most studied object in the Solar system, aside from Earth. It is markedly smaller than the Earth (with an equatorial radius of just 3396 km, compared to Earth's 6378 km), with mass just 0.107 times that of our planet. It is also significantly less dense than the other terrestrial planets, with a bulk density of 3.934 g/cm$^3$, which recent work has suggested might be the result of the planet being starved of material during its formation as a result of the migration of Jupiter (e.g. Brasser et al., 2016), as we discuss in section four. Unlike the other planets visible with the unaided eye, which typically appear to have a vaguely yellowish hue, Mars appears a striking red – the result of the iron oxide that dominates the planet's surface.

During the 19$^{th}$ Century, telescope observations of Mars began to reveal a world that displays some similarities to the Earth. Like Earth, Mars has polar caps whose size waxes and wanes with the passing seasons. Whilst Earth's polar caps are dominated by water ice, we now know that those on Mars are a combination of a 'bedrock' made of water-ice and carbon dioxide ice, with their expansion and contraction caused primarily by seasonal deposits of carbon dioxide ice – the result of the winter temperatures at the poles being far colder than those on the Earth. Those observations also revealed dark areas on Mars' surface, which some considered to potentially be areas of vegetation (e.g. Hollis, 1908), though we now know that they are instead highland regions, and bedrock scoured clear of Mars' ubiquitous reddish dust. Towards the end of the 19$^{th}$ Century, speculation on the possibility that there could be widespread life on Mars – potentially even advanced, technological life – reached fever pitch (e.g. Flammarion, 1894; Comstock, 1902; Lowell, 1908), with speculation fuelled by the observation of 'canali' (or channels) on the planet's surface. The concept of the Martian Canals was born, primarily through a mistranslation of the Italian 'canali' (e.g. Green, 1879; Maunder, 1888; Lowell, 1895; Hamilton, 1916). As telescopes improved, however, it gradually became accepted that these 'canali' were actually optical illusions/artefacts (e.g. Evans & Maunder, 1903) – though debate over the topic continued to rage through to the dawn of the space age (e.g. Webb, 1955, and references therein). Nevertheless, through the first half of the twentieth century, it became clear that the modern Mars is far from an ideal location for widespread, complex life (e.g. Tombaugh, 1950, who wrote "*It appears likely that Mars has always had a thin atmosphere, very little water, and a very dry climate*"). The first images returned of Mars from space drove the final nail into that particular coffin – revealing an arid, cratered, cold and desolate world (e.g. Chapman, Pollack & Sagan, 1968), and confirming the lack of canals on the planet's surface (e.g. Sagan, 1975).

Over the past few decades, the idea that Mars might once have supported life, or at least supported conditions suitable for the development of life, has played an important role in directing our ongoing exploration of the red planet. The first spacecraft to successfully visit Mars was the American *Mariner 4*, which flew rapidly past the planet in July 1965, returning a small amount of data and imagery (e.g. Fjeldbo, Fjeldbo & Eschleman, 1966; Chapman, Pollack & Sagan, 1968; Siscoe et al., 1968). The first mission to orbit Mars, the American *Mariner 9*, in November 1971 (e.g. Hanel et al., 1972; McCauley et al., 1972; Conrath et al., 1973; Sagan et al., 1973), was quickly followed by the first successful soft landing on the surface of the red planet, by the Russian mission *Mars 3* (which failed some 20 seconds after reaching Mars' surface). The *Viking* missions, in the mid-1970s, were a huge success – both *Viking 1* and *Viking 2* comprised a Mars orbiter and a lander – with the two landers surviving and operating on the surface of Mars for several years (six for *Viking* 1, and three for *Viking* 2). The *Viking* landers provided our first in-situ measurements of Mars surface, yielding a vast amount of useful data (e.g. Hess et al., 1977; Toulmin et al., 1977; Adams, Smith & Johnson, 1986, and many others). The landers carried a suite of three biological experiments, designed to search for any evidence of life in the soils at their landing site – the results of which were inconclusive, at best (e.g. Klein et al., 1976; Levin & Straat, 1976; Klein, 1977, 1978).

After the success of the *Viking* landers, our exploration of Mars underwent a hiatus of almost two decades, before undergoing a renaissance in the second half of the 1990s. In the 21$^{st}$ Century, Mars has been peppered with a series of landers and rovers, with new missions departing to the red planet with every opposition (with the mean synodic period of Mars being approximately 780 days). The Mars rovers *Spirit* and *Opportunity* were a particular highlight of the 21$^{st}$ Century exploration of Mars. The two rovers landed on Mars in January 2004, for a mission initially scheduled to last just 90 Sols (1 Sol is one Martian day, 24h 37m). Both rovers vastly outlived their expected lifetimes – *Spirit* remained active for more than six years, whilst *Opportunity*'s journey across Mars was finally brought to an end as a result of a catastrophic loss of power resulting from an intense Martian dust storm on June 10$^{th}$, 2018, after a remarkable 5352 Sols. In that time, *Opportunity* covered a distance of more than 45 kilometres – meaning that it holds (by a comfortable distance) the record for the greatest distance travelled over the surface of another world. At the same time, a series of Mars orbiters have returned exquisite images of the planet's surface – with the result that the surface of Mars has been more closely studied than that of the Earth (given the challenges inherent in imaging the bottom of the Earth's oceans).

The result of all that work is that Mars is better studied and characterised than any other object in the Solar system (except the Earth). The earliest missions to Mars revealed that the planet has essentially no magnetic field (e.g. Smith et al., 1965), though parts of the planet's crust retain evidence of ancient magnetism (Acuña et al., 1999; Langlais, Purucker & Mandea, 2004), evidence supported by measurements of the Martian meteorite ALH84001 (Weiss et al., 2002). Based on that evidence, it seems likely that the young Mars had a molten, iron-rich core, which allowed the planet to generate a strong magnetic field. Unlike the Earth, which has maintained its core dynamic until the current day, Mars lost its dynamo, and thence its magnetic field, after a few hundred million years – potentially as a result of the planet cooling more rapidly than the Earth (e.g. Stevenson, 2001, 2003; Williams & Nimmo, 2004). The surface of Mars shows that, unlike the Earth, the planet does not undergo plate tectonics. Indeed, the lack of plate motion has allowed a cluster of vast shield volcanoes to grow above a hotspot, forming the largest volcanoes in the Solar system (the largest of which, Olympus Mons, is almost 26 km tall, as measured from its base, and has a surface area only marginally smaller than that of France; e.g. Mouginis-Mark, 2018). For more information on Mars' volcanic history, see Werner (2009), and references therein.

There is abundant evidence that Mars was once warm and wet, with an ocean potentially filling the planet's vast northern lowlands (e.g. Helfer, 1990; Clifford & Parker, 2001; Carr & Head, 2003; Rodriguez et al., 2015). This in turn suggests that the planet's atmosphere must once have been much thicker and warmer than that we see today (e.g. Ramirez, 2017, and references therein). It seems likely that the lack of a global magnetic field for the bulk of Mars' evolution has allowed Mars' atmosphere to be denuded by the influence of the Solar wind. The lack of plate tectonics on the red planet has also played a role in the loss of its original thick atmosphere, since without plate tectonics, the planet lacks a mechanism by which any gas chemically trapped in the planet's surface rocks could be recycled (e.g. Tomkinson et al., 2013). For a more detailed discussion on the loss of Mars' atmosphere, we direct the interested reader to Horner & Jones (2010c), and references therein. Interestingly, recent radar observations of Mars' south polar cap have revealed that there may exist a permanent reservoir of liquid water on modern Mars – in the form of large lakes buried beneath the ice of Mars' southern polar cap (Orosei et al., 2018). It seems likely that future Mars mission will look to explore this in greater detail, since such deeply buried lakes in Antarctica are known to teem with life (e.g. Christner et al., 2014). If life ever did find a foothold on Mars, then it seems plausible that it could still survive, buried deep beneath the Martian polar ice.

Tied to these ideas of Mars' ancient oceans, it is worth noting that Mars' exhibits a remarkably varied range of surface features – from the vast Valles Marineris (a valley over 4000 km long, 200 km wide, and up to 7 km deep) to the vast volcanoes of the Tharsis bulge. Perhaps the most interesting feature of Mars' global topology, however, is known as the Martian Dichotomy. Aside from the vast impact basin Hellas, Mars' southern hemisphere consists of highlands, which are heavily cratered, and considered to be ancient terrain. The northern hemisphere, by contrast, consists of smooth lowlands, with few if any scars. The Dichotomy is show below, in Figure 9, based on data from the Mars Orbiter Laser Altimeter on *Mars Global Surveyor*. That dichotomy may well be the scar left behind by a cataclysmic impact on the planet (e.g. Andrews-Hanna, Zuber & Banerdt, 2008), as we discuss in more detail in section 4.1.1.

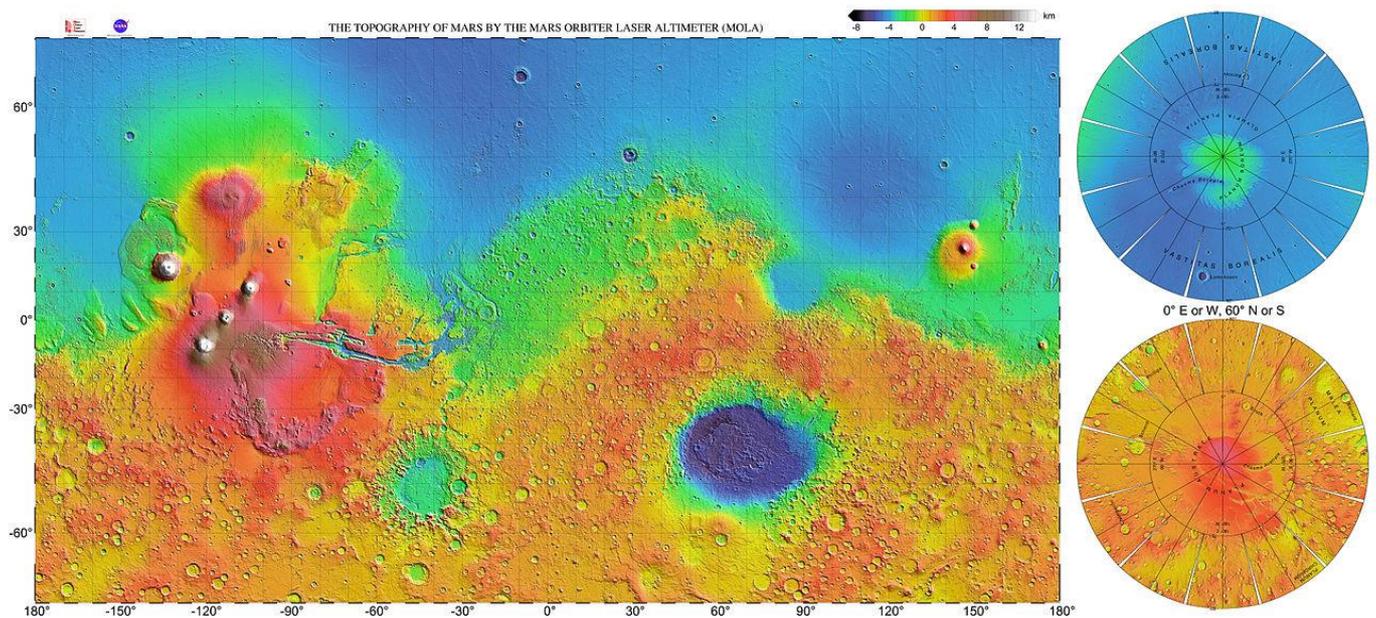

*Figure 9*: *The topography of Mars, based on data taken by the MOLA instrument on Mars Global Surveyor. The shield volcanoes of the Tharsis bulge can be seen close to the equator, to the left, with the Hellas basin the deep round feature in the southern hemisphere between longitudes 60 and 90 degrees. The most striking feature, however, is the Martian Dichotomy – the difference between the smooth northern lowlands and the ancient, cratered, southern highlands. Image credit NASA/JPL/USGS; image is public domain.*

The robotic exploration of Mars is scheduled to continue over the coming years. Of particular interest is the NASA *InSight* mission, which landed on Mars in November 2018. *InSight* carries a sensitive seismometer, and a thermal probe, which together will provide a wealth of fresh insights on the nature of Mars' interior – from the thickness of the planet's crust, and new information on the radius and density of the planet's core (e.g. Folkner et al., 2018), to the rate at which heat flows outward from the planet's interior (e.g. Spohn et al., 2018), and even new data on the frequency with which Mars experiences impacts from cometary and asteroidal bodies (e.g. Teanby, 2015; Dauber et al., 2018).

Early highlights from the *InSight* mission include the discovery that the local magnetic field around *InSight*'s location is an order of magnitude stronger than that estimated from instruments in orbit around Mars, which is thought to be the result of magnetised rocks buried beneath the surface near the lander, and the finding that that field is time variable, likely as a result of activity high in the planet's atmosphere (e.g. Banerdt et al., 2020; Johnson et al., 2020). Perhaps the highest profile results from *InSight* to date have come from the SEIS instrument, designed to measure Mars' seismic activity (Lognonné et al., 2019). That instrument has proved hugely successful in detecting Marsquakes, having catalogued several hundred to date. The largest of those quakes occurred in the Cerberus Fossae region, revealing that the faulting observed in that area remains geologically active today (e.g. Witze, 2019; Banerdt et al., 2020; Giardini et al., 2020), and it will be fascinating to see to what degree that activity continues in the coming years. For all that Mars is by far the best studied celestial body, it seems certain that the coming decade will deliver a wealth of new surprises, as our ongoing exploration allows us to learn ever more about the red planet.

*Jupiter and Saturn: The Gas Giants*

The gas giant planets Jupiter and Saturn contain most of the planetary mass in Solar system (at 318 and 95.2 $M_{Earth}$, respectively; see Table 1). Although they are similar in diameter (with Saturn's equatorial diameter being just 17% smaller than that of Jupiter), Jupiter is more than three times the mass of Saturn, and as a result, it's bulk density is markedly higher (1.326 g cm$^{-3}$ vs. 0.687 g cm$^{-3}$). In addition to comprising the bulk of the mass of the Solar system (aside from the Sun), Jupiter and Saturn also comprise the majority of the angular momentum budget of the entire system - a fact that points to the importance of the giant planets and their evolution on the architecture of planetary systems.

Both planets maintain systems of rings composed of rocky and ice particles of varying size. The presence of rings around Saturn has been known since the seventeenth century (as described in section 4.1.1) – in part as a result of the planet's axial tilt of almost 27°[30]. When they are close to edge on, the rings disappear from the view of all but the largest telescopes – and if Saturn's axial tilt were as low as that of Jupiter (a meagre 3.13°), it seems quite likely that they would have evaded such early telescopic discovery. In contrast to Saturn's magnificent ring system, Jupiter's rings were not discovered until the *Voyager 1* spacecraft visited the planet in March 1979 (see Miner et al., 2007, and references therein). Recent observations have revealed that such ring systems may well be transient companions for the giant planets, providing evidence that Saturn's ring may well become entirely depleted as a result of an ongoing process of "ring rain" within the next 300 million years (O'Donoghue et al., 2019). The question of whether Saturn's rings are primordial or transient has implications for the origin of the system, as we discuss in sections 4.1.1 and 4.1.3.

Both Jupiter and Saturn host large numbers of natural satellites (with, between them, 161 of the 205 planetary satellites discovered to date), some of which display active volcanic (i.e., Io) or cryovolcanic (i.e., Enceladus) activity. Over the past five decades, Jupiter and Saturn have been investigated in exquisite detail by a series of spacecraft: *Pioneer 10* and *11* (e.g. Carlson & Judge, 1974; Smith et al., 1974; Gehrels et al., 1980; Null et al., 1981), *Voyagers 1* and *2* (e.g. Broadfoot et al., 1979; Smith et al., 1979, 1981, 1982), *Galileo* (e.g. Carlson et al., 1996; Niemann et al., 1998; Gautier et al., 2001; Vasavada & Showman, 2005), *Ulysses* (e.g. Balogh et al., 1992; Stone et al., 1992; Grun et al., 1993), *Cassini-Huygens* (e.g. Porco et al., 2005; Doherty et al., 2009; Fletcher et al., 2010; Kanani et al., 2010), *New Horizons* (e.g. Baines et al., 2007; Gladstone et al., 2007; Reuter et al., 2007; Spencer et al., 2007), and *Juno* (e.g. Bolton et al., 2017; Connerney et al., 2017; Wahl et al., 2017).

Jupiter and Saturn are composed of mostly hydrogen and helium at a roughly Solar composition ratio, but both planets are enriched in heavy elements relative to Solar (see Atreya, 2016, for a discussion). After diatomic hydrogen, methane and ammonia are the two most abundant molecular species in the atmospheres of Jupiter and Saturn. Global cloud layers consisting of ammonia ice, ammonium hydrosulfide ($NH_3SH$), and water ($H_2O$) are present in both atmospheres, although at different pressure levels between 0.1 and 10 bars (see e.g. Ragent et al., 1998; West et al., 2009). Jupiter's global clouds are clearly visible when imaged in true (or enhanced colour) images, such as that shown in Figure 10 (taken by the *Juno* spacecraft). The clouds in Jupiter's and Saturn's atmospheres have been used to measure these planets' zonal winds, although the generation of those winds and their connection to deeper layers of the atmosphere or interior is still an area of active research (e.g., Kong et al., 2018; Galanti et al., 2019). Jupiter and Saturn both have warm stratospheres, within which temperature inversions are driven by the deposition of solar radiation. This radiation drives photochemistry in both atmospheres that generates hazes (e.g., Irwin et al., 1998; Fouchet et al., 2009), which can be seen to reflect solar radiation near 2 microns in Saturn (as can be seen in Figure 11, below). Above the region dominated by neutral species, the extended atmospheres of both Jupiter and Saturn include ionospheres and magnetospheres. These magnetospheres are dynamic (e.g., Mitchell et al., 2009, Connerney et al., 2017) and produce auroral emission (visible in Figure 11) caused by interactions with the solar wind as well as material originating on the surfaces, or in the interiors, of the planet's satellites (e.g., Vasavada et al., 1999; Clarke et al., 2009; Mura et al., 2019).

---

[30] It has been suggested that the tilt of Saturn's spin axis could well be the result of the giant planet having accreted, at a late stage in its formation, a planetesimal several times more massive than the Earth, in an oblique collision – a theory we discuss in more detail in section 4.1.1.

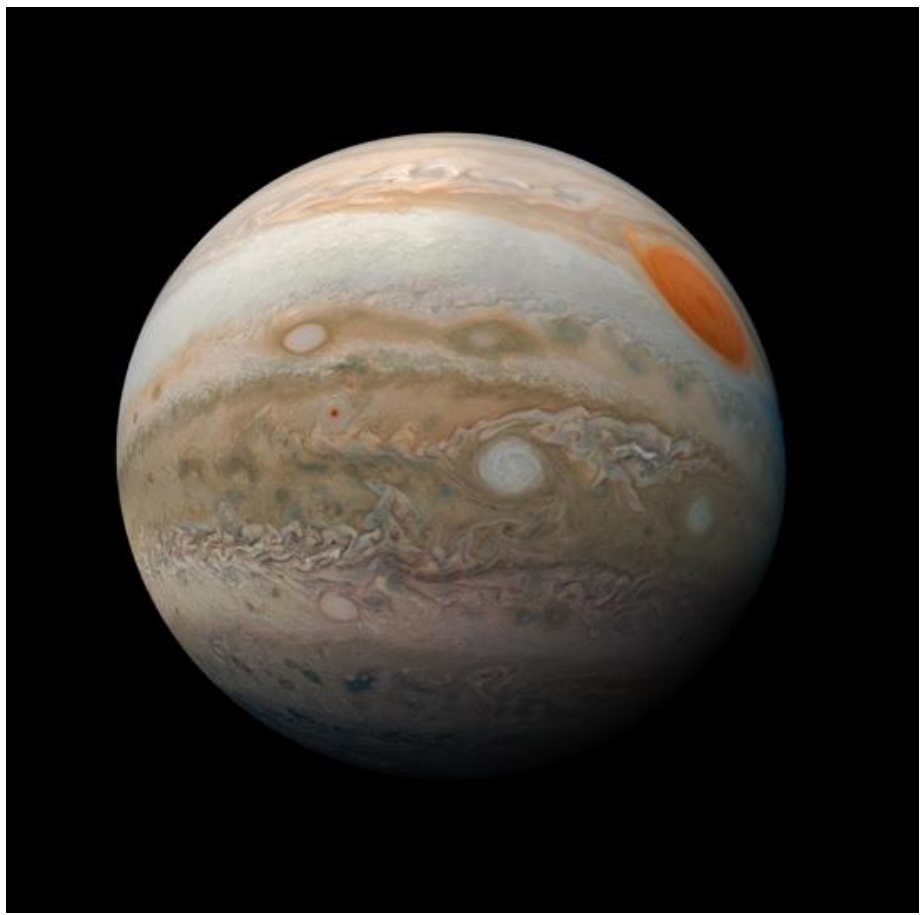

*Figure 10: Colour-enhanced image of Jupiter taken by the JunoCam onboard the Juno Spacecraft. The cloud bands and various storms are clearly visible. Image credit: NASA/JPL-Caltech/SwRI/MSSS/Kevin M. Gill (https://www.jpl.nasa.gov/spaceimages/details.php?id=PIA22946); ; image is public domain.*

Both Jupiter and Saturn exhibit differential rotation, with their equatorial regions rotating more rapidly than the poles[31]. As a result of their rapid rotation (with periods of order ten hours), the two planets are noticeably oblate[32]. Within the interiors of Jupiter and Saturn, molecular hydrogen transitions to liquid metallic hydrogen ($H^+$) at a pressure of roughly 2 Mbar (e.g., Fortney et al., 2018). The depth at which this transition occurs in differs between the two planets, as a result of the significant difference in their mass. Deeper still, Jupiter and Saturn contain cores of rock and ice. Previous measurements of the nondimensional moments of inertia of the two giant planets have led to the suggestion that Saturn is more centrally condensed than Jupiter (e.g. Helled, 2011, Guillot and Gautier 2015). Indeed, recent high-precision measurements of the gravitational moments of Jupiter from the *Juno* spacecraft have revealed that Jupiter's core is diluted, and potentially extends to a large fraction of the planet's radius (Wahl et al., 2017). Such diffusion between the core and the envelope of giant planets has been predicted by previous high-pressure equation of state experiments (Wilson & Militzer, 2012), but there are also suggestions that this dilution of the Jovian core could be evidence that Jupiter was once victim to a giant collision (e.g. Liu et al., 2019; see section 4.1.1). Jupiter's core is thought to contain between seven and twenty-five Earth masses of heavy elements (Wahl et al., 2017), whilst Saturn's core is thought to contain between five and twenty $M_{Earth}$ of such material (Fortney et al., 2018) – findings that offer strong support to their idea that the two planets formed through a process of core accretion[33].

---

[31] Jupiter's polar regions rotate with a period of around 9h 55m 44s, whilst its equator rotates once every 9h 50m 30s. The most fundamental spin period for the planet as a whole is considered to be the System III period of 9 h 55m 29.71s, which corresponds to the rotation of the planet's magnetosphere as measured by radio astronomers (Dessler, 1983). In the case of Saturn, the difference between polar and equatorial rotation rates is even greater – 10h 38m at the poles vs 10h 14m at the equator.

[32] Jupiter's equatorial and polar radii are 71492 km and 66854 km, respectively, whilst those of Saturn are 60268 km and 54364 km, respectively. The resulting oblatenesses of the two planets are 6.5% and 9.7% – far greater than Earth's 0.34%.

Both gas giants are still in the process of cooling from their formation, and this thermal radiation can be readily detected with observations at 5 microns (see Figure 11). However, research suggests that the thermal evolutionary histories of Jupiter and Saturn are markedly different. Homogeneous evolutionary models tend to accurately reproduce the current luminosity of Jupiter, but substantially underpredict that of Saturn (see e.g. Fortney & Nettlemann, 2010 for a review on this topic). This discrepancy has presented a longstanding, critical issue to the fundamental understanding of the formation and evolution of the entire class of gas giant objects. Helium phase separation and subsequent "rain out" onto Saturn's core is often used to invoke an extra source of energy within Saturn (e.g., Stevenson & Salpeter, 1977a, 1977b). Over forty years after the introduction of this theory, results from the *Cassini* mission seem to have provided verification (Koskinen & Guerlet, 2018). However, modelling and observational efforts will likely continue to explore this topic into the future.

---

[33] It is now widely accepted that the gas giant planets did not form at precisely their current locations, but instead underwent a certain amount of migration as they grew. A number of different scenarios have been proposed describing that migration – from relatively sedate migration over small distances, to chaotic and 'jumpy' migration, driven by encounters with massive planetesimals, and even scenarios where Jupiter migrates first inward (to around 2 au from the Sun), then back out to its current location. The bulk of the evidence for the proposed migration is found in the distribution of the Solar system's small body populations – and we discuss the various theories proposed to explain that distribution in detail in section 4.3. It is even considered possible that the giant planets ejected a planet-mass object from the region in which they formed to an orbit beyond that of Neptune – a theory we discuss in section 4.5.2.

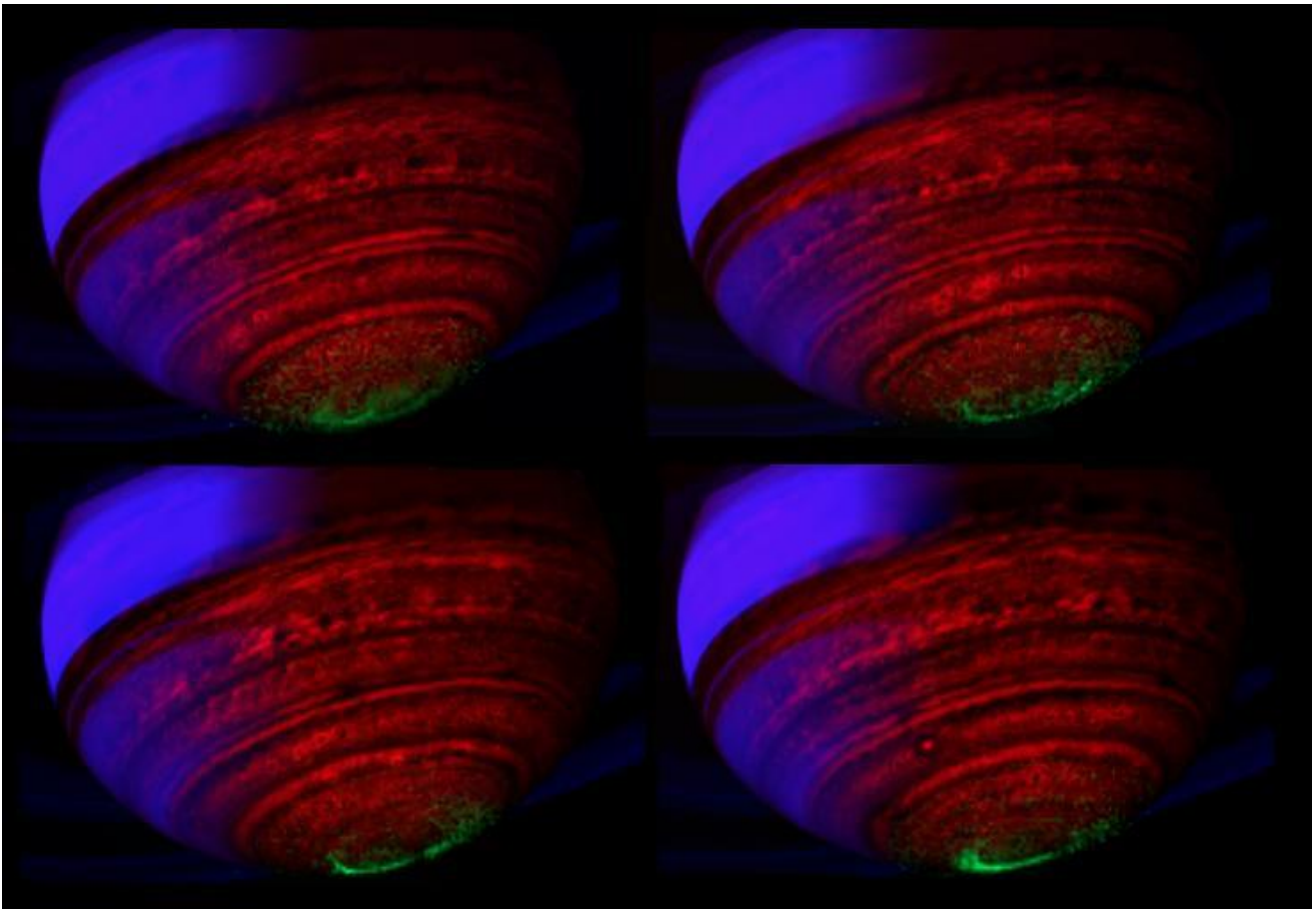

*Figure 11: Saturn in false colour as viewed from the Visual and Infrared Mapping Spectrometer onboard the Cassini Spacecraft. The blue represents reflected solar radiation at 2 microns, the green represents auroral emission between 3 and 4 microns, and the red represents deep thermal emission at 5 microns. Image credit: NASA/JPL/ASI/University of Arizona/University of Leicester (https://photojournal.jpl.nasa.gov/catalog/PIA13403); image is public domain.*

### *Uranus and Neptune: The Ice Giants*

A whole generation of space scientists were inspired when *Voyager 2* flew by Neptune in August 1989, sending us the first images of its dark blue disc. Four years earlier, the same spacecraft had flown past Uranus, and still over 30 years later remains the only spacecraft to have visited our Solar system's ice giants. Many of the observations made by *Voyager 2* remain to date the best information we have to constrain our understanding of the ice giants. Based on *Voyager 2* results, both Uranus and Neptune have been the subjects of books within the University of Arizona's Space Science Series – 'Uranus' (Bergstralh et al., 1991) and 'Neptune and Triton' (Cruikshank et al., 1995). However, over the last 30 years, Earth and near-Earth observations have made a significant impact to our understanding of these worlds. Observations from facilities such as the Very Large Telescope, the Atacama Large Millimeter/sub-millimeter Array (ALMA), and the *Hubble Space Telescope*, have shown both the ice giants to be much more dynamic worlds than the *Voyager 2* data had suggested.

Why the name 'Ice Giants'? This moniker has come from the average densities of Uranus and Neptune, which 'weigh' in at 1.27 and 1.65 g/cm$^3$ respectively. This early observation led to the conclusion that these planets were markedly enriched with the 'heavier' elements of oxygen, carbon, sulfur and nitrogen compared to the composition of Jupiter and Saturn, whose overall makeup reflects more closely that of our Sun (as discussed above). The widely accepted view is that these elements will be in the form of ices; with $H_2O$, $CH_4$, $H_2S$ and $NH_3$ combined making up ~70% of the mass of both Uranus and Neptune, and that it is the physical properties of these materials that will govern the interactions on these planets.

Though Uranus and Neptune are often collectively written about as 'the Ice Giants', it should be pointed out that there are some large differences between these worlds. The starkest of these is the negligible heat flux exhibited flowing from Uranus' interior, as observed by *Voyager 2*, making its upper layers the coldest planetary location within our Solar system. Added to this is the 98° obliquity of Uranus, which results in the planet experiencing extreme seasonal change. In contrast, Neptune has very strong self-luminosity (2.61 times more the solar influx; Pearl & Conrath, 1991), and has been continually observed to display distinct meteorological features. As a result, it has been suggested that the two planets represent end-members of a 'Neptune and sub-Neptune' class of astronomical object (Fulton and Petigura, 2018).

Uranus and Neptune both host a suite of satellites, many of which are discussed in subsequent sections. Uranus's larger, regular satellites (including Titania, Oberon, Ariel and Umbriel) have orbits which are coplanar to Uranus' rotation, tilted to the plane of the rest of the Solar system[34]. Thirteen of these regular satellites have, to date, been discovered orbiting interior to the orbit of Miranda, the innermost of the five moons known to orbit the planet at the dawn of the Space Age. Ten of those satellites were discovered during the *Voyager 2* flyby of the planet (Smith et al., 1986; Jacobson, 1998), with an eleventh, Perdita, being discovered more than a decade later using data obtained during the flyby (e.g. Karkoschka, 2001). The other two new regular satellites (Cupid and Mab) were discovered in 2003 using the *Hubble Space Telescope* (Showalter and Lissauer, 2003). In the last three decades, ground-based observations have revealed a secondary satellite system orbiting Uranus, comprising (at the time of writing) nine much smaller irregular satellites largely moving on retrograde orbits (see e.g. Gladman et al., 1998b, 2000; Sheppard et al., 2005)[35].

Neptune's satellite system is dominated by its largest moon, Triton, which stands out as it is the only large satellite within our Solar system to move on a retrograde orbit about its planet. It is now considered that the most likely explanation for the origin of Triton is that it was once a binary trans-Neptunian object, moving on an orbit not unlike that of the dwarf planet Pluto, and that it was gravitationally captured by Neptune (Agnor and Hamilton, 2006) as a result of an encounter with the planet that tore the binary asunder. That capture event is thought to have had catastrophic implications for the rest of Neptune's satellites at the time. We discuss the origin of Triton in more detail in section 4.1.3.

Both Uranus and Neptune rotate with similar periods, roughly intermediate between those of the gas giants and the Earth. Uranus rotates once every 17 hours, 14 minutes, whilst Neptune spins slightly faster, with a period of 16 hours, 6 minutes. These similar rotation periods, coupled with the limited available measurements of the two planet's gravity fields, have led to a common interior structure being proposed for both planets. This is based on a three-layer model (Guillot, 1999), comprising of an upper layer of ices mixed with molecular hydrogen and helium, that transitions to an icy mantle of water, methane and ammonia - potentially mixed with silicates. The very centre of the ice giants is thought to comprise a dense core of silicates and metals, though some have suggested that carbon could condense to diamond under these conditions (Benedetti et al., 1999).

Our abiding image of Uranus is of a pale turquoise gas world devoid of the tumultuous storms that are enduring on the other gas giants. This imagery, taken by *Voyager 2* as it flew past Uranus in 1985 is perhaps unrepresentative, as it is of the planets' Northern hemisphere summer. Due to its extreme obliquity, Uranus is subject to extreme seasons, and during a given hemisphere's winter the majority of it will not see the Sun for 20 years. On the progression to equinox in 2007, a noticeable increase of cloud activity was observed in Uranus' atmosphere (de Pater et al., 2015), and this continues to be monitored in the hope it will reveal much about Uranus' atmospheric dynamics. Neptune, in contrast, has shown constant atmospheric activity since being first observed by *Voyager 2* on its 1989 flyby. This activity is mostly observed in the infra-red and at the planet's mid-latitudes, however a bright equatorial storm tracked by professional and amateur observers through 2017-2018 is challenging our emerging understanding of Neptune's atmosphere (Molter et al., 2019).

---

[34] The extreme tilt of Uranus and its regular satellite system is often considered to be evidence that the giant planet was involved in a giant collision, either during its formation or soon after, with an Earth-sized protoplanet – a theory we discuss in more detail in section 4.1.1.

[35] Such irregular satellites are found orbiting each of the giant planets and are thought to have been captured by their host planets through a variety of mechanisms during the latter stages of planet formation – as we discuss in more detail in the opening to section four, and in section 4.1.3.

Both Uranus and Neptune exhibit magnetic fields that are anomalous to those observed elsewhere in the Solar system. Unlike the magnetic fields of Earth, Jupiter and Saturn; those of Uranus and Neptune are not dipolar or axially symmetric. It was suggested (Hubbard and MacFarlane, 1980) that to generate fields of this nature a type of thin-shell dynamo would need to arise. Calculations have shown that this is possible (Stanley and Bloxham, 2004), but for this to occur a convective region in the interior must occur. This is at odds with many of the current interior models of the two planets, which have assumed purely conductive interiors for Uranus and Neptune.

With the great success of the *Galileo*, *Cassini*, and *Juno* missions in exploring the Solar system's gas giants, there is a current groundswell of support for a similar flagship mission to explore Uranus and Neptune. Our current lack of detailed information about our own ice giants is particularly disappointing given the great number of potential 'ice-giant' type planets that have been detected orbiting other stars in recent years (as we discuss in section 5). Whilst those planets orbit at much smaller radii than Uranus and Neptune, the best laboratories for the detailed study of such planets are, at least in the relatively near future, those in our own planetary system. If we are to understand these far away worlds at all we need first to well constrain our own neighbours Uranus and Neptune.

# 4 THE FORMATION AND EVOLUTION OF THE SOLAR SYSTEM

Our understanding of the formation and evolution of our Solar system has been shaped by the almost overwhelming amount of information we have about it. Prior to the discovery of the first exoplanets, all models of planetary formation were based on our own Solar system, which is now actually thought to be somewhat unusual as planetary systems go (e.g. Wittenmyer et al., 2011b, 2016; Bryan et al., 2019). For example, "super-Earths" (planets of ~1.2 – 1.9 Earth radii) are the most common type of planet as revealed by the *Kepler* mission (Howard et al., 2012, Zhu et al., 2018, Hsu et al., 2019), yet are conspicuously absent in our Solar system.

Current thinking holds that our Solar system formed from a collapsing giant molecular cloud. It has been suggested that that collapse may have been triggered by the shock wave from a nearby supernova (e.g. Cameron & Truran, 1977; Boss & Keiser, 2010; and many others). Key evidence for the influence of a nearby supernova comes from the study of the composition of meteorites, which reveal the existence of short-lived radioactive nuclei in the material from which they formed. As Boss & Valhara (2000) state:

*"Because the half-lives of these nuclides are so short, this evidence requires that no more than about a million years elapsed between their nucleosynthesis and their inclusion in cm-sized solids in the solar nebula. This abbreviated time span can be explained if these nuclides were synthesized in a stellar source such as a supernova, and were then transported across the interstellar medium by the resulting shock wave, which then triggered the gravitational collapse of the presolar molecular cloud core."*

Essentially, then, the idea is that the same explosion both triggered the collapse of the pre-solar cloud, and polluted it with large amounts of heavy elements. It should be noted, however, that there are some problems with the invocation of a single nearby supernova as the sole source of these short-lived radioactive nuclei, as is discussed in review by Pfalzner et al. (2015). Instead, those authors describe how any contribution to the Solar system's budget of radioactive nuclei resulting from nearby supernova(e) was likely supplemented by an additional dose of $^{26}$Al, injected through the wind of a massive nearby star (e.g. Arnould et al., 2006; Tatischeff et al., 2010). In order that the Solar system's birth cluster be large enough that there be a reasonable likelihood of such pollution, Pfalzner et al. (2015) suggest that the cluster was relatively massive - comprising at least 2000 stars.

An alternative explanation for the unusually high abundance of $^{26}$Al in the early Solar system was proposed by Dwarkadas et al., 2017, who note that, whilst the protoplanetary nebula from which the Solar system formed was enriched in $^{26}$Al, the abundance of $^{60}$Fe (compared to $^{56}$Fe) appears to have been somewhat depleted compared to the norm in the galaxy at the time. To explain this discrepancy, Dwakadas et al. propose that the formation of the Solar system was not triggered by a nearby supernova, but was instead triggered by the outflow from a nearby Wolf-Rayet star. If the Solar system formed on the edge of a bubble blown clear by such a star, it would naturally be inoculated with an appreciable quantity of $^{26}$Al, without simultaneously being polluted by $^{60}$Fe.

In contrast to these ideas of a peculiar/rare confluence of events leading to the formation of the Solar system, Young (2016) argues that no such unusual processes (nearby supernovae, or AGB stars) need to be invoked to explain the Solar system's initial budget of short-lived radionuclides – instead suggesting that '*The radiochemistry of the early Solar system is therefore unexceptional, being the consequence of extensive averaging of solids from molecular clouds*'. More recently, Bartos & Marka (2019) have proposed that the early Solar system was inoculated by short-live radio isotopes as a result of a relatively nearby binary neutron-star merger. Had such an event occurred in the ~100 Myr prior to the Solar system's formation, it could account for the abundances of radioactive 'r-process' isotopes in the early Solar system – isotopes with halflives of less than 100 Myr. In particular, they suggest that single merger event could be responsible for the bulk of the early Solar system's curium and plutonium, and proposed that that event might have occurred approximately 80 million years before the Solar system formed, at a distance of approximately 300 parsecs.

As the proto-Solar nebula collapsed, it is thought to have flattened out into a dynamically cold protoplanetary disc, with the great bulk of the material moving on near-circular, low-inclination orbits. Within the disc, the temperature varied greatly – with the inner regions, near our proto-Sun, fiercely hot, while the outer reaches were cold. Beyond a distance of ~3 au, temperatures were sufficiently low for water to form icy grains, which meant that beyond this "ice-line", significantly more solid material was available to speed the accretion of the planets (e.g. Dodson-Robinson et al., 2009). Over time, the solid material in the nebula experienced collisions, causing the agglomeration of planetary embryos[36]. The more material that was available, and the greater the density of that material, the faster these embryos grew. Beyond the ice-line, this core-accretion process occurred sufficiently quickly that the cores were able to become massive enough that they could capture gas from the proto-planetary nebula, leading to them undergoing "runaway growth" (e.g. Pollack et al., 1996). In this model, the growth of the giant planets (Jupiter, Saturn, Uranus and Neptune) can be broken down into three key stages.

> 1) The accretion of solid material from the disc starts very slowly, but as the mass of the planetesimal increases, the rate at which material is accreted undergoes a runaway growth, then falls away again once the planet has swept up all the material in its immediate "feeding zone" (e.g. Lissauer, 1987). This process involves the formation of a number of 'embryos' or 'oligarchs' - the largest objects in the feeding zone - that can undergo violent giant collisions with one another. Eventually, one mass comes to dominate a given region - an object often described as a 'protoplanet'.

> 2) Once the feeding zone is cleared, the protoplanet undergoes a very slow continued accretion, as small amounts of material (both gas and solids) are captured from the edges of the feeding zone. This stage tends to determine the evolutionary timescale of giant planet formation (e.g. Pollack et al., 1996).

> 3) Eventually the mass of the protoplanet reaches a critical value (roughly ten times the mass of the Earth, e.g. Mizuno, 1980), at around the time its gaseous and solid mass are equivalent. At this point, it undergoes a process of runaway gas accretion, which only ceases once the gas in the disc is cleared out by the Sun – the end of the lifetime of the proto-planetary disc (typically believed to occur around 2 – 3 Myr after the start of the formation process, though some discs are thought to survive for up to ~10 Myr, or more; e.g. Li & Xiao, 2016; Murphy, Mamajek & Bell, 2018, whilst those in clusters might be more short-lived; e.g. Concha-Ramírez et al., 2019).

A more recently developed hypothesis for the early part of this process is the so called 'Pebble Accretion' model (Lambrechts & Johansen 2012; Levison et al., 2015; Chambers, 2016). One of the issues with the 'classical' formation theory detailed above is that the rate at which material is accreted into planetesimals is often too slow to produce a massive enough object to act as the nucleus of a gas giant planet before the gas disc dissipates. Historically, this has proven to be a major flaw in the classical accretion process (Goldreich et al., 2004). A solution to this problem has planetesimals accrete 'pebbles', centimetre to metre sized objects, from the debris disc (Lambrechts & Johansen, 2012). If the pebbles form in conjunction with the scattering of the smaller planetesimals in the disc (Levison et al., 2015), then the cores of gas giants can form in a realistic number and timeframe. For a detailed overview of current theories of terrestrial planet formation, we direct the interested reader to Izidoro & Raymond, 2018.

---

[36] At sub-cm sizes, the collisional accretion of dust grains is only possible for very low collision velocities, as a result of the extremely low strength of such grains. Whilst the precise details of accretion for grains in the micron- to centimetre- scale remain unclear, it is certainly the case that that accretion was greatly aided by the effects of gas-drag within the proto-planetary disc, which would act to rapidly damp the motion of those small grains, ensuring that their typical collision velocities were very low. In laboratory experiments, dust growth has been demonstrated up to millimeter sizes through mutual collisions, but the centimetre size barrier is insurmountable due to mutual velocities and low surface adhesion resulting in bouncing rather than sticking between the largest grains (e.g. Blum, 2010; Windmark 2012b; Testi, 2014). However, if larger grains are already present in the disc further growth of small dust grains beyond the centimetre size barrier is a rapid process (~$10^4$ years; Windmark, 2012a). This barrier to growth can also be overcome by assuming some 'stickier' dust (e.g. some fraction of $H_2O/CO_2$ ice, or the presence of an organic mantle to a grain; e.g. Homma et al., 2019) in their compositions, or a distribution of collision speeds (e.g. Windmark et al., 2012b).

Once the gas has been removed from the system, the growth of the giant planets all but stops, with the only ongoing mass gain coming from collisions with dynamically unstable small bodies (such as the Shoemaker-Levy 9 impact on Jupiter, in 1994; e.g. Zahnle & Mac Low, 1994; Hammel et al., 1995; Noll et al., 1995). As the giant planets interacted with the small body reservoirs (in particular the trans-Neptunian disc), the transfer of material from those reservoirs around the Solar system (with, ultimately, the bulk being ejected from the system[37], or colliding with one or other of the planets) caused the planets to continue the migration that would have been observed during their formation, albeit at a much, much slower rate. Indeed, some authors (e.g. Tsiganis et al., 2005; Gomes et al., 2005) have suggested that such migration could lead to a period of significant destabilisation of the outer Solar system, and potentially have been the cause of the putative Late Heavy Bombardment of the terrestrial planets (e.g. Bottke & Norman, 2017). On the other hand, it has recently been suggested that the instability of the giant planets probably occurred much earlier than previously thought, perhaps within the first 100 Myr of Solar system evolution (Kaib & Chambers, 2016; Quarles & Kaib, 2019). It has even been suggested that Uranus and Neptune formed closer to the Sun than Saturn, and were ejected to their current orbits in that process (e.g. Thommes et al., 2002), or that the Solar system initially contained at least one additional ice giant that was ejected during such a period of system instability (e.g. Nesvorny & Morbidelli, 2012; Batygin, Brown & Betts, 2012; Cloutier, Tamayo & Valencia, 2015). Such a planet could still lurk in the Solar system's outer reaches, a proposal that has recently driven wide discussion of, and a number of searches for, the proposed 'Planet Nine' (e.g. Batygin & Brown, 2016a; Beust, 2016; Brown & Batygin, 2016, 2019; Batygin & Morbidelli, 2017; Batygin, Adams, Brown & Becker, 2019)

An even more speculative idea is that Jupiter undertook a dramatic excursion through the inner Solar system, with its orbital radius decreasing from > 5 au to < 2 au before increasing again to its current position within the Solar system. This model, which has become known as the 'Grand Tack' (e.g. Walsh et al., 2011, Nesvorny, 2018), has been invoked to explain the mixing of icy and rocky bodies in the asteroid belt. In addition, it offers a plausible explanation for the relatively low mass of Mars , suggesting that the alleged mass-deficit is the result of proto-Mars having been a stranded protoplanetary embryo that was starved of material as a result of the dynamical clearing of material from its zone of capture by the migrating Jupiter (e.g. Brasser et al., 2016).

Regardless of the fine details of the formation of the outer Solar system, there is significant evidence that the giant planets migrated over significant distances (both inwards and outwards; e.g. Malhotra, 1995; Levison et al., 2007; Lykawka, 2012; Nesvorny & Vokrouhlicky, 2016; Nesvorny, 2018) to reach their current locations. The bulk of that evidence comes from the distribution of the Solar system's small bodies, which will be discussed in some depth in section 4.3.

The formation of the terrestrial planets, far interior to the ice line, is thought to have been a slower and more stochastic process. Without the extra solid mass afforded by ices in the outer Solar system, none of the embryos in the inner regions grew sufficiently rapidly to become massive enough to capture a large gaseous atmosphere prior to the removal of the proto-planetary nebula by the youthful Sun, or to become super-Earths. Instead, the terrestrial planets are thought to have grown more slowly, with the final stages of that formation taking between 10 – 100 Myr to reach their conclusion (e.g. Chambers, 2004). The initial growth proceeded much as described above, for the giant planets, with a process of pebble accretion aiding the growth of a number of embryos which became the dominant large bodies in the inner reaches of the proto-planetary nebula (e.g. Izidoro & Raymond, 2018[38]).

---

[37] It is certain that other planetary systems are also continually ejecting objects to interstellar space - both during their formation, and their long-term evolution. As a result, it was inevitable that we would eventually observe one of those objects passing through the Solar system. The first such interstellar vagabond, 1I/'Oumuamua, was discovered in 2017 (e.g. Meech et al., 2017; Jewitt et al., 2017; Bannister et al., 2017; Fitzsimmons et al., 2018), with the second, the cometary 2I/Borisov, being discovered and passing through perihelion in 2019 (e.g. Fitzsimmons et al., 2019; Jewitt & Luu, 2019; Opitom et al., 2019; Guzik et al., 2020; Hallatt & Wiegert, 2020; McKay et al., 2020). It is likely that the next generation of surveys will discover many more in the years to come.

Once formed, these embryos dynamically interacted with one another, with giant collisions between embryos being the primary route by which they gained mass. This highly stochastic process eventually resulted in the terrestrial planets we observe today, each of which have features that have been explained as being the result of giant collisions between embryos towards the end of planetary formation (e.g. Dormand & McCue, 1987; Benz et al., 1986, 1987, 1989, 2007; Raymond et al., 2014; Canup & Salmon, 2018; Lykawka & Ito, 2019).

The formation of the planetary satellites appears to have occurred through a number of routes. For the regular satellites of the giant planets, current theories suggest that those planets developed their own small sub-nebulae, within which the satellites formed in much the same manner as the terrestrial planets within the main proto-planetary nebula (e.g. Mosqueria & Estrada, 2003a, b; Sasaki, Stewart & Ida, 2010; Mosqueria, Estrada & Turrini, 2010). Some recent studies take this theory one step further, and have generations of satellites forming in the sub-nebulae of the outer planets, migrating inwards, and being devoured by their hosts. Only the final generation of satellites formed in this way would have survived to the current day (e.g. Canup & Ward, 2009).

By contrast, the irregular satellites of the outer planets are thought to have formed elsewhere and then been captured, either through collisions (e.g. Goldreich et al., 1989; Woolfson, 1999; Koch & Hansen, 2011), three-body encounters between the host planet and binary planetesimals (e.g. Agnor & Hamilton, 2006; Vokrouhlický et al., 2008), three-body encounters involving two of the giant planets and the captured object (e.g. Nesvorný et al., 2007), or through gas-drag (e.g. Ćuk & Burns, 2004). No one mechanism is sufficient to explain all the observed properties of all irregular satellites (e.g. Jewitt & Haghighipour, 2007), and so these objects are thought to be representative of the different processes that occurred during the final stages of planetary formation. The Moon is different again, thought to have formed when the proto-Earth was involved in a giant collision with a Mars-sized embryo, towards the latter stages of its accretion (e.g. Benz et al., 1986, Reufer et al., 2012). In recent years, a similar process has been suggested to explain the origin of Mars' two unusual satellites, Phobos and Deimos - albeit invoking a somewhat less catastrophic collision (e.g. Craddock, 2011; Canup & Salmon, 2018).

The formation of our Solar system left behind large populations of small bodies, whose orbits and composition provide valuable data that are used to help untangle the detail of the formation process (e.g. Malhotra, 1993; Gomes, 1997; Lykawka & Mukai, 2008; Minton & Malhotra, 2009; Lykawka, 2012; Nesvorny & Vokrouhlicky, 2016). In the coming subsections, we present detailed reviews of many of the processes thought to have occurred during planetary formation, together with those that continue to the current day, and highlight a number of theories put forward to explain the various features of our system as a whole.

## 4.1 GIANT AND CATACLYSMIC IMPACTS

When we look around the Solar system today, many objects bear witness to the giant collisions that occurred during the latter stages of planetary formation. From the impact that created the Moon (e.g. Benz et al., 1986, Reufer et al., 2012) to that which has been invoked to explain the capture of Neptune's giant moon Triton (e.g. Goldreich et al., 1989; Woolfson, 1999), it seems that cataclysmic impacts and collisions between planet-sized bodies were a regular occurrence as the planets formed, and played a pivotal role in shaping the Solar system as we know it.

---

[38] For a detailed overview of current theories of terrestrial planet formation, we direct the interested reader to Izidoro & Raymond's excellent chapter in the Handbook of Exoplanets, published in 2018.

## 4.1.1 GIANT IMPACTS AND THE PLANETS
A number of the planets in our Solar system have features that are best explained as having resulted from giant impacts thought to have occurred during the final stages of planet formation.

*Mercury*

The planet Mercury has long been recognised as being both unusually dense, and unexpectedly small (as discussed in section three). Its uncompressed mean density of ~5.3 g/cm$^3$ (Cameron et al., 1988) is significantly greater than the value for the next densest planet (the Earth, with ~4.45 g/cm$^3$; Lewis, 1972). This increased density must be the direct result of Mercury containing significantly more iron, as a fraction of its total mass, than any of the other terrestrial planets. Indeed, Benz et al. (2007) state that the expected silicate-to-iron ratio of Mercury is typically thought to lie in the range 30:70 – 50:50, though some authors have suggested that the true silicate content could be even less than the lower extreme stated here, if the interior contains a significant amount of sulphur (e.g. Harder & Schubert, 2001).

Over the years, a number of models have been put forward to explain the strange properties of Mercury, but the leading contender (e.g. Benz et al., 1988; Benz et al., 2007; Chau et al., 2018) involves the collision of a primordial "super-Mercury" with a smaller, but still planet-sized object, which occurred towards the latter stages of planetary formation. Through the use of smoothed-particle hydrodynamics (hereafter SPH) modelling, Benz et al. (2007) suggest that, prior to the impact, the large proto-planet moving on an orbit similar to that occupied by our Mercury was at least twice the mass of the current planet, and was made of the same materials as the other terrestrial planets – a mix of silicates and iron. At the time of the impact, the proto-planet was differentiated, such that the bulk of the planet's iron had settled to the core, which was surrounded by a mantle rich in silicates. At this point, the proto-planet was struck by another proto-planetary body, and severely disrupted. The outer layers were stripped away, and dispersed into the Solar system by the force of the collision, leaving behind the bulk of the core. Over time, some small fraction of the ejected material re-accreted, but the great majority was lost, leaving the iron-rich, silicate-poor, over-dense and under-sized husk that we observe today.

More recent work by Chau et al. (2018) delves deeper into the nature of Mercury's origin, investigating a variety of collision scenarios, from a single giant impact to a series of smaller collisions. They find that it is possible to replicate Mercury's current composition by invoking a range of initial scenarios - a high speed head-on collision with a very massive impactor, a glancing 'hit-and-run' collision between a massive proto-Mercury (4-5 times the mass of the current planet) and another body, and even a series of smaller, less catastrophic collisions (as is also discussed by Asphaug & Reufer, 2014). In a similar vein, Jackson et al. (2018) investigated the leading giant impact scenarios that have been proposed to explain the origin of Mercury. They suggest that all models that invoke a single, high velocity impact should be disfavoured, and that a multiple hit-and-run scenario seems to be the most likely explanation for the formation of Mercury.

Recent studies investigating the formation of the terrestrial planets through *N*-body simulations have found it hard to replicate the initial conditions that would be needed for the collisional scenarios described above. However, those models were also unable to explain Mercury's current mass, orbit, and high core mass fraction (Clement et al., 2019b; Lykawka & Ito, 2019) - leaving the full picture of Mercury's formation and evolution open to further debate. What remains clear, however, is that the best explanation for Mercury's peculiarities is an extreme collisional past, which serves as a reminder that the final stages of terrestrial planet formation were a chaotic and violent time.

*Venus*

The second planet out from the Sun, Venus, also has a few unusual features (e.g. Kane et al., 2019b). Whilst the majority of the planets spin (relatively rapidly) in the same direction as they orbit the Sun (counter-clockwise, as viewed from above the Sun's north-pole, Venus actually spins, very slowly, in the other direction. With respect to the background stars (the sidereal day), Venus takes some 243 days to spin once on its axis (a period longer than that of the planet's orbit, 225 days). However, because it spins in the opposite direction, the Solar day (the time between successive sunrises or sunsets) is around 117 days. Although, in the past, suggestions were made that this unusual spin had its origin in an ancient collision between Venus and another body (e.g. Dormand & McCue, 1987; Dormand & McCue, 1993), the preferred current theories for the unusual spin invoke other mechanism. Dobrovolskis & Ingersoll (1980) proposed that Venus' current slow spin could principally be the result of atmospheric torques upon the planet's surface, driven by the heating of the planet's dense lower atmosphere by Solar insolation[39]. Such a torque would, over long time periods, have led to the gradual slow-down of Venus' rotation, resulting in a spin state that they suggest may be "*a steady state among tides in the atmosphere, tides in the solid body, and possibly the influence of the Earth*". More recently, Correia et al. (2001a, b) proposed that the chaotic dynamical evolution of Venus' spin under the gravitational influence of the other planets might explain its peculiarities. Neither of these mechanisms would require a giant collision involving Venus at some point in the past, though of course it should be noted that neither would rule out the occurrence of such a collision.

Interestingly, the idea that a giant impact might have played some role in the formation of Venus has not totally gone away. Over the years, a number of studies have suggested that the interior of Venus is significantly drier than that of the Earth. Zolotov et al. (1997) state that the bulk-silicate-Venus "*plausibly contains only 1% of the amount of water in the bulk silicate Earth*", a deficiency that is widely accepted in the Solar system science community, and one that has been used to explain the current lack of widespread plate tectonics on that planet (e.g. Ward & Brownlee, 2000; O'Neil et al., 2007). In order to explain the dehydration of Venus, Davies (2008) suggests that the planet may once have experienced a giant collision with a large planetary embryo. This, he suggests, would have helped the planet to outgas the bulk of the water it accreted as it formed, resulting in it being significantly dryer than the other terrestrial planets. More recent work (Gillmann, Golabek & Tackley, 2016) revealed that a large, late impact could readily facilitate the loss of Venus' water. However, if the timing were different, a similar impact could instead help to generate a dense, volatile-rich atmosphere for the planet - leading to the kind of hot-house conditions observed on the planet today. Indeed, Lykawka and Ito (2019) suggest that Venus would likely have experienced a number of giant impacts through the latter stages of its formation. Venus' desiccated nature might, instead, reflect differences in the delivery of water to the terrestrial planets (as we will discuss in more detail in section 4.5), or may simply be the result of the dehydration of the planet's surface (such that all water out-gassed is lost, stripped from the atmosphere by the Solar wind, effectively drying the planet from the outside in; e.g. Kulikov et al., 2006; Dubinin et al., 2011), it is certainly interesting that the idea of such a giant collision keeps recurring in studies of that planet.

*Earth*
Possibly the most famous theorised giant impact is that which is thought to have created the Earth's moon (e.g. Benz et al., 1986, 1987, 1989; Stephenson 1987; Cameron & Benz 1991, Canup & Asphaug, 2001; Wada et al., 2006; Halliday, 2008; Reufer et al., 2012). It is proposed that a large, Mars-sized object (often called Theia) struck the proto-Earth at some point up to ~100 Myr after the planets began to form, with recent work suggesting that the impact most likely occurred whilst the Earth's surface was still covered by a global magma ocean (Hosono et al., 2019). The low-velocity, oblique impact severely disrupted both the Earth and the impactor, emplacing a large amount of debris into orbit around the Earth - debris that primarily comprised the upper layers of the Earth and the impacting body. This debris then accreted to form the Moon, at that time located just beyond the Roche limit (the distance at which tidal forces would disrupt a strengthless body).

---

[39] It is worth recollecting, here, that Venus is thought to have experienced a 'runaway greenhouse effect' in the past, allowing it to attain and maintain its current superheated surface conditions. The young Venus may well have had a climate superficially similar to that of the Earth. As the Sun grew more luminous, the planet warmed until it reached a crucial tipping point – the evaporation of water from the planet's oceans accelerated. The increased levels of water vapour in Venus' atmosphere acted to strengthen the greenhouse effect – driving the temperature still higher. Eventually, the oceans were gone, which served to further accelerate the greenhouse warming – by removing the oceans, the water-driven weathering that would act to remove other greenhouse gasses from the atmosphere ceased. But those gasses continued to build up in the atmosphere, as a result of the planet's volcanic activity. The result? Venus eventually warmed to the point that its surface is hot enough to melt lead – far from the 'Earth-like planet' it may once have been.

Since its formation, the Moon has receded from the Earth as a result of their tidal interaction, which has also acted to slow the rotation of the Earth, and has locked the Lunar rotation such that one face perpetually points towards Earth (a 1:1 spin-orbit resonance). The theory explains why the Earth and Moon have significantly different bulk densities. Had the Earth and Moon simply accreted together, or were the Moon captured having formed independently, it would be expected that they would both have roughly similar fractions of iron and silicates. Instead, the Earth is observed to have a large iron core, and a density of some ~5.5 g/cm$^3$ (corresponding to an uncompressed density of ~4.45 g/cm$^3$, as discussed above), whilst that of the Moon is just 3.346 g/cm$^3$ (Table 3.13, Wieczorek et al., 2006). It also explains the high angular momentum of the Earth-Moon system, and the particularly large mass of the Moon relative to that of the Earth - both factors which suggest that the system had an unusual formation.

In order to explain the low collision velocity required by SPH simulations to form the Moon, Belbruno & Gott (2005) suggest that the impactor formed as a Trojan-companion to the Earth, librating around our planet's $L_4$ or $L_5$ Lagrange point. Such a configuration can be stable for a lengthy period of time (indeed, both Jupiter and Neptune still host significant populations of Trojans at the current day, as will be discussed later, particularly in section 4.3.2), so it is perfectly reasonable to assume that such a body could survive and continue to grow for a hundred million years prior to escaping from the resonant orbit (potentially through dynamical interaction with other growing embryos, or distant perturbations from other planets) and colliding with the Earth. If the impactor formed at the same heliocentric distance as the Earth, this would fit well with the observed oxygen isotope abundances in the Moon and Earth, which are essentially identical (e.g. Wiechart et al., 2001; Young et al., 2016). Had the impacting body formed elsewhere in the Solar system, it would have brought material with measurably different abundances into play, and there would be no guarantee that both Earth and Moon would receive just the right dose that the two bodies, post-impact, would look the same. The idea of a local origin for Theia found further support in the work of Quarles & Lissauer (2015), who performed a series of integrations of an inner Solar system that contained, initially, five terrestrial planets. They found that a variety of scenarios could lead to a Moon-forming impact, with Theia starting life on a variety of orbits in the general vicinity of 1 au, without requiring Theia to form in 1:1 resonance with the Earth. The scenarios the resulted in planetary systems most similar to our own (with the terrestrial planets on orbits with relatively low eccentricities) came when the proto-Earth and Theia began with similar semi-major axes.

Whilst the search for the most promising impact scenario for the formation of the Moon remains an active topic of research, recent work has identified subtle but statistically significant differences between the oxygen isotope abundances of the Earth and Moon - at the ~12 parts per million level (Herwatz et al., 2014) - potentially revealing the isotopic signature of Theia for the very first time.

The debris that was ejected from the Moon-forming collision above the escape velocity of the Earth-Moon system would have taken a long time to disperse. A significant amount of that debris would have collided with the Earth and the other terrestrial planets over the millions of years following the collision (~20% falling back to Earth, and ~17% accreted by Venus, for example, e.g. Jackson & Wyatt, 2012), and the impact would have resulted in a markedly increased debris and dust environment in the inner Solar system that would have lasted for tens of millions of years (e.g. Wyatt & Jackson, 2016).

*Mars*
Mars, too, has been proposed as the recipient of at least one giant impact. A key feature of that planet is the clear dichotomy between the terrain in the Southern Highlands, and that in the Northern Lowlands. The origin of the feature has long been the subject of debate, with models suggesting that it might be the result of either a giant impact, or convection and overturn in the planet's mantle (e.g. Wilhelms & Squyres, 1984; McGill & Squyres, 1991; Roberts & Zhong, 2006). In order to address this question, Andrews-Hanna et al. (2008) studied the morphology of the feature, examining satellite measurements of Martian gravity and the topography of the boundary between the two distinct zones. They were able to remove the effects of the Tharsis bulge, a volcanic region of crustal uplift featuring the Solar system's largest volcanoes (including Olympus Mons), and revealed that, once that feature was removed, the crustal dichotomy itself could be modelled as a large elliptical region that, in turn, is best explained as being the scar left over from a giant impact that occurred during the latter stages of planetary formation.

Following that work, Marinova et al. (2008), performed detailed SPH modelling to study whether it was possible that an impact sufficiently large to create such a vast basin (the largest impact basin known to date in our Solar system) would actually leave such a scar (since the energy involved in such a collision could easily melt the entire crust of the planet, obliterating any evidence that it had happened). They present a number of scenarios that appear to fit the observations, suggesting that the basin-forming impact likely involved an oblique (rather than face-on) low velocity collision (6 - 10kms$^{-1}$) between the planet and an impactor of diameter 1,600 - 2,700km. Given the extreme age of the feature, such an explanation seems perfectly reasonable, as one would expect there to be a significant number of such potential impactors left over at around the epoch that planet formation was coming to a close, and so this collisional origin for the crustal dichotomy is currently considered the leading explanation[40]. It has even recently been suggested that the peculiar Martian moons, Phobos and Deimos, could have been produced in such a giant impact – a theory which may solve the long standing question of the origin of those peculiar tiny satellites (e.g. Craddock, 2011).

The most recent studies of such a collision suggest that, rather than the impactor being of comparable size to the Red Planet, the formation of Mars' two satellites might only require a collision with an impactor some $10^{-3}$ times Mars' mass (Canup & Salmon, 2018). Such a low relative mass would suggest an impactor diameter between that of the asteroids Vesta (~525 km) and Ceres (946 km). Interestingly, that model suggests that such an impact could have created several small satellites, accreted from a circum-planetary disc resulting from the initial collision. In that scenario, Phobos and Deimos would have been the outermost of those new satellites, with the others destroyed relatively quickly (on timescales less than a million years) after spiralling back into Mars as a result of their tidal interaction with the planet. The authors note that the scale of impact required for the formation of Phobos and Deimos would potentially be compatible with the formation of Mars' largest known impact basins - Utopia (diameter ~3,300 km) and Hellas (diameter ~2,300 km).

*Jupiter*
One of the key predictions of the Core Accretion model is that the giant planets should have relatively dense core of solid material - with masses of several to ten times that of the Earth (e.g. Mizuno, 1980; Bodenheimer & Pollack, 1986; Pollack et al., 1996; Inaba, Wetherill & Ikoma, 2003; Hubickyj, Bodenheimer & Lissauer, 2005). It is only once the giant planets had accreted that much solid matter that their gravitational pulls would be sufficient to begin to capture the hydrogen and helium that made up the bulk of the protoplanetary nebula. However, it has historically proven challenging to determine the true scale and nature of the cores of Jupiter and Saturn (e.g. Guillot, Gautier & Hubbard, 1997; Saumon & Guillot, 2004; Hori et al., 2008; Nettlemann, 2011; Lozovsky et al., 2017). NASA's *Juno* mission (e.g. Matousek, 2007; Bolton et al., 2017) has attempted to provide an answer to that question for the giant planet Jupiter. *Juno* moves on a highly eccentric and inclined orbit that, at perijove, skims Jupiter's cloud tops. Moving on such an orbit has allowed *Juno* to probe the structure of Jupiter's deep interior - mapping the distribution of mass deep within the giant planet (e.g. Militzer et al., 2016; Folkner et al., 2017; Wahl et al., 2017).

---

[40] Indeed, it seems likely that the Earth would have experienced several impacts of a similar scale to this Martian 'giant impact', which was several orders of magnitude less energetic than that between the proto-Earth and Theia, which led to the formation of the Moon. Zahnle et al. (2007) present a fascinating review of the impact such collision would have on the early Earth, in the time after the Moon forming impact, whilst O'Neill et al. (2017) discuss the role that large impacts (of objects with diameters greater than 500 km) could have played in the genesis of Earth's global system of plate tectonics.

The first results from *Juno* have found strong evidence for a massive core deep within Jupiter - containing between 7 and 25 times the mass of the Earth (e.g. Wahl et al., 2017). Surprisingly, however, those results also reveal that Jupiter's core is somewhat larger (and therefore less dense, or more '*dilute*') than had been expected (e.g. Wahl et al., 2017; Debras & Chabrier, 2019). To explain this otherwise anomalous mass distribution, Liu et al. (2019) recently proposed that Jupiter's diluted, extended core, may well be the result of a giant head-on collision between the young proto-Jupiter and a massive planetary embryo, itself around ten times the mass of the Earth. Such an impact could shatter the proto-Jupiter's dense, compact, massive core, dispersing its material through the lower reaches of the planet, and resulting in the degree of dilution observed in the *Juno* data. As more data comes in from the ongoing *Juno* mission, it seems likely that the precision with which Jupiter's inner regions can be mapped will increase still further, and it will be interesting to see whether those results continue to match well with the violent collision hypothesis proposed by Liu et al. (2019).

*Saturn*
The first evidence that Saturn's spin axis is tilted, compared to the plane of its orbit, came with the discovery of its rings, in the 17th Century. Galileo's observations of Saturn in 1610 suggested that the planet was triple – he wrote (in translation) "*This discovery is that Saturn is not single but a composite of three, which seem to touch each other and never change their relative position and never move among themselves nor change: they are placed in a line parallel to the Zodiac; the one in the middle being about three times larger than the two lateral ones, and being situate in this manner, oOo.*"[41]. By the end of 1612, however, the two smaller bodies accompanying Saturn appeared to vanish – much to Galileo's consternation. Forty-seven years later, Huygens (1659) published the solution to Galileo's quandary. Based on his own observations, made with a telescope he constructed with his brother, Huygens discovered Saturn's giant satellite, Titan, and revealed that the 'smaller bodies' were in fact a thin ring of material, in orbit around the planet. The ring and the satellite moved in the same plane, tilted to the plane of Saturn's orbit by more than 20 degrees. The cause of Galileo's vanishing 'companions', therefore, was the tilt of Saturn's spin axis – which caused the rings to wax and wane over the years, as Saturn moved around the Sun. During the summer and winter for Saturn's northern hemisphere, Saturn's rings were face on – clearly visible, and the source of the optical illusion of a 'triple planet' observed by Galileo. Around the Saturnian equinoxes, however, the rings approach edge-on, and all but disappear as seen from Earth.

In many ways, Saturn's rings are the most obvious indication of a planet's axis tilt in the Solar system – since they can readily be seen even through binoculars, and they reveal that the giant planet's axis is tilted by almost 27 degrees to the plane of its orbit. Lissauer & Safronov (1991) considered the degree to which accretive collisions between protoplanets and large planetesimals would impact the final spin of the planet formed through those collisions. They found that such collisions would, on average, serve to hasten the spin of the youthful planet, as well as acting to increase their obliquities. Their results added weight to the idea the Moon could have formed in a giant collision between the proto-Earth and another body, and supported the idea that Mercury was a collisionally shattered husk. Building on those ideas, they considered whether such impacts could be the cause of Saturn's observed obliquity. They suggested that, if Saturn's obliquity were the result of its collisional history, during its formation, then it most likely must have accreted at least one planetesimal several times more massive than the Earth.

The origin of Saturn's obliquity remains to be definitively determined, however. Ward & Hamilton (2004) noted an alternative mechanism that could drive Saturn's spin axis from a negligible tilt to its current value. They noted that the precession period for Saturn's spin axis is strikingly similar to the precession period of Neptune's orbital plane (which is slightly tilted compared to the Solar system's invariable plane, at an inclination of 0.72 degrees). They note that Neptune's orbit precesses with a period of ~1.87 million years – a period which would, in the past, have been faster. Under the assumption that, when the Edgeworth-Kuiper belt was more massive, Neptune's orbit would have precessed faster than

---

[41] Excerpt from "Galileo's work on Saturn's Rings", Partridge & Whitaker, 1896.

Saturn's spin axis, the authors noted that, as Neptune's orbit precessed ever more slowly, it would eventually reach the same period as Saturn's spin axis. They proposed that Saturn's axial tilt would have been pumped to its current value as a result of the planet's spin moving into this secular spin-orbit resonance with Neptune's orbital precession – and note that, during occupation of that secular resonance, Saturn's axis might well have liberated with an amplitude of up to 31 degrees. In a follow-up work, Hamilton & Ward (2004) used numerical integrations to demonstrate that such resonant excitation of Saturn's axial tilt could occur, and that it would not be disrupted by the perturbations exerted upon Saturn by the other planets in the Solar system.

More recently, Brasser & Lee (2015) considered the same mechanism, in the context of the migration of the giant planets, to see whether the obliquities of Saturn (tilted by ~27 degrees) and Jupiter (tilted by just ~3 degrees) were compatible with a variety of migration models. They found that it was challenging to explain the obliquities of both Jupiter and Saturn as a result of migration – with even their best simulations only yielding appropriate values for the orbital and obliquity evolution of the two giant planets some 0.3% of the time. Whilst it was possible to tilt Saturn's axis as a result of a resonance crossing during migration, then, such a hypothesis was hard to fit together with the other constraints on the final architecture of the Solar system. As a result, the true story of Saturn's obliquity remains an open question – and it seems likely that future observations (similar to those being obtained by the *Juno* mission, currently in orbit around Jupiter) may be required to shed light on the true origin of the ringed planet's tilt. Indeed, during the last few months of the *Cassini* mission, the spacecraft moved on an ever more eccentric orbit, with a pericentre moving closer and closer to the giant planet. That evolution facilitated *Juno*-like measurements, offering a wealth of new information on Saturn's interior structure, from the fact that Saturn's differential atmospheric rotation must extend to at least 9000 km depth (Iess et al., 2019), to a confirmation that the results are consistent with Saturn having an increased central density (indicative of the presence of a core; e.g. Movshovitz et al., 2020). As researchers continue to analyse the wealth of data returned by *Cassini*, it seems likely that the interior structure of the planet will become more clear cut, and that the answer to the question of Saturn's tilt may finally find a definitive answer.

*Uranus*
The discovery of Uranus by Sir William Herschel, in 1781, marked a watershed in Solar system astronomy. Prior to that date, our Solar system was thought to have just six planets, all of which had been known since prehistory. The new planet was observed widely, and enthusiastically, by astronomers around the globe, and Herschel followed his discovery, six years later, with the detection of the first two of the planet's moons (Herschel, 1787). In the letter through which he announced the discovery of these satellites, Herschel states:

*"Their orbits make a confiderable angle with the ecliptic; but to affign the real quantity of this inclination, with many other particulars, will require a great deal of attention, and much contrivance: for, as aftimations by the eye cannot but be extremely fallacious, I do not expect to give a good account of their orbits till I can bring fome of my micrometers to bear upon them; which, thefe laft nights, I have in vain attempted, their light being fo feeble as not to fuffer the leaft illumination, and that of the planet not ftrong enough to render the fmall filk-worm's threads of my delicate micrometers vifible.".*

Herschel's surprise at the inclination of these satellites' orbits was understandable - our Moon orbits the Earth within around 5º of the ecliptic (our planet's orbital plane). Similarly, the regular satellites of the other planets that were known in Herschel's day (4 around Jupiter, 5 around Saturn) all orbit a degree or less from the equatorial plane of their host planets, which, in turn, lie close to the plane of the ecliptic. The fact that this new planet's moons instead orbit far from that plane was therefore understandably surprising. It turns out, in fact, that the spin-axis of Uranus is tilted almost perpendicular to the plane of its orbit, with an axial tilt of approximately 98º. The planet's regular satellites and ring system orbit about the planet's equator, and so the entire system is tilted over dramatically compared to the other planetary systems. In order to explain this, it has been suggested that the planet was struck, during or after its formation, by an Earth-sized protoplanet (e.g. Safronov, 1966; Cameron, 1975; Slattery et al., 1992;

Brunini, 1995; Parisi & Brunini, 1997; Kegerreis et al., 2018). Such a collision would tip the rotation axis of the planet over from that with which it initially formed. The planet's regular satellites, which orbit very close to the plane of its equator, presumably formed after the giant collision that tilted the planet, either from material ejected from the impactor, or from the planet itself as a result of the collision (Mousis, 2004). Although several alternative explanations for the planet's tilt have been suggested over the years (such as the dynamical influence of an unusually massive Uranian satellite, which would have since escaped the system; Boué & Laskar, 2010, or the tidal capture and disruption of a massive object; Singer, 1975), the giant impact hypothesis seems a remarkably good fit to the known data.

**4.1.2 GIANT IMPACTS, THE ASTEROID BELT, AND THE TNOS**
The evidence of the violent collisional history of our Solar system is not limited to the planets themselves. Obviously, the larger the object, the larger the collision must be to significantly disrupt it. It is no surprise, then, that when we look at the small body populations of the Solar system, examples abound of objects that at some point suffered giant collisions. Indeed, it is likely that those known are only the tip of the iceberg. In this subsection, we present just a few exemplar cases that reveal the violent history of the Solar system's small body reservoirs.

The asteroid belt is a collisionally-ground reservoir, which is generally thought to contain only a small fraction of the mass it had at the birth of the Solar system (as discussed by e.g. Weidenschilling, 1977; O'Brien et al., 2007; Morbidelli et al., 2009; Bottke et al., 2015)[42]. Within the belt, tens of asteroid families have been identified, thought to be the results of catastrophic collisions between asteroids (e.g. Marzari et al. 1998, Milani et al., 2014; Nesvorny et al, 2015; Cellino et al., 2019). Unlike the giant impacts that are thought to have affected the planets, the collisional grinding of the asteroid belt is an ongoing process, with new families having been created at all points in the Solar system's history (e.g. the Karin family, a sub family of the older Koronis family, which is thought to have formed just ~5.8 Myr ago; Nesvorný et al., 2002). Over time, the members of these families diffuse through the belt, to a greater or lesser extent, depending on a variety of dynamical and non-gravitational effects, to the extent that some of the families that must have been created by the oldest collisions can no longer be detected (e.g. Hanuš et al., 2019). Figure 12 shows the distribution of known collisional families across the asteroid belt.

The X-type asteroids in the inner main belt have recently been shown to comprise, among their number, at least two such widely dispersed collisional families (Delbo et al, 2019). The eldest of these, whose largest member may be the asteroid (689) Zita, has a wide distribution of orbital elements, with members spread across the entirety of the inner belt. That family was only identified after the authors considered the correlation between the size of a given asteroid and the rate at which it will migrate through the belt under the influence on non-gravitational forces (e.g. Vokruihlický et al, 2006, Milani et al, 2014, Spoto et al., 2015, Bolin et al., 2017). The wide distribution of the members of the Zita family indicates that that family may even have formed prior to the early planetary instabilities invoked by the Grand Tack and Nice Models of planetary migration (Delbo et al., 2017). The second family identified in the inner belt by Delbo et al (2019) is the Athor family, after asteroid (161) Athor, has an estimated age of approximately 3 Gyr age, and whilst still widely dispersed, clusters more closely in element space than the Zita family, potentially as a result of its formation coming after periods of instability in the outer Solar system.

Some of the fragments ejected in these asteroid-asteroid collisions are ejected onto orbits that evolve, under the influence of both non-gravitational forces (see section 4.9) and distant gravitational perturbations from the planets, to regions of instability in the belt, and are then injected into the inner Solar system, to become the near-Earth asteroids (e.g. Morbidelli, 1999; Morbidelli et al., 2002; Morbidelli & Vokrouhlický, 2003; Bottke et al., 2015; Granvik et al., 2017). Such material forms a significant fraction of the ongoing meteoritic flux at the Earth, and certain classes of meteorites have been strongly linked to specific asteroid families, or even individual asteroids (e.g. the basaltic achondrite meteorites and asteroid 4 Vesta, Binzel & Xu, 1993; Asphaug 1997; Drake, 2001; Marchi et al., 2012).

---

[42] Though it has been suggested that the belt originally formed with much less mass than classical models of Solar system formation would suggest (e.g. Hayashi, 1981), or may even initially have been completely empty (e.g. Raymond & Izidoro, 2017).

It has been proposed (Bottke et al., 2007) that the fragmentation of the parent body of the Baptistina family, around 160 Myr ago, was the source of the object that collided with the Earth 65 Myr ago, creating the Chicxulub impact crater (e.g Hildebrand et al., 1991; Morgan et al., 1997; Dressler et al., 2003), and potentially causing the Cretaceous/Tertiary (K/T) mass extinction (e.g. Alvarez et al., 1980; Smit, 1999; Schulte et al., 2010), in which the dinosaurs went extinct. It should be noted, however, that the direct link between the extinction and the impact is still the subject of some debate (e.g. Reddy et al., 2009), with more recent work supporting the idea that the extinction was in fact the result of a double-whammy - the combined effects of an enormous ongoing volcanic extrusion of flood basalts, forming the Deccan Traps in India, and the impact that created the Chicxulub impact crater (e.g. Peterson, Dutton & Luhmann, 2016). It is even possible that the scale of the eruption that formed the Deccan Traps was influenced by the Chicxulub impact, with Renne et al. (2015) finding that as much as 70% of the total extruded material was erupted within 50,000 years of the impact.

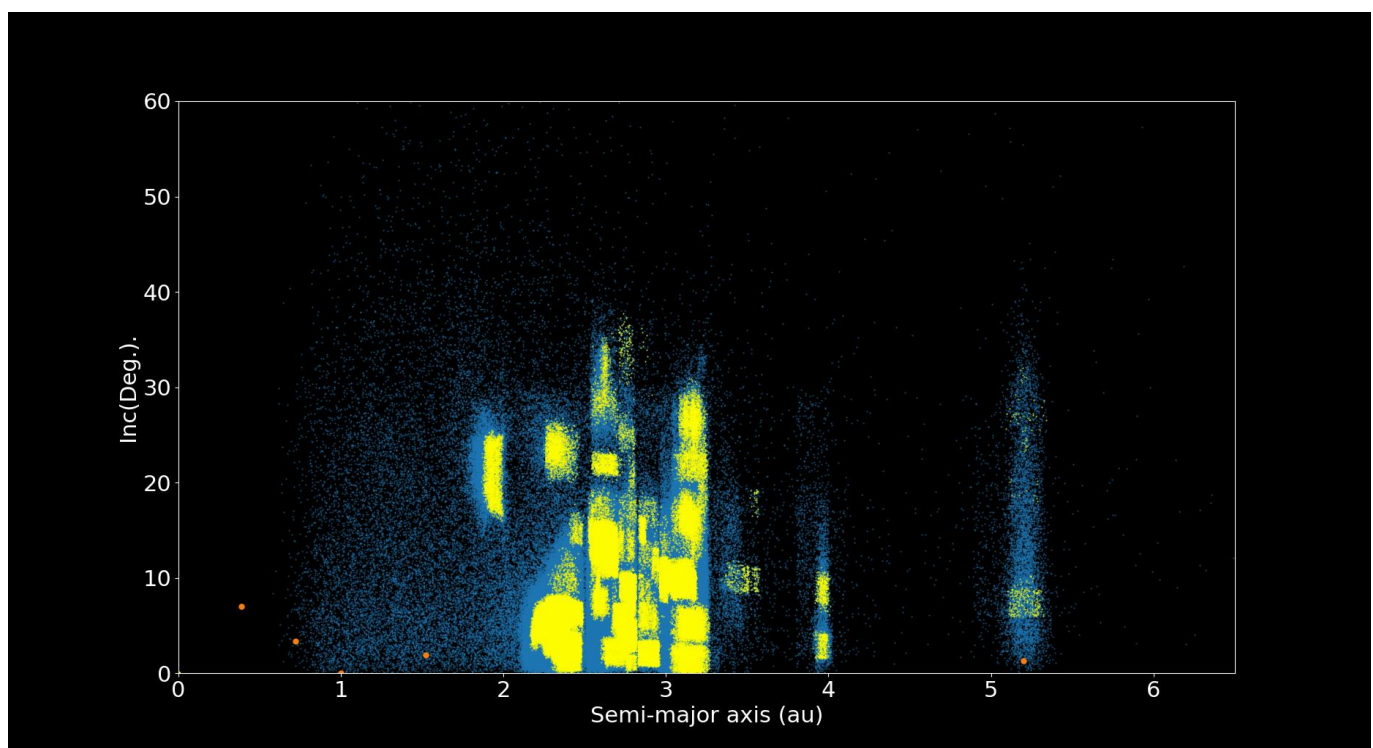

*Figure 12: The distribution of collisional families in the inner Solar system. The blue dots show the orbital elements of all known asteroidal bodies in the inner Solar system. Overlaid on that distribution, in yellow, are shown all asteroids identified as belonging to one or other of the system's myriad collisional families - with such membership determined based on the work of Nesvorný et al. (2015). The five orange dots show the orbital elements of the five innermost planets, Mercury, Venus, Earth, Mars and Jupiter, for scale.*

Even though the known trans-Neptunian population is far smaller than that of the asteroid belt (an artefact of its greater distance from the Sun - we remind the reader that the estimated population of the trans-Neptunian region is far greater than that of the asteroid belt; e.g. Durda & Stern, 2000; Gomes et al., 2008), it is already known to contain a number of examples of giant collisional processes. The dwarf planet Pluto is accompanied by five satellites (Charon, Hydra, Nyx, Kerberos and Styx; Showalter & Hamilton, 2015), the orbits of which are best explained by a giant-impact formation mechanism like that invoked for the Earth-Moon system (e.g. McKinnon 1989; Canup 2005; Stern et al, 2006; Canup, 2011) - a theory supported by the more recent observations performed by the New Horizons mission, which visited the dwarf planet in 2015 (e.g. McKinnon et al., 2017; Sekine et al., 2017)[43].

Eris, another of the trans-Neptunian region's dwarf planets, has a satellite, Dysnomia, which may also have been produced in this fashion (Greenberg & Barnes, 2008), while a number of the smaller trans-Neptunian objects have similar satellite systems which might have formed in this way (e.g. the Orcus-Vanth system, Brown et al., 2010). However, recent observations of both the Orcus-Vanth and Eris-Dysnomia systems have cast some doubt on the origin of their satellites (Brown & Butler, 2018). Those observations reveal that the two satellites in question (Vanth and Dysnomia) have low albedos, similar to those observed for other objects of a similar size in the Edgeworth-Kuiper belt - rather than the high albedos that might be expected had those moons formed through accretion from an icy disc around their parent bodies resulting from a giant collision. As Brown & Butler (2018) note, further study is needed to definitely determine the origin of the satellites orbiting these two large trans-Neptunian objects.

The dwarf planet Haumea has two satellites, thought to have formed in a giant impact (e.g. Brown et al., 2006), and its own ring system (Ortiz et al., 2017; Winter, Borderes-Motta & Ribeiro, 2019). In addition, it is the largest object in the first detected collisional family in the trans-Neptunian region (e.g. Brown et al., 2007; Ragozzine & Brown 2007; Schlichting & Sari, 2009; Leinhardt, Marcus & Stewart, 2010; Lykawka et al., 2012; Volk & Malhotra, 2012; Villenius et al., 2018; Proudfoot & Ragozzine, 2019). It is almost certain that, over the coming years, more collisional families will be found in the trans-Neptunian region (Marcus et al., 2011). However, since the orbital velocities of objects in that region are far smaller than the orbital velocities of the asteroids, the ejecta from any given collision will disperse far more widely through the region (since the ejecta velocities will be a far greater fraction of the orbital velocity)[44]. In addition, as Marcus et al. (2011) note, the slow collisional velocities may well be "*insufficient to disrupt the largest (and most visible) bodies*" – with collisions on such objects likely resulting in either accretion, or a hit-and-run scenario for large impact angles (as described in Agnor & Asphaug, 2004). This, coupled with the fact that TNOs are significantly fainter than main belt asteroids, means that such detections will be significantly more difficult – and it is actually quite remarkable that the first TNO family should have been found so soon after the discovery of the first members of the population.

### 4.1.3 GIANT IMPACTS AND PLANETARY SATELLITES

Even within the satellite systems of the Solar system, there is ample evidence for the key role played by giant impacts in shaping the evolution of the system. A number of the satellites bear the scars of impacts that were close to totally disrupting the body - the largest crater on Saturn's moon, Mimas, for example, Herschel, is around a third of the diameter of the moon, and the collision which formed it left fractures that can be found all around the satellite. Were the impactor slightly more massive, or travelling a little more quickly, it could well have totally disrupted the moon (e.g. Beatty et al., 1981).

Staying within the Saturnian system, giant impacts have been proposed as the source of the unusual ridge which encircles another of Saturn's moons, Iapetus. The ridge, which runs along the satellite's equator, is one of the most unusual looking features of the Solar system, leaving the moon resembling a cricket ball with a raised seam. In this case, however, the seam is over 10km high, and some 20km wide, and stretches almost completely around the circumference of the satellite. Over the years, a number of explanations have been suggested for the ridge, ranging from it being a relic of the satellite having once been significantly oblate (e.g. Castillo-Rogez et al., 2007), to the collapse of a circum-Iapetan ring (Ip, 2006; Stickle & Roberts, 2018), or even the contraction of the moon's interior as it cooled following its formation (e.g. Sandwell & Schubert, 2010). Dombard et al. (2010, 2012) have proposed that the feature could be the result of a giant impact upon the satellite. They suggest that a giant impact upon the satellite led to the formation of a moon, in orbit around Iapetus, in much the same way that models suggest our own Moon, and Pluto's Charon (among others) were formed. Such an object would accrete on an orbit close to the equatorial plane of the satellite, and then undergo tidal decay, with its orbit shrinking until it passed within the Roche limit of Iapetus, at which point the tides raised upon it by the large satellite would tear it into fragments, which would then slowly de-orbit to the surface of Iapetus at subsonic speeds. This would allow sufficient material to fall to create the ridge whilst simultaneously explaining the fact the ridge lies on the satellite's equator. More recently, Leleu, Jutzi & Rubin (2018) suggest that the merger of two similarly-sized objects could explain not only Iapetus' unusual equatorial ridge and oblate shape, but that similar processes could well be behind the peculiar shapes of several of Saturn's other innermost satellites - Pan, Atlas and Prometheus.

---

[43] Interestingly, although data from the *New Horizons* mission supports the collisional formation mechanism for the Pluto system, it has also revealed that Pluto and Charon have fewer small ($\lesssim$ 13 km diameter) craters than one would expect based on current estimates of the population of small trans-Neptunian objects, a result that cannot be explained solely on the basis of geological resurfacing (Singer et al., 2019). This, in turn, has been taken to suggest that there may be fewer such objects than previously thought - a result which might also explain the relatively pristine surface of 2014 MU69 (Stern et al., 2019), which was visited by *New Horizons* on January 1st, 2019

[44] This might seem somewhat counter intuitive, since the orbital velocities in the Edgeworth-Kuiper belt are much slower than those in the Asteroid belt, which should result in lower mean collision velocities. However, as Marcus et al. (2011) note in their introduction, "*collisions typically result in a collection of objects with a velocity dispersion comparable to the parent body's escape velocity (e.g., Durda et al., 2007) ... the comparable size of the typical ejection velocity and the background velocity dispersion causes Kuiper Belt families to spread across huge swaths of the trans-Neptunian region, significantly diluting the dynamical clumping of families that is so prominent in the asteroid belt.*".

A giant collision leading to the disruption of a massive, close-in Saturnian satellite has also been proposed as the origin of the planet's splendid ring system (e.g. Ip, 1988). More recently, Charnoz et al. (2009) take this idea one step further, suggesting that Saturn's rings were formed during the putative Late Heavy Bombardment (which we discuss in more detail in section 4.2.1). In that work, they suggest the bombardment would also have led to significant disruption of the closer in Saturnian satellites, with all smaller than Mimas being destroyed, and then re-accreting after the bombardment finished - which in turn suggests that, if that theory is to be believed, the giant Herschel crater could be the relic of an impact during that epoch that almost managed to disrupt Mimas. Instead of invoking a giant collision, Canup (2010) suggests that the rings could instead have been formed by the tidal stripping of an inward-spiralling Titan-sized satellite, during the final stages of the planet's formation. If such a satellite were to migrate inwards through a circum-planetary disc surrounding Saturn, Canup found that Saturn could strip the mantle and crust from such an object, whilst leaving its core relatively intact, fated to spiral inwards and be devoured by the planet. Given the size of such a satellite, it would be reasonable to expect it to be fully differentiated, and hence the material stripped away in this manner would be a good match to the observed composition of Saturn's rings - which are thought to be between 90 and 95% water ice. However, recent results based on analysis of data obtained during the final phase of the Cassini mission have cast doubt on both these proposed origins for the Saturnian ring system. Rather than being an ancient feature, dating back to the Solar system's youth, that data instead suggests that the rings are far younger than previously thought - having formed in just the last 10 - 100 million years (Iess et al., 2019). Indeed, it seems likely that the rings are actually a transient feature - with observations suggesting that they may well decay within the next 300 million years (e.g. O'Donoghue et al., 2019). Such a recent origin is still entirely compatible with the idea that the rings were created in a catastrophic collision - and the transient nature of the Saturnian ring system seems in keeping with the fact that the other three giant planets, Jupiter, Uranus and Neptune, each also have ring systems - albeit ones far less spectacular than those that gird Saturn.

In orbit around Uranus, the small satellite Miranda proved the highlight of the Voyager 2 flypast for many astronomers. Unlike the other Uranian satellites imaged, which were heavily cratered, pictures of Miranda revealed an incredibly unusual terrain - as can be seen in Figure 13.

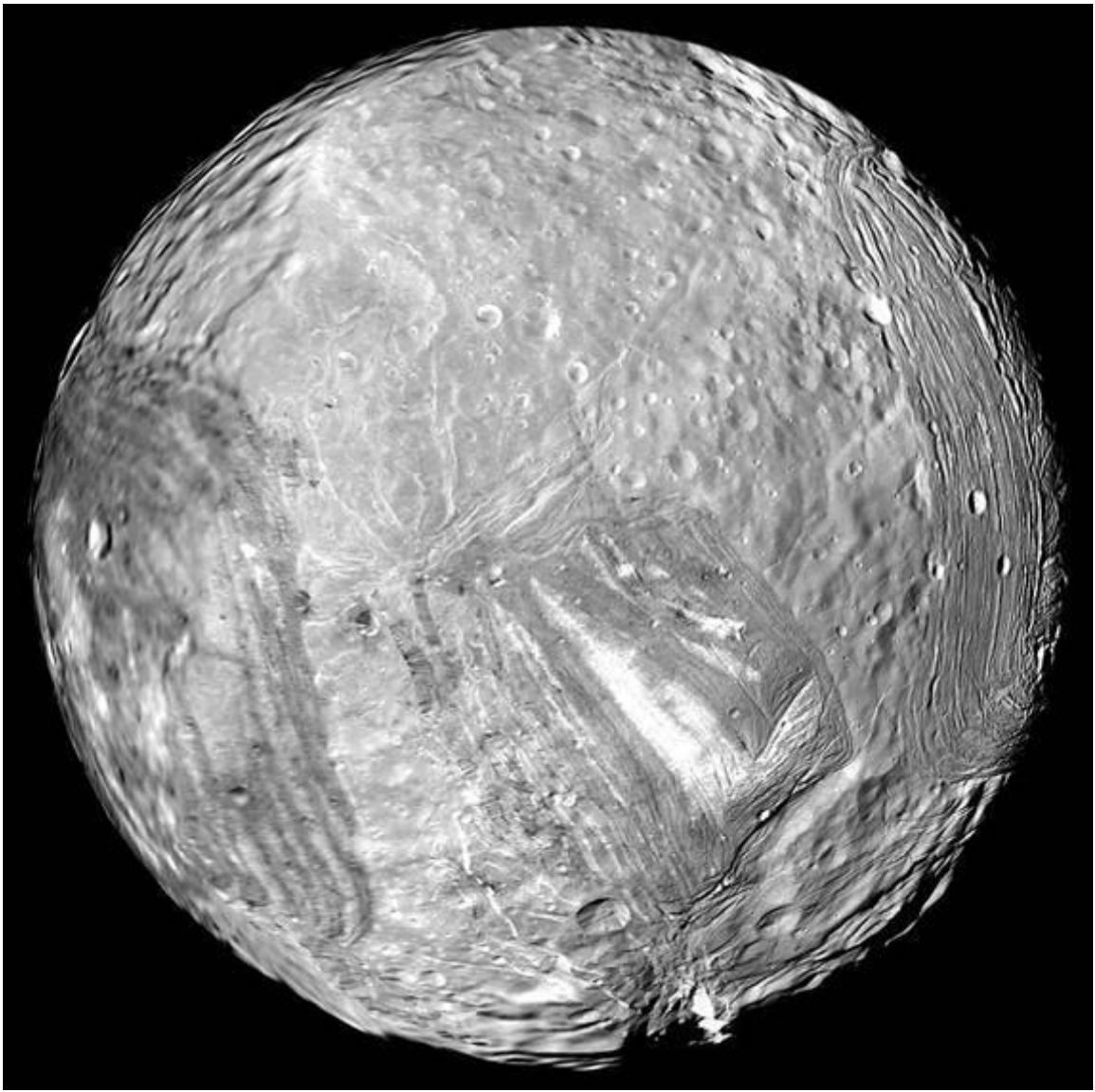

*Figure 13: Uranus' icy moon Miranda, as imaged by the Voyager 2 spacecraft on 24 January 1986. Image courtesy of NASA/JPL-Caltech; image is public domain. At the bottom of the image can be seen Verona Rupes, vertical cliffs that may stand 20 km tall. Just below the centre is the Inverness Corona - a feature often called the Chevron, whilst the two other unusual coronae terrains, Arden and Elsinore, are located on the left and right-hand sides, respectively.*

Some of the features looked as though large "blocks" had been dropped into the planet, with adjacent regions appearing greatly distinct from one another. Some of the surface appeared "normal", with the usual craters, while other regions exhibited stretch marks, and enormous cracks stretched across other parts of the satellite. One particularly fascinating feature was the region known as *Verona Rupes*, a giant vertical cliff face that may be as much as twenty kilometres in height -- the tallest cliff in the Solar system. In order to explain these features, it has been suggested that Miranda was formed through the re-accretion of debris left over when an earlier moon in its orbit was collisionally shattered (e.g. McKinnon, 1988; Marzari et al., 1998; Movshovitz et al., 2015). Although this theory remains the most favoured in discussions of Miranda's formation, some authors (e.g. Marcialis & Greenberg, 1987; Janes & Melosh, 1988; Pappalardo et al., 1997; Hammond & Barr, 2014) have suggested that other mechanisms (such as convection-driven resurfacing) could create the observed features, removing the need for their cataclysmic origin.

The families of irregular satellites in orbit around the outer planets are another example of the effects of giant collisions. In much the same way as the asteroid families mentioned in section 4.1.2, distinct families of irregular satellites can be found around the giant planets, each of which is thought to be linked to the collisional disruption of a larger parent object, followed by gentle orbital dispersion (e.g. Sheppard & Jewitt, 2003; Nesvorný et al., 2004), a finding supported by recent cladistical analysis of the satellite systems (Holt et al, 2018). Despite this collisional erosion of the irregular satellite population, it has been noted that the number of irregular satellites larger than a given size appears to be roughly constant between the outer planets (e.g. Jewitt & Sheppard, 2005). Jewitt & Haghighipour (2007) present a detailed review of our understanding of these unusual objects, and describe the various models that have been proposed for their origin in some detail. Future observations of these satellites, and the discovery of further members of the population, will not only help to constrain the mechanism by which they were captured, but will also help to constrain the exact processes involved in the formation of their host planets themselves.

Neptune's gigantic irregular satellite Triton is an oddity among the planetary satellites. The seventh largest satellite in the Solar system (with a diameter of 2706 km), it is by far the largest and most massive irregular satellite known, with a diameter an order of magnitude larger than Saturn's Phoebe, the next largest irregular. It also orbits remarkably close to its host planet, compared to the other irregulars, with an orbital radius of $3.5 \times 10^5$ km placing it slightly closer to Neptune than our Moon is to the Earth (by contrast, the next closest irregular satellite known is Uranus' Francisco, with an orbital radius of just over $4 \times 10^6$ km). Triton's unusual properties mark it as having a markedly different origin to the other irregular satellites, and many theories have been suggested over the years to explain its current orbit. It is noteworthy that, aside from Triton, Neptune lacks any sizable satellites – of the thirteen moons in the Neptunian system, Triton contains over 99% of the total mass. The second largest of Neptune's moons, Proteus, has dimensions of just 424x390x396 km (Stooke, 1994), and the total mass of the regular satellite system of Neptune is far, far smaller than that of any of the other giant planets. In addition, the orbit of the irregular satellite exterior to Triton, Nereid, is remarkably eccentric.

All this evidence points to an unusual and possibly catastrophic origin for Triton (e.g. Banfield & Murray, 1992). It is widely acknowledged that Triton is most likely a captured object, rather than one which formed in its current orbit, but the mechanism by which it was captured is still under some debate. Early theories (e.g. Lyttleton, 1936) attempted to link the origin of Triton to that of Pluto. This was based on the observational similarities of the two objects, the fact that Pluto's orbit crosses that of Neptune, and on estimates that both objects were of order $1/10^{th}$ the mass of the Earth. After a lengthy discussion of hypothetical encounters between Neptune, Pluto and Triton, Lyttleton (1936) highlighted a scenario by which such an encounter could capture Pluto to a prograde orbit around Neptune while simultaneously causing the direction of Triton's orbit to reverse, and also become prograde. Noting that dynamics is a time-reversible process, Lyttleton then proposed that Triton and Pluto could initially have been large prograde satellites of the giant planet. At some point, such an encounter, operating in reverse, could have occurred, leading to the ejection of Pluto to its current orbit, and could have been the origin of Triton's current retrograde orbit. The idea that Pluto was once a Neptunian satellite has played an important role in many theories of the origin of Triton – from the proposed disruption of the Neptunian satellite system by an encounter with a massive trans-Neptunian planet (Harrington & van Flandern, 1979), to suggestions that Triton was an unbound interloper that was captured by Neptune through a close encounter with proto-Pluto (suggesting also that the approach was so close that the proto-Pluto was disrupted in the process, leading to the formation of its largest moon, Charon; Farinella et al., 1979). It should be noted, however, that none of these theories are currently considered as serious answers to the question of either Triton or Pluto's origin – but they still serve to highlight how our knowledge of Pluto and Triton has changed over the past century, and how those changes have informed our theories on their origins.

As our knowledge of both Triton and Pluto increased, it was realised that their masses were so low that such scenarios were implausible, unless the two objects had been involved in an actual physical collision (e.g. McKinnon, 1984). Once this was realised, studies instead typically concentrated on mechanisms through which Triton could be captured alone – although some authors (e.g. Goldreich et al., 1989; Woolfson, 1999, 2013) continue to consider the statistically improbable idea that Triton was captured as the result of such a giant collision. The model currently accepted by the majority of researchers, however, considers that Triton was likely a binary dwarf planet (like Pluto, or Eris) that approached Neptune sufficiently closely that it was captured by the planet as a result of the disruption of the binary pair – with the other member of the system continuing on an unbound orbit (e.g. Agnor & Hamilton, 2006). Rufu & Canup (2017) considered the effect that such a capture event would have on a pre-existing satellite system, and found that the interactions between a newly-captured Triton and a satellite system comparable to that seen orbiting Uranus could readily result in a system like that seen around the modern-day Neptune. In addition, they note that the interactions involved in such a scenario offer a mechanism by which the orbital eccentricity of Triton could be reduced on a timescale short enough to prevent the ejection of satellites like Nereid, which might otherwise be lost from the system as a result of scattering by Triton should it's orbit solely be circularised by tidal interactions with Neptune.

**4.2 THE IMPACT RATE OF SMALLER BODIES IN THE SOLAR SYSTEM**
At sizes smaller than those at which the impactor would destroy the target body, impacts become ever more frequent. Whilst the epoch of giant collisions, at least from the point of view of the planets, passed in the early days of the Solar system, smaller impacts have continued right up to the current day. Although such impacts are no doubt of interest when one considers the potential habitability of Earth-like planets (e.g. Horner & Jones, 2008a, 2010c, 2011, 2012), the effects of such impacts, taken individually, would be so minor as to be undetectable in other planetary systems. Indeed, the first confirmed impacts to be observed on any Solar system body only came in the last thirty years[45] – the Shoemaker-Levy 9 impacts on Jupiter in 1994 (e.g. Hammel et al., 1995, Ausphag & Benz, 1996), the two further impacts upon that planet observed in 2009 and 2010 (e.g. Sánchez-Lavega et al., 2010; Orton et al., 2011), and the impact of meteoroids observed on the surface of the Moon (e.g. Bellot Rubio, Ortiz & Sada, 2000; Ortiz et al., 2002; Madiedo et al., 2014, 2019).

Despite this, the study of these impacts in our Solar system has provided a wealth of information on the processes by which the objects within formed and evolved. Studies of the chemistry and origin of meteorites reveal the timescales on which asteroids accreted and differentiated (e.g. Whitby et al., 2000; Kleine et al., 2002; 2005), as well as providing strong evidence that our Solar system was seeded with short-lived radionuclides by a nearby supernova at some point during its formation (e.g. Lattimer et al., 1978; Hoppe et al., 1996; Yin et al., 2002; Mostefaoui et al., 2005). In addition to these results, which clearly provide valuable data to be incorporated in models of planetary formation, the two main areas of small-impact research that probably have the most impact on our studies of exoplanets are the proposed Late Heavy Bombardment of the terrestrial planets, and the suggested role of giant planets in shielding terrestrial planets from potentially catastrophic fluxes of small bodies.

**4.2.1 THE LATE HEAVY BOMBARDMENT**
The theory of the Late Heavy Bombardment suggests that, approximately 700 million years after the formation of the Solar system, there was a dramatic increase in the rate at which the terrestrial planets were bombarded by asteroids and comets. Evidence for the theory first came from the analysis of impact melt dates in Apollo Lunar samples (Tera et al., 1974) that showed a predominance of dates in the range 3.8 – 4.0 Gyr which were interpreted as a "Terminal Lunar Cataclysm". This cataclysm hypothesis was, however, criticized as "a misconception" by Hartmann (1975). The controversy has continued ever since.

Supporters of the cataclysm hypothesis (e.g. Ryder, 1990, 2002; Stoffler & Ryder, 2001) derive dates for the best-studied major Lunar impact basins in the very tight range from 3.85 Gyr (for Imbrium) to 3.92 Gyr (for Nectaris). Since relative dating of Lunar basins based on stratigraphic analysis (Wilhelms, 1987) places 11 of the major basins at ages between Nectaris and Imbrium, this implies a very short period of intense bombardment. They also argue (Ryder, 2002) that extrapolating the impact rates back to the Moon's formation, as a steady decline, would imply accretion of more than a Lunar mass of material.

---

[45] It should be noted that, in the 1970s, there were suggestions that an impact was observed on the Moon in 1178, forming the 20 km diameter crater Giordano Bruno (see e.g. Hartung, 1976; Calame & Mulholland, 1978). However, that idea has since fallen into disfavour, with recent data suggesting the crater is far older, having most likely formed between one and ten million years ago - e.g. Morota et al., 2009; Basilevsky & Head, 2012). We discuss this in more detail in section 3.8.1.

However, an alternative view is that the impact history of the Moon is consistent with a steady decline of impact rates with time. Advocates of this view (e.g. Neukum et al., 2001; Hartmann, 2003) argue that the Lunar surface record obtained from the Apollo and Luna missions (limited to the equatorial near-side) is strongly biased by major events such as the Imbrium impact, and that the record of older impacts has been largely erased by subsequent events.

To remove the selection effects that result from the limited coverage of the Lunar surface, Cohen et al. (2000) analysed Lunar meteorites. They found a range of impact melt ages with no sharp impact spike, but with none earlier than 3.92 Gyr, which was argued as supporting the cataclysm hypothesis. Support for the cataclysm hypothesis has also been reported from studies of other Solar system bodies. Kring & Cohen (2002) used analysis of meteorites to determine that main belt asteroids were heavily cratered ~3.9 Gyr ago. The only ancient Martian meteorite (ALH 84001) shows a shock event at 3.92 Gyr (Turner et al., 1997) and a thermal event at ~3.9 Gyr has been recorded in Hadean zircons on Earth (Trail et al., 2007). While these results are consistent with a cataclysmic bombardment throughout the inner Solar system, the evidence for a strong spike in impacts rests on the Lunar record.

A key to the cataclysm interpretation is the ages for the older Lunar basins, Serenitatis and Nectaris, which are given as 3.89 Gyr and 3.92 Gyr by Stoffler & Ryder (2001). However, both these ages are controversial. Spudis et al. (2011) has concluded that the Apollo 17 impact melts used to date Serenitatis are not, in fact, from Serentatits which means the basin could in fact be much older. Norman et al. (2010) argue that the Apollo 16 Descartes terrain often considered to be ejecta from Nectaris are in fact ejecta from the Imbrium basin, allowing a much older age for Nectaris. Older ages for Nectaris and Serenitatis would weaken the case for a Lunar cataclysm. The controversy may not be resolved until future Lunar missions are carried out that would enable better age determinations for the older basins, e.g. the proposed MoonRise sample return mission to the South Pole-Aitken basin (Jolliff et al., 2012).

So while it is well established that there was a high impact rate on the Moon at around 3.8-3.9 Gyr, it remains somewhat uncertain as to whether this represents a cataclysmic spike, the tail of a steady decline or some more complex history, at least as far as the Lunar sample record is concerned. Bottke et al. (2007) have analysed the problem from the point of view of Solar system dynamics. They argue that the occurrence of major impact events as late as Imbrium and Orientale (the last major Lunar basin forming impacts at ~3.8 Gyr) is inconsistent with being the tail of a steadily declining population of objects left over from the formation of the Solar system. This is true whatever is assumed about the initial population of objects, since with a large population of objects, collisions between the objects themselves depletes the population of the larger impactors.

Building on that work, the excellent review by Bottke & Norman (2017) describes the current state of the art of research into the true nature of the Late Heavy Bombardment, and the overall decay in the scale and frequency of impacts across the inner Solar system during the first billion years of its life. They find that the evidence for a '*strong version of the Terminal Cataclysm hypothesis*' is now far weaker than it was in the past, but note that both the Imbrium and Orientale basins were formed during the short time interval spanned by that hypothesis. Based on the latest data, they propose a hybrid, two-phase model for the late bombardment of the inner Solar system. First, in the very early stages of planet formation, there was a very intense phase of bombardment, lasting until ~4.4 Gyr before the present day. That phase was driven by the cleanup of left-over planetesimals from the formation of the terrestrial worlds. A second, prolonged phase of intense bombardment followed somewhat later, beginning between 4.2 and 4.0 Gyr before the present day - a phase that included the impacts that created the Imbrium and Orientale basins, and might well have extended for well over a billion years (e.g. Bottke et al., 2012). Rather than being the result of the clean-up phase of terrestrial planet formation, that secondary period of intense bombardment was the result of the migration, and potential dynamical instability, of the Solar system's giant planets - as we will discuss in more detail in section 4.3.2. In the conclusion of their work, Bottke and Norman suggest several regions for future research that should help to shed further light on the nature of the Late Heavy Bombardment, and it seems likely that the coming decade will yield fresh insights into the processes that sculpted the face of Solar system as we see it today.

**4.2.2 DO GIANT PLANETS OFFER SHIELDING TO TERRESTRIAL WORLDS?**
As described earlier, there are three main types of potentially hazardous objects in our own Solar system, each of which finds its origin in one of the three main reservoirs of dynamically stable objects. The near-Earth asteroids originate within the Asteroid belt (e.g. Binzel et al., 1992; Bottke et al., 2000; Morbidelli et al., 2002; Dunn et al., 2013), between the orbits of Mars and Jupiter, and are injected into the inner Solar system, after being created in collisions between larger bodies, primarily by distant perturbations from the planets Jupiter and Saturn. The short-period, or Jupiter-family, comets are continually replenished from the Centaur population (e.g. Horner et al., 2003, 2004a, 2004b; Tiscareno & Malhotra, 2003; Volk & Malhotra, 2008; Grazier, Castillo-Rogez & Horner, 2018; Grazier, Horner & Castillo-Rogez, 2019; Peixinho et al., 2020), which are themselves sourced from objects in the trans-Neptunian region (e.g. Holman & Wisdom, 1993; Duncan & Levison, 1997; Emel'yanenko, Asher & Bailey, 2005; Volk & Malhotra, 2008), though a significant contribution to that population may originate amongst the Neptunian and Jovian Trojan clouds (e.g. Marzari, Farinella & Vanzani, 1995; Horner & Evans, 2006; Horner & Lykawka, 2010a; Horner, Lykawka & Müller, 2012). The third population of potential impactors, the Oort cloud comets, are sourced from the Öpik-Oort cloud (e.g. Öpik, 1932; Oort, 1950), and are typically thrown into the inner Solar system by the gravitational influence of passing stars and the galactic tide (e.g. Biermann, Huebner & Lust, 1983; Heisler, Tremaine & Alcock, 1987; Matese & Lissauer, 2004; Feng & Bailer-Jones, 2015; Vokrouhlický, Nesvorný & Dones, 2019), although it has been suggested that some fraction might well be sent inwards as a result of perturbations by a Jovian-mass Solar companion (e.g. Matese, Whitman & Whitmire, 1999; Horner & Evans, 2002; Matese & Whitmire, 2011), as we will discuss in more detail in section 4.6.2.

In the case of the near-Earth asteroids and short-period comets, despite the presence of relatively populous reservoirs of parent objects (the asteroid belt, and trans-Neptunian disc, respectively), the number of objects being moved from orbits in those regions to orbits in the inner Solar system (where they can imperil the Earth) would be minimal were it not for the perturbative influence of the giant planets. For this reason, it is immediately apparent that the presence of giant planets can play a significant role in determining the frequency with which debris rains down upon the terrestrial planets. In other words – the presence of a massive debris disc around a given star should not be taken, *per se*, as evidence that the frequency of collisions on a planet in a different part of that planetary system would be prohibitively high for life to develop. If no giant planets were present in that system, then no mechanism would exist to effectively transfer material from that disc to a potentially habitable planet elsewhere in the system, and so one might instead expect the impact regime on such a world to be relatively peaceful, despite the presence of a massive disc.

Despite the clarity of such logic, it has historically been widely taught that the planet Jupiter acts as a giant shield, significantly lowering the rate at which asteroids and comets collide with the Earth. Without Jupiter, the argument goes, significant impacts would be sufficiently frequent on the Earth that the development of life would be stymied, or even prevented absolutely (e.g. Ward & Brownlee, 2000). The origin of this idea is unknown, but it likely has its roots around the time that craters on the surface of the Earth were finally acknowledged as being the result of impacts – an idea that was not fully accepted until the 1960s (e.g. Chao, Shoemaker & Madsen, 1960; Shoemaker, 1960; Shoemaker & Chao, 1961; Bjork, 1961).

Given that weathering removes all but the largest craters on relatively short geological timescales, it was obvious that most craters could not be ancient features, but had instead been created by impacts in the relatively recent past. At that time, very few short-period comets, and even fewer near-Earth asteroids were known. The obvious culprits, therefore, were the long-period comets, objects sourced from the Oort cloud. Such comets are only tenuously bound to the Sun, with their aphelia located at enormous distances (typically tens of thousands of au), and as such only need to experience a very small perturbation in order to become unbound, and be ejected from the Solar system forever. In most cases, the planet that ejects comets in this way is Jupiter, and so therefore it seems an obvious conclusion that, if the planet were not there, far more of these objects would survive to pass through the inner Solar system multiple times, and therefore the impact rate at the Earth would be much higher. The idea that Jupiter has shielded us is now firmly entrenched in both the public and scientific consciousness. By extension, it is often argued that giant planets are a necessity in order to shield Earth-like planets from impacts for those planets to be habitable (as in, for example, Greaves, 2006, who suggests that the lack of a Jupiter-type planet in the Tau Ceti system has resulted in a significant amount of debris being present that would threaten any Earth-like planets therein).

Given the pervasiveness of this idea, it is somewhat surprising that, until recently, very little work had been done to examine it in any depth. The first study of any note into the question of planetary shielding was carried out by Wetherill, 1994, who concluded that, in planetary systems containing only "*failed Jupiters*" (i.e. planets which only manage to reach the mass of Uranus or Neptune, rather than that of a Jupiter), the impact flux experienced by an Earth-like planet would be of order a thousand times higher than that experienced in our own Solar system. This would be a direct result of the reduced efficiency with which small objects would be ejected from such a system, allowing them to remain on planet-crossing orbits for a much longer period of time. However, in order for a given object to pose a threat to an Earth-like planet, it must first be emplaced on an orbit that crosses that of the planet. The dynamical lifetime of such objects is relatively short, and so in order for small bodies to pose an on-going threat, their numbers must continually be replenished – a question which was not addressed in that early work.

Laasko et al. (2006) approached the question of planetary shielding from a different angle. Using numerical integration, they examined the effect of the position and mass of a Jovian planet on the rate of ejection of particles that were initially placed on eccentric orbits that crossed the classical habitable zone from the start of the integration. Using our Solar system as a test case for their method, they found the surprising result that Jupiter "*in its current orbit, may provide a minimal amount of protection to the Earth*". Unfortunately, again, their work did not consider the transfer of material from non-threatening to dangerous orbits, but it did provide the first hints that something might be a little amiss with the canonical view of Jupiter's role.

Our current understanding of the various populations of bodies that can threaten the Earth is vastly different to that which was present around the time that the idea the Jupiter is a shield caught on. Rather than the long-period comets being the main source of impact threat to the Earth, it is now believed that the Near-Earth Asteroids constitute the great bulk of the impact risk, with the comets contributing just a few percent of the total risk (e.g. Chapman & Morrison, 1994; Bottke et al., 2002). Whilst it is still true that Jupiter plays a significant role in removing the near-Earth asteroids from threatening orbits, it is also well known that it is the main source of fresh asteroidal material in the inner Solar system (e.g. Morbidelli et al., 2002). In the case of the short-period comets (often known as the Jupiter-family), the situation is equally confused – the great majority of such comets make their way onto Earth-crossing orbits as a result of Jupiter's perturbations, and so any conclusions as to the true level of shielding offered by the giant planet must take into account the efficiency with which it creates new threats, as well as the ease with which it removes them.

In order to study this problem, a series of detailed dynamical studies have been carried out (Horner & Jones 2008a, 2008b, 2009, 2010b, 2012; Horner et al., 2010a) that examine the influence of Jupiter on the three populations of potentially hazardous objects in our own Solar system – namely the near-Earth asteroids, the short-period comets and the long-period comets. The results are startling. In the case of the near-Earth asteroids, it turns out that the Earth actually experiences a greater number of impacts with our Jupiter than it would with no Jupiter at all (see Figure 14, below). However, if Jupiter were just a quarter of its current mass, the impact rate of near-Earth asteroids would be far higher than in the other two cases.

This rise and fall of the impact rate as the mass of Jupiter increases is found to be a direct result of the changing location of the $\nu_6$ secular resonance, which can drive asteroids trapped within into the inner Solar system on astronomically short timescales. In our own Solar system, this resonance, although strong, is located almost at the inner edge of the asteroid belt, where very few objects can fall prey to its perturbative influence. When Jupiter is less massive, however, the resonance is located farther from the Sun, and is also somewhat broader, and can therefore influence a greater fraction of the total mass of the asteroid belt. The "worst case" scenario occurs when Jupiter is around a quarter of its current mass (similar in mass to the planet Saturn), at which point the resonance is both broad and strong, and is located in the middle of the asteroid belt, ideally located to disturb the orbits of the maximum number of asteroids[46].

For the short-period comets, a similar pattern emerges, albeit for different reasons - as can be seen in Figure 14. At low Jupiter masses, the planet is unable to either efficiently eject comets from the Solar system or emplace them on Earth-crossing orbits. At high masses, the planet can very efficiently emplace comets on Earth-crossing orbits, giving them the opportunity to collide with our planet. However, it can also easily remove them from the inner Solar system and eject them, so any given comet will most likely have relatively few opportunities to hit the Earth before being removed again. In both these cases, then, the impact flux experienced by Earth is relatively low. Once again, there is a "worst case" scenario, however. Remarkably, this "worst case" again occurs when the Jupiter-like planet has around a quarter of the mass of our Jupiter. At that mass, the planet can efficiently emplace comets on to Earth-crossing orbits. However, it is still of sufficiently low mass that it is very difficult for it to eject those comets from the Solar system entirely. As such, any comets placed on Earth-crossing orbits by the giant planet at that mass will have plenty of opportunities to collide with the Earth before they are removed, and so the impact flux reaches a peak. Beyond that peak, as the planet's mass grows larger, objects are ever more efficiently ejected, causing the time spent on Earth-crossing orbits to fall, thereby reducing the impact rate.

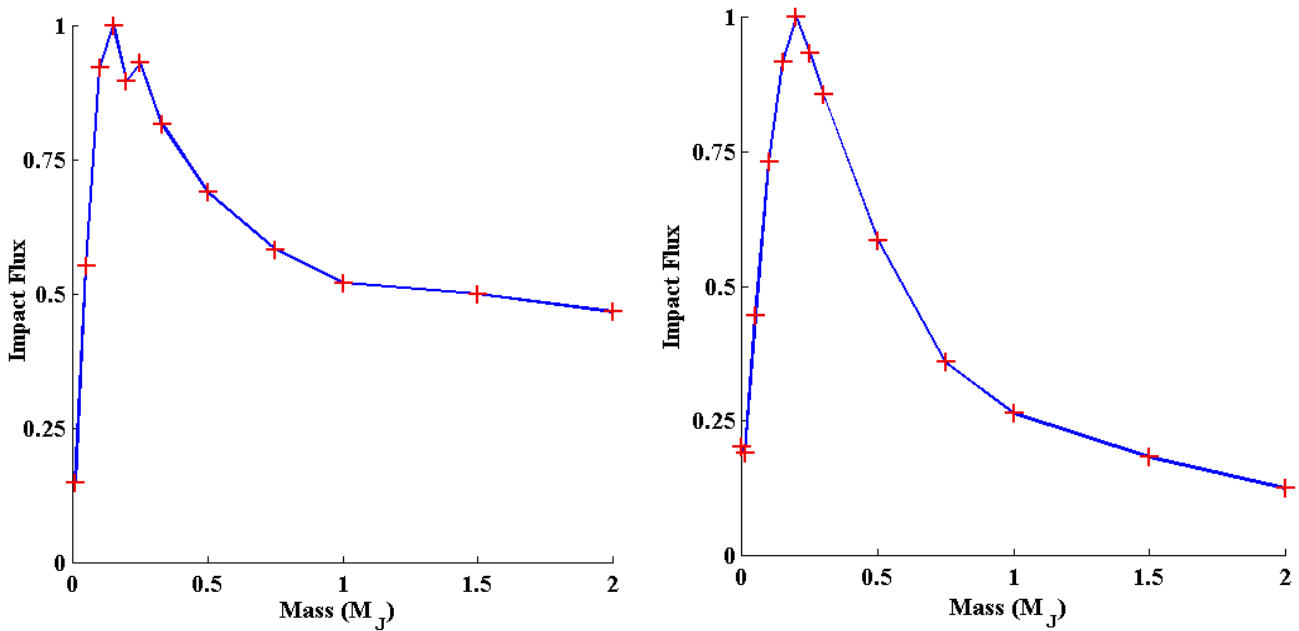

*Figure 14*: *The impact rate experienced by the Earth as a result of near-Earth asteroids (left) and Jupiter-family comets (right), as a function of Jupiter's mass, based on the simulations described in*

---

[46] Were Jupiter smaller, the asteroid belt would likely look very different to how it appears today – the result of the influence of the broad area of instability caused by the $\nu_6$ secular resonance. The results of the simulations described in Horner & Jones (2008) suggested that the end result might well have been a bifurcated belt – with both inner and outer belts separated by a wide gap, in semi-major axis – a scenario very different from the belt we observe in our own Solar system.

*Horner & Jones (2008a, 2009). In both cases, the flux has been normalised to the highest flux experienced by the Earth through the course of those simulations. For both the near-Earth asteroids and Jupiter-family comets, the impact flux Earth would experience without Jupiter is less than a quarter of that we see in the simulations that feature Jupiter as we know it in the Solar system. The impact flux due to both population peaks for a 'Jupiter' somewhat less massive than our Solar system's Saturn, before falling away again - although in both cases, the flux experienced in our Solar system still exceeds that experienced when Jupiter is absent from those simulations.*

In the case of the long-period comets, it seems that the long-held belief holds true – Jupiter does indeed act as a shield. As its mass increases, it becomes ever more efficient at perturbing loosely-bound comets in such a way that their orbits become unbound, and they head out of the system, never to return. A more detailed study of the effect of giant planets on the long-period comet flux (Lewis, Quinn & Kaib, 2013) studied the evolution of cometary orbits from the initial population of the Oort Cloud by the giant planets through to their eventual re-injection and ejection from the inner planetary region. That work once again highlighted the role of giant planets (of mass comparable to, or greater than, Saturn) in ejecting potentially Earth-crossing comets from the system, ensuring that the chance of them colliding with a terrestrial planet is minimised. Once again, for long-period comets, it seems that Jupiter (and Saturn) truly act to reduce the impact threat posed to the Earth.

Although a significant amount of further research into the question of the shielding offered by giant planets is needed before any definitive answer is reached as to the true balance between their roles as friend and foe, it is clear from these studies that, at the very least, the situation is significantly more complicated than had previously been thought (e.g. Grazier, 2016; Grazier, Castillo-Rogez & Horner, 2018). Rather than being the benign and beneficial influence on our planet's development that has long been assumed, it seems that Jupiter might actually be more of an enemy than a friend!

**4.3 THE MIGRATION OF THE GIANT PLANETS**
The idea that giant planets migrate during the process of their formation and evolution is now well accepted. In exoplanetary systems, key evidence for this behaviour includes the hot Jupiters (e.g. Mayor & Queloz, 1995; Johnson et al., 2006; Vogt et al., 2010; Winn et al., 2010, Bryan et al., 2012; Becker et al., 2015), and a number of dynamically compact systems with planets trapped on mutually resonant orbits (e.g. Lissauer et al., 2011, Johnson et al., 2011, Robertson et al., 2012a, b; Wittenmyer, Horner & Tinney, 2012; Gillon et al., 2017). Such migration is the most common explanation for all giant planets discovered orbiting closer to their host stars than the location of the ice line, with the idea being that the planet first accreted beyond that point in its host system before careening inwards during the latter stages of its formation (e.g. Kuchner & Lecar, 2002; Masset & Papaloizou, 2003; Crida & Morbidelli, 2007; Beaugé & Nesvorný, 2012) - although in recent years, alternative mechanisms have been proposed to explain at least some of the known short-period giant planets (e.g. Batygin, Bodenheimer & Laughlin, 2016; Dawson & Johnson, 2018; Bailey & Batygin, 2018). It has even been suggested that such inward migration might often lead to the in-fall of one (or more) giant planets to the atmosphere of a youthful star, polluting that star's atmosphere with the heavy elements accreted, and so artificially inflating the observed metallicity (e.g. Pasquini et al., 2007). A planet migrating in such a fashion might also drag with it vast quantities of volatile material from beyond the ice-line, leading to the accretion of "water-worlds", which might potentially be habitable, in the giant planet's wake (e.g. Fogg & Nelson, 2007a, 2007b, 2009; Izidoro, Morbidelli & Raymond, 2014; Darriba et al., 2017).

Within our own Solar system, too, evidence abounds for the post-formation migration of the giant planets. However, whilst currently migration scenarios involving exoplanets typically only discuss inward migration, within our own Solar system there is a significant body of evidence that argues for the outward migration of Neptune, and potentially Uranus, leading to the dynamical relaxation of the outer Solar system. Indeed, it is currently thought that Neptune migrated over a distance of at least 7 au, and potentially as much as 15 au outward after its formation, doubling the radial extent of the planetary component of our Solar system.

## 4.3.1 EVIDENCE FOR MIGRATION - THE SMALL BODIES

The stable reservoirs of small bodies within the Solar system contain a great deal of information pertaining to the formation and migration of the giant planets, with their current distributions the direct result of the gravitational sculpting they experienced during that period. The main belt asteroids, for example, are distributed across a wide range of orbital eccentricities and inclinations – plots of their locations in *a:e* and *a:i* space (as shown in Figure 2) reveal a great deal of intricate structure within the belt, from the gravitationally cleared locations of MMRs with Jupiter to the separation of a small population interior to the main body of the belt from the rest of the population. The locations of the key mean-motion and secular resonances which have sculpted the belt are nicely portrayed in the first three figures of Nagasawa et al., 2000, who discuss whether the current level of excitation observed within the asteroid belt could be the result of secular resonances sweeping through the region as the Solar nebula became depleted.

In other studies, the role of such sweeping resonances as a result of planetary migration is considered pivotal in creating the excited distribution we observed today (e.g. Gomes, 1997, Minton & Malhotra, 2009). Indeed, Minton & Malhotra (2011) used the distribution of the eccentricities of the main belt asteroids to place constraints on the rate at which Saturn migrated. At the current epoch, the $\nu_6$ secular resonance, driven by Saturn, lies at the inner edge of the asteroid belt (for orbits with zero inclination), and delineates the high-inclination boundary for main-belt asteroids at greater heliocentric distances (see e.g. Fig 3 of Nagasawa et al., 2000). In order for the asteroid belt to have survived the inward migration of Saturn, as the secular resonance swept through it, Minton & Malhotra (2011) propose that Saturn must have migrated at a speed of at least 0.15 au $\text{Myr}^{-1}$ – although they note that this migration limit is a function of Saturn's orbital eccentricity. If Saturn's eccentricity was smaller at the time of its migration, then slower migrations could have occurred without overly disrupting the asteroid belt.

The Jovian Trojans, too, are thought to be a direct result of planetary migration (Morbidelli et al., 2005; Lykawka & Horner, 2010; Nesvorný, Vokrouhlický & Morbidelli, 2013; Pirani et al., 2019) – the source of their dynamically hot distribution (with some objects exhibiting inclinations up to 55 degrees ((83483) 2001 $SC_{93}$) and eccentricities as high as 0.272 ((5144) Achates)). Interestingly, models describing the capture of the Jovian Trojans during planetary migration suggest that objects would be captured with a wide range of dynamical stability, from those that are stable on timescales far longer than the age of the Solar system to those that remain Trojans for just tens, or hundreds, of years. As such, even four billion years after their capture, objects should still dynamically bleed from the Jovian Trojan clouds to the rest of the outer Solar system, as the less stable objects are gradually "shaken free" by the distant perturbations of the planets (e.g. Levison, Shoemaker & Shoemaker, 1997; Di Sisto et al 2014; Di Sisto, Ramos & Gallardo, 2019; Holt et al., 2020). Recent dynamical simulations have revealed that up to a third of the members of the Jovian Trojan swarms could be unstable (e.g. Di Sisto et al 2014), whilst at least one Jovian Trojan exhibits dynamical instability on Gyr timescales (Horner, Lykawka & Müller, 2012). That object, (1173) Anchises, was found to have a dynamical half-life of order 619 million years, which is entirely compatible with it being an unstable Trojan captured during planet formation, rather than a recent entrant to the cloud. The Jovian Trojan population might even contain the key to tying down the timing of the migration of the giant planets, with recent simulations of the binary Trojans Patroclus-Menoetius suggesting that the migration of the giant planets must have occurred very early in the Solar system's history, prior to the Late Heavy Bombardment (Nesvorný et al., 2018). That work suggests that the Patroclus-Menoetius binary formed in a thick planetesimal disc beyond the orbit of Neptune, and argues that it must have been ejected from that disc, and captured to Jupiter's Trojan population, at a very early stage in the planet formation process, otherwise it would have been disrupted through encounters within that disc.

The Neptunian Trojans, like the Jovian Trojans, have an orbital distribution far too dynamically excited to be considered pre-formed objects. In the past decade, a number of studies have been carried out into the formation and evolution of these objects, revealing that they were most likely captured during Neptune's outward migration (e.g. Lykawka et al., 2009, Nesvorny & Vokrouhlicky 2009; Horner & Lykawka, 2010b, Horner & Lykawka, 2012; Parker, 2015; Gomes & Nesvorný, 2016). Just as for the Jovian Trojans, such a scenario would suggest that there remain unstable members of the Neptunian Trojan population at the current day, gradually bleeding from that population to contribute to the Centaur population, the dynamically unstable parents of the Jupiter Family comets (e.g. Horner & Lykawka, 2010a, d). Indeed, recent studies have revealed that two of the known Neptunian Trojans may well be such objects, namely 2001 $QR_{322}$ (Horner & Lykawka, 2010c) and 2008 $LC_{18}$ (Horner et al., 2012a), with two additional objects (2004 $KV_{18}$ and 2012 $UW_{117}$) proving in actuality to be recently captured interlopers to that population (Horner et al., 2011a; Wu, Zhou & Zhou, 2018).

The orbital distribution of objects in the trans-Neptunian region, too, carries a wealth of information on the precise nature of that planet's migration (e.g. Chiang & Jordan, 2002; Hahn & Malhotra, 2005; Murray-Clay & Chiang, 2005; Nesvorný, 2015; Gomes et al., 2018; Volk & Malhotra, 2019). The Plutino population, trapped within the 3:2 Neptunian MMR, are believed to have been captured and transported as that resonance migrated outward with the planet. Once objects were trapped in the resonance, their orbital eccentricities and inclinations were pumped as they were pushed outwards by the moving resonance, resulting in the distribution observed today (e.g. Malhotra, 1993, 1995; Lykawka & Mukai 2008; Levison et al., 2008). The farther Neptune migrated, the more extreme the inclinations and eccentricities acquired would be, and so it is in theory possible to determine, from the Plutino distribution, the precise scale of Neptune's outward motion. That outward migration also played a significant role in sculpting the dynamically stable Edgeworth-Kuiper belt (e.g. Chiang & Jordan, 2002; Hahn & Malhotra, 2005; Petit et al., 2011; Nesvorný 2018; Volk & Malhotra, 2019), driving the observed asymmetric distribution of objects trapped in Neptune's 2:1 MMR (Murray-Clay & Chiang, 2005), and resulting in the excited yet stable population we observe today (e.g. Gladman et al., 2012).

**4.3.2 MIGRATION AND THE LATE HEAVY BOMBARDMENT**
In section 4.2.1, we discussed the possibility that the inner Solar system was subject to a period of enhanced impact rates, known as the Late Heavy Bombardment, several hundred million years after the formation of the Solar system. Whilst there remains some debate over the true nature and duration of that bombardment, there remains evidence that the impact rate was markedly elevated during that time - with the formation of Imbrium and Orientale basins at around 3.8-3.9 Gyr seemingly inconsistent with the idea that the impact rate was solely due to the steady depletion of a population of impactors dating from the Solar system's formation (e.g. Bottke et al., 2007; Bottke & Norman, 2017).

Given the ample evidence for the migration of the giant planets detailed in the previous sub-section, it is natural to wonder whether such migration could help to explain the origin of the Late Heavy Bombardment. In 2005, a series of papers exploring that reasoning proposed what became known as the 'Nice Model' (e.g. Gomes et al., 2005; Tsiganis et al., 2005; Morbidelli et al., 2005). That model suggests that the four giant planets formed in a more compact architecture than that we see today[47]. As those planets interacted with the debris left behind from the planet formation process, they migrated, leading to a very gradual relaxation of their orbits - causing them to slowly spread apart. Eventually, after a few hundred million years, that migration caused their orbits to become dynamically unstable, leading to a period of chaotic evolution that destabilised the outer Solar system - and in the process led to the injection of large amounts of debris to the inner Solar system, causing the Late Heavy Bombardment.

In the Nice Model, the authors suggest that the four giant planets formed in a compact group, with Jupiter and Saturn located closer together than the location of their mutual 2:1 mean motion resonance, and the ice giants packed in relatively closely beyond the orbits of the two largest planets. Beyond the orbit of the outermost ice giant (which could either be Uranus or Neptune, depending on the particulars of the integration considered), there was a massive, thick disc of planetesimals - stretching outward from a distance of approximately 15 au from the Sun (see e.g. Figures 1 & 2 of Gomes et al., 2005). Over a period of several hundred million years, the orbits of the four giant planets relaxed, ever so slowly. The outermost continually stirred the disc of debris beyond its orbit, scattering material sunwards, and migrating outwards in response (as a result of the conservation of angular momentum). The objects injected to the realm of the giant planets were then scattered in the same manner as the modern Centaurs, experiencing stochastic close encounters with the giant planets before finally being ejected from the Solar system. The two planets whose orbits lay in the centre of the four-planet sandwich only migrate slightly as a result of this, as they can hand objects both inward (to the next sunward planet's control) or outward (to the next planet out) - although both exhibit slight outward migration due to the net exchange of angular momentum through objects passed inward to Jupiter. Jupiter, on the other hand, migrates inward. When Jupiter throws objects into the inner Solar system, there is no mechanism by which they can readily be decoupled from the giant planet's influence[48], and so they will eventually once again encounter the giant planet. But when Jupiter ejects objects from the Solar system, they clearly never return. As a result, Jupiter gradually loses angular momentum, in exchange with the ejected objects, and migrates sunwards.

---

[47] Indeed, in work that presaged the development of the Nice Model, Thommes, Duncan & Levison (1999, 2002) proposed a model that suggested that Uranus and Neptune formed between the orbits of Jupiter and Saturn, to be scattered outward to their current orbits as a result of encounters with proto-Jupiter and proto-Saturn.

Eventually, this gradual migration leads to a cataclysmic period of instability in the orbits of the giant planets. Jupiter and Saturn first approach, then cross, the location of their mutual 2:1 mean-motion resonance. This destabilises the orbits of the two ice giants, causing them to evolve chaotically, leading to the ejection of at least one of those planets into the bulk of the massive disc in the Solar system's outer reaches. That planet would then disrupt that disc, scattering debris in all directions, and sending a large amount of icy material inwards, to impact upon the planets in the inner Solar system - all the while experiencing significant dynamical friction from those encounters that would serve to rapidly circularise its orbit, pulling it away from the possibility of further encounters with the other ice giant.

In addition to the destabilisation and disruption of the massive debris disc beyond the orbit of the outermost planet, this period of chaotic evolution would also markedly disrupt and deplete the asteroid belt - again showering the inner Solar system with potentially deadly debris. At the same time, the destabilisation would allow the emplacement of large quantities of icy material to the outer reaches of the main belt - including, potentially, the ice-rich dwarf planet Ceres (e.g. McKinnon, 2008; Levison et al., 2009). Eventually, the orbits of the outer planets would settle into the new, stable, relaxed architecture we observe today. The trans-Neptunian region and asteroid belt would have been significantly denuded of material, and the inner planets would have experienced a delayed period of intense bombardment, triggered by the instability in the outer planet's orbits - the Late Heavy Bombardment. Furthermore, the theory does an elegant job of explaining the capture of the Jovian Trojans (e.g. Morbidelli et al., 2005), plundered from the debris liberated from the asteroid and trans-Neptunian belts, and captured to Jupiter's 1:1 mean-motion resonance during the period when that resonance was itself destablised by the resonance-crossing of the two largest planets. As Jupiter and Saturn moved away from mutual 2:1 mean-motion resonance, any objects temporarily captured to Jupiter's 1:1 MMR would be 'frozen in', locked into that resonance as it returns to its former extremely stable state. Whilst other models exist that can also explain the origin of the Jovian Trojan population (e.g. Pirani et al., 2019), the Nice Model remains a fascinating insight into the possibilities of delayed dynamical instabilities in planetary systems, and serves as an important reminder that, just because a planetary system appears to be dynamically stable at the current epoch, that does not necessarily mean that it always has been, or always will be.

In the years since the Nice Model was proposed, a wide variety of variants and alternate theories have been proposed, all invoking the dramatic migration and interaction of the outer planets to explain the various features of the modern Solar system. Levison et al. (2008) investigate the impact that the chaotic evolution described in the Nice Model would have on the formation and structure of the Edgeworth-Kuiper belt. They find that, in order to explain why Neptune stopped its outward migration at 30 au, the trans-Neptunian disc must have been truncated at between 30 and 35 au from the Sun - otherwise the outward motion would have continued until the planet reached the outer edge of the disc (e.g. Gomes et al., 2004). Such truncation of the Solar system's protoplanetary disc has been proposed and studied before, in the context of the sharp outer edge of the Edgeworth-Kuiper belt (e.g. Melita et al., 2002; Levison, Morbidelli & Dones, 2004; Lykawka & Mukai, 2008), but those works typically invoked an outer truncation distance closer to ~50 au, rather than the 30-35 au suggested by Levison et al. (2008). Morbidelli et al. (2010) also expanded on the work of the Nice model, showing how the stochastic 'jumping' of Jupiter and Saturn that would result from chaotic close encounters with a Neptune mass planet during the period of instability could explain both the depletion and current structure of the asteroid belt. Follow up work (Nesvorný, Vokrouhlický & Morbidelli, 2013) showed how such a scenario could result in the current orbital distribution of the Jovian Trojans, and note that it could even help to explain the origin of the observed asymmetry between the populations of the leading and trailing Jovian Trojan clouds.

---

[48] At least, within the scope of their simulations. In reality, a fraction of the material flung inwards by Jupiter would be lost - through collision with the terrestrial planets, collision with the Sun, and fragmentation. That said, it seems improbable to suggest that the amount of material lost this way would come close to balancing that which would eventually be ejected from the Solar system by the giant planet - and so the end result (that Jupiter would migrate inwards) remains the same.

More recently, still more dramatic versions of a chaotic early Solar system have been proposed. Some scenarios (e.g. Batygin, Brown & Betts, 2012; Nesvorný & Morbidelli, 2012;) invoke the idea that the outer Solar system might once have contained more than the four giant planets we observe today, and suggest that the additional planet(s) that were once there were either ejected entirely from the system, or were flung onto long-period, distant orbits (e.g. Bromley & Kenyon, 2014, 2016; Trujillo & Sheppard, 2014; Cloutier, Tamayo & Valencia, 2015; Nesvorný 2018), as we discuss in more detail in section 4.5.

A particularly interesting model of early Solar system chaotic behaviour is known as the 'Grand Tack Hypothesis', which is based on the idea that a pair of inwardly-migrating giant planets can become trapped in mutual mean-motion resonance, as a result of the outer of the two migrating more rapidly, and catching the innermost up. In certain circumstances, interactions between those two planets and the protoplanetary disc within which they are migrating can result in the direction of their migration reversing, with the pair marching back outwards towards their initial formation locations (e.g. Masset & Snellgrove, 2001; Morbidelli & Crida, 2007).

In the Grand Tack scenario, Jupiter and Saturn formed somewhat closer to the Sun than their current locations (with Jupiter at ~3.5 au) before initially migrating inwards, towards the Sun. After Jupiter had migrated some ~2 au inwards, to around 1.5 au, it and Saturn changed the direction of their migration after becoming trapped in mutual mean motion resonance. Thereafter, the two planets moved back outwards to their current locations (Walsh et al., 2011). This 'Grand Tack' (Walsh et al., 2012) of the giant planets would have the effect of both stirring and truncating the solid material in the inner parts of the protoplanetary disc - creating conditions similar to the truncated inner proto-planetary disc invoked by Hansen (2009) to explain the origin of the terrestrial planets. It would also act to significantly deplete the asteroid belt, and would also aid in the delivery of volatile material to the inner planets (e.g. O'Brien et al., 2014).

A key and widely studied advantage of this theory is that it offers a mechanism by which the growth of Mars can be truncated, helping to explain why the Red Planet is so much less massive than the Earth or Venus (e.g. Walsh et al., 2011; Brasser 2013; Kobayashi & Dauphas, 2013; Izidoro et al., 2015b; Brasser et al., 2017)[49]. More recent work (Brasser et al., 2016) suggests that the innermost point of Jupiter's grand tack was likely somewhat more distant than previously thought - around 2 au, rather than the 1.5 au suggested by Walsh et al., 2011. In the simulations carried out by Brasser et al. (2016), a tack inward to 1.5 au was found to be strongly incompatible with the formation of the most massive terrestrial planet (the Earth) occurring at 1 au, even though such a tack was capable of truncating Mars' formation as required. By halting Jupiter's tack at a greater heliocentric distance, those authors were more readily able to replicate the modern inner Solar system.

---

[49] It should be noted, here, that other theories exist to explain Mars' small size, suggesting that the protoplanetary disc was depleted in material where Mars formed (e.g. Izidoro et al., 2014; Raymond & Izidoro, 2017); one mechanism by which such depletion could occur is that a secular resonance with the planet Jupiter swept through the region in which Mars was to form, prior to its formation. The passage of that resonance would excite debris in the region of Mars' current orbit, causing the available mass for planet formation to be depleted as a result of enhanced particle fragmentation due to increased mutual collision speeds (Bromley & Kenyon, 2017). Equally, Levison et al. (2015) suggest that pebble accretion models might offer an alternative explanation for Mars' small mass.

One problem with the Grand Tack hypothesis, however, is that the migration and associated disruption of the asteroid belt would occur long before the time of the Late Heavy Bombardment. It should, however, be noted that the Grand Tack model and the Nice Model are not necessarily mutually exclusive. The Nice Model teaches us that a period of instability amongst the giant planets can occur after a lengthy period of quiescence - so it is, of course, possible that the giant planets experienced both a period of evolution in the earliest stages of the Solar system's formation that matched the behaviour proposed in the Grand Tack model, and then, later, experienced a delayed period of instability, resulting in the injection of the ice giants (Uranus and Neptune) to a massive trans-Neptunian disc. However, as yet, the two models have not been unified, and there remains extensive debate on the true narrative of planetary migration and the origin of the Late Heavy Bombardment.

In that light, it is interesting to note that a number of studies have suggested that a Nice-model like dynamical instability in the orbits of the giant planets may well have occurred within the first 100 million years of the Solar system's history (e.g. Agnor & Lin, 2012; Kaib & Chambers, 2016; Clement et al., 2018; Nesvorný, 2018; Quarles & Kaib, 2019; Mojzsis et al., 2019), with some suggesting that the giant planets had reached their current architecture prior to the end of terrestrial planet formation (e.g. Kaib & Chambers, 2016). Such studies have attempted to address the potential incompatibility between the late giant planet migration proposed by the Nice Model and the observed orbits of the terrestrial planets, noting that, had the instability between the giant planets occurred after the terrestrial planets were fully formed, then it should have significantly impacted and disrupted their orbits (e.g. Brasser, Walsh & Nesvorný, 2013). This situation is far less problematic if the instability happens at an earlier epoch - and several studies have shown that such an instability could provide an additional mechanism by which the growth of Mars could be truncated (e.g. Clement et al., 2018, 2019a), whilst also offering a potential explanation for the origin of Mercury's unusual orbit (Roig, Nesvorný & DeSouza, 2016; Clement et al., 2019b), and being able to well reproduce the observed depletion and structure of the asteroid belt (Clement, Raymond & Kaib, 2019). The true history of the migration of the Solar system's giant planets remains to be determined, and it will be interesting to see how future studies address the apparently dissonant evidence provided by the Solar system's small body populations and impact record.

**4.4 THE ORIGIN OF WATER ON THE EARTH, AND THE HYDRATION OF THE TERRESTRIAL PLANETS**
One of the ingredients considered most vital for the development and ongoing survival of life is water (e.g. Mckay, 1991, 2014; Rampino & Caldeira, 1994; Chyba & McDonald, 1995; Ward & Brownlee, 2000; and many others). In fact, the presence of water is considered so important that the classical definition of the "Habitable Zone" around a star is simply the range of distances from the star over which an Earth-like planet would be able to have stable liquid water at some point on its surface (e.g. Hart, 1979; Kasting, Whitmire & Reynolds, 1993; Horner & Jones, 2010a; Kopparapu et al., 2013, 2014, 2016; Zsom et al., 2014; Agnew et al., 2017).

The role of water is considered to go beyond simply providing a medium in which life can develop, and being the key solvent for its survival. It is thought, for example, that if the Earth was desiccated, plate tectonics would not occur, which would remove a key route through which gasses that are removed from the atmosphere by weathering, life, and other processes, are recycled from the surface (e.g. Wallace & Hobbs, 2006, section 2.3). It is even argued that the lack of plate tectonics might lead to the collapse of the dynamo that produces the magnetic field that protects our planet (as is the case for both Venus and Mars; e.g. Nimmo & McKenzie, 1998; Nimmo, 2002; Breuer & Spohn, 2003; O'Neil, Jellinek &

Lenardic, 2007). For a more detailed review of the various influences on planetary habitability, we direct the interested reader to the detailed review of that topic, Horner & Jones, 2010a.

Given that water is such an important part of life on Earth, it is somewhat surprising that the origin of that water is still under heavy debate. Our planet formed well sunwards of the location of the ice-line in the Solar system, and so temperatures were far too high for water to condense from the Solar nebula to be directly accreted in situ. It seems obvious, then, to conclude that the terrestrial planets should have accreted solely from dry rocks. It has been proposed, however, that a significant amount of water could be trapped in the form of hydrated silicates in that region, which would allow water to be accreted concomitantly with the planet's silicate budget. The apparent paradox of our wet planet, and its hot origins, has prompted significant amounts of debate, and has led to a number of different models being proposed to detail the hydration of the terrestrial planets. These can be broken down into three main types – *endogenous hydration*, *early exogenous hydration* and *"late veneer" exogenous hydration*.

Models detailing the endogenous hydration of the terrestrial planets suggest that the great bulk of their water was accreted from their local environment, with only a small contribution being added from sources farther from the Sun. In such a scenario, the planets are thought to have accreted the bulk of their water in the form of hydrated silicates, which have been shown to be stable at far higher temperatures than those at which water can exist alone as a solid (e.g. Drake, 2004, 2005; Asaduzzaman, Muralidharan & Ganguly, 2015). One problem with such scenarios, which on the surface seem to provide a promising avenue by which terrestrial planets throughout the universe could gain significant amounts of water during their formation, is that there is a well-established correlation between the water content of meteorites that fall on the Earth and the region of the asteroid belt in which they originated (with those that originated farther out in the belt having a greater water content than those with origins in its inner reaches). In particular, Morbidelli et al. (2000) detail results for a class of meteorites known as the Enstatite chondrites, which are thought to originate in the innermost region of the asteroid belt. They highlight the fact that these meteorites are the driest of all known meteorites in the Solar system (containing only 0.05-0.1% water, by mass), which suggests that the rocky material which formed interior to the ice-line was unable to hold much in the way of water, and therefore that the amount of water concomitantly accreted during the formation of the terrestrial planets from local materials most likely was insufficient to explain the Earth's water budget.

Because of these problems tying the Earth's water to the materials from which it formed, most researchers now believe that exogenous models of hydration are the best explanation for our water budget. Such scenarios postulate that, rather than having a local reservoir of hydrated material, the water budget of the terrestrial planets was instead sourced from objects that formed much farther from the Sun, beyond the ice-line, and were then flung inwards to impact on the planets. These models are further broken down into two main types – early accretion models, which suggest that the water was pumped into the inner Solar system (typically from objects that formed in the outer reaches of the asteroid belt, just beyond the ice line) whilst the planets were themselves in the process of accreting (e.g. Morbidelli et al., 2000; Petit et al., 2001; Marty, 2012). Some of these models require the bulk of the water to have been delivered in a few stochastic giant impacts (similar to those described in section 4.1; e.g. Marty, 2012), a process which has been invoked to explain the fact that the terrestrial planets appear to have significantly different water budgets (with, for example, the Earth being significantly more hydrated than Mars). It has even recently been suggested that the bulk of Earth's water was delivered in the planetary collision that gave birth to the Moon - with Earth's water being delivered by the impacting planet, Theia (Budde, Burkhardt & Kleine, 2019). On the other hand, mechanisms have also been proposed that could deliver water in the form of a continuous early rain of icy pebbles, ranging in size from millimetres to tens of centimetres in diameter, that formed through pebble accretion beyond the ice-line, then migrated inward following the ice-line as it, in turn, briefly moved in to heliocentric distances smaller than that of the Earth's orbit (e.g. Sato, Okuzumi & Ida, 2016).

The "late-veneer" exogenous hydration models, by contrast, suggest that the hydration of the terrestrial planets occurred at some point after their bulk accretion was completed (e.g. Chyba, 1987; Owen & Barnun, 1995). One potential event which could have been the source of this hydration is the Late Heavy Bombardment (e.g. Oberbeck & Fogleman, 1989; Wells et al., 2003), which may well have led to a prolonged influx of material from both the outer Asteroid belt, and the trans-Neptunian disc, through the inner Solar system (e.g. Gomes et al., 2005). Such a bombardment, even though it would consist of many impacts of smaller bodies, rather than a few giant impacts (as in the early exogenous scenarios) could easily lead to the terrestrial planets having different water budgets, particularly if the impactors were sourced from both the Asteroid belt and the outer reaches of the Solar system, as discussed by Horner et al. (2009). Equally, the Grand Tack model, in which Jupiter and Saturn careen inward through the asteroid belt, and then migrate back out to their current locations, could efficiently deliver material from beyond the snow-line to the terrestrial planets (e.g. O'Brien et al., 2018).

One way in which researchers have attempted to differentiate between the various methods suggested for the delivery of water to the Earth is through the study of the ratio between deuterium and hydrogen in water (the D:H ratio). A number of studies (e.g. Drouart et al., 1999; Dauphas, Robert & Marty, 2000; Mousis et al., 2000; Mousis, 2004; Horner, Mousis & Hersant, 2007; Horner et al., 2008; Yang, Ciesla & Alexander, 2013) of the formation of our Solar system have suggested that the D:H ratio in the water accreted to icy bodies should vary significantly as a function of heliocentric distance within the Solar nebula – with the lowest values being found near the ice-line (in the outer Asteroid belt), and the largest values occurring in the region beyond Neptune. Clearly, then, one should expect that water delivered from the asteroid belt would have a measurably different D:H signature than that sourced from the outer regions of the Solar system - a result supported by observed large D:H values in comets (e.g. Balsiger et al., 1995; Bockelee-Morvan et al., 1998; Meier et al., 1998; Villanueva et al., 2009), and the significantly lower values measured in meteorites (e.g. Fig. 2 of Drouart et al., 1999). One would expect that it would be a relatively simple procedure to compare the different values expected for different objects to that measured in the Earth's oceans, and therefore have an independent test of the origin of the water stored therein. However, the issue is complicated by the fact that the D:H ratio measured in the Earth's water is not solely the product of its origin, but rather has been significantly modified over geological timescales by a variety of processes. This also greatly hinders comparisons between the deuteration of terrestrial, venutian, and martian water. For simplicity, many authors assume that the initial D:H ratio of water on these planets was identical (e.g. Krasnopolsky et al., 1998; Lécuyer et al., 2000; Gurwell, 1995) and then use the observed differences in the deuteration at the current day to draw conclusions on the processes that have occurred on those planets in the intervening time.

Such difficulties are one of the key reasons that the true origin of the Earth's water has not yet been universally agreed, although, at the current time, the most widely accepted scenario seems to be that the bulk of the water was sourced from far beyond the Earth's orbit, with at least some contribution coming as a "late veneer" during the Late-Heavy Bombardment (e.g. Morbidelli et al., 2012).

The situation is further complicated by the fact that Venus is markedly drier than the Earth at the current epoch, whilst Mars is believed to have once been both warm, and wet (although most of that water is likely still locked up in permafrost across the planet). It seems likely that both those planets have lost significant amounts of water through it being stripped from their atmospheres (e.g. Jakosky & Phillips, 2001; Hodges, 2002; Lammer et al., 2003; Valeille et al., 2010). In addition, a number of studies have suggested that both Mars and Venus were dessicated as a result of the Solar system's early impact regime (e.g. Davis, 2008; Kurosawa, 2015; Rickman et al., 2019a). Whilst the final narrative of the inner Solar system's hydration remains to be determined, it seems increasingly likely that the giant collisions that dominated the latter stages of terrestrial planet formation may have played a vital role in ensuring the disparate volatile budgets of Venus, Earth and Mars (e.g. Lykawka & Ito, 2019).

## 4.5 PLANET-MASS OBJECTS BEYOND NEPTUNE

Ever since Neptune was discovered as a direct result of its perturbations on the orbit of Uranus, the idea that there might be one (or more) planetary body orbiting farther out has repeatedly come into vogue. The fortuitous discovery of Pluto, in 1930, was a direct result of a search for such a body – a search that simply led to the detection of the first trans-Neptunian object. In the early 1980s, studies of the frequency of mass extinctions on the Earth, allied to the discovery of the Chicxulub impact crater in Mexico, believed to be tied to the extinction of the dinosaurs (e.g. Alvarez et al., 1980; Hildebrand et al., 1991; Schulte et al., 2010), led to the idea that such extinctions were occurring with a period of approximately 26 million years (e.g. Raup & Sepkoski, 1984, 1986; Schwartz & James, 1984). In order to explain the observed periodicity, a number of competing theories sprang up, including the idea that our Sun might have a distant companion, most likely a brown or red dwarf, moving on a highly eccentric orbit that would, at perihelion, bring it through a sufficiently dense region of the inner Oort cloud that it would trigger a severe shower of comets, leading to impacts on the Earth and a mass extinction (e.g. Davies, Hut & Muller, 1984; Whitmire & Jackson, 1984; Alvarez & Muller, 1984). In the years that followed, however, the theory of "Nemesis" quickly fell by the wayside, with a number of studies casting doubt on the validity of the hypothesis (e.g. Bailey, 1984; Clube & Napier, 1984a; Hills, 1984; Hut, 1984; Torbett & Smoluchowski, 1984). However, theories of planetary-mass objects beyond the orbit of Neptune continue to crop up in order to explain a variety of features of our Solar system, from the population of detached objects beyond Neptune's orbit to observed aphelion asymmetries among dynamically new long period comets. Here, we detail a variety of those theories, showing how observations of the dynamical structure of the outer Solar system can allow us to make predictions about what may lie at greater distances.

### 4.5.1 A PLANET IN THE OORT CLOUD - SOURCE OF THE LONG-PERIOD COMET DISTRIBUTION?

Whilst the idea of "Nemesis", a companion star to the Sun responsible for periodic mass extinctions on Earth, has fallen out of favour, there remains beguiling evidence that suggests that the Sun may be accompanied by a large planet, moving in the Oort cloud. Dynamically new long-period comets, making their first pass through the inner Solar system having been perturbed from their cold storage in the Oort cloud, should come in from all directions in approximately equal quantities – their aphelia should be isotropically distributed across the sky. In the early 1980s, studies of the distribution of cometary aphelia revealed evidence of the Biermann comet shower (e.g. Biermann, Huebner & Luster, 1983) – a clustering of cometary aphelia in a small area of the sky that is accepted as being evidence of the effect of the passage of a star through the Oort cloud a few million years ago. That passage disrupted the orbits of comets in the region affected by the passing star, causing a shower of new comets to fall inward to the inner Solar system. The observed clustering represents the tail-end of the shower event.

It is not just close-encounters with low mass stars that can inject significant numbers of comets to the inner Solar system. Fouchard et al. (2011) performed simulations of the effect of a wide variety of stellar encounters, and found that the influence of far more distant, massive stars, can be significant. Indeed, they note that single massive stars could cause what they term "comet drizzles" - lengthy periods (of 100 Myr, or more) of enhanced cometary flux from the Oort cloud to the inner Solar system. Whilst the impact of close, low-mass stars would be limited to their tracks through the outskirts of the cloud, that of more distant, massive stars can impact the whole of the cloud, moving cometary nuclei onto orbits that are more likely to be injected into the inner Solar system by the influence of the galactic tide. It is by enhancing the population of nuclei in this 'tidally active zone' that massive stars can generate such prolonged enhancements to the observed long-period comet flux, as those comets proceed to drizzle into the inner Solar system for tens of millions of years.

More recent work (Feng & Bailer-Jones, 2015) notes that, whilst the majority of new long-period comets are injected into the inner Solar system by the influence of the galactic tide, a non-negligible fraction (>5%) are flung inwards by encounters with passing stars. They also identify past encounters with the stars Gliese 710 and HIP 14473, which they suggest may have contributed up to 8 and 6%, respectively, of the current long-period comet flux. In the coming years, our knowledge of historical stellar close encounters should increase markedly, as a result of the data being returned by the *Gaia* space telescope (e.g. Gaia Collaboration et al., 2016, 2018; Evans et al., 2018), and it seems almost certain that such work will lead to the identification of the parent of the Biermann shower, and the discovery of additional showers within the ever-growing catalogue of long-period comets (e.g. Rickman et al., 2012; Bailer-Jones, 2018; de la Fuente Marcos, de la Fuente Marcos & Aarseth, 2018).

In the late 1990s, two groups independently suggested that another signal was becoming visible in the distribution of new cometary aphelia, as the catalogue of known dynamically new comets was increasing. Murray (1999) suggested that the observed aphelia of the long-period comets showed an excess at distances between 30,000 and 50,000 au, and that the aphelia within that region were preferentially aligned along a great circle inclined at ~30 degrees to the plane of the ecliptic. At the same time, Matese et al. (1999) noticed a similar pattern, claiming an excess of aphelia on a different great circle, this time centred around a galactic longitude of ~135 degrees. Both groups suggested that the alleged asymmetries could be explained by the presence of a planet of around three times the mass of Jupiter, orbiting within the Oort cloud, and perturbing comets therein toward the inner Solar system. After an exhaustive study of the biases that affect cometary observations, Horner & Evans (2002) found that the signal proposed by Murray (1999) did not stand up to detailed investigation, being merely the product of the aforementioned biases. However, they found that such biases could not explain the statistically significant feature observed by Matese et al. (1999). In the decade since that work, the number of known comets with well determined orbits has increased significantly, leading Matese & Whitmire (2011) to revisit the earlier work. In that work, they find that the current catalogue of known dynamically new comets contains a strong signal showing the perturbations of the galactic tide upon the Oort cloud. On top of this, they find that the population of these objects contains:

*"an ≈ 20% impulsively produced excess. The extent of the enhanced arc is inconsistent with a weak stellar impulse, but is consistent with a Jovian mass solar companion orbiting in the [Outer Oort cloud]".*

On this basis, it seemed that the evidence for our Sun having a Jovian companion (an object that would perhaps be classified as an evolved T- or Y-dwarf, if detected floating freely in space, or in orbit around another star – e.g. Luhman, Burgasser & Bochanski, 2011; Liu et al., 2011; Burgasser et al., 2012) was growing firmer as more observations were made. However, if such an object exists, then one would expect that it would be discovered during the analysis of data obtained by the Wide-field Infrared Survey Explorer, WISE (Wright, 2007), since such an object should be relatively bright at the infrared wavelengths studied by that mission. On that basis, Luhman (2014) performed an all-sky search for such an object using images taken by WISE across multiple epochs. Their work found no such object down to a W2 magnitude of 14.5, which they note suggests that no Saturn-mass object exists within 28,000 au of the Sun, nor any Jupiter-mass objects within 82,000 au of our star - though they note that using alternative models for the brightness of Jupiter-mass objects could bring that survey limit down to a range of 26,000 au for an object the size of Jupiter.

It now seems likely that there is no Jupiter-mass companion, lurking in the Oort Cloud - instead, the origin of the observed asymmetries in the long-period comet flux may be the influence of the Galactic Tide itself (e.g. Feng & Bailey-Jones, 2014). In the coming years, the number of newly discovered long-period comets should rise considerable, and analysis of their orbits may likely yield a final, definitive answer on the nature and origin of the observed aphelion asymmetries.

**4.5.2 A TRANS-NEPTUNIAN PLANET AS AN EXPLANATION FOR THE STRUCTURE OF THE TRANS-NEPTUNIAN POPULATION**
In the first few years of the 21st Century, observers found growing evidence that the outer edge of the Edgeworth-Kuiper belt, at about 47 or 48 au from the Sun, was relatively abrupt (e.g. Gladman et al., 1998a; Jewitt, Luu & Trujillo, 1998; Chiang & Brown, 1999; Allen, Bernstein & Malhotra, 2001; Trujillo

& Brown, 2001; Gladman et al., 2001) - a feature that became known as the 'Kuiper Cliff' (e.g. Chiang & Brown, 1999; Chiang et al., 2003). Such a sharp cut-off was unexpected - indeed, some theories had even suggested that the region beyond 50 au should contain more objects, and more large objects, than the region now recognised as the Edgeworth-Kuiper belt (e.g. Stern, 1996; Stern & Colwell, 1997). At the same time, studies of debris discs around nearby stars had revealed them to extend to much greater astrocentric distances around their hosts than the extent of the Sun's Edgeworth-Kuiper belt (e.g. Smith & Terrile, 1984; Backman & Paresce, 1993; Dominik et al., 1998; Greaves et al., 1998; Holland et al., 1998; Augereau et al., 1999; Jourdain de Muizon et al., 1999; Schneider et al., 1999; Kenyon & Bromley, 2001; see also Matra et al., 2018).

A number of possible explanations were put forward in an attempt to explain the origin of the Kuiper Cliff. In some scenarios, the proto-planetary disc from which the Solar system formed was truncated at ~30 au, and the current trans-Neptunian objects were swept outward to their current orbits as Neptune migrated to its current orbit (e.g. Ida et al., 2000; Levison & Morbidelli, 2003; Gomes, Morbidelli & Levison, 2004). To explain such truncation, a close encounter (with a pericentre of a hundred or a few hundred au) between the Sun and a nearby star during the Solar system's formation in a relatively dense stellar cluster was suggested (e.g. Ida, Larwood & Burkett, 2000; Kobayashi & Ida, 2001; Levison, Morbidelli & Dones, 2004; Kobayashi, Ida & Tanaka, 2005), and the discovery of the dwarf planet (90377) Sedna, in 2003 (Brown, Trujillo & Rabinowitz, 2004), appeared to add some weight to that hypothesis (e.g. Kenyon & Bromley, 2004; Morbidelli & Levison, 2004). It soon became clear, however, that such a solution could not explain all of the observed features of the belt (e.g. Melita, Larwood & Williams, 2005).

An alternative explanation for the sculpting of the outer Edgeworth-Kuiper belt was the presence of an additional planet, beyond the outer edge of the belt. Brunini & Melita (2002) considered the influence that a Mars-mass object at ~60 au would have on the primordial Edgeworth-Kuiper belt, which they modelled as extending far beyond the Kuiper Cliff. They found that such a planet would be able to carve a gap in such a disc with an inner edge at ~50 au - a good match to the location of the Cliff. In a similar manner, Lykawka & Mukai (2008) found that the influence of a planet several times the mass of Mars, ejected by Neptune to an eccentric orbit trapped in an N:1 resonance with the giant planet (where N is an integer > 1), could explain the sculpting of the disc's outer edge, the distribution of the excited Classical/Scattered Disc objects, the formation of distant Resonant objects (beyond 50 au), and the presence of the detached objects. They suggested that, if the planet still lurks in the system's outer reaches, it would move on an orbit with inclination between ~20 and 40 degrees, and a semi-major axis ~100-200 au, placing it within the reach of detection by contemporary surveys. de La Fuente Marcos & de La Fuente Marcos (2014) went one step further - following the discovery of the detached, Sedna-like object 2012 VP$_{113}$ (Trujillo & Sheppard, 2014), they suggested that the most likely explanation for the existence of such objects was the existence of at least two trans-Neptunian planets, whose influence was required to shepherd the orbits of the detached objects that had been discovered at that time, building on the single super-Earth proposed by Iorio (2014) to explain the same new discoveries.

The idea of a trans-Neptunian planet has never really gone away, but it has gained further exposure in recent years, following the discovery of 2012 VP$_{113}$. A series of papers proposed the existence of a distant, Neptune-mass planet (popularly known as 'Planet Nine') in order to explain an apparent clustering in the orbits of a number of extremely distant trans-Neptunian objects with $a > 250$ au, including (90377) Sedna and 2012 VP$_{113}$ (e.g. Trujillo & Sheppard, 2014; Batygin & Brown, 2016a, b; Brown & Batygin, 2016), a result that gained support by the discovery of additional objects moving on distant, detached orbits within the apparent cluster (Sheppard & Trujillo, 2016). Bromley & Kenyon (2016) considered a scenario in which such a planet initially formed in the vicinity of Jupiter and Saturn, and was scattered outwards onto a highly eccentric orbit. They found that, if the protoplanetary disc around the young Sun featured an extended gaseous disc (as is observed for protoplanetary discs around other stars), then the planet could have undergone orbital circularisation at a heliocentric distance greater than 100 au as a result of dynamical friction with the gas of that disc. Once the gaseous disc around the Sun was removed, such a planet could sculpt the orbits of objects well beyond Neptune, creating the observed clustering of the extremely distant TNOs (Batygin & Morbidelli 2017). Bailey, Batygin & Brown (2016) studied the manner in which such a distant planet would perturb the orbits of the four giant planets, and found that its influence could be sufficient to explain the observed misalignment (of approximately 6 degrees) between the spin axis of the Sun and the invariable plane of the Solar system.

Other explanations have, of course, been advanced to explain the observed clustering in the orbits of these distant objects - including recent work revisiting and reviving the idea that the structure of the Solar system's outer reaches might be the result of a stellar fly-by during the system's youth (Pfalzner et al., 2018). Perhaps the most convincing, however, is the suggestion that the observed clustering may be the result of the observational biases that are inherent in the surveys that have discovered those objects (Shankman et al., 2017). In that work, the authors examined the distribution of distant Solar system objects discovered by the Outer Solar System Origins Survey (OSSOS), which discovered eight such objects using the Canada-France-Hawaii Telescope between 2013 and 2017. Shankman et al. demonstrated that, whilst their discoveries might appear to fit the paradigm of a clustered set of orbits, in line with the Planet Nine theory, once the biases that afflicted the survey were properly accounted for, the orbital distribution of their discoveries "*is consistent with being detected from a uniform underlying angular distribution*". On the other hand, Brown (2017) analysed the biases involved in observations of Edgeworth-Kuiper belt objects, finding that such biases were unlikely to explain the alignment of TNOs located beyond $a = 230$ au.

The 'Planet Nine' theory has received widespread attention, and has caught the imagination of astronomers and the wider public alike, even spawning a citizen science search program, through the Zooniverse website[50] - a search that has yet to bear fruit. Similar searches using data from the WISE and NEOWISE survey (Fortney et al., 2016; Meisner et al., 2018) have failed to find any evidence of the proposed planet - but if it does exist, it seems likely that it will be detected in the coming years.

## 4.6 THE FORMATION AND EVOLUTION OF THE OORT CLOUD
Stretching halfway to the nearest star, the Oort cloud is one of the great mysteries of the Solar system. Whilst we know it is there, as a result of the continual flux of dynamically new comets maintaining the long-period comet population (e.g. Öpik, 1932; Oort, 1950), it is too distant to study directly. Instead, we must learn what we can from the long period comets it sends our way.

As we discussed earlier, those comets are injected from the Oort cloud through the combined effects of the galactic tide, and perturbations from passing stars (e.g. Feng & Bailer-Jones, 2015, and references therein), In coming years, it is likely that the combined efforts of the *Gaia* space telescope (Gaia Collaboration et al., 2016, 2018; Evans et al., 2018), surveying the motions and distances of approximately 2% of all stars in the galaxy, and the next generation of all sky surveys (such as the Large Scale Synoptic Telescope, LSST; LSST Science Collaboration et al., 2009; Ivezić et al., 2019), which will discover vast numbers of new comets, will reveal evidence of additional past encounters between the Sun and passing stars, to add to the Biermann comet shower (Biermann, Huebner & Lust, 1983). Within a decade, we may even get our first close-up view of a dynamically new comet, passing through the inner Solar system for the first time in more than four billion years, as a result of the recently announced Eurpean Space Agency mission 'Comet Interceptor'[51], scheduled for launch in 2028.

But what of the Oort Cloud's origins? It is clear that it can not have formed in situ - instead, it is thought to have been created as a side-effect of the accretion of the giant planets, and their subsequent clearing of the outer Solar system. As the giant planets accreted, they were embedded in a disc containing icy planetesimals of all shapes and sizes. Close encounters between the planets and the planetesimals would scatter the latter, and as the mass of the planets grew, so too did the efficiency with which such scattering events could eject those planetesimals to highly eccentric, and even unbound, orbits.

---

[50] https://www.zooniverse.org/projects/marckuchner/backyard-worlds-planet-9
[51] http://www.cometinterceptor.space/

The farther an object strays from the Sun, the weaker the Sun's gravitational influence upon it will be. Once a comet is sufficiently distant, the influence of nearby stars and the galactic tide will therefore be strong enough to significantly alter the orbit of that comet. As a result, there are three possible outcomes for a planetesimal ejected by the giant planets to a high eccentricity orbit, that takes them well beyond the orbit of Neptune. The first is that they are either ejected with such great velocity that they move on a hyperbolic orbit, never to return, or are flung sufficiently far (on a nominally bound orbit) that the influence of nearby stars and the galactic tide can decouple them from the Sun's influence. Such objects, clearly, will not become members of the Oort Cloud. At the other extreme, you have those comets whose aphelia are not sufficiently far from the Sun for passing stars and the galactic tide to significantly perturb them. Those objects will return to the regime of the giant planets unperturbed, no doubt to experience further scattering. Between the two extremes, you have a scenario whereby an ejected object obtains sufficient distance from the Sun for its orbit to be strongly perturbed by the influence of the galactic tide and passing stars, without becoming so distant that those perturbations remove it from the Solar system entirely. It is these objects that give birth to the Oort Cloud. The perturbations from the galactic tide and nearby stars act to lift the perihelia of those objects out of the region of the Solar system dominated by the planets, placing them on orbits where they can remain for billions of years, held in cold storage.

Over the billions of years that follow, the orbits of comets in the outer reaches of the Oort Cloud will slowly become randomised, as a result of their continual stirring by the galactic tide, and periodic nudges from passing stars. From a distribution that would initially resemble the disc from which those comets were sourced (i.e. relatively low inclinations with respect to the plane of the Solar system), the Oort Cloud will eventually obtain its current roughly spheroidal shape, with the inclinations of the comets therein wholly randomised.

The outer boundary, beyond which objects would have been effectively removed from the Sun's influence, would have been closer to the Sun if our birth cluster was particularly massive and dense (with more stars forming closer to the Solar system), and at a greater distance if the Solar system formed in a relatively loose cluster or stellar association. As noted at the start of section 4, evidence gathered from meteorites suggests that our birth cluster was relatively massive, containing at least 2,000 stars (e.g. Pfalzner et al., 2015). Such a tight cluster is invoked in order to explain the abundance of $^{26}$Al in the early Solar system, which it is argued was provided by the wind of a nearby massive star.

Dones et al. (2004) provide an excellent review of theories describing the formation and evolution of the Oort Cloud. They note that, on the basis of dynamical simulations, Uranus and Neptune likely dominated the emplacement of cometary material to the Oort Cloud - see their section four for a full overview of studies of the Oort Cloud's formation prior to that date. However, more recent work has reopened the question of the primary source of material to the primordial Oort Cloud. Brasser et al. (2012b) performed detailed numerical simulations of the formation of the Oort, taking account of the influence of the other stars in the Sun's birth cluster. They found that the inner Oort cloud was primarily populated by comets ejected by Jupiter and Saturn, with the pericentres of those comets then being raised by encounters with passing stars to decouple them from the planetary domain of the Solar system. As such, the question of the 'planetary parents' of the Oort cloud remains somewhat in doubt, and the true answer might well depend on the precise nature of the cluster in which the Sun formed, as well as the chaotic migration history of the giant planets. For a more detailed discussion of the Oort cloud's formation, we direct the interested reader to the recent review by Dones et al. (2015), which builds on their earlier work to provide a detailed overview of the formation of all of the Solar system's reservoirs of cometary body.

As discussed in section 4.4, it has been proposed that the level of deuteration trapped in the water of the Solar system's icy bodies could act as a tracer of their formation locations within that disc (e.g. Drouart et al., 1999; Dauphas, Robert & Marty, 2000; Mousis et al., 2000; Mousis, 2004; Horner, Mousis &

Hersant, 2007; Horner et al., 2008; Yang, Ciesla & Alexander, 2013). As such, Horner, Mousis & Hersant (2007) suggested that measurements of the D:H ratio in incoming comets could be used to test the various theories of their formation, showing how a variety of scenarios for the formation regions of cometary bodies would manifest as different distributions of deuteration in the nuclei we observe today. Following that logic, Kavelaars et al., 2011 compared the D:H enrichment measured in Saturn's icy satellite Enceladus with the levels of deuteration measured for a number of long-period comets. They found that the comets displayed similar levels of deuteration to the icy moon, suggesting that they had formed at roughly equivalent locations in the protoplanetary nebula. Given that the established wisdom is that the Oort Cloud was probably primarily populated by Uranus and Neptune, those authors considered this to be evidence that the majority of material in the Oort Cloud formed at a similar heliocentric distance to Saturn, which in turn would suggest that Saturn formed close to Uranus and Neptune. Such a scenario is exactly what was proposed by Thommes, Duncan & Levison (1999, 2002), and that formed the cornerstone of the Nice Model (Gomes et al., 2005; Tsiganis et al., 2005; Morbidelli et al., 2005).

Because of the tenuous grasp with which the Sun holds on to comets in the Oort cloud (particularly the outer reaches), the cloud should continually be bleeding objects to interstellar space, stripped from the Sun by the combined effects of passing stars and the galactic tide. It is fair to assume that other stars behave likewise - and a result, interstellar space should contain vast numbers of free-floating comets and asteroids, some of which, by chance, must occasionally pass through the inner Solar system. The first such object, 1I/'Oumuamua, was discovered in 2017 (e.g. Meech et al., 2017; Jewitt et al., 2017;' Bannister et al., 2017; Fitzsimmons et al., 2018) - and should another such object be discovered moving on a favourable path whilst the Comet Interceptor mission stands ready at Earth's L2 Lagrange point, that object would likely be considered an ideal target for the mission. In other words, in the relatively near future, that mission means that there is a chance that we may get our first 'up-close-and-personal' view of a comet or asteroid from another planetary system.

Just as passing stars perturb our Oort cloud, so too will the Sun perturb the Oort clouds of its stellar neighbours. Particularly whilst the Sun was still young, and moving within its birth cluster, this process of mutual Oort cloud stirring would have been pronounced, and one could readily envision a cluster awash with comets, torn from one parent star, co-moving within the cluster. For this reason, Levison et al. (2010) performed numerical simulations to investigate the degree to which comets could be transferred from one star to another - with the idea that our Oort cloud could contain a significant number of comets captured from other stars before the dispersal of our birth cluster. The authors conclude that a significant fraction of comets in the Oort cloud were indeed captured from our Sun's birth neighbours - with such comets potentially making up the vast majority (>90%) of nuclei within the cloud.

Whilst the capture of comets to the cloud might have been efficient while the Sun resided in its birth cluster, recent work by Hanse et al. (2018) has shown that the situation became markedly different once the Sun began wandering the galaxy alone, having left that cluster behind. They consider the capture of comets from field stars (and, by extension, the transfer of Solar comets to those stars) as those stars fly past the Sun. They find that capture events are relatively rare, since the typical encounter velocities of passing stars are far greater (by several orders of magnitude) than the orbital velocities of comets in the Oort cloud. They note that the transfer of any 'exocomets' to the Oort cloud will typically only happen during unusually slow ($\lesssim 0.5$ km s$^{-1}$) and close ($\lesssim 10^5$ au) encounters - and as a result, such exocomets likely make up just a small fraction of the mass of the cloud (between $10^{-4}$ and $10^{-5}$). In the same work, those authors were able to model how the Oort cloud is stripped of comets by stellar encounters, suggesting that such encounters have likely caused the Oort cloud to lose between 25% and 65% of its initial mass through the age of the Solar system.

As we discussed in section 4.5, comets from the Oort cloud have been invoked as the cause of periodic mass extinctions on Earth since the early 1980s. In order to explain the suspected ~26 Myr periodicity in mass extinctions reported by Raup & Sepkoski (1984, 1986), several authors discussed the possibility that the Sun might have an undetected binary companion, a dim red dwarf star called 'Nemesis', whose motion through the Oort cloud sent periodic showers of comets to imperil life on Earth (e.g. Davies, Hut & Muller, 1984; Whitmire & Jackson, 1984; Alvarez & Muller, 1984) - an idea that quickly fell into disfavour (e.g. Bailey, 1984; Clube & Napier, 1984a; Hills, 1984; Hut, 1984; Torbett & Smoluchowski, 1984).

Schwartz & James (1984) proposed a different mechanism by which the Oort cloud could repeatedly send comet swarms towards the Earth, noting that the ~26 Myr periodicity in extinctions found by Raup & Sepkoski (1984) was remarkably similar to the period with which the Solar system oscillates about the plane of the galaxy. Whilst it is typical to imagine the orbital motion of the Sun around the galaxy as being roughly circular (with an orbital period of ~225 Myr), the true story is somewhat more complicated. As it moves around the galactic centre, the Sun also undergoes a periodic vertical oscillation back and forth through the plane of the galaxy. When the Sun is located above that plane, the mass of stars, gas and dust within it exerts a downward force, pulling our star (and therefore the Solar system) down towards the plane. As a result, the system accelerates, falling downwards until it passes through the galactic plane and out the other side. Once it is below the plane of the galaxy, the system now feels the same restoring force pulling upwards, causing its vertical motion to slow, stop, and then reverse. The result is that the Sun's motion around the galactic centre is somewhat reminiscent of an automated fairground carousel ride, with the Sun oscillating up and down as it moves around the galactic centre.

Schwartz & James (1984) proposed that it was this motion that was to blame for the periodic extinctions noted by Raup & Sepkoski (1984). The vertical oscillations of the Sun result in the Solar system passing through the plane of the galaxy repeatedly, with those passages occurring with a frequency comparable to the periodicity proposed for Earth's mass extinctions. As the system passes through the galactic midplane, encounters with stars, and with giant molecular clouds become more likely, increasing in turn the likelihood that the Earth will collide with a comet flung inwards by such an encounter. The idea was revisited by Randall & Reece (2014), who invoked the influence of a smooth disc of dark matter, located at the midplane of the galactic disc, in order to explain how the Oort cloud could always be guaranteed to trigger a comet shower - something that had proved problematic when the only influences on the cloud were considered to be encounters with stars and giant molecular clouds.

An alternative theory connecting the Sun's motion through the galaxy was proposed by Leitch & Vasisht (1998), who suggested that several of Earth's largest extinction events coincided with periods at which the Sun was moving through one or other of our Galaxy's spiral arms. The spiral arms of the Milky Way (and of other spiral galaxies) are regions within which there is a greater concentration of dust, gas, and star formation - and as a result, a greater stellar density. When the Sun moves through one of those arms, then, the likelihood of close encounters with stars (which could inject comets to the inner Solar system) is higher. At the same time, the chance of an extinction event caused by a nearby supernova (the death of one of the galaxies most massive stars) is also greatly enhanced, since those stars are almost exclusively found within the spiral arms. Gilman & Erenler (2008) and Filipovic et al. (2013) revisited this idea, using updated maps of the Milky Way's structure to attempt to determine whether any correlation between such spiral-arm crossings and mass extinctions could be found.

It is fair to say, however, that the jury is still out on whether mass extinctions on Earth are truly periodic, and even if they are, many of the theories proposed to explain such periodic extinctions are considered by many to be fanciful or unfeasible. Nevertheless, it is clear that comets from the Oort cloud have posed, and will continue to pose, an impact threat to the Earth - and the origin and evolution of the Oort cloud will remain a hot topic for research for many years to come.

**4.7 DUST AND DEBRIS IN THE SOLAR SYSTEM**
When we look out at the Solar system, we see dust everywhere. On any clear night, we observe dust ablating in Earth's atmosphere, producing meteors, and their brighter siblings fireballs and bolides (e.g. Jenniskens, 2006). With the unaided eye, we can also observe the Zodiacal Light and Gegenschein (e.g. Roosen, 1970; Leinert, 1975), both the result of sunlight scattering from dust in the plane of the Solar system. Whenever an active cometary nucleus is sufficiently close to the Sun, the sublimation of its surface ices will result in the continual ejection of dust grains - which will be pushed away from the Sun to produce those comet's spectacular dust tails. And when those comets fragment (as they often do; e.g. Crovisier et al., 1996; Sekanina, 2000; Jenniskens & Lyytinen, 2005; Jenniskens & Vaubaillon, 2007), or go into outburst (e.g. Sekanina et al., 1992; Dello Russo et al., 2008; Montalto et al., 2008; Ye & Clark, 2019), large quantities of dust are released - occasionally even resulting in spectacular meteor showers, as that dust encounters Earth (e.g. Asher, 1999; Wiegert et al., 2005; Jenniskens & Vaubaillon, 2005; Jenniskens, 2016).

Asteroids, too, are a ready source of dust. The near-Earth asteroid 3200 Phaethon (the parent of the Geminid meteor shower; e.g. Gustafson, 1989; Williams & Yu, 1993) behaves as a 'rock comet', shedding dust near perihelion as a result of surface weathering, caused by the extreme temperatures to which it is subject through the course of its orbit (e.g. Jewitt & Li, 2010; Li & Jewitt, 2013). More often, asteroids shed dust as a result of collisional processes - from the small scale sputtering caused by the continual hypervelocity impacts of dust on their surface, through to the rare but catastrophic collisions that shatter whole objects, creating vast quantities of dust in the process (e.g. Sykes & Greenberg, 1986; Sykes, 1990; Reach, 1992; Durda & Demott, 1997; Nesvorný et al., 2002; Spahn et al., 2019), and even through activity like that of comets (e.g. Hsieh, Jewitt & Fernández, 2004; Hsieh & Jewitt, 2006; Jewitt, Yang & Haghighipour, 2009; Jewitt, 2012; Agarwal et al., 2017; Chandler et al., 2019).

Farther from the Sun, too, dust is abundant - from the dusty debris orbiting several of the Solar system's small bodies (e.g. Stansberry et al., 2004; Braga-Ribas et al., 2014; Ortiz et al., 2015; Ortiz et al., 2017) to the ring systems around the giant planets (e.g. Elliot, Dunham & Mink, 1977; Smith et al., 1979; Goldreich & Tremaine, 1982; Smith et al., 1989). In short, dust is a pervasive presence in the Solar system.

**4.7.1 THE TEMPORAL EVOLUTION OF DUST IN THE SOLAR SYSTEM - STEADY STATE OR TIME-VARIABLE?**
Given the wide variety of sources for that dust, it is natural to assume that the amount of dust in the system is roughly in steady state - in other words, to assume that we do not observe the system from a privileged, unusual epoch in the system's history, at which dust levels are elevated above the norm. However, that assumption may well be invalid. As we discussed earlier, evidence is growing that Saturn's spectacular ring system is a relatively recent feature of the Solar system (Iess et al., 2019), and that those rings may fade and decay over the next few hundred million years (O'Donoghue et al., 2019). Dust in the Solar system is relatively short lived (for reasons we will discuss in section 4.9), and as such, absent any sources to replenish the system's dust budget, one would expect the Solar system to become almost dust free on a relatively short timescale. To put this another way - should an event happen to inject large quantities of dust and debris to the Solar system (such as the collision of two large asteroids, or the decay and fragmentation of a giant cometary nucleus), the dust created from that event would decay over time.

The Solar system today contains at least three features that might be considered evidence of such stochastic events, each of which has contributed to the fact that the current dust and debris budget of the system can be considered to be enhanced over the potential 'steady state' background that would be generated by the ongoing collisional sputtering of dust from the asteroid and trans-Neptunian belts, and from the outgassing of comets.

The first of these, as described above, is the existence of Saturn's spectacular ring system. If the work of Iess et al. (2019) and O'Donoghue et al. (2019) holds true, those rings can be considered a transient phenomenon, rather than being considered to be a permanent fixture of the system. Whilst the rings may decay on timescales of hundreds of millions of years, the timescales of their formation and decay could be taken to suggest that the other giant planets, whose ring systems are currently minor in comparison to those of Saturn, might once have hosted spectacular, fully fledged rings. It would be interesting to discover for what fraction of the Solar system's history one or other of the giant planets has displayed

Saturn-like rings - and it might well be the case that future observations of Jupiter- and Saturn-analogue exoplanets might help us to determine whether such rings are the temporal exception, or the norm.

The second unusual feature is the plethora of 'sungrazing comets', moving on near-identical orbits that cause them to skim the Sun's surface at perihelion (e.g. Marsden, 1967). Thanks to the work of the SOHO spacecraft, in recent years, the number of known sungrazing comets has grown dramatically, with that spacecraft alone discovering several thousand such objects - most of which either disintegrated as they approached the Sun, or never survived the rigours of their perihelion passage (e.g. Battams & Knight, 2017).

Of particular interest among the sungrazing comets are the Kreutz family (after Kreutz, 1888), which number among them some of the most spectacular comets of the last millenium, visible, at their brightest, in broad daylight. These include the Great Comets of 1106, 1843, 1882, and, more recently, comet C/1965 S1 Ikeya-Seki. Members of the Kreutz family move on near-identical, retrograde, orbits, inclined to the plane of the Solar system by ~140 degrees. As more members have been discovered, it has been possible to group them into several sub-families, each of which features slightly different orbital elements (e.g. Marsden, 1967, 1989).

In order to explain the vast number of Kreutz sungrazers, a process of ongoing fragmentation has been proposed, with the idea that all of the observed sungrazers are fragments of a vast parent body, possibly 100 km (or more) in diameter, that first fragmented at least two millennia ago (e.g. Marsden, 1968, 1989; Sekanina & Chodas, 2004, 2007). This theory is supported by the fact that the great majority of family members fragment as they pass close to the Sun. So, for example, the Great Comet of 1106 fragmented as it passed by the Sun, and the Great Comet of 1882, and comet C/1965 S1 Ikeya-Seki were the first returns of two of the larger fragments created in that event (e.g. Hasegawa & Nakano, 2001).

The different sub-groups within the Kreutz family are thought to be objects that all tie to a particular fragmentation event, farther back in time - with each subsequent fragmentation giving birth to a new sub-sub family, and so on. Whilst it is natural to assume that those fragmentation events would happen at, or around, perihelion, when the combination of thermal and tidal stresses on the nucleus are the greatest (e.g. Bailey, Chambers & Hahn, 1992; Sekanina & Chodas, 2012), there is growing evidence that some Kreutz fragmentation events happen far from the Sun, most likely as a result of thermal stress from the penetration of heat from the previous perihelion passage to the comet's interior (e.g. Sekanina, 2001; Sekanina & Chodas, 2004, 2012).

Based on the current spread of the sub-groups, it has been suggested that the origin of the Kreutz sungrazing family might be tied to a spectacular comet observed in the year -371 (i.e. 371 BC), documented by Aristotle and Ephorus (e.g. Marsden, 1989), or the bright comet of -214 (i.e. 214 BC) (e.g. Sekanina & Chodas, 2007). In either case, it is suggested that the parent of the Kreutz sungrazer population was a true behemoth of a comet - with a nucleus of at least 100 km in diameter. Whilst such a large nucleus might at first sound unfeasible, there is evidence that at least two of the great comets in the historical record had nuclei of comparable size - these being C/1729 P1 Sarabat (which was visible with the naked eye for several months, despite a perihelion distance of over 4 au) and C/-43 K1 (often known as Caesar's Comet)[52], whilst comet C/1996 O1 Hale-Bopp, which was visible with the unaided eye for a record-breaking 18 months, had a nucleus of diameter ~60 kilometres in diameter (e.g. Fernández, 2002).

Looking forward, it is virtually certain that we will see more spectacular members of the Kreutz sungrazing family, as the ongoing fragmentation and decay process will likely continue for several thousand years, or more. Indeed, some studies predict the arrival of another cluster of bright Kreutz members in the coming decades (e.g. Sekanina & Chodas, 2007; Sekanina & Chodas, 2012; Sekanina &

---

[52] Descriptions of the apparitions of these two comets can be found in Volume 1 of Kronk's Cometography (Kronk, 1999)

Kracht, 2013), of which the comet C/2011 W3 Lovejoy may have been the first substantial member (Sekanina & Chodas, 2012).

In addition to the famous Kreutz sungrazing group, there are at least three other families of sungrazing comets, identified as a result of the plethora of discoveries made by the SOHO spacecraft. These families - the Marsden, Meyer and Kracht groups (e.g. Jones et al., 2018) - are also thought to be relics of the fragmentation of other large sungrazing objects. Indeed, it seems likely that both the Marsden and Kracht groups are part of a broad swath of debris throughout the inner Solar system that includes the highly eccentric short-period comet 96P/Macholz, and several meteor streams (including the delta Aquariids, the daytime Aqauriids, and the Quadrantids; e.g. McIntosh, 1990; Gonczi, Rickman & Froeschle, 1992; Ohtsuka, Nakano & Yoshikawa, 2003; Sekanina & Chodas, 2005; Jenniskens, Duckworth & Grigsby, 2012).

All this suggests that such families may well be a relatively common occurrence through the Solar system's history - a finding supported by dynamical studies of the long term evolution of cometary orbits, which suggests a significant fraction might eventually evolve to sungrazing orbits (e.g. Bailey, Chambers & Hahn, 1992; Bailey et al., 1996). The sungrazing comet population is therefore expected to be continually in flux, with new families being born as a natural consequence of cometary evolution. As the number of known sungrazers continues to climb in the coming years, it seems likely that other families may become apparent, relics of more ancient cometary fragmentation episodes. For more information on sungrazing comets, we direct the interested reader to the excellent review by Jones et al. (2018).

The third of the Solar system's unusual dust features is known as the 'Taurid Stream', a swath of debris in the inner Solar system that is so broad that it encounters all four terrestrial planets. Through the northern hemisphere late summer and autumn, the stream produces the Northern and Southern Taurid meteor showers on Earth[53], and is also responsible for the active daylight Beta Taurid meteor shower, in June and July each year. In other words, the Taurid stream is so broad that the Earth spends fully one third of each year traversing it. And it is not just the Earth that encounters the Taurid stream - so broad is the swath of debris that all four terrestrial planets encounter it. On Mercury, impacts from Taurid meteoroids have been linked to an annually repeated excess of Ca in the planet's exosphere (e.g. Christou, Killen & Burger, 2015), whilst both Venus and Mars spend a significant fraction of their orbits moving through the broad stream, at both its ascending and descending nodes (e.g. Christou, 2010).

The Taurid stream contains (at least) one active comet - 2P/Encke, which has long been considered to be the parent of the meteor showers produced by the stream (e.g. Whipple, 1951; Kresak, 1978; Babadzhanov, Obrudov & Makhmudov, 1990). Whipple (1951) notes that comet Encke must '*originally have been massive to have persisted so long in a short-period orbit*' - since the breadth of the stream suggests a grand old age. In the early 1990s, several studies examined the Taurid stream, and their relationship to comet Encke, suggesting that the stream is at least 20,000 years old (e.g. Babadzhanov, Obrubov & Makhmudov, 1990; Steel, Asher & Clube, 1991). Steel, Asher & Clube (1991) note that the stream's initial progenitor must have been a giant comet that was captured into a particularly short period orbit[54] before undergoing a process of ongoing fragmentation. That process generated a significant fraction of the dust we observe in the Zodiacal cloud, with Comet Encke being the largest, and most active, remaining fragment.

The idea that the Taurid stream is the result of an ongoing process of cometary disintegration found support with the discovery of a number of near-Earth asteroids moving within the Taurid complex (e.g.

---

[53] (active from 10th September - 20th November, and 20th October - 10th December, respectively, according to the 2020 edition of the International Meteor Organisation's Meteor Shower Calendar, which can be found at https://www.imo.net/files/meteor-shower/cal2020.pdf ; accessed 29/6/19.
[54] Comet 2P/Encke has the shortest orbital period of any of the 'traditional' comets, at 3.3 years, by some distance. The only comets with shorter orbital periods are the Main Belt Comets.

Asher, Clube & Steel, 1993; Babadzhanov, 2001; Porubčan, Kornoš & Williams, 2006; Babadzhanov, Williams & Kokhirova, 2008). To generate such a broad stream of debris, the diameter of the parent object has been estimated as being at least ~100 km (e.g. Asher & Clube, 1993; Clube et al., 1996), with a number of authors suggesting it was most likely a former Centaur, captured to a short-period orbit (e.g. Clube & Napier, 1984b; Horner et al., 2003; Napier, 2015) after a close encounter with Jupiter. Asher & Clube (1997) suggest that the parent of the stream may have been captured to the inner Solar system as much as 50,000 years ago, becoming trapped in a near-sungrazing orbit in 7:2 mean-motion resonance with Jupiter. Over the millennia that followed, the nucleus of that giant comet would gradually fragment as a result of repeated tidal stress, before experiencing a final, cataclysmic disruption as a result of a very close encounter with one of the Solar system's planets (potentially Mercury), approximately 5,000 years ago.

Given the large amounts of material contained within the Taurid stream, it has long been considered of interest as the source of potential impactors at the Earth. On June 30th, 1908, a large meteoroid exploded over Siberia, levelling almost 2200 square kilometres of forest - an impact thought to be the largest on Earth for at least a thousand years. Based on calculations of the radiant point for that impactor performed by Zotkin & Tsikulin (1966), Kresak (1978) suggested that the impactor was a member of the Beta Taurid meteor shower, and that it might therefore have been a fragment of Comet Encke, a finding supported by updated radiant calculations presented in Zotkin & Chigorin, (1991). Asher & Steel (1998) showed that a genetic link between the two objects was entirely feasible - suggesting that the two bodies could have separated as recently as 10 kyr ago - a timescale well within the longer lifetime of the Taurid stream as a whole. Hartung (1976) suggested that the impact crater Giordano Bruno, on the Moon, was formed in an impact on 18th June[55] 1178, based on observations reported to Gervase of Canterbury by '*five or more men who were sitting there facing the moon*'. This, in turn, led to the suggestion that the cratering event was linked to the Beta Taurid shower, and comet Encke (e.g. Hartung, 1993), though more recent work (Morota et al., 2009) has found that the crater is far too old for this to be the case - putting its age in the range 1 - 10 Myr.

In the early 1980s, there was wide discussion of the idea that mass extinctions on Earth occurred periodically, as a result of cometary impacts sourced from the Oort cloud (e.g. Raup & Sepkoski, 1984, 1986; Schwartz & James, 1984). Building on that narrative, Clube & Napier (1984b) suggested that, for the last 20,000 years, the Earth has been experiencing a 'spike' in impact flux, related to the disintegration from a Chiron-sized (i.e. D ~200 km) comet - the progenitor of Encke and the Taurid stream. They suggested that the debris released in that disintegration had a marked impact on the environment of the Earth during that time - driving climatic variations that helped to cause the last of Earth's glaciations. Asher & Clube (1993) built on that work, suggesting that the influence of the Taurid stream could have contributed to intermittent periods of cooling at roughly ~200 year intervals over the past millennium, at times when Earth would have encountered particularly dense patches of debris, trapped within the 7:2 mean-motion resonance with Jupiter. Clube et al. (1996) linked the historical evolution of the Taurid swarm with a variety of ancient catastrophes on Earth - '*One could regard otherwise enigmatic events in history such as the sudden collapse of the Indus Valley Civilization of Mohendojaro and of the Old Kingdom in Egypt ... , both occurring at ~2500 BC, as fitting well with the precession of the primary orbital nodes ... and the (Taurid) cometary collision picture in general.*'. They again tied the cooling that triggered the last ice age to the influence of impacts from the Taurid stream - along with tying the stream to events in the Old Testament of the Bible, and the end of the Roman Empire.

This theory, which became known as 'Cometary Catastrophism', still occasionally rears its head (e.g. Napier, 2010, 2015) - and recent years have seen a growing interest in the study of a relatively dense core to the Taurid stream, as noted in Asher & Clube (1993), trapped in 7:2 mean-motion resonance with Jupiter. That stream undergoes periodic close approaches to the Earth, causing an uptick to Taurid activity, and enhanced numbers of bright fireballs from the stream (as was observed in 1998, 2005, and 2015; e.g. Olech et al., 2017; Spurný et al., 2017). In June 2019, the Earth was forecast to experience its closest approach to the centre of that resonant swarm since 1975 - albeit at a distance large enough that any impact risk was expected to be negligible. Clark, Wiegert & Brown (2019) proposed to take advantage of this close approach to search for near-Earth objects moving in the swarm, detailing the optimal observation strategy to take advantage of the encounter. At the time of writing, no results have been published detailing the success (or otherwise) of this strategy, suggesting that any such work may have proven unsuccessful. It remains likely, however, that in the coming years, new asteroidal members of the Taurid swarm will be discovered, with such finds likely growing in number as we approach our next encounter with the centre of the swarm, in the early 2030s.

---

[55] or July - in the paper, Hartung gives the date first as 18th July, and later as 18th June

It is clear, on the basis of these three distinct features of the Solar system (Saturn's rings, the Kreutz sungrazing comets, and the Taurid stream), that the amount of dust and debris in the Solar system can vary dramatically, and its seems highly unlikely that we are currently seeing the Solar system's steady-state dust background. Instead, we are likely in a period of enhanced dust levels in the inner Solar system, with the amount of dust gradually decaying following the disintegration of the Taurid progenitor, tens of thousands of years ago, and the Kreutz progenitor, in the past few thousand years. It might well be the case that for much of the Solar system's evolution, the amount of dust present was lower than that we see today - with long periods of quiescence interspersed with short periods of enhanced dust loading, as giant comets fragment, or asteroids collide.

**4.7.2 CHONDRITES: A WINDOW INTO THE FORMATION OF THE SOLAR SYSTEM**
Primitive meteorites contain a direct record of the environment and the physical processes that took place in the early Solar system. Studies of those meteorites complement analytical and numerical models of Solar system formation, as well as observational studies of current star formation, but putting this record in its astronomical context is not trivial. Chondrite meteorites, in particular, are regarded as the 'building blocks' of the Solar nebula, out of which the Solar system formed (Wood, 1962). They are the most common type of meteorites (about 90% of observed falls[56]), and are subdivided into ordinary, carbonaceous, and enstatite classes according to mineralogical, chemical and oxygen isotope systematics (Weisberg et al., 2006; Krot et al., 2007; Scott & Krot, 2005b, 2014).

The *Hayabusa* sample return mission to the asteroid 25143 Itokawa (e.g. Yano et al., 2006) showed that ordinary chondrites are related to the S-class asteroids (Nakamura et al., 2011; Yurimoto et al., 2011), the second most common class of asteroids. Two other missions of this type are underway (*Hayabusa* II to asteroid 162173 Ryugu; Tsuda et al., 2013, and *OSIRIS-REx* to asteroid 101955 Bennu; Lauretta et al., 2017). These are C-type asteroids; potentially linked to the carbonaceous chondrites (e.g. Burbine et al., 2002; McSween, 2011). These meteorites (particularly the CI sub-type) have elemental abundances close to the Solar photosphere (e.g. Lodders et al., 2009), and are regarded to be a good representation of the composition of the early Solar system (e.g. Desch et al., 2018). They also contain large amounts of water (approximately 12-13 wt%; Alexander et al., 2013), which is thought to be consistent with their having accreted an initial water-to-rock mass ratio closer to the Solar system primordial value (~1.2; Krot et al., 2015) than is the case for other well characterised meteoritic material.

Chondrite meteorites are, essentially, cosmic 'breccias' aggregated from the ambient material in the early Solar system. They provide unique information on the timescales for the formation of the first solids, as well as on the dynamical, thermal and chemical properties of the Solar nebula (e.g. Jones et al., 2000; Wadhwa & Russell 2000; Krot et al., 2009). They are composed of three separate constituents (e.g. Alexander et al., 2007; Davies et al., 2014): Chondrules, Calcium-Aluminium rich Inclusions (CAI) and Matrix. Chondrules are millimetre-sized sub-spherical inclusions of magnesium-silicate minerals, predominantly olivine and pyroxene. They are a major component of chondrites (up to 80% by volume, depending on the chondrule class; Connolly 2005; Scott & Krot, 2014). Their properties – texture, shape and mineralogy -- indicate that they were molten prior to their incorporation into the meteorite parent bodies, with melting temperatures of order 1800 K (Petaev & Wood, 2005; Scott & Krot, 2005a). They are thought to have formed rapidly (in a process lasting minutes to hours), with rapid cooling rates of ~10 – 1000 K/hr (e.g. Jones et al., 2000; Ciesla, 2005). Moreover, compound chondrules and the presence of dusty rims seem to indicate that some of them may have experienced multiple processing events.

---

[56] Meteorites can be divided into two categories, based on the circumstances of their recovery. 'Falls' are those meteorites whose flight through the atmosphere (as a bright fireball) is observed, and which are located on the ground just hours or days after the appearance of their parent fireball. 'Finds' are much more common - meteorites located on the surface of the Earth having lain there for years, decades, or even longer. Essentially, we see 'falls' fall to Earth, whilst 'finds' are found fortuitously long after they reached our planet. 'Falls' are of particular scientific value, since, if recovered quickly enough, they can be considered to be pristine, and unaffected by chemical weathering/alteration on the surface of our planet. For this reason, global camera networks are now being established (Devillepoix et al., 2020), such as the Australian Desert Fireball Network (e.g. Bland et al., 2012; Howie et al., 2017; Devillepoix et al., 2018; Sansom et al., 2019), to maximise our chances of observing 'falls', and being able to locate the meteorites they produce in a timely fashion.

CAI are regarded as the oldest and most refractory Solar system solids, with their formation dated to 4.567 Gyr (Amelin & Ireland, 2013). They are thought to have experienced even higher temperatures than chondrules, of order 1700 – 2000 K (Rubin, 2000; Richter et al., 2002; Grossman, 2010) over timescales of hours to days. CAI can be up to 13% by volume of chondrites, and exhibit sizes in the range of a few micrometres to centimetres. Their compositions suggest that they condensed out of a solar-composition gas (Grossman, 1972). Puzzlingly, CAI and chondrules are embedded in a fine-grained mixture (matrix) of silicates, oxides, organic material and presolar grains, which appears never to have experienced temperatures beyond ~400 K. Moreover, even though the proportion and composition of chondrule and matrix material can vary among chondrites types, the overall chemical composition of each chondrite group is approximately constant, indicating a level of complementarity between these components (e.g. Bland et al., 2005; Zanda et al., 2006; Hezel & Palme, 2008; Krot et al., 2009; Desch et al., 2010). In turn, this suggests a close connection between the high-temperature process leading the formation of chondrules, and the preservation of the (complementary) cold matrix in the nebula.

The prevalence of chondrules and CAI in meteorite samples suggests that they formed via a general process, closely associated with the star and planet formation mechanism (e.g. Wood, 1962; Ciesla et al., 2003; Scott & Krot, 2005a; Morris & Desch, 2010). Their formation timescales are broadly consistent with the inferred protostellar disc lifetimes ($10^6$–$10^7$ years; Calvet et al., 2000; Hartmann et al., 2016); but CAI are thought to have formed over a shorter period (< 0.1 to ~ 0.3 Myr) during the embedded phase of Solar formation (Bizzarro et al., 2004; Scott & Krot, 2005b), whereas chondrules may either have formed at that time (Bizzarro et al., 2004) or during a period ranging up to 4 Myr later (Kita et al., 2005; Russell et al., 2005; Scott & Krot 2005a). However, the mechanism(s) responsible for their thermal processing has remained poorly understood for many decades. The presence of presolar grains in meteorite samples (Anders & Zinner, 1993; Nittler, 2003), geochemical evidence for a poorly mixed nebula (Clayton, 2005; Jacobsen & Ranen, 2006), and direct evidence from astronomical observations all point to a generally cold protoplanetary disc environment, with insufficient temperatures to melt precursors at the radial distances they sample (~3 au; e.g. Cassen, 2001). Proposed mechanisms have included shock waves (Hood & Horanyi, 1993; Connolly & Love, 1998; Ciesla et al., 2003; Morris & Desch, 2010, Boley et al., 2013; Mann et al., 2016), collisions between asteroids (Lichtenberg, 2018) and processing close to the Sun followed by ejection to the outer Solar system (e.g. via X-winds; Shu et al., 1996).

Processing via shocks appears to be broadly consistent with some chondrule thermal histories inferred from laboratory experiments (Desch et al., 2005; Morris & Desch, 2010). However, the origin of these shocks is not fully determined. On the other hand, processing in the proximity of the early Sun is problematic, as dust may be absent at such distances (Muzzerolle et al., 2003) and it may be difficult to explain the chondrule-matrix complementary composition.

Another mechanism has been proposed (Salmeron & Ireland, 2012a,b), which invokes formation in a laterally extended, magnetocentrifugal wind accelerated from the disc surfaces (Blandford & Payne, 1982; Pudritz et al., 2007; Königl & Salmeron 2011; Frank et al., 2014). Young stars often exhibit powerful outflows of material that are accelerated away along the polar axis of the star – disc system, and current observations suggest that they may extend to radial distances ~ a few au from the protostar (e.g. Anderson et al., 2003; Coffey et al., 2004, 2007; Bjerkeli et al., 2016). This mechanism is attractive because thermal processing occurs where meteorite parent bodies were formed (the disc) and it naturally explains basic chondrule/chondrites properties (e.g. their tight size range, efficient processing, peak temperatures, compound chondrules, and complementary composition). However, many more properties would need to be explored in the context of this formation scenario, and further modelling is warranted. This is a rapidly evolving field, and harmonising astronomical observations, evidence from remote sensing missions, and studies of meteorites and sample returns is an important objective of early Solar system research, which will drive further theoretical, experimental and numerical work into the future.

**4.8 SMALL BODIES WITH RINGS - CHARIKLO, CHIRON AND HAUMEA**
Over the past decade, narrow ring systems and other orbiting material have been discovered orbiting several of the Solar system's small bodies (as detailed below, in Table 3). Such gravitationally bound material is thought to be able to persist for long periods of time, even in the absence of cometary activity from the host object, and may take the form of orbiting fragments (essentially moonlets), arcs or rings.

The first definite evidence for material orbiting distant Solar system small bodies came with the discovery of bona-fide satellites, such as Charon, Nix, Hydra, Styx and Kerboros around Pluto. Given the suggested collisional origin of those satellite families, it seems likely that their formation was preceded by the existence of temporary ring systems orbiting their parent bodies, as the material ejected from those bodies accreted to form its new moons.

In the last decade, orbital fragments (moonlets) have been detected around the Centaurs 29P/Schwassmann-Wachmann 1 (Stansberry et al., 2004; Gunnarsson et al., 2008; Womack et al., 2017) and (60558) 174P/Echeclus (Rousselot, 2008; Fernández, 2009). More excitingly, however, occultation observations of the dwarf planet Haumea and the Centaurs 10199 Chariklo and 2060 Chiron have resulted in the detection of narrow ring systems in orbit around them - just as, in 1977, the rings of Uranus were discovered during observations of a stellar occultation by the planet from the Kuiper Airborne Observatory (e.g. Elliot, Dunham & Mink, 1977).

The first set of rings to be discovered orbiting an object other than one of the giant planets were confirmed around the Centaur 10199 Chariklo (Braga-Ribas et al., 2014). That work made use of a network of telescopes distributed across South America to observe the occultation of a faint (R=12.4 mag) star by the Centaur. Such occultation observations are a great boon to researchers, as they allow the true size and shape of an object such as Chariklo to be determined on the basis of the durations of the occultation as observed from a variety of locations, in addition to information on any locations from which the occultation does not occur. By piecing together the various occultation 'chords' observed in this way, it is possible to reconstruct the projected size and shape of the object in question. In the case of Chariklo, the sites that observed the occultation recorded the occulted star dimming and brightening on two occasions before the true occultation began, then repeating that behaviour after the conclusion of that main occultation event. On the basis of their observations, Braga-Ribas et al. found that Chariklo hosts two narrow, dense rings, with orbital radii 391 and 405 km and widths ~7 km and ~3 km, respectively.

Once the rings around Chariklo had been announced to the world, Ortiz et al. (2015) returned to earlier observations of the Centaur 2060 Chiron, which had exhibited a similar behaviour during its own occultation of a faint background star. The observed pre- and post-occultation dimming of Chiron had originally been explained as being the result of jet-like features around the Centaur's nucleus (Ruprecht et al., 2013), which has long been known to exhibit cometary activity throughout the entirety of its orbit (e.g. Luu & Jewitt, 1990; Meech & Belton, 1990; Bus et al., 2001). Instead of being the result of such jetting, Ortiz et al. (2015) proposed that Chiron, too, hosts a dense, narrow ring, some 10 km wide, with an orbital radius of approximately 324 km.

More recently, observations of a stellar occultation by the dwarf planet 136108 Haumea, on 21 January 2017 revealed that, in addition to its satellites, Hi'iaka and Namaka, that dwarf planet, too, hosts a dense, narrow ring (Ortiz et al., 2017; Winter, Borderes-Motta & Ribeiro, 2019). The orbital radius of that ring is far wider than those around Chariklo and Chiron (which is not unexpected - indeed, the diameter of Haumea is so great that the rings of the two Centaurs would lie within its solid body), at ~2,287 km. Its width is also significantly larger than those around the two Centaurs, estimated at ~70 km. Interestingly, the ring appears to be coplanar with the orbit of Haumea's outer moon, Hi'iaka, and the equatorial plane of the dwarf planet - suggesting a potential common origin for the moons and rings that is entirely in keeping with the idea that they were created in a significant impact event. Additional evidence for such an event can be found in the extended Haumea collisional family (e.g. Brown et al., 2007; Ragozzine & Brown 2007; Schlichting & Sari, 2009; Leinhardt, Marcus & Stewart, 2010; Lykawka et al., 2012; Volk & Malhotra, 2012; Villenius et al., 2018), as discussed in section 4.1.2.

The exact origin of the rings around the two Centaurs remains unclear, but theories include debris from cometary activity (Pan & Wu, 2016; Wierzchos et al., 2017), a collision between the small body and another body, or between an orbiting satellite and another body (Melita et al., 2017), the tidal disruption of an orbiting satellite (El Moutamid et al., 2014), and the tidal disruption of the Centaur itself due to a close encounter with a planet (Hyodo et al., 2016). At this time, no single origin theory can be discounted, and observations of the two Centaurs only hint at possible origin scenarios. 2060 Chiron has been known to display cometary activity at heliocentric distances beyond the water-ice sublimation distance of approximately 3 au (Bus et al., 1991; Womack & Stern, 1999; Jewitt, 2009; Womack, 2017), which might well help to maintain its ring system. To date, 10199 Chariklo has not been observed to display such activity, but it is certainly possible that it may have done so in the relatively recent past (Wood et al., 2017). So, ring formation by cometary activity remains a possibility.

Given the frequency with which Centaurs are expected to experience close encounters with the giant planets, Wood et al. (2017, 2018) performed detailed $n$-body simulations to determine whether those encounters could be close enough to either disrupt the rings, or to help generate them (through tidal disruption of their parent bodies). Those simulations revealed that encounters that are close enough to create or disrupt rings around Centaurs should be very rare - sufficiently rare, in fact, that it seems unlikely that either Chariklo or Chiron would have experienced such an encounter during their time in the Centaur region. In other words, the results of those simulations argue against the idea that the rings of either body were created by a close encounter with a giant planet. Furthermore, the lack of disruptive encounters during the lifetimes of those Centaurs means that a primordial origin for the rings (or, at least, an origin that predates the transfer of their hosts to the Centaur region) cannot be ruled out.

In the coming years, it seems highly likely that further occultation observation programs will reveal additional rings orbiting more of the Solar system's small bodies - indeed, given the scarcity of such observations for those small bodies in the past, the fact that three ring systems have already been discovered may argue that such rings are in fact relatively commonplace - revealing once again that our knowledge of the minutiae of the Solar system's small body populations remains incomplete, and serving as a reminder that there is always something new to discover!

| Object | Orbital Radius (km) | Widths (km) | Ref |
|---|---|---|---|
| Chariklo | 390.6 ± 3.3 (inner)<br>404.8 ± 3.3 (outer) | 7.17 ± 0.14 (inner)<br>3.4 $^{+1.1}_{-1.4}$ (outer) | [1] |
| Chiron | 324 | 10 | [2] |
| Haumea | 2,287 | 70 | [3] |

*Table 3:* *The properties of the rings around 10199 Chariklo, 2060 Chiron and 136108 Haumea.*
*[1] Braga-Ribas et al. (2014); [2] Ortiz et al. (2015); [3] Ortiz et al. (2017).*

**4.9 THE REMOVAL OF MATERIAL FROM SMALL BODY RESERVOIRS**
As was discussed earlier, the Solar system is littered with debris moving on dynamically unstable orbits. In the absence of replenishment, these small body populations would quickly become depleted, on timescales of just a few million years, as a result of collisions with the Solar system's planets (e.g. Zahnle & Mac Low, 1994; Hammel et al., 1995; Sánchez-Lavega et al., 2010; Hueso et al., 2018) or the Sun (e.g. Michels et al., 1982; Farinella et al., 1994; Jones et al., 2018), fragmentation and disintegration (e.g. Crovisier et al., 1996; Sekanina, 2000; Jenniskens & Lyytinen, 2005; Jenniskens & Vaubaillon, 2007), and ejection from the system as a result of close encounters with one or other of the system's giant planets (e.g. Levison & Duncan, 1994, 1997; Bottke et al., 2002; Horner et al., 2004b).

Given the number of objects we observe in the various unstable populations, it is obvious that they must be continually replenished with fresh material, injected from those populations in the Solar system that are nominally dynamically stable. In this manner, the near-Earth asteroids are held to be primarily sourced from the asteroid belt (e.g. Wetherill, 1988, Morbidelli et al., 2002; Morbidelli & Vokrouhlický, 2003); the short-period comets from the trans-Neptunian region (e.g. Duncan & Levison, 1997; Levison & Duncan, 1997; Di Sisto & Brunini, 2007; Lowry et al., 2008; Volk & Malhotra, 2008, 2013), with contributions from the inner Oort cloud (e.g. Emel'yanenko, Asher & Bailey, 2005; Brasser et al., 2012a; de la Fuente Marcos & de la Fuente Marcos, 2014) and the Jovian and Neptunian Trojans (e.g. Horner & Lykawka, 2010a; Horner, Lykawka & Müller, 2012; Di Sisto, Ramos & Gallardo, 2019; Holt et al., 2019); and the long-period comets from the Oort cloud (e.g. Öpik, 1932; Oort, 1950).

At the same time, vast quantities of dust are released into the Solar system by cometary activity, and by collisional processes throughout the system - evidence of which can be seen on any clear night, in the form of meteors, the Zodiacal Light (e.g. Leinert 1975; Leinert et al., 1981; Reach et al., 2003), and the Gegenschein (e.g. Brandt & Hodge, 1961; Wolstencroft, 1967; Roosen, 1970; Buffington et al., 2009; Ishiguro et al., 2013). Since dust is continually produced, the fact that the Solar system is not substantially dustier is clearly the result of the efficient clearing of dust, by a variety of processes.

In this section, we describe the non-gravitational mechanisms by which planetary systems are kept clear of dust, and those that can help to facilitate the liberation of objects from the dynamically stable regions of the system, to join the various populations of dynamically unstable object.

**4.9.1 EFFECTS THAT REMOVE DUST**
Three major processes dominate the lifetimes of dust grains in a planetary system; Poynting-Robertson drag, radiation pressure, and collisional destruction (e.g. Wyatt & Whipple, 1950; Burns, Lamy & Soter 1979; Backman & Paresce, 1993; Lagrange et al., 1995, Moro-Martín & Malhotra, 2003; Krivov, 2010; Wyatt et al., 2011). All of these processes act on timescales much shorter than the stellar lifetime, implying that the dust that permeates our Solar system, and the debris discs observed around other stars, are not primordial (e.g. Backman & Paresce 1993; Wyatt 2008; Moro-Martin 2013). An unseen population of larger asteroidal and cometary bodies (planetesimals) are invoked as the mass reservoir for the ongoing production of dust grains we observe around other stars (Artymowicz & Clampin, 1997; Dominik & Decin, 2003). Analogously, the sources of dust in the Solar system include the collisional grinding of asteroidal bodies and material ejected from active comets (e.g. Nesvorný et al., 2010; Koschny et al., 2019).

**Radiation Pressure**

The dynamics of small dust grains in the Solar system are dictated by a combination of the incident stellar radiation field and the gravitational potential of the Sun. The strength of influence that radiation pressure has on a given dust grain can be described by the ratio of the radiation pressure force $F_{rad}$ to the stellar gravitational force $F_{grav}$ acting on it, defined as the $\beta$ parameter (Burns et al., 1979; Krivov 2010). The $\beta$ ratio is calculated as follows:

$$\beta = \frac{F_{rad}}{F_{grav}} = \frac{3L_* Q_{pr}}{16\pi c G M_* s \rho}$$

Radiation pressure is proportional to the stellar luminosity, $L_*$, the grain cross-section, and the efficiency $Q_{pr}$ with which momentum is transferred from photons to the dust, averaged over the stellar spectrum. The gravitational force, on the other hand, is proportional to the stellar gravitation parameter, $GM_*$, as well as the grain mass, and hence, to its volume and its density, $\rho$. Since both gravitation and radiation intensity are $r^{-2}$ laws, the value of $\beta$ is dimensionless, and is fixed for a dust grain of given radius, $s$, and density, $\rho$. Dust grains effectively see a reduced gravitational potential from the central star due to this incident radiation. As a result, for dust grains released by objects[57] moving on circular orbits, or around the perihelion passage of objects moving on eccentric orbits, the influence of radiation pressure drives dust grains onto more eccentric orbits than those occupied by their parent bodies.

For sufficiently large and opaque grains, the radiation pressure efficiency, $Q_{pr}$, is constant near unity. As a result, the $\beta$ ratio becomes inversely proportional to the grain radius, $s$. Radiation pressure is less and less important as grain size increases, becoming insufficient to perturb the orbital motion for particles larger than ~100μm (e.g. Pawellek et al., 2019). Equally, very small dust grains are inefficient absorbers and scatterers of incident stellar radiation. Hence, in the Rayleigh regime, where grains are much smaller than the dominant stellar photons, $Q_{pr}$ is reduced, and becomes proportional to the grain radius, $s$. As a result, the $\beta$ ratio levels off to a lower, constant value, at these small sizes. The exact size range over which radiation pressure is important depends on both the properties of the host star and the composition of the dust.

The grain size at which $\beta = 0.5$ is often called the blow-out limit. For dust originating from a larger parent body on a circular orbit moving at Keplerian velocity around the host star, grains with $\beta$ values < 0.5 will remain bound to the star. In such a scenario, grains for which $\beta > 0.5$ will become unbound from the host star upon release from their parent body. They are blown out of the system on hyperbolic orbits, leaving on orbital timescales, vacating the terrestrial region on a timescale of years, and leaving the Edgeworth-Kuiper belt on a timescale of centuries. Around the Sun, the blow-out size of compact silicate dust is ≈ 1 μm, whereas around an A-type star the blow-out size of identical dust is ≈ 10 μm (Backman & Paresce, 1993).

For dust originating from larger bodies moving on eccentric orbits, the limit between bound and unbound depends on the point in their parent body's orbit at which the dust grains are released (Kresák, 1976):

---

[57] Dust grains can be released through a variety of processes, such as cometary activity, spallation, large scale impacts/collisions, surface expansion and contraction under heating (as is the case for the Geminid meteor shower parent, (3200) Phaethon, e.g. Jewitt & Li, 2010; Li & Jewitt, 2013), and even as a result of the rapid rotation of their parent bodies.

$$\beta_{lim} = \frac{r}{2a}$$

where *r* is the distance at which the grain is released from its parent body, and *a* denotes the semi-major axis of that object. For grains released closer to perihelion, the excess kinetic energy results in a blow-out limit that falls in the range 0 < $\beta_{lim}$ < 0.5 (dependent on the exact motion at the time). By contrast, for grains released closer to an object's aphelion, the limiting value for dust grain blow-out falls in the range 0.5 < $\beta_{lim}$ < 1, and initially, for low *β* values, emission near aphelion will actually cause dust grains to move on orbits more circular than that of their parent (as seen in Figure 15, below). Indeed, for particles release at aphelion, a *β* value that is identical to the orbital eccentricity of the parent body will result in a dust grain moving on a circular orbit.

For cometary parent bodies, where orbital eccentricities, *e*, are often very high, and dust production occurs primarily in the months around their perihelion passage, $r_{emission}$ ≈ *q* = *a* (1 - *e*), and the limiting *β* ratio can be approximated with $\beta_{lim}$ ≈ (1 - *e*)/2. For orbital eccentricities in excess of 0.99, which is typically the case for long-period comets, the lower size limit for bound grains can thus exceed the circular blowout limit by a factor ≳ 100. This distinction is critical when it comes to the discussion of dust behaviour in other planetary systems. Studies of debris discs often discuss the 'blowout limit' for the dust we observe in those systems – which brings with it an implicit assumption that the parent bodies are moving on circular, or near circular orbits. In reality, however, this blowout limit is actually an upper limit, since grains ejected by objects moving on eccentric orbits would be blown out of a system when identical grains emitted from a circular orbit would not be removed. This is particularly relevant when one considers observations of systems containing exo-comets (e.g. *β* Pictoris; Kiefer et al., 2014). The stability and distribution of dust grains released from parent bodies on circular and eccentric orbits are illustrated in Figure 15.

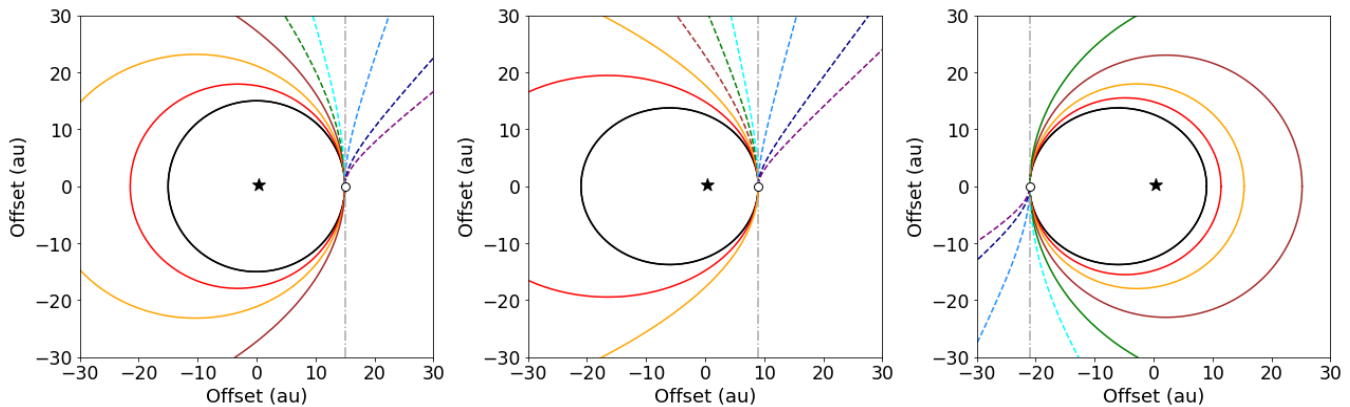

*Figure 15: Orbits for dust grains emitted from a parent body moving on an orbit with a semi-major axis of 15 au. The left panel shows a scenario where the parent body moves on a circular orbit (i.e. e = 0), whilst the central and right-hand panels show the behaviour when the parent body's orbit has an eccentricity of 0.4. The middle panel shows the behaviour of dust emitted at perihelion, whilst the right hand shows emission at aphelion. In each panel, we plot, in colour, the resulting orbits of the dust for a variety of values of β (specifically 0.0, 0.15, 0.3, 0.45, 0.60, 0.75, 1.5, 3.0, and 4.5), using the formulation outlined in Kresák, 1976. The solid lines show the orbits of bound grains (those moving on orbits with e < 1). The dashed lines show dust grains moving on hyperbolic orbits (i.e. those with e > 1). The dotted lines show those grains for which the β value is so great that the outward force due to radiation equals or exceeds the inward force due to gravity. The vertical grey dot-dash line marks the dividing line between cases where the acceleration due to radiation pressure is less than that due to gravity, and those where radiation pressure exceeds the effect of gravity. In the case of emission from an object moving on an eccentric orbit, the limiting β that determines whether a grain will be bound or unbound depends on the point of release in the parent body's orbit around the Sun. The location of the parent body of the dust at the time of emission is denoted by the 'o' symbol, and the Sun by the '\*".*

It is challenging in the extreme to get an overview of the dust halo of the Solar system because we are buried deep within it. As a result, whilst significant work has studied the evolution of dust in the context of meteor showers and the zodiacal light (as described in section 4.7.1), the discovery of debris orbiting other stars has proved a great boon to researchers, since it allows the study of the entirety of the debris in a given planetary system at a single time. The presence of extended haloes of sub-blowout sized grains (the radii of which may extend up to several microns) has been observed in the optical and near-infrared scattered light images of debris disc-host stars (e.g. Smith & Terrile, 1984; Schneider et al., 2014; Ren et al., 2019). For example, multi-wavelength observations of HD 181327 revealed a difference in debris ring location between optical and millimetre wavelengths that is consistent with radiation pressure induced dust grain size segregation. The millimetre wavelength observations also revealed a broad halo component (or unresolved second belt) to the disc (Marino et al., 2016).

It would be expected that any debris disc with a host star luminous enough to drive dust out of its system[58] should exhibit such a halo structure, and it has been suggested that up to half of the emission from debris discs at infrared wavelengths could be produced by sub-micron dust grains, for those stars luminous enough to drive the ejection of dust in this way (Thébault & Kral, 2019).

We note here that not all dust haloes observed around other stars are the result of radiation pressure. For example, the young M-dwarf star AU Mic has a bright debris disc surrounded by a significant extended halo. However, as an M-dwarf, the star is clearly insufficiently luminous for that halo to be the result of radiation pressure driven escape. Instead, the presence of a halo is attributable to the strong stellar wind from the young M-dwarf star (Schüppler et al., 2015; Boccaletti et al., 2018). Similar to the pressure induced by stellar radiation, the pressure exerted by stellar wind particles is mostly directed radially away from the star, with the intensity dropping as $r^{-2}$. Stellar wind pressure is proportional to, and varies with, the stellar mass loss rate and the wind velocity. In the Solar system, typical wind pressures are by more than three orders of magnitude lower than radiation pressure (e.g. Burns et al., 1979).

In combination with the influence of Poynting-Robertson drag (as described below) and stellar wind forces, radiation pressure induces a radial distribution of dust grains within a debris disc that is dependent on both the size of the particles and their (initial) orbits (e.g. Burns et al., 1979; Strubbe & Chiang 2006; Krivov, 2010). Whilst in the Solar system, we can see individual objects as they contribute to the system's dust budget, determining the source of dust in extrasolar debris discs is more challenging. To accurately trace the location of the dust-producing planetesimals within those discs it is therefore necessary to observe larger dust grains which are less strongly influenced by these radiation forces. Modelling of debris discs from mid-infrared to millimetre wavelengths suggests that dust seen at wavelengths in the far-infrared and beyond traces the location of the underlying planetesimal belt (Pawellek et al., 2019).

**Poynting-Robertson Drag**
The Poynting-Robertson effect (Poynting, 1904; Robertson, 1937) induces the in-spiralling of a dust grain orbiting the Sun through that grain's interaction with a 'head wind' of solar photons. This head wind is the result of the orbital motion of the dust grains, relative to the mostly radial stellar radiation field. Whilst a dust grain at rest would just be affected by the radiation pressure described above, any orbital motion leads to an additional force proportional to $-v/c$, where $v$ denotes the grain's orbital velocity, and $c$ is the speed of light. The momentum transfer that causes both radiation pressure and the Poynting-Robertson effect happens when the solar photons are absorbed or scattered by the grain. Subsequent re-emission, if anisotropic, can then lead to additional forces on macroscopic objects such as the Yarkovsky effect (which we describe in more detail in section 4.9.2, below).

---

[58] Below a certain luminosity, a star simply cannot accelerate dust grains from a circular orbit to escape velocity purely by radiation pressure alone. For stars on the main sequence, this threshold is typically thought to lie at around mid-K spectral type, though the precise location depends on both the dust composition and density.

This perceived 'head wind' causes a loss of orbital energy, resulting in the gradual decay of the dust grain's orbit. The Poynting-Robertson effect acts to reduce both the semi-major axis *and* the eccentricity of the dust grain's orbit. Grains therefore migrate towards the Sun from the initial orbits into which they were released until they eventually reach distances where the material sublimates. This happens at a few Solar radii for silicate dust and at several astronomical units for volatiles (e.g. Kobayashi et al., 2011). For grains moving on roughly circular orbits, the timescale for loss via Poynting-Robertson drag is given by

$$t_{PR} = \frac{ca^2}{\beta GM_*} = \frac{16\pi c^2 s \rho a^2}{3L_* Q_{pr}}$$

Here, $t_{PR}$ is a function of the grain radius, *s*, density, *ρ*, and the radiation pressure efficiency $Q_{pr}$, combined with the orbital radius of the grain, *a*, around a star with luminosity $L_*$ (Wyatt & Whipple, 1950; Burns, Lamy & Soter, 1979). The timescale is proportional to grain size and makes Poynting-Robertson drag most effective on µm-sized dust grains, for which it causes grains to spiral inward from 1 au to a few Solar radii in roughly one thousand years. Smaller grains are quickly blown out of the system by direct radiation pressure. For dust grains initially moving in the region of the Edgeworth-Kuiper belt, the timescale increases by a factor of approximately one thousand (to a million years, or so). Similarly, increasing the grain size to millimetres would again yield an increase of the timescale by a factor of a thousand.

Whilst direct stellar wind pressure has little influence on dust dynamics and lifetimes in most systems, wind *drag* is more effective. The radial velocity of the wind particles (~100 to 1000 km/s) is roughly three orders of magnitude smaller than that of the photons. The dust's azimuthal motion and the momentum transfer from the corresponding 'head wind' resulting from the stellar wind particles becomes more significant; the additional factor $c/v_{wind}$ makes wind drag comparable in efficiency to radiation-induced Poynting-Robertson drag. Gustafson (1994) estimates that wind drag contributes about one quarter of the total drag, at times of average Solar wind strength. Around stars with stronger stellar winds, wind drag can dominate (e.g. Reidemeister et al., 2011; Schüppler et al., 2015).

**Collisional Destruction**
The impact craters that cover most Solar system objects bear witness to the importance and frequency of mutual collisions (as described in depth earlier in this manuscript). Relative speeds on the order of kilometres per second mean that such collisions often end the lives of dust grains. At the same time, collisions among bigger objects produce fresh dust. A cascade of collisions transfers mass from planetesimals down to observable dust. Once the material is ground down to sufficiently small sizes, radiation pressure and Poynting-Robertson drag move the dust grains out of the system or towards the Sun, respectively. Objects in regions and of sizes for which the loss timescale due to mutual collisions is shorter than the direct removal timescale due to radiation pressure and drag forces are called collisionally dominated bodies.

For a dust grain moving on a Keplerian orbit, its collisional lifetime, $t_{coll}$, will be broadly defined by its orbital period, $t_{orb}$, and the cross-section surface density, *σ* (cross section per disc surface area; also known as normal geometrical optical depth, *τ*), of potentially harmful colliders in the disc through which it moves (e.g. Burns, Lamy & Soter, 1979; Backman & Paresce, 1993; Wyatt & Dent, 2002), such that:

$$t_{coll} = \frac{t_{orb}}{2\pi\sigma} \cdot \frac{2i}{v_{rel}/v_k} \sim \frac{t_{orb}}{2\pi\sigma} \propto \frac{a^{3/2}}{\sigma M_*^{1/2}}$$

where *i* is the orbital inclination, $v_{rel}$ the typical relative velocity, and $v_K = (GM_*/a)^{0.5}$ the Keplerian speed. The fraction containing the inclination and the velocities is of the order of unity, and through substitution of the Keplerian speed and the circumference of the orbit (= $2\pi a$) for $t_{orb}$, we obtain the relation in terms of the stellar mass, $M_*$, orbital radius, *a*, and surface density, *σ*.

A collision is commonly considered catastrophic if the largest remaining fragment has at most half the mass of the colliders. The critical energy required to disrupt and disperse an object increases with increasing size. The critical specific energy (the energy per unit mass) decreases up to object radii ∼100 m, and then increases for larger objects which are strengthened by self-gravity (e.g. Benz & Asphaug, 1999; Leinhardt & Stewart, 2012). The critical impactor size required for a catastrophic collision also increases with object size, and as a result, the rate at which such collisions happen decreases. The lifetime of an object against collision increases with increasing grain size, albeit not as strongly as the Poynting-Robertson timescale. Below a critical grain size, Poynting-Robertson drag acts on shorter timescales than collisional disruption, whilst above that critical size, collisions dominate. As a result of the different dependencies in $t_{coll}$ and $t_{PR}$, the critical size at which $t_{coll} = t_{PR}$ is smaller in denser or more distant regions (e.g. Kuchner & Stark, 2010). Observations of the interplanetary dust bands and counts of microcraters suggest that the critical size is around ∼100 μm (corresponding to ∼ $10^{-5}$ g) in the inner Solar System (e.g. Fechtig & Grün 1975; Le Sergeant d'Hendecourt & Lamy 1980; Grün et al., 1985; Grogan et al., 2001).

Whilst in the Solar system, there is a clear distinction between dust grains large enough to be collisionally dominated and those small enough to be driven by radiation pressure and Poynting-Robertson drag, the same is not true of the debris we have observed beyond the Solar system. With increasing disc surface density (or mass), the critical size at which $t_{coll} = t_{PR}$ decreases until it becomes comparable to or smaller than the blowout limit. At that point, all dust is either bound and dominated by collisions or moving on unbound orbits. At the current epoch, our instrumentation is simply not sensitive enough to detect true analogues of the Solar system's dust environment. As a result, all of the currently known debris discs are collisionally dominated, as a direct result of observational biases that favour the detection of bright, dense and cool dust discs (see e.g. Matthews et al., 2014; Hughes et al., 2018).

Detections of near-Edgeworth-Kuiper belt analogues (∼10✕ $F_{EKB}$) around Sun-like stars have been made at far-infrared wavelengths in continuum emission by *Herschel* (Eiroa et al., 2013, Sibthorpe et al., 2018), and have allowed us to begin to place our Solar system's dust environment in the context of its peers. These detections approach the predicted limits on Edgeworth-Kuiper belt flux of 1.2 mJy at 70 microns (Vitense et al., 2012) set by *IRAS* and *COBE* at 60 μm, favouring an upper limit < 1 MJy/sr (Aumann & Good, 1990; Backman, Dasgupta & Stencel, 1995), and in-situ dust measurements from *New Horizons* (Han et al., 2011). Those observations suggest that the Edgeworth-Kuiper belt lies around the mid-point of dust disc brightness based on an extrapolation from the observed systems (Moro-Martín et al., 2015), but still lies within the collisionally dominated regime.

In the typical situation where a collisional cascade constantly replenishes and removes material, the size-frequency distribution of dust grains can be approximated by a power law, $dN \propto s^{-p} ds$, with index $p \approx 3.5$ (Dohnanyi, 1969; O'Brien & Greenberg, 2003), where $dN/ds$ is the number of particles per unit grain radius. For the smaller grains that are produced in collisions but whose lifetimes are shortened by Poynting-Robertson drag, the size distribution flattens to an index in the range $2.5 < p < 3.0$ (Wyatt et al., 2011; Reidemeister et al., 2011). Such flattening is observed for the dust distribution in the inner Solar system (e.g. Fechtig & Grün, 1975).

The combination of a collisional cascade and a sharp lower size cutoff set by radiation pressure creates a distinctive wavy pattern in the size distribution of dust grains, deviating from the single power law. The small grains that are unbound by radiation pressure are significantly underabundant with respect to those that are still small, but big enough to stay bound. The latter, barely bound grains, therefore lack a reservoir of potentially destructive colliders; their lifetimes and abundance are increased. In turn, yet bigger grains suffer more strongly from collisions with these barely bound grains; they are again underabundant, and so on (Campo-Bagatín, 1994). The wavelength, amplitude, and grain size to which this pattern extends depend on the details of the collision physics, including material properties and impact velocities (e.g. Thébault et al., 2003; Krivov et al., 2006). The wavy size distribution can translate to observable waves in the spectral energy distribution (Thébault & Augereau, 2007; Kim et al., 2018). When orbital eccentricities induced by radiation pressure exceed those inherited from the belt of parent bodies, collision velocities and rates are increased, the abundance decreased. For parent belts with very low dynamical excitation ($e \lesssim 0.01$), this can cause an additional depletion of the barely bound grains up to grain radii *s* corresponding to $\beta(s) \gtrsim e$, resulting in a more tenuous halo (Thébault & Wu, 2008).

In a collisional cascade, the destruction of two asteroidal sized bodies in a system produces large numbers of small dust grains, causing a temporary brightening of the observed debris disc. Such events occur stochastically, and the brightness increase is short lived ($t_{coll} \sim 10^4$ yr, Rieke et al., 2005, Jackson et al., 2014). The dust grains produced in such a collision are swiftly removed from the system by the subsequent collisional cascade and dynamical evolution through radiation forces as described above. The presence of very bright asteroid belt analogues orbiting other stars (with $L_{dust}/L_* > 10^{-3}$, $T_{dust} > 200$ K; approximately ten thousand times more luminous than our own asteroid belt) is believed to be evidence of ongoing collisional processes, possibly analogous to the giant impacts seen in the Solar system's youth (as described in section 4.1; see e.g. Song et al., 2005; Lisse et al., 2009; Fujiwara et al., 2009, 2010; Melis et al., 2010). The rapid evolution of such bright asteroid belt analogue debris discs, including both stochastic variability on periods of months to years and monotonic dissipation over timescales of years to decades, has been observed in a handful of systems (Melis et al., 2012; Meng et al., 2005, 2012, 2015; Su et al., 2019), and suggests that the violent processes that shaped our Solar system are far from unique.

### 4.9.2 EFFECTS THAT REMOVE LARGE OBJECTS

The great majority of near-Earth asteroids are thought to be sourced from the asteroid belt. The primary routes by which these asteroids are transferred from the belt to the inner Solar system feature interaction with mean-motion and secular resonances with the giant planets (e.g. Wetherill, 1988; Morbidelli et al., 2002). However, those mechanisms are so effective, on astronomical timescales, that they rapidly clear any debris from their regions of influence in the belt. To maintain a steady flux of new material to the inner Solar system through those mechanisms, then, fresh material must first be transferred into those regions from elsewhere in the belt. To address this problem, Morbidelli & Vokrouhlický (2003) investigate the influence of a non-gravitational force known as the Yarkovsky effect on the main belt asteroids, finding that that effect could cause the migration of asteroids through the belt to the regions of instability that direct those asteroids into the inner Solar system.

The diurnal Yarkovsky effect, explained in detail in Bottke et al., (2006), and the references therein, involves the absorption and re-emission of photons on an orbital body. Photons of light from the Sun hit an object and are absorbed. As the object is rotating, and its surface has a significant, non-zero, thermal inertia, the infrared radiation of the photon happens at a different angle to the absorption. This produces a small net force. In a prograde rotator, the force has a component in the orbital direction that creates a net increase in orbital rate (e.g. Bottke et al., 2006). The consequence of this small force is an increase in semimajor axis over time. For a retrograde rotator, the result is a decrease in orbital momentum, resulting in a decrease in semi-major axis. The magnitude of the force depends upon how close a body is to the Sun, the obliquity of the body's spin axis with respect to the orbital plane, and the body's thermophysical characteristics (e.g. Bottke et al., 2006). Mathematically, this is represented in equation:

$$\left(\frac{da}{dt}\right)_{diurnal} = -\frac{8}{9}\frac{\alpha\Phi}{n}W(R'_\omega,\Theta_\omega)cos\gamma$$

The radiative pressure coefficient ($\Phi$) is dependent upon the radius of the object (*R*) and the Solar radiation flux (*F*).

$$\Phi = \frac{\pi R^2 F}{(mc)}$$

The Stefan-Boltzmann constant (α = 1 − A) comprises the bond albedo of the object (A). The thermal component, (W (R$_\omega$ ′, Θ$_\omega$)) is dependent on the radius scaled for penetrative depth (R$_\omega$ ′ = R/$l_v$) and the thermal parameter (Θ$_\omega$). The thermal parameter (Θ$_\omega$) is to account for the thermal properties of the body.

The area to mass ratio of the object has a major effect on how the Yarkovsky effect influences an object. A large object has a large mass to area ratio, reducing the effect of such a small force. If an object is too small, the thermal gradient over the object lessens and the radiative difference becomes minimal. It is generally accepted that the size range where the diurnal Yarkovsky effect has the most influence is between one metre and approximately 10km (e.g. Bottke et al., 2006).

In addition to the diurnal Yarkovsky effect, a seasonal effect has been described (Bottke et al., 2006). Unlike the diurnal Yarkovsky effect, the seasonal effect can only slow the object, reducing the semi-major axis.

$$\left(\frac{da}{dt}\right)_{seasonal} = \frac{4}{9}\frac{\alpha\Phi}{n}W(R_n, \Theta_n)sin^2\gamma$$

From the above equation, due to the radiative flux (*F*), it can be seen that the Yarkovsky effect has the most effect in the inner Solar system. Simulations have shown that the Yarkovsky effect has an effect on the evolution of the near-Earth Asteroids (e.g. Morbidelli & Vokrouhlický, 2003; Farnocchia et al,. 2013), Martian Trojans (e.g. Ćuk et al,. 2015) and the main asteroid belt (e.g. Nesvorný & Bottke, 2004). Further out in the Solar system, the impact of the Yarkovsky effect on the Jovian Trojans has been found to be negligible on objects larger than 100m (Wang & Hou, 2017), with objects more distant than Jupiter likely not experiencing the effect to any significant degree.

The Near-Earth asteroid population is of great interest as a result of the potential hazard it poses to the Earth (see section 4.2.2), and the Yarkovsky effect will play an important role in altering the orbits of these objects, causing the degree to which they can be considered to be hazardous objects to change over time. As a result of the effect's influence on the orbital evolution of NEAs, there has recently been a concerted effort to detect the Yarkovsky effect on individual members of the NEA population, with over 40 confirmed detections to date (e.g. Vokrouhlický et al, 2015; Durech et al., 2018), including the OSIRIS-REx target, (101955) Bennu (Deo & Kushvah, 2017; Hergenrother et al., 2019).

Over the past two decades, the influence of the Yarkovsky effect on the migration of asteroids across the asteroid belt has been widely studied, and asteroidal drift under the influence of the Yarkovsky effect has become a key metric in the identification of asteroid collisional families, and the determination of their ages (e.g. Bottke et al, 2001; Spoto et al, 2017; Bolin et al,. 2017, 2018; Delbo et al, 2019; Paolicchi et al., 2019).

# 5. APPLICATION TO KNOWN EXOPLANETARY SYSTEMS

The thought that the Solar system might not be unique, and that there might be planets orbiting other stars, is one that has been debated for at least several hundred years. Famously, the Dominican Friar Giordano Bruno argued, in the third dialogue of his 1584 work 'De l'Infinito, Universo e Mondi', that the stars we see in the night sky are suns, accompanied by their own planets, and that the cosmos contains an infinite number of suns and an infinite number of planets[59] (Bruno, 1584). He even suggested that the planets can not be seen because they are too small and too faint, compared to their host stars[60] - an issue that has proven painfully true over the centuries since!

The first steps towards the detection of planets orbiting other stars came in the 18$^{th}$ and 19$^{th}$ Centuries, with discoveries that would presage the most successful exoplanet detection techniques. In 1783, two years after the discovery of Uranus, the young British astronomer John Goodricke presented an explanation for the peculiar periodic dimmings of the bright star Algol (Goodricke, 1783). Goodricke realised that an unseen companion orbited Algol[61], and that the unseen secondary eclipsed the primary every 2.86 days. Whilst Algol's variability had been known for thousands of years (as it is easily seen with the unaided eye; e.g. Jetsu et al., 2013), Goodricke was the first (that we know of) to correctly identify the cause of the periodic dimming - a discovery that would lay the groundwork for the modern exoplanet transit technique.

In the 1840s, two additional discoveries were made that set the scene for the eventual dawn of the Exoplanet Era. In 1844, Friedrich Wilhelm Bessel announced the discovery of unseen companions to the bright stars Sirius and Procyon, on the basis of periodic variations in the proper motions of those stars (Bessel, 1844)[62]. Just two years later, in 1846, the planet Neptune was discovered as a direct result of its gravitational perturbation of the motion of Uranus (e.g. Adams, 1846; Galle, 1846; Le Verrier, 1846).

In all of these cases, observations of one object (Algol, Sirius, Procyon and Uranus) enabled the discovery of another, unseen body, as a result of the unseen object's influence on the target of the observations. In other words, all of the new objects were first discovered indirectly, as a result of their causing otherwise inexplicable changes in the behaviour of another body. It was obvious that the same techniques could be used to detect planets orbiting other stars - but the problem, as Bruno had realised, was that planets are far smaller than their host stars - and so their influence on those stars would be too small to detect.

In the middle of the 20th Century, the discovery of a number of planets orbiting nearby red dwarf stars, including Barnard's Star, Lallande 21185, and 61 Cygni, were announced as a result of a long-term astrometric survey carried out using the Sproul Observatory refracting telescope (e.g. Strand, 1943, van de Kamp & Lippincott, 1951; van de Kamp, 1963, 1969a,b). The discoveries were announced based on observed periodic astrometric wobbles from the target stars, which were explained on the basis that those stars had planet-mass unseen companions (in much the same way that Bessel, a century beforehand, had inferred the presence of Sirius B and Procyon B). No other observers were able to replicate his results, however - and it seems that the observed 'wobbles' were actually the result of changes to the optical properties of the Sproul refractor caused by the periodic cleaning of its main objective lens (e.g. Gatewood & Eichhorn, 1973, Hershey, 1973).

---

[59] *'Ivi innumerabili stelle, astri, globi, soli e terre sensibilmente si veggono…'*; taken from http://www.ousia.it/content/Sezioni/Testi/BrunoDeInfinitoUniverso.pdf , accessed on 14$^{th}$ July 2019

[60] *'La raggione è, perché noi veggiamo gli soli, che son gli più grandi, anzi grandissimi corpi, ma non veggiamo le terre, le quali, per esserno corpi molto minori, sono invisibili…'*; taken from http://www.ousia.it/content/Sezioni/Testi/BrunoDeInfinitoUniverso.pdf, accessed on 14$^{th}$ July 2019

[61] *'If it were not perhaps too early to hazard even a conjecture on the cauſe of this variation, I ſhould imagine it could hardly be accounted for otherwiſe than either by the interpoſition of a large body revolving around Algol, or ſome kind of motion of its own, whereby part of its body, covered with ſpots or ſuch like matter, is periodically turned towards the earth…'*, from Goodricke, 1783.

[62] These were the first two 'white dwarf' stars to be discovered - Sirius B and Procyon B.

The dawn of the Exoplanet Era finally arrived in the late 1980s, with the discovery of planets orbiting the stars Gamma Cephei A and HD 114672 (Campbell et al., 1988; Latham et al, 1989). At the time, however, the planets were thought to be brown dwarfs, too massive to count as true planetary detections. Those discoveries were followed, however, by the detection of three planets orbiting the pulsar PSR 1257+12 (the first two being announced in Wolszczan & Frail, 1992; the third in Wolszczan, 1994), a single planet orbiting the pulsar PSR B1620-26 (Thorsett, Arzoumanian & Taylor, 1993), and then the announcement, in 1995, of the 'first planet orbiting a Sun-like star' - 51 Pegasi b (sometimes known as Bellerophon; Mayor & Queloz, 1995).

Prior to the detection of those first planets, it was widely expected that planetary systems around other stars (if they existed) would strongly resemble our own Solar system, having formed in a disc of material around their young suns (e.g. Swedenborg, 1734; Laplace, 1796; Edgeworth, 1949; Lissauer, 1993). However, those first planets were nothing like those we might have expected, and forced a radical overhaul of our theories of planetary formation and evolution.

In the first three decades of the Exoplanet Era, we have learned that planets are ubiquitous in the cosmos - that, just as Bruno said, the myriad stars in the night sky all host planets of their own. We have also learned that planetary systems are far more diverse than we could have imagined. We have found planets moving on highly eccentric, comet-like orbits (e.g. Naef et al., 2001; Tamuz et al., 2008; Wittenmyer et al., 2017b), and others that are denser than osmium (e.g. Siverd et al., 2012; Johns et al., 2018) or more tenuous than fairy floss (e.g. Masuda, 2014). Some planetary systems are incredibly tightly packed (e.g. MacDonald et al., 2016, Gillon et al., 2017), and others feature 'hot Jupiters' on highly inclined or even retrograde orbits (e.g. Bayliss et al., 2010; Addison et al., 2013; Esposito et al., 2014).

Despite the remarkable diversity observed in exoplanet systems, the Solar system remains our key touchstone for the understanding of these alien worlds. Of the several thousand planetary systems that have been discovered to date, the only one that we can study in intricate detail is our own[63]. For that reason, in this section, we describe exoplanets in the context of Solar system science, showing how our knowledge of our planetary system has helped to shape our understanding of those that orbit other stars.

**5.1 DEMOGRAPHICS OF EXOPLANETS**
Since the dawn of the Exoplanet Era, over 4,000 exoplanets have been identified in approximately 3,000 systems[64], with many more candidate planets and systems discovered by NASA's *TESS* and *Kepler* missions awaiting confirmation using ground-based facilities. The vast majority of these systems were identified through transit photometry, which looks for the almost imperceptible 'wink' as an unseen companion passes across the face of the star as viewed from the Solar system. The *Kepler* spacecraft (e.g. Borucki et al., 2010; Howell et al., 2014) proved to be the most productive planet search project to date, performing what was essentially the first census of the Exoplanet Era. Between its primary and K2 mission phases, *Kepler* led to the discovery of 2730 planets, with a further 2945 still awaiting confirmation[65].

Despite the apparent wealth of exoplanets discovered in the last thirty years, our current understanding of the exoplanetary systems we have discovered remains far less than our knowledge of the Solar system's planets before the launch of the first interplanetary missions in the 1960's. Our knowledge, in the main, is limited to knowing the planet's minimum mass (if discovered using the radial velocity technique) or its size (using the transit method). A small fraction of the planets discovered to date have been observed directly - but even then, we can usually gain only limited understanding of the planet's nature (with the mass of such planets estimated based on an assumed age and their observed brightness; e.g. Marois et al., 2008[66]). Beyond these basic characteristics, little more is known for the vast majority of known exoplanets. In some cases, a bulk density can be calculated (for those planets that have been observed using both the transit and radial velocity techniques; e.g. Siverd et al., 2012; Masuda, 2014; Johns et al., 2018, and many others). In systems containing multiple planets, some of which transit their hosts, it is possible to obtain improved mass and orbit measurements, and even to uncover the presence of additional planets, through the analysis of *transit timing variations* (TTVs) – which result from the effect of one planet's gravity perturbing the orbit of another (e.g. Agol et al., 2005; Holman & Murray, 2005; Lithwick et al., 2012; Becker et al., 2015; Gajdoš et al., 2019; Hamann et al., 2019; Lam et al., 2020). Specific molecular species have been identified in the atmospheres of a small number of planets through multi-wavelength transit photometry (e.g. Charbonneau et al., 2002; Knutson et al., 2014; Khalafinejad et al., 2017) or direct measurement of the planet's spectrum (e.g. in the case of HR8799 c; Oppenheimer et al., 2013; Wang et al., 2018). But, in the main, our knowledge of the planet's physical nature is limited to either a minimum mass, or a size - nothing more[67].

---

[63] In much the same way, the Sun remains the only star we can study from close range - and so has formed the touchstone of our understanding of other stars, and the Space Weather environments around them. Whilst a detailed discussion of the Sun is beyond the scope of this work, we direct the interested reader to the recent review by Airapetian et al. (2019) for more details.

[64] As of 28th July, 2019, the NASA Exoplanet Archive (https://exoplanetarchive.ipac.caltech.edu/) details 4025 confirmed exoplanets in 2991 systems.

[65] Values obtained on 28th July, 2019, from the NASA Exoplanet Archive's 'Exoplanet and Candidate Statistics' page, at https://exoplanetarchive.ipac.caltech.edu/docs/counts_detail.html

Prior to the panoply of exoplanet discoveries beginning in the late 1980's, speculation on the types and abundance of planets within the galaxy was mostly inspired by planetary formation theory, which was rooted in the Solar system. The Sun, an average star, hosts a set of planets that seemed to be loosely arranged in their orbits by size and mass, with rocky (terrestrial) inner planets, and more distant gaseous giants. It seemed reasonable to conclude that other Sun-like stars would like have similar groups of planets, arranged in similar ways, with rocky worlds interior to gaseous ones.

The first exoplanets to be discovered around Sun-like stars were far different to what was expected following this line of thought. 51 Pegasi b (Mayor & Queloz, 1995) is comparable in mass to Jupiter, but orbits its host at a distance of just over 0.05 au, with a period of 4.23 days. In the years that followed, such 'hot Jupiters' dominated exoplanet discoveries - a finding that led to significant innovation in our models of planet formation (e.g. Dawson & Johnson, 2018, and references therein). The variety of exoplanets that have been found has revealed an unexpected diversity - in other words, not all planetary systems resemble our own.

With that said, as the sample size of known exoplanets has increased, this conclusions we had made on planet formation based on the sample of one system (the Solar system) have somewhat stood the test of time. Our broad understanding (planet formation in a disc around a young star) has proven accurate, and we now know that planets are ubiquitous - most, if not all stars host planets. However, the sheer diversity of these planets—in size, mass, and orbital architecture—has proven to be far beyond the scope of Solar system-based predictions for the demographics of exoplanets.

Figure 16 shows the dramatic rise in the number of known exoplanets, organized by detection technique. The two most productive methods are transit and radial velocity (RV) techniques. Here, we will summarise the current state of knowledge in exoplanet demographics as determined through transit and RV surveys. Figure 17 shows the distribution of known exoplanets as a function of their discovery method, their mass, and their orbital period. It is immediately apparent that the transit and RV methods of exoplanet detection are sensitive to different subsets of the population of exoplanets. Both techniques fall victim to significant observational biases, and are limited by the available observational temporal baseline, which effectively determines the maximum orbital period at which a given survey they can detect planets. The transit method is heavily biased towards detecting planets that move on short period orbits, and most effectively finds planets with larger radii (as they produce a significantly more obvious transit signature in the light curves of their host stars). Similarly, the radial velocity method is biased towards finding the most massive planets, with an efficiency that is limited towards short orbital periods by the available observational cadence, and towards long periods by the temporal baseline over which observations have been obtained.

---

[66] Though it should be noted that this may soon change - Gravity Collaboration et al (2019) report the direct detection of HR 8799 e by optical interferometry, and describe the determination of the mass of that planet ($10_{-4}^{+7}$ times the mass of Jupiter) on the basis of the application of synthetic spectra to their observational data.

[67] Those planets detected through Transit Timing Variations are a notable exception to this general rule of thumb. In such cases, the actual mass (rather than the minimum mass) can be directly determined.

When interpreting the sample of known exoplanets, especially as displayed in plots such as Figure 17, it is critical to remember that what is known (and not yet known) about exoplanet demographics is shaped by the biases of the techniques used to detect exoplanets. These biases must be carefully accounted for to extract robust demographic information from the sample of detected exoplanets. To give just one example - when one examines Figure 17, it would be natural to think that hot Jupiters were abundant, and that most planetary systems feature at least one such scorching giant. This is, however, the result of the aforementioned observational biases. Simply put - both the transit and radial velocity techniques are strongly biased towards the detection of hot Jupiters - they are the easiest of all planets to find, and so one would expect them to dominate exoplanet demographics, particularly in the early days of the field. Indeed, the true occurrence rate of hot Jupiters is low - with studies based on data taken by *Kepler* and *TESS* consistently returning rates of less than 1% (e.g. Howard et al., 2012; Fressin et al., 2013; Wang et al., 2015; Zhou et al., 2019).

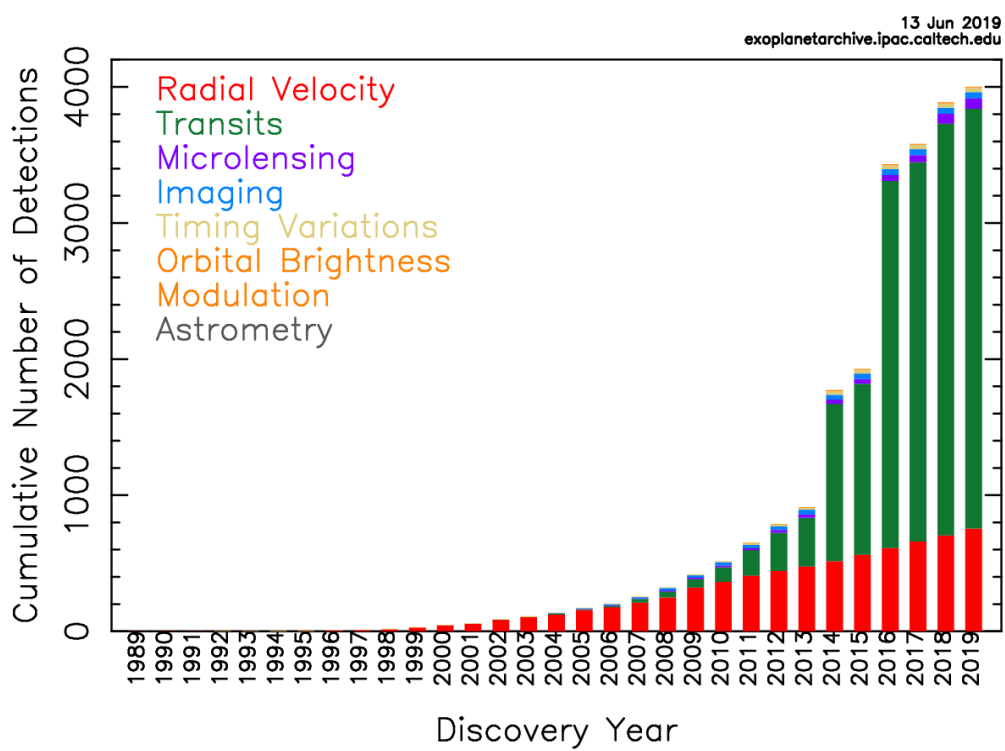

*Figure 16*: *Exoplanet detections as a function of year, coloured by the detection technique. Steep increases in the number of transiting exoplanets in 2014 and 2016 are due to observations from the Kepler mission. Image courtesy of the NASA Exoplanet Archive*[68], *accessed on 13th June, 2019.*

---

[68] https://exoplanetarchive.ipac.caltech.edu/

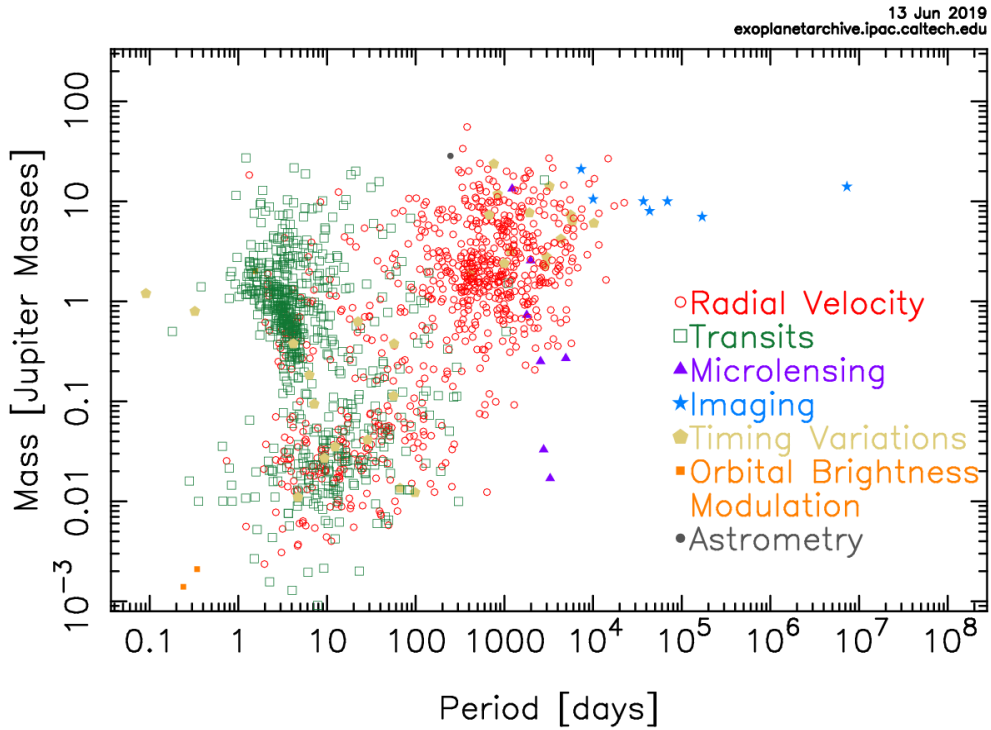

*Figure 17: Distribution of known exoplanets in mass-period space coloured by detection technique. The different sensitivities of the transit and radial velocity methods are demonstrated by the largely separate samples of green and red points. The masses used are those published in the literature – and as a result, the data for radial velocity planets shows the minimum mass compatible with the observations (i.e. m* sin *i), whilst the only transiting planets included are those for which a direct mass measurement has been made. Interpretations of the exoplanet demographics from figures such as this must be made with caution, as the observational biases from each detection technique are present, and the mass of any given planet is subject to the degree of uncertainty with which its host star has been parameterised. Image courtesy of the NASA Exoplanet Archive[69], accessed on 13$^{th}$ June, 2019.*

Numerous ground- and space-based transit surveys are responsible for having found most of the presently known exoplanets. One of the most notable efforts is the *Kepler* mission (Borucki et al., 2010), which acquired a nearly continuous four-year observational baseline for over 150,000 stars. *Kepler* identified over 4,000 exoplanet candidates (Akeson et al., 2013, Thompson et al., 2018), of which more than 2,300 have been confirmed or statistically validated (Rowe et al., 2014; Morton et al., 2016). Thanks to the observational baseline and photometric precision of *Kepler*, the orbital periods of these exoplanets range from less than a day to several hundred days and their radii are as small as ~1 Earth radius (Twicken et al., 2016).

Hsu et al. (2019) provided a thorough analysis of the vetting and detection efficiency of the final data release for the primary *Kepler* mission (DR25) using updated stellar parameters from the *Gaia* DR2 release. They found that a typical FGK star hosts $1.07^{+0.10}_{-0.08}$ planets with radii in the range 1.0 – 16 $R_{Earth}$ and orbital periods spanning 4 - 128 days. Hsu et al. (2019) also estimated that planets with sizes 0.75 - 1.5 $R_{Earth}$ and orbital periods of 237 - 500 days have an occurrence rate for FGK stars of < 27%. This value remains solely an upper limit, as a result of the low detection efficiency and high false alarm probability for planets in that region of parameter space. At even longer periods, an automated planet search through the *Kepler* data determined that, on average, Sun-like stars have 2.0 ± 0.7 planets between 0.1 and 1 Jupiter radii moving on orbits with periods between 2 and 25 years (Foreman-Mackey et al., 2016). Interestingly, the exquisite precision of the *Kepler* photometry has enabled the investigation of specific sub-groups of the exoplanet population for which there are no analogues in the Solar system. The radius distribution of short-period (<100 days) exoplanets contains a statistically significant decrease in occurrence between 1.5 and 2 Earth radii (Fulton et al., 2017). This feature is currently the subject of intense investigation, but is generally thought to be related to the erosion of planetary atmospheres by stellar radiation.

---

[69] https://exoplanetarchive.ipac.caltech.edu/

At longer orbital periods (hundreds of days), the reliability of candidates from transit surveys decreases and RV confirmation is typically employed to ensure the robustness of planet occurrence rate statistics. Santerne et al. (2016) combined *Kepler* photometry with long-baseline RV observations from the SOPHIE spectrograph to find that the total occurrence rate of giant planets (with transit depth between 0.4 and 3%) within an orbital period of 400 days is 4.6 ± 0.6%. At wider orbital separations, the efficiency of transit detections decreases sharply and long-baseline RV surveys provide much of the exoplanet demographic information. Between 0.3 and 7 au, giant exoplanets with mass in the range 0.3 to 13 Jupiter masses have an occurrence rate of $6.2^{+2.8}_{-1.6}$% (Wittenmyer et al., 2016b).

A primary goal of many studies has been to relate stellar properties to exoplanet occurrence rates. Since stars and planets form from the same protostellar material, it seems reasonable to expect that the properties of the host star would influence those of any planets. Such investigations were restricted to the realm of theory prior to the discovery of exoplanets, but now that planets have been discovered around a wide variety of stars, such studies are becoming ever more important.

In the early days of the exoplanet era, RV surveys of F-, G-, and K-type stars identified a positive correlation between the occurrence rate of giant planets within several au and stellar metallicity (Santos et al., 2003; Fischer and Valenti 2005). This is now commonly known as the "planet metallicity relation" for exoplanets, and is interpreted as providing strong support for the core accretion theory of giant planet formation (Pollack et al. 1996). Subsequent work has found that this relation likely extends to later-type (lower mass) stars (Johnson et al., 2010; Neves et al., 2013).

Planet detection efforts aimed specifically at M dwarf stars have also uncovered interesting relationships between exoplanet demographics and stellar properties. In general, small planets are found to be much more prevalent around M dwarfs. According to Dressing & Charbonneau (2015) and Gaidos et al. (2016), the average M dwarf has roughly 2.2-2.5 exoplanets with radii between Earth and Neptune moving on orbits with periods of ~200 days or less. M dwarfs also favour the formation of "compact multiple" systems of exoplanets, featuring several Earth-size planets on tightly-packed short-period orbits, with the scale of the entire planetary systems comparable to that of the Galilean moons orbiting Jupiter (e.g., Muirhead et al., 2015; Ballard & Johnson 2016). Whilst low mass planets orbiting late-type stars seem to be abundant, the prevalence of giant planets orbiting such stars is notably smaller than that observed around more Sun-like FGK stars (e.g., Cumming et al., 2008). In other words, it seems that the smaller the star, the smaller will be its largest planets.

## 5.2. THE SOLAR SYSTEM AS AN EXOPLANET SYSTEM

As noted in section 5.1, the architecture of the Solar system dominated theories regarding formation processes and orbital evolution until the discovery of other systems. Over time, it has become clear that the range of possible architectures, although not a continuum, is incredibly vast (Winn & Fabrycky, 2015). One way to place the Solar system within the broader context is to consider the appearance and planetary signatures of our Solar system if it were to be observed as an exoplanet system.

| Planet | Semi-major axis (au) | Orbital Period (days) | K (m/s) | Transit Depth (ppm) | Transit Probability (%) |
|---|---|---|---|---|---|
| Mercury | 0.387 | 87.97 | 0.008 | 12 | 1.20 |
| Venus | 0.723 | 224.70 | 0.086 | 76 | 0.64 |
| Earth | 1.000 | 365.25 | 0.089 | 84 | 0.47 |
| Mars | 1.524 | 686.97 | 0.008 | 24 | 0.31 |
| Jupiter | 5.20 | 4332.6 | 12.48 | 10551 | 0.09 |
| Saturn | 9.54 | 10759 | 2.76 | 7498 | 0.05 |
| Uranus | 19.19 | 30689 | 0.30 | 1348 | 0.02 |
| Neptune | 30.1 | 60182 | 0.28 | 1266 | 0.02 |

*Table 4*: *The semi-major axes and orbital periods for the Solar system's planets, and the radial velocity semi-amplitude (K), transit depth, and transit probability for those planets, as they would be seen by an external observer. We remind the reader that the value of K given above is the maximum amplitude that would be observed, in the case of an orbit that was oriented along the observer's line of sight. In cases where the orbit was tilted to that line of sight, only the line-of-sight component of the motion would be observed, resulting in a smaller signal.*

Shown in the above table are the semi-major axes and orbital periods for the planets in the Solar system. Also included are the radial velocity semi-amplitudes (m/s) for the planets, under the assumption that the Solar system is viewed close to edge-on. The typical precision of current RV surveys is ~1 m/s, but next generation instruments are poised to be deployed (Plavchan et al., 2015). The "NN-explore Exoplanet Investigations with Doppler spectroscopy" (NEID) instrument aims to have total per measurement error budget of ~30 cm/s (Halverson et al., 2016). The "Echelle SPectrograph for Rocky Exoplanets and Stable Spectroscopic Observations" (ESPRESSO) instrument is anticipated to achieve a per measurement precision of ~10 cm/s (Pepe et al., 2014). Thus, although current RV surveys are only sensitive to Jupiter and Saturn analogues (subject to their having a sufficiently long observational baseline), Earth/Venus analogues may be detectable in the coming years. Note that although the semi-amplitude of Uranus and Neptune are large compared to the terrestrial planets, their exceptionally long orbital periods mean that the detection of their analogues continues to pose a significant challenge.

Also shown in the above table are the predicted transit depths (in parts per million, ppm) and transit probabilities for the eight Solar system planets. The transit depth of Mercury lies at the very threshold of the photometric precision of the *Kepler* mission (Borucki et al., 2010), and whilst *Kepler* did detect planets with diameters smaller than or comparable to Mercury (e.g. *Kepler*-37 b Barclay et al., 2010; *Kepler*-444 b Campante et al., 2015), those planets moved on orbits with periods far shorter than that of Mercury (~13.4 and ~3.6 days, respectively). Such short orbital periods allowed large numbers of transits to be observed during *Kepler*'s primary mission, which greatly increased the ease with which those planets could be detected.

However, the major challenge in the detection of Solar system analogues lies in the transit probability, which follows an inverse relationship with the semi-major axis (Kane & von Braun 2008). Simply put, the more distant the planet from its host, the lower the likelihood of it being observed to transit from any given external location. Depending on the duration and window function of the survey (von Braun, Kane

& Ciardi, 2012), most of the Solar system's planets would, at best, be detected as single transit events by current and past surveys. Such single transit events are far from optimal, and pose significant challenges with regards to their follow-up and validation (Osborn et al., 2016; Villanueva et al., 2019; Yao et al., 2019). As a result, the bias of survey completeness towards short orbital periods means that the full context of true Solar system analogues is an on-going process that will rely on combinations of exoplanet detection techniques to fully realize.

A further critical aspect of our Solar system as an analogue for exoplanetary systems is the realisation that in-situ data for exoplanets is unlikely to be achieved for several generations (and possibly significantly longer). Hence, all inferences regarding the connection between planetary atmospheric data and surface processes relies upon the in-situ data obtained within the Solar system from which models applicable to exoplanets may be constructed. The primary terrestrial sources of such atmospheric data and geological profiles are Venus, Earth, Mars and several of the moons of the giant planets, most particularly Titan.

Furthermore, it is important to study the evolution of the Solar system's planets through time, since they occupy a specific evolutionary state at the present epoch (Del Genio et al., 2019). The most relevant Solar system bodies to the study of terrestrial planets are Venus and the Earth since they represent the primary target demographic of exoplanet searches in the context of habitability (e.g. Kane et al., 2019b). To illustrate the importance of the study of our own planets, we note that the abundance of oxygen in the Earth's atmosphere was significantly lower than the present day for most of our planet's history (Lyons & Reinhard, 2009), whilst recent studies have revealed that Venus could have retained its liquid surface water up until ~1 Gya (Way et al., 2016). Searches for potential Venus analogues will play an important role in understanding the bifurcation of atmospheric evolution between Venus and Earth in our Solar system (Kane et al., 2013, 2014). More detailed studies of the Venusian atmosphere are required in order to fully characterise runaway greenhouse environments for terrestrial exoplanets (Schaefer & Fegley 2011; Ehrenreich et al., 2012; Kane et al., 2018). Likewise, the Solar system's giant planets are critically important laboratories for deriving atmospheric and interior models of giant exoplanets as a function of composition, age, and insolation flux (Mayorga et al., 2016; Withers & Vogt 2017; López-Puertas et al., 2018).

As mentioned previously, analogues of the Solar system's ice giants are particularly challenging for us to detect at present. Despite this, data from studies of the Solar system's ice giants are being utilised for exoplanet models (Arridge et al., 2014), and scaled down versions of the Solar system may make the detection of ice giant planets beyond the snow line more accessible (Kane, 2011). Overall, the expanding exploitation of the synergies between planetary science and exoplanetary science is an essential step in reliably inferring the derived properties of exoplanets, most particularly inferences regarding the surface conditions of terrestrial exoplanets.

Over the course of its history, the Solar system has exhibited large variability both in the structures within the system and the visible emission from the attendant planets and debris belts that would be detectable by an outside observer. As a result, how the system would have looked to an external observer has changed dramatically over the past 4.5 billion years.

At the earliest epoch, around 3 – 10 Myr after the Sun arrived on the main sequence, the gaseous proto-planetary disc would still be in the final stages of dissipation, or have already dispersed. At that time, the outer Solar system would be dominated by the giant planets, which would still be warm from accreting their gaseous envelopes, and would therefore be radiating brightly in the near infra-red (much as is the case with the four giant planets orbiting HR8799 e.g. Marois et al., 2008, 2010). At this stage the primordial debris belts would most likely still be dynamically cold, producing little dust through collisions, and therefore be faint. In the inner Solar system, the terrestrial planets would not yet be fully formed, and the inner regions of the Solar system would therefore be full of small, short-lived dust particles, as a chaotic pinball of proto-planetary embryos collided and aggregated to one day form the four terrestrial planets. At infrared wavelengths, the warm dust produced in this manner would be detectable as a bright infrared excess at near and mid-infrared wavelengths, perhaps with silica emission features as a result of the colliding embryos producing high temperature, gaseous SiO, as seen around HD 15407 (Fujiwara et al., 2012) and HD 172555 (Lisse et al., 2009; Johnson et al., 2012).

In the Nice Model for the evolution of the Solar system, it was suggested that the migration of Saturn and Jupiter resulted in the two planets crossing a mutual mean motion resonance approximately 700 Myr after the Solar system formed. The result of that resonance crossing was a massive dynamical instability, which disrupted the orbits of the giant planets and the Solar system's debris belts. The orbits of the outer four planets become chaotic, and the quiescent primordial debris belts were shattered, scattering cometary and asteroidal material far beyond their point of formation by the rearrangement of the giant planets onto their current orbital configuration. The model suggests that this rearrangement resulted in the Late Heavy Bombardment, delivering a drenching of water rich cometary material to the terrestrial planets. Some of the scattered small bodies were captured into the Jovian and Neptunian Trojan populations (e.g. Morbidelli et al., 2005, Lykawka & Horner, 2010), whilst the migration of Jupiter sculpted the Asteroid belt, gutting the inner system and removing ~90% of its original mass. If the Solar system did indeed experience a period of dynamical instability such as that proposed in the Nice Model, then the collisions driven by this proposed period of extreme instability would have resulted in the production of a large density of small dust grains, which would in turn produce an appreciable, although short lived ($< 10^4$ yrs), brightening of the Sun's debris disc by several orders of magnitude (Wyatt et al., 2007; Booth et al., 2009; Raymond et al., 2011). Dust-producing collisions occur stochastically over the lifetime of a planetary system, such that the levels of dust (and hence observed infra-red excess) in the system will peak and decay many times.

Currently, at an age of 4.5 Gyr, the Sun's two principal debris belts are very faint, with fractional luminosities, $L_{dust}/L_{star}$ ~$10^{-7}$ for the inner Solar system (Backman, Dasgupta & Stencel, 1995 – the definition of 'Zodi') and ~$1.2\times10^{-7}$ for the Edgeworth-Kuiper belt (Vitense et al., 2012)[70]. The peak of the continuum emission from the Edgeworth-Kuiper belt (the Solar system's cold disc) is 0.5 mJy at a wavelength of approximately 40 – 50μm, as seen from a distance of 10pc - a flux that is less than 1% of the contribution from the Solar photosphere at those wavelengths (Vitense et al., 2012). By contrast, the peak of the continuum emission from the zodiacal light falls at approximately 20 μm (Reach et al., 2003), at a flux level of 0.1 mJy, as seen from a distance of 10 pc (compared to a stellar flux at 2 Jy at the same wavelength)[71].

These values are lower than those observed for the debris discs detected around nearby Solar type stars (Montesinos et al., 2016; Sibthorpe et al., 2018), which typically exhibit fractional luminosities for cold debris discs of the order $10^{-4} – 10^{-5}$ and warm discs (e.g. Fujiwara et al., 2013; Kennedy & Wyatt 2013) up to levels of a few percent of the stellar flux (e.g. BD+20 307, Zuckerman et al., 2008; HD15407, Fujiwara et al., 2012). The known menagerie of debris discs around other stars therefore represent brighter, and more massive (and/or more collisionally active), analogues to the Solar system's Edgeworth-Kuiper or Asteroid belts.

Current detection of circumstellar debris dust is limited by instrument sensitivity, and many fainter discs remain to be detected – the Edgeworth-Kuiper belt is believed to have a disc brightness about average for its age (Moro-Martin et al., 2015), but is as yet undetectable. With this caveat, far-infrared observations by *Herschel* identified cool debris discs around ~15% of nearby, Sun-like stars (Trilling et al., 2008;

---

[70] These values relate to the total excess flux emitted by the discs at long-wavelengths, as would be seen from a distant star. Since almost all that excess comes at infrared wavelengths, these values are often called the '*fractional infrared luminosity*' (e.g. Perryman, 2018, p342).

[71] Due to the faint nature of the zodiacal light, few studies have considered the detection of the Solar system's zodi from another stellar system. The values here are taken from page four of a talk given by Chris Koresko at MPIA in 2003, recovered online at http://www.mpia.de/MIDI-RB/Contributions/Koresko.pdf on 3rd April 2020. The plot on which the values are based credits Kuchner, 2012.

Eiroa et al., 2013; Montesinos et al., 2016; Sibthorpe et al., 2018), with a higher incidence of ~ 30% around more massive A-type stars (Chen et al., 2006; Thureau et al., 2014).

The combination of volume-limited far-infrared surveys with exoplanet survey results for the same sample of stars makes a direct statistical comparison of the properties of stars with planets (down to a few Earth masses) and/or debris discs (with dust brightness levels ~ 10x the Edgeworth-Kuiper belt) possible for the first time. This led to the determination that stars with detected planets have an enhanced incidence of circumstellar debris (Matthews et al., 2014). Searching for more refined correlations, such as with planetary mass or stellar metallicity, has led to the identification of weak trends (Wyatt et al., 2012; Marshall et al., 2014a; Meshkat et al., 2017), but such analyses are hampered by the heterogeneity of the underlying disc and planet data sets, making direct determination of the impact of such properties on the occurrence of planets/debris difficult (Moro-Martín et al., 2015; Wittenmyer & Marshall, 2015).

Spatially resolved imaging of the scattered light and/or the continuum emission from circumstellar debris has revealed the architectures of these planetary systems in greater detail. In this regard the *Hubble Space Telescope*, Gemini/GPI, and VLT/SPHERE at optical/near-infrared wavelengths, and ALMA operating at radio wavelengths have been revolutionary. Most obviously, the complex dynamical structure exhibited by β Pictoris' debris disc, believed to be induced by its directly imaged Jovian-mass companion, clearly exhibits the interplay between the planetary and planetesimal components in a system (e.g. Smith & Terile 1984; Golimowski et al., 2006; Dent et al., 2014; Matra et al., 2017, 2019).

At millimetre wavelengths, other systems have exhibited an equally intriguing variety of structures thought to be the result of disc-planet interaction, such as eccentric debris rings (MacGregor et al., 2017; Faramaz et al., 2019), gaps in broad planetesimal belts (Marino et al., 2018, 2019), multiple planetesimal rings (MacGregor et al., 2019), haloes of millimetre-sized dust grains from dust filtering (Marino et al., 2017; MacGregor et al., 2018) and scattered planetesimal populations (Marino et al., 2017; Geiler et al., 2019). At optical and near-infrared wavelengths, high angular resolution coronagraphic imaging reveals similar structures to those seen at millimetre wavelengths, including the presence of haloes of small dust, asymmetries, and eccentric rings (Schneider et al., 2014).

Initial successful efforts toward scattered light imaging of debris discs (building on the detection of the disc around β Pictoris; Smith & Terrile, 1984) were dominated by the *Hubble Space Telescope* (Schneider et al., 2014; Choquet et al., 2018; Ren et al., 2019) and the reanalysis of archival observations (Choquet et al., 2016). More recently, advances in high contrast adaptive optics have led to a renaissance in ground-based disc imaging, with high profile results including the polarimetric imaging of HR 4796A (Perrin et al., 2015), the resolution of many young discs (e.g. Hung et al., 2015; Millar-Blanchaer et al., 2016; Chauvin et al., 2018; Hom et al., 2020) including new detections (Sissa et al., 2020), the identification of asymmetries (Draper et al., 2016) and complex substructures (Olofsson et al., 2018). The information gleaned from the dust distribution and properties in these two wavelength regimes is complementary, providing constraints on the dust grain scattering albedo and size (e.g. Choquet et al., 2018; Marshall et al., 2018).

Determining the distribution and intensity of the scattered light levels around other stars has been of particular interest due to its implications for future exoplanet direct imaging instruments (e.g. Roberge et al., 2012). This exo-zodiacal light would constitute a confusing background to any attempts at the direct imaging of exoplanets in the habitable zone of their host stars. The direct measurement (by photometry or spectroscopy) of exo-zodiacal light would be stymied by the brightness of the star. Interferometric methods have therefore been employed at optical and infrared wavelengths to infer exo-zodiacal dust levels through the measurement of the visibility deficit revealing the presence of an extended component (i.e. circumstellar dust) to the total emission from a stellar system.

Using the Keck interferometer, limits at the level of 100x to 1000x Zodi were obtained for tens of stars, with a few detections for the brightest discs (Mennesson et al., 2014). Measurements with VLTI surveyed around 90 stars at near-infrared wavelengths spanning a broad range of stellar spectral types, finding a hot dust incidence comparable to that of cool debris discs detected by *Spitzer* and *Herschel* but uncorrelated with the presence of a known outer debris belt (Absil et al., 2013; Ertel et al., 2014). Most recently, the Hunt for Observable Signatures of Terrestrial Systems on the Large Binocular Telescope Interferometer searched for dust close to the habitable zone of 38 stars, finding that the presence of habitable zone dust was correlated with cool dust, and that most Solar-type stars were not very dusty, with a median of 3 Zodis and a 95% upper limit of 27 Zodis based on their N-band measurements (Kennedy et al., 2015; Ertel et al., 2018, 2020).

## 5.3 DEBRIS DUST IN EXOPLANETARY SYSTEMS

Almost all stars begin their existence with a circumstellar disc most commonly identified through excess emission at infrared wavelengths (Williams & Cieza 2011; Ribas et al., 2014; Wyatt et al., 2017). The effects of grain growth (at the earliest stages) and collisional evolution lead to a predictable monotonic decay in disc brightness over time, assuming a steady-state evolution (e.g. Wyatt et al., 2007; Kains, Wyatt & Greaves, 2011; Marshall et al., 2011; Löhne et al., 2012). The dust masses in circumstellar discs can be traced from the sub-millimetre emission, revealing a drop in the observed mass at around 1-10 Myr in age (e.g. Panić et al., 2013; Holland et al., 2017), comparable to the approximate lifetime of protoplanetary discs as seen through observation of accretion and gas tracers (Ribas et al., 2014; Richert et al., 2018). More massive and compact (smaller semi-major axis) discs evolve more quickly due to the shorter collisional timescales for dust-producing planetesimals, leading to the same overall brightness decay over time, which follows an approximate $1/t$ relation. A bright disc is not necessarily a massive disc[72], except potentially at the earliest phase of its evolution.

---

[72] Since the brightness of the disc is linked to the amount of dust, rather than the mass of larger planetesimals - so an excited low-mass disc (with high collision rates) could appear equally bright as a high-mass, dynamically cold disc (with low collision rates). Both scenarios generate equal amounts of dust, and so result in equally bright discs. In other words, in understanding the nature of a disc from its brightness, there remains a degeneracy between the collision rate and disc mass.

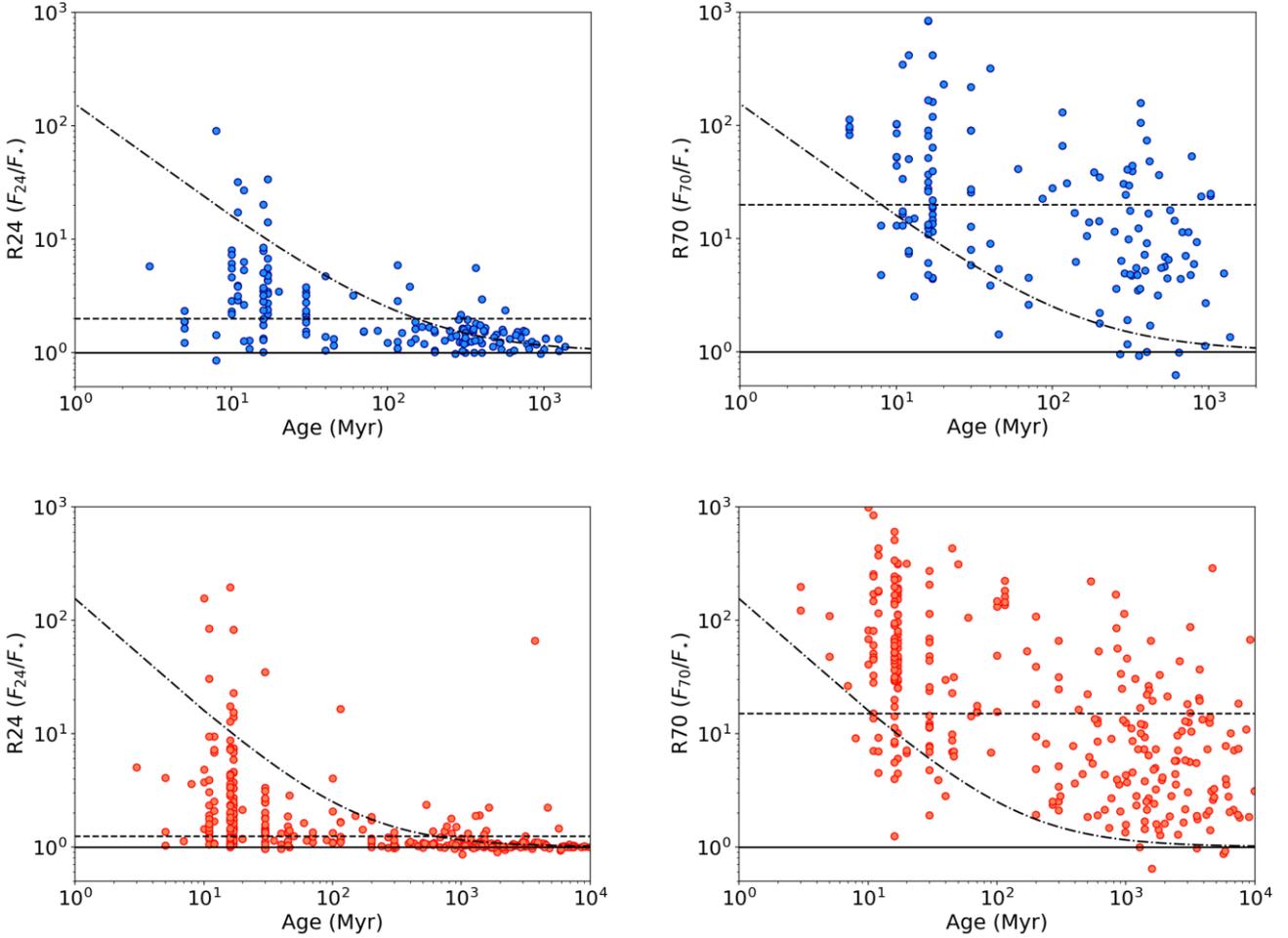

*Figure 18: Here we show the evolution of infrared excess for warm (Asteroid belt) and cold (Edgeworth-Kuiper belt) analogues as a function of time for A stars (blue, top row) and FGKM stars (red, bottom row). Excess emission is characterised by the ratio of total flux to stellar photospheric flux as measured by the Spitzer MIPS instrument at 24 and 70 um. Photometric measurements were taken from the Spitzer IRS debris disc catalog (Chen et al., 2014). The horizontal solid line denotes $R_{wav} = F_{wav}/F_* = 1$ (i.e. no excess emission). The dashed lines denote the boundary above which 'high' excess systems exist R = 2/20 for A-stars at 24/70 um, R = 1.25/15 for G-stars at 24/70 um; (Wyatt et al., 2007; Kains et al., 2011). The dot-dash curved line is the envelope of 24 um excesses ($150 t_{age}^{-1}$) as determined by Rieke et al. (2005).*

Whilst the evolutionary trend with time holds for the vast majority of debris discs, there exist several well-known anomalies, such as BD+20 307 (Song et al., 2005) and q$^1$ Eri (Liseau et al., 2008, 2010), that are brighter (higher fractional luminosity) than expected given the stellar age (see Figure 18). The interpretation of these abnormally bright discs is that the observed emission is, wholly or partially, the result of stochastic event(s) that have caused an enhanced rate of dust production at the present epoch, rather than being the result of those discs being unusually massive.

The mostly likely mechanism behind this enhanced dust production is thought to be either self-stirring by the largest planetesimals within the disc (e.g. Kenyon & Bromley 2008; Krivov & Booth 2018), or the migration of a planet through the planetesimal belt (e.g. Mustill & Wyatt, 2009). The process by which self-stirring could result in excess dust production is subject to many assumptions (such as the size and strength of the planetesimals in the disc), leading to orders-of-magnitude uncertainty in the timescale for the mechanism to act. At the most extreme end of systems with enhanced collision rates are bright ($L_d/L_*$ > 0.01) exo-Asteroid belts that display evidence of rapid time variability or decay at near- or mid-infrared wavelengths (e.g. Melis et al., 2012, Meng et al., 2015), ruling out an in-situ, steady-state origin for the observed dust. At far-infrared wavelengths, the detection of spatially resolved emission from overly bright debris discs around young stars (e.g. Moór et al., 2015, Vican et al., 2016) places a constraint on the disc evolution, revealing several systems that are consistent with either self-stirring or planet migration influencing the observed disc architecture. Given that these discs are anomalously bright, the stirring event should have only been a (relatively) recent occurrence in these systems.

For example, a recent detection of CO gas in the debris disc around HD 32297 reveals that the observed debris disc could be coeval with the gas disc, which in turn suggests that it might therefore be only a few thousand years old (Cataldi et al., 2020). This bright debris disc system is young, but otherwise unremarkable in its extent or architecture. As a result, this discovery suggests that the tacit assumption that the discs we observe around other stars are the same age as the host star and present as a result of (broadly) steady state evolution is not always robust.

A handful of debris discs have mid-infrared spectra with features that reveal the composition of their constituent dust. As a result, the origin of that dust, and the nature of its production, can be directly determined. The debris disc orbiting HD 172555 has SiO gas in its spectrum, consistent with high temperature, shocked material produced in a hypervelocity impact (Johnson et al., 2012), whereas that around HD 109085 reveals a mixture of primordial (icy, amorphous) and processed (crystalline) material consistent with an origin of the parent bodies in the outer belt of that system (Lisse et al., 2013).

As we become better able to study the structure and distribution of dust in exoplanetary systems, it will become possible to model the processes that cause that structure to occur. For example, the debris dust orbiting HR 8799 has been found to lie in two distinct reservoirs, one interior to, and the other exterior to, the star's four known giant planets. Much as is the case in the Solar system, the structure of debris belts in the HR 8799 system is clearly driven by the gravitational sculpting resulting from the influence of those four planets (e.g. Su et al., 2009; Contro et al., 2016; Goździewski & Migaszewski, 2018, Geiler et al., 2019). By contrast, the sharp inner edges to the dust distributions of two young stars (HR 4796A and HD 141569) have been proposed as being the result of the photophoretic driving of dust through a particularly tenuous dispersing gas disc (e.g. Krauss & Wurm, 2005; Hermann & Krivov, 2007) - a process that has also been invoked to explain the origin of high-temperature inclusions in the Solar system's asteroidal and cometary bodies (e.g. Mousis et al., 2007a, b).

Amongst known debris discs, Poynting-Robertson drag (as discussed in section 4.9.1) has a minimal effect on the radial distribution of dust grains (Wyatt, 2006). These systems are dense enough (~10 to 1000 times the mass of the Edgeworth-Kuiper belt) that the dust lifetime is dominated by mutual collisions rather than by migration. For fainter discs, Poynting-Robertson drag can be an important dust removal mechanism, but such systems lie beyond our present capacity to detect. The proposed *Spica* infrared/sub-millimetre mission, scheduled for launch and operation in the mid-2030's, would provide observations of many exact Edgeworth-Kuiper belt analogues (Roelfsema et al., 2018).

In the coming years, new facilities will enable us to identify the origin, composition, and dynamics of the dust-producing bodies that comprise debris discs with greater precision than has been achieved to date. Such data will not only help us to better understand and characterise the debris discs in question - it will also play an important role in our efforts to address the more esoteric questions of habitability in these planetary systems in the longer term, since it will offer the potential for us to quantify the impact regimes that would be experienced by potential rocky habitable zone planets, following methods initially developed to study the evolution and history of our own planetary system (e.g. Horner & Jones, 2008a, 2009, 2010a,c).

## 5.4 COMPARATIVE ANALYSIS OF PLANETARY SYSTEMS
A primary scientific aim of exoplanet searches is to determine the degree to which our Solar system represents a typical outcome of planetary system formation and evolution. To this end, much attention and effort is lavished on the quest to detect Earth-like planets in Earth-like orbits. Indeed, this was the chief aim of the *Kepler* spacecraft mission (Borucki et al., 2010). Space-based transit surveys such as *Kepler*, *CoRoT*, and *TESS* (Barge et al., 2008, Ricker et al., 2015) are particularly adept at detecting such small planets. The photometric sensitivity of these missions reaches about 30 parts per million (ppm) for suitably quiet stars, whilst the flux decrement due to an Earth-size planet transiting a Solar-type star is 80 ppm. In the four years of the *Kepler* prime mission, a handful of potential Earth- analogue candidates were identified moving on orbits within their host stars' habitable zones (Kane et al., 2016). Whilst the radial velocity (RV) method has been a successful technique for detecting exoplanets for decades (Campbell et al. 1988, Cochran & Hatzes 1994, Butler et al. 1996), the RV signal of an Earth-mass planet in a 1 au orbit is a dauntingly diminutive 8 cm/s, which remains below the current best precision of a few tens of cm/s for instruments such as HARPS and ESPRESSO (Fischer et al., 2016).

Of equal importance, but perhaps capturing less public interest, is the presence of outer giant planets analogous to Jupiter and Saturn, moving on orbits with periods of more than 10 years. For these planets, the transit method is of minimal use, both because the probability of a planet transiting its host scales inversely to the planet's semimajor axis, and because the transit of such a distant planet would be a singular ~1 day event in any given 10+ year period. Fortunately, however, the RV method is eminently capable of detecting Jupiter analogues: our Jupiter imposes an RV signal of 12 m/s in a 12 year period on the Sun, and RV surveys have achieved precisions of, or better than, 3 m/s for at least the last 25 years (Fischer et al., 2014; Endl et al., 2016). From the long-running RV surveys, there is a growing consensus that about 3-6% of Solar-type stars host a "Jupiter analogue," typically defined as being a giant planet orbiting beyond the ~3 au snow line (Wittenmyer et al., 2006, 2011b, 2016b; Rowan et al., 2016, Zechmeister et al., 2013). As the temporal baseline of legacy RV surveys lengthens, true Saturn analogues may be detectable; the RV signal of Saturn is 3 m/s in an almost 30 year orbit. Given that the orbital periods of the ice giants Uranus and Neptune are 84 and 165 years, respectively, and that their maximum RV amplitudes would be just 0.3 and 0.28 m/s, the RV method is not, currently, likely to deliver true analogues for those planets any time soon, and so to find such planets, we must turn to other detection techniques.

Direct imaging probes a different but complementary region of parameter space to the RV and transit methods, being most sensitive to giant planets moving on orbits at tens of au from their host stars. To date, the favourable contrast ratios of young planetary systems have biased direct imaging detections toward systems younger than ~100 Myr. The HR 8799 system (Marois et al., 2008, 2010, 2012; Wang et al., 2018) has been considered a benchmark "scaled up Solar system analogue", with four giant planets of mass 5-7 Jupiter masses orbiting between 14-68 au, as well as two distinct debris belts (potentially analogues of the Asteroid and Edgeworth-Kuiper belts; e.g. Su et al., 2009; Contro et al., 2016; Goździewski & Migaszewski, 2018). As mentioned in section 5.2, scaled down versions of the Solar system may also assist in the detection of ice giant analogues via RVs and transits (Kane, 2011).

As we noted earlier, the combination of transit and RV observations enables the density of certain exoplanets to be determined. In the main, these results have been used to draw conclusions on the potential composition of the planets in question, but it has also recently been used to highlight the significant disparity between the densities of two almost identically-sized planets in the Kepler-107 system. Both Kepler-107 b and c have radii around 1.5-1.6 times that of the Earth, and so it would be natural to assume that RV observations would show them to have similar masses. However, such observations instead reveal that the bulk density of Kepler-107 c is more than twice that of Kepler-107 b (12.6 g cm$^{-3}$ vs 5.3 g cm$^{-3}$). As discussed by Bonomo et al. (2019), the best explanation for this disparity is that Kepler-107 c was once the victim of a giant, mantle-stripping collision - as was the case for Mercury in the Solar system (e.g. Benz et al., 2007; see also section 4.1.1, above). These results demonstrate that we are now entering an era where mass and size measurements of exoplanets can begin to give information not only on the composition of those planets, but also on their collisional history. Given the widespread evidence of giant impacts in the Solar system, this result is perhaps not particularly surprising - but it is worth stressing that, without the evidence from our own backyard, planets such as Kepler-107 c would probably be hugely confusing additions to the exoplanet catalogue. Given the uncertainties inherent in calculating the densities of transiting exoplanets (particularly the smallest), it seems unlikely that we would be able to definitively identify the Earth and Mercury as having relatively high densities compared to the other terrestrial worlds, were we observing the Solar system from the outside. However, in the coming years, the precision with which we can make such measurements will improve as new facilities come on-line. In the future, we may well therefore reach a point where such forensic studies can be carried out for planetary systems similar to our own.

Currently there are in excess of two hundred known satellites orbiting the eight planets within the Solar system (as detailed in section 2, in Table 1), most of which orbit the two largest planets in our system, with Jupiter hosting 79 known moons and Saturn hosting 82 known moons. Whilst none of those moons are large enough that we would be able to confidently detect them were they placed in orbit about the known exoplanets, the sheer number of moons in the Solar system, particularly the large number orbiting the Jovian planets, indicate a high probability that moons will also exist orbiting exoplanets. As a result, there is a general consensus in the scientific community that exomoons do exist (e.g. Williams et al. 1997; Kipping et al., 2009; Heller 2012; Heller & Pudritz 2015; Zollinger et al., 2017). However, whilst recent work suggested the first preliminary detections of exomoon signatures (Teachey et al., 2017; Teachey & Kipping 2018), those results remain heavily debated, and as a result, no exomoons have been confirmed to date (Heller et al., 2019; Kreidberg et al., 2019).

The lack of confirmed detections should not, however, be taken as evidence that such moons do not exist. Instead, it is, at least in part, the result of such moons remaining beyond the limits of our detection ability, and it is likely that future surveys with more powerful tools will confirm that, like planets, exomoons are ubiquitous. It may, however, be that we have been looking in the wrong places. Astronomers speculate that exomoons are likely ubiquitous based primarily on our knowledge of the Solar system's outer planets, yet, to date, we have not looked for moons around planets that are true analogues to the Solar system's gas and ice giants. Despite extensive efforts to detect moons around exoplanets with orbital distances between 0.1 and 1 au (e.g. Kipping et al., 2012), the wait for a confirmed detection goes on. This dearth can potentially be explained by the instability of exomoons in systems that experience intense stellar tides (e.g. Barnes & O'Brien, 2002) or inward migration (Spalding et al., 2016). Long-period transiting exoplanets may indeed be better candidates for studies of exomoons than their short-period counterparts.

One of the key challenges inherent to the detection of exomoons is the impact of the orbital motion of the moon around the planet as the planet orbits its star. It has been suggested that a potential method by which exomoons could be detected is through their effect on the transits of their host planet across its star. There are two ways in which such variations could manifest and be detected - through transit duration variations (TDV) and/or transit timing variations (TTV) (Brown et al., 2001; Kipping et al., 2012, 2013). In addition, as the planet transits its host star, the moon may also be seen to transit. However, as the planet orbits the star, the moon also orbits the planet, and so the position of the moon in its orbit about the planet will change each time the planet transits the star. The signature of the moon will therefore vary its position within the transit of the planet depending on its orbital phase, as shown in Figure 19, below. The problems caused by this variability in the timing of the moon's transit relative to that of its host planet are further exacerbated by the possibility that there may be occasions when the moon is lined up perfectly with its host planet, and therefore transiting or passing behind the planet as the planet is in transit. Such a scenario will mean that the transit signature of the moon would disappear for that particular transit, leaving a signal that would be indistinguishable from a solitary, moonless planet.

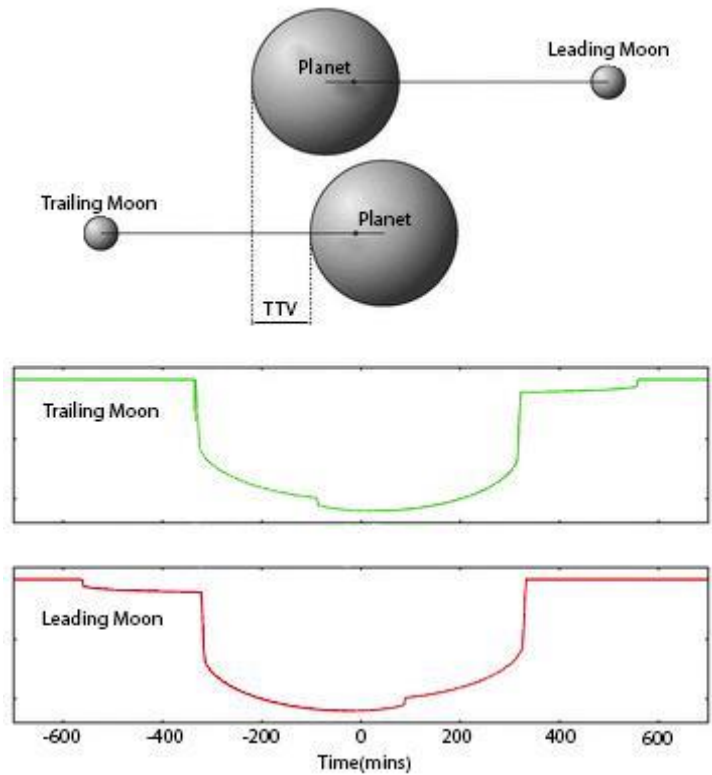

*Figure 19: The transit of a moon may be detected as a planet-moon system passes in front of its host star. As the planet orbits the star, the moon also orbits the planet, and so will produce a different signature depending on its orbital position as it transits the star. The panels above show the expected planet-moon transit signatures for both a moon trailing behind the planet as it transits (green) and a moon leading the planet as it transits (red). Note that if the moon is transiting or eclipsed by the planet when the planet passes in front of the star, then the signature of the moon will disappear, and the planet will appear to be moonless.*

The time at which the planet transits can also be affected by the existence of an exomoon. The moon-planet system orbits about the center of mass of the two objects (their barycentre). In the absence of any perturbations from other planets in the system, the barycentre of the planet-moon system will follow a Keplerian orbit around the star. At those times when the moon trails the planet, the planet must therefore be leading the barycentre in its orbit around the star (since the planet and moon must always be directly opposite one another across the barycentre). As a result, when the moon trails the planet, the planet will be seen to transit its host star a little earlier than would otherwise be expected. Equally, at times when the moon is leading the planet in their orbit around their host star, the planet must therefore lag behind, trailing the barycentre. As a result, the planet will be observed to transit the star later than would otherwise be expected. Thus the variance of the transit timing of the planet can be an indication of the existence of an exomoon. However this method is also used to determine the existence of additional planets in orbit around the star causing a similar gravitational effect on the transiting planet. As a result, caution is needed when using this method for exomoons. Other potential methods that have been suggested for exomoon detection include microlensing (Han and Han 2002; Bennett et al. 2014), pulsar timing variation (Lewis et al., 2008), radio emission modulations (Noyola et al., 2014) and direct imaging (Cabrera & Schneider 2007). Hill et al., 2018 found that instruments would need the capacity to resolve a separation of 10 - 4400 $\mu$ arcseconds in order to directly image and resolve a planet-moon system. Direct imaging could also reveal the existence of a moon through the transit of a moon or of its shadow across a bright planet (Heller 2016).

The current focus of exomoon searches on planets at relatively small orbit radii is the direct result of the fact that the majority of the most promising techniques for the detection of exomoons rely on the detection of transits of the moon's host planet. Such transit observations are strongly biased towards short period planets, with transits growing more unlikely as the orbital radius of the planet increases. Two giant exoplanets moving on orbits well beyond 1 au that are possible candidates for in-transit exomoon detection are Kepler-421b (Kipping et al., 2014) and Kepler-167e (Kipping et al., 2016). The ephemerides of these planets have been found to be free of transit timing variations (Dalba & Muirhead 2016, Dalba & Tamburo 2019), enabling the accurate prediction of future transits that may be searched for photometric signals of exomoons.

Over the past few years, the study of exomoons has become a topic of great interest, with particular focus on the possibility that such moons might increase our chances of finding potentially habitable worlds beyond the Solar system. Studies on the occurrence rates of giant planets have found that they are less likely to be found in the habitable zone of their star than smaller terrestrial planets. However, if each giant planet hosts more than one moon, then habitable zone exomoons may well be more numerous than in the habitable zone terrestrial planets (Hill et al., 2018).

One reason that exomoons are considered particularly interesting in the context of the search for life beyond Earth is that they would likely offer a variety of sources of energy to a potential biosphere, rather than relying on the flux received from the host star - an idea fuelled by our observations of Io and Europa in the Jovian system. The reflected light and emitted heat from the host planet could help to ensure a clement climate on such a moon, whilst tidal heating could drive volcanic activity, providing both additional energy and nutrients to any life that developed there (Heller & Barnes 2013; Hinkel & Kane 2013). These combined heating effects act to increase the size of the habitable zone around the host star of such moons, creating a wider temperate area in which they could maintain conditions that are amenable to liquid surface water (Scharf 2006). At the same time, occultations of the host star by the giant planet could help to reduce the incoming energy flux for satellites moving close to the inner edge of the habitable zone – preventing such a moon from overheating. This interplay between tidal and radiative heating from the host planet and the diminution of the incoming stellar flux during occultation events is potentially complex, with the inclination of the moon's orbit, and the presence of additional moons in the system likely to play an important role. For an exomoon orbiting an exoplanetary Uranus, for example, occultation events would occur seasonally, with the satellite spending most of its time in full starlight. At the same time, the orbital eccentricity of the planet itself will doubtless impact upon the potential habitability of its moons. For a planet like BD+14 4559 (whose orbit is shown below, in Figure 20) that spends part of its orbit outside the outer edge of the habitable zone, the extra tidal and radiative energy delivered by the host planet could potentially enable any potentially habitable exomoon to maintain its habitability during this time, though it should also be noted that such a moon would likely lack the time to freeze over during these excursions beyond the nominal outer edge of the habitable zone.

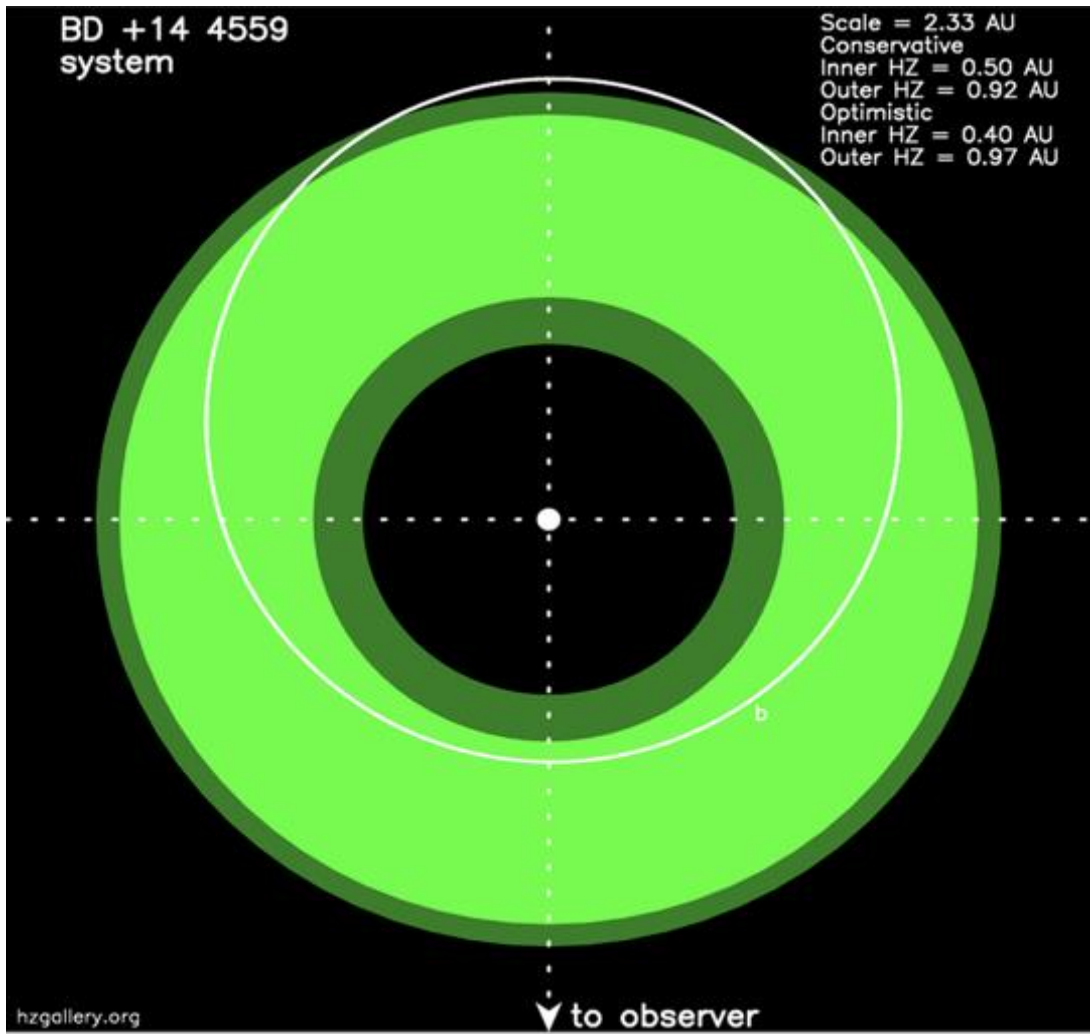

*Figure 20: BD+14 4559 is a giant planet that resides mostly within the habitable zone of its star. Due to the extra energy sources provided to an exomoon, any moon orbiting this planet may still be able to maintain habitable conditions during the part of the planet's orbit when it exits the outer edge of the planetary habitable zone. (Image Source: The Habitable Zone Gallery[73]; described in Kane & Gelino, 2012)*

In addition, as we have seen in our own planetary system, large exomoons offer the possibility for liquid water to be found far beyond the boundary of the habitable zone, at least as it has traditionally been defined. Moons like Europa, Enceladus, and Titan have become targets of great interest for astrobiological research, despite the fact that they lie many au beyond the outer edge of the Solar system's habitable zone. Whilst such moons would clearly be of great interest were they found orbiting Jupiter-analogue exoplanets, it is likely that they would be poor targets for the search for life elsewhere. Whilst it is reasonable to assume that Europa is a habitable world, for example, it is not fair to consider that it would be *detectably* habitable. Europa's ocean is buried beneath a kilometres thick ice sheet - and even in our own backyard, we cannot yet say whether that ocean contains life. For this reason, the focus on exomoon habitability, in a detectable/measurable sense, will remain on those satellites of giant planets that themselves orbit within (or close to) the habitable zone of their host star - such satellites would offer not only the possibility of being considered potentially habitable worlds, but also the possibility that their habitability could be remotely probed, using the next generation of astronomical telescopes and instrumentation.

---

[73] http://www.hzgallery.org/index.html; accessed 23rd April 2020

## 5.5 SUPER-EARTHS AND MINI-NEPTUNES

In the Solar system, Earth and Neptune bound a clear division between the terrestrial and gaseous planets. Separated in size by a factor of four, no known Solar system object has a radius larger than Earth's but smaller than Neptune's. It was natural, therefore, to expect that exoplanetary systems might follow this same behaviour - rocky, terrestrial worlds close to their host stars would be relatively small (as in the Solar system), whilst the more distant giant planets would all be large - Neptune-sized or bigger.

It has become obvious, however, that this is not the case. One of the great surprises to come out of the *Kepler* mission was that, rather than being rare, planets between the size of Earth and Neptune are actually abundant. *Kepler* found such planets in droves - as can be seen in Figure 21, below. Now, it is widely known that *super-Earths*—likely terrestrial worlds with radii slightly larger than the Earth—and *sub-Neptunes*—likely planets with thick gaseous atmospheres, reminiscent of the Solar system's ice giants, with radii slightly smaller than Neptune—are by far the most common of the planets that we can currently detect in the galaxy. Some authors have even suggested that such planets might be the most abundant of all - outnumbering their smaller *and* larger siblings - but at the same time, logic would dictate that there are likely to be more small planets than large ones in the cosmos (just as, in the Solar system, there are more small bodies than larger ones, and just as there are more small, low mass stars than massive stars). The true size distribution of planets smaller than the *super-Earths* remains to be uncovered in the coming decade - but regardless of the degree to which they are the most common planets in the cosmos, the *super-Earths* and *sub-Neptunes* pose a fascinating conundrum, showing where our ability to infer planetary properties based on the Solar system breaks down.

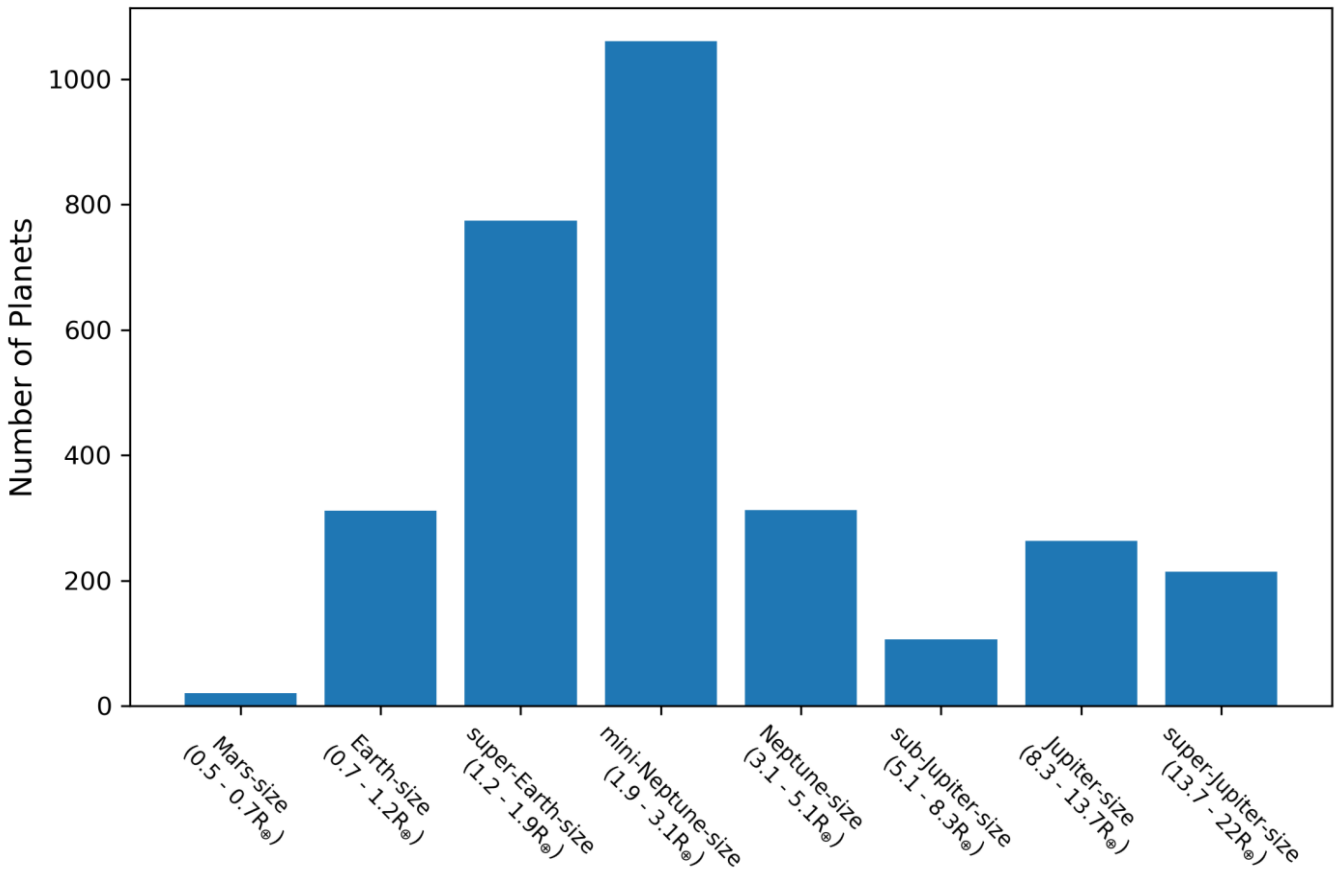

*Figure 21*: *Known transiting planets by size. This histogram demonstrates how readily the primary Kepler mission discovered super-Earth and mini-Neptune exoplanets. Even when corrected for observing*

*biases, exoplanets with radii in between that of the Earth and Neptune seem to be the most prevalent planets in the galaxy. Data from the NASA Exoplanet Archive[74] (accessed July 9th, 2019).*

Prior to the era of *Kepler*, the surprising abundance of super-Earth and mini-Neptune exoplanets was already becoming clear, with the detection of such planets being achieved by several RV surveys. For planets moving on orbits with periods less than fifty days, it is clear that smaller planets are more common than giants - with the occurrence rate of exoplanets on such orbits rising by approximately a factor of ten as one moves from Jupiter-mass objects to those only a few times more massive than the Earth (Mayor et al., 2009; Howard et al., 2010; Wittenmyer et al., 2011a). The sensitivity of *Kepler* to exoplanets in this region of parameter space enabled a more detailed investigation of super-Earth and mini-Neptune prevalence. Accounting for both false positives and completeness, Fressin et al. (2013) found the occurrence rates of both super-Earths and mini-Neptunes with periods less than 85 days to be ~23%, higher than any other size planet in any region of parameter space considered in their study.

Understanding the occurrence rate of super-Earths and mini-Neptunes is of critical importance to theories of planet formation. Early population synthesis models that simulated planet formation via core accretion anticipated an increasing planet occurrence rate with decreasing planet mass for objects orbiting at several au (e.g., Ida & Lin 2004; Alibert et al., 2005; Mordasini et al., 2009). However, the observed prevalence of super-Earth and mini-Neptune exoplanets at orbital distances within 1 au challenged the theoretical models. During formation, planets whose mass grew to several times that of Earth were expected to either spiral into the host star via Type-II migration or rapidly accrete a gaseous envelope to become a gas giant planet (Ida & Lin 2008; Schlaufman et al., 2009; Alibert et al., 2011) - a process of runaway growth that should result in planets rapidly passing through the super-Earth/mini-Neptune mass range. Producing a super-Earth or mini-Neptune planet would therefore require the unlikely dispersal of the gas disc at a particular time, truncating that process of runaway growth. The mere existence of so many super-Earth and mini-Neptune exoplanets in the galaxy demonstrated that this theory was flawed, and has led to numerous suggested mechanisms to explain how these exoplanets may have formed (e.g., Ida & Lin 2010; Rogers et al., 2011, Hansen & Murray 2012; Chiang & Laughlin 2013; Cossou et al., 2014).

Super-Earths and mini-Neptunes occupy a fascinating region of the mass-radius parameter space whereby vastly different interior and atmospheric compositions are a priori plausible for a given planet. These planets can have substantial fractions of silicate rocks, metals, ices, and gases (e.g., Valencia et al., 2007; Rogers & Seager 2010). A subset of super-Earth and mini-Neptune exoplanets that are particularly amenable to transit and RV characterization have had their radii and masses measured precisely (e.g., Marcy et al., 2014, Weiss & Marcy 2014). Substantial effort has been made to explore the atmospheres and interiors of this subset of well characterized exoplanets. Lopez & Fortney (2014) used planetary evolution models to suggest that, for mini-Neptunes, the planet's radius can be used as a proxy for its hydrogen-helium envelope fraction. Wolfgang et al. (2015) employed a probabilistic mass-radius relationship model to explore the intrinsic astrophysical scatter present in the relation between super-Earth and mini-Neptune mass and radius, whilst Rogers (2015) demonstrated that most 1.6 Earth-radius exoplanets have densities that are too low to be explained by just iron and rock, suggesting that these exoplanets are not terrestrial in nature. Beyond the small group of studies listed here, the atmospheres and interiors of these transitionary worlds are the subject of intense ongoing investigation.

Given the ubiquity of super-Earth and mini-Neptune exoplanets in the galaxy, one must wonder: *why is the Solar system devoid of this class of planet?* In other planetary systems, the presence of cold, long-period giant exoplanets has been found to be strongly correlated to the existence of inner super-Earth-sized companions (e.g., Zhu & Wu 2018; Bryan et al., 2019). One potential explanation for the lack of such a planet in the inner Solar system is that Jupiter could have acted as a barrier to Uranus, Neptune, and Saturn, halting their inward migration and preventing them from becoming super-Earth or mini-Neptune class planets (Izidoro et al., 2015a). Equally, given proper disc conditions, super-Earths may have existed in the Solar system previously, clearing the material inside of Mercury's orbit before colliding with the Sun (Martin & Livio 2016). Yet another explanation could be that the Solar system *does* contain a super-Earth or mini-Neptune planet that has so far eluded detection (see Batygin et al., 2019 for a review). Additional explorations of the small and large bodies nearby and across the galaxy will likely shed light on this peculiar aspect of the Solar system.

---

[74] https://exoplanetarchive.ipac.caltech.edu

## 5.6 EXOPLANETARY ATMOSPHERES

Exoplanetary atmospheres are a window to an exoplanet's composition, formation history, evolution, and even habitability. The proximity of the Solar system planets has afforded us the luxury of being able to study their atmosphere up-close and personal. Unlike exoplanets, which cover just a single pixel through the largest telescopes, the Solar system's planets are easily resolved, allowing the behaviour of their atmospheres to be studied in detail. The advent of planetary space exploration, in the 1960s, meant that we could get closer still - sending orbiters to observe the planets continuously, and even launching probes that could dive into their atmospheres, and, for the terrestrial planets, land on their surfaces. This has allowed us to investigate the atmospheres of the planets in our Solar system to a level that will most likely not be possible for an exoplanet in the foreseeable future. Almost all of the information we can gather pertaining to an exoplanet's surface, formation, and evolution will be filtered through distant observations of its atmosphere.

Not long after the discovery of the first exoplanet came the first theoretical predictions regarding exoplanet atmospheric characterization. Seager & Sasselov (2000) proposed that measurements of the wavelength dependence of an exoplanet's transit depth would reveal variations in atmospheric opacity. This could, in turn, be used to infer the properties and composition of an exoplanetary atmosphere through a technique known as transmission spectroscopy. This technique was first successfully used to identify sodium in the atmosphere of HD 209458 b based on four transits observed by the *Hubble Space Telescope* (Charbonneau et al., 2002).

Since the initial characterisation of the atmosphere of HD 209458 b, several other techniques for exoplanet atmospheric characterization have been developed. The observation of the occultation of short-period, highly irradiated exoplanets by their host stars (secondary eclipses[75]) led to the first detection of photons directly emitted by an exoplanet atmosphere (Deming et al., 2005; Charbonneau et al., 2005). The observation of secondary eclipses enables the calculation of planetary Bond albedo and atmospheric temperature. Similarly, Knutson et al. (2007) conducted infrared observations of half of the orbital phase of the hot Jupiter HD 189733 b, effectively mapping the temperature distribution across planetary longitude. Phase curve maps, such as those presented in that work, have provided valuable information about energy transport and heat redistribution in the atmospheres of the hottest exoplanets.

Exoplanet atmospheres are also being characterised through high dispersion ground-based spectroscopy, whereby spectral lines of an exoplanet are detected separately from stellar lines thanks to the time varying radial motion of the exoplanet (e.g., Snellen et al., 2010). This technique has lead to confident detections of carbon monoxide and water vapor in exoplanet atmospheres (e.g., Brogi et al., 2012, Birkby et al., 2013), has enabled the estimation of day-night wind velocities on exoplanets, and has provided validation for 3D exoplanet atmosphere models (Kempton et al., 2014).

At greater orbital separations, direct imaging has also become a viable technique for the atmospheric characterisation of young, self-luminous exoplanets. The HR 8799 system contains three directly imaged exoplanets that have been observed for over a decade, revealing critical information about the atmospheres and orbital dynamics of young giant planets (e.g., Marois et al., 2008, 2010). However, the immense planet-star contrast and the fundamental limits on angular resolution set by diffraction add difficulty to the direct observation of smaller or cooler (i.e., more mature) exoplanets at visible and near-infrared wavelengths.

---

[75] when the planet passes behind its host, as seen from Earth.

Amongst the various methods of exoplanet atmospheric characterization, transmission spectroscopy is presently the most productive means of probing exoplanet atmospheres. Although the entire sample of exoplanets that have been subject to transmission spectroscopy observations are hotter and have shorter-period orbits than any object in the Solar system, substantial efforts have been made to simulate the transmission spectra of the Solar system's planets. For Venus, Barstow et al. (2016) generated synthetic spectra using a radiative transfer model to determine that the cloudiness of a Venus-analogue exoplanet would preclude atmospheric characterisation from an observatory such as the *James Webb Space Telescope*, even if the hypothetical planet orbited a small star only 10 parsecs away. Pondering the Earth as an exoplanet has become a cottage industry within exoplanetary science. With appropriately high-signal observations, transmission spectroscopy of an Earth-analogue exoplanet could reveal the presence of key biosignature gases as well as other species such as ionised calcium (e.g. Pallé et al., 2009). Importantly, mid-transit observations of an Earth-twin would be hindered by atmospheric refraction, which would mask atmospheric information below ~10 km (e.g. García Muñoz et al., 2012).

Transmission spectra have also been synthesized for Jupiter, Saturn, and Titan. Irwin et al. (2014) used spectroscopic observations and a radiative transfer model to simulate the transmission spectrum of Jupiter between 0.4 and 15 μm. The transmission spectrum of Jupiter as measured in a single transit would yield accurate constraints on the vertical profiles of temperature and hydrocarbon abundance in Jupiter's stratosphere (Irwin et al., 2014). Dalba et al. (2015) demonstrated that occultations of the Sun by Saturn observed by the *Cassini* Spacecraft in the near-infrared could be used to derive a transmission spectrum between 1 and 5 μm. The spectrum displayed multiple absorption features due to methane, as well as features from the hydrocarbon byproducts of Saturn's photochemistry (i.e. acetylene, ethane, and higher-order aliphatic hydrocarbons). Saturn's transmission spectrum also displayed a flat continuum between 1 and 5 μm caused by atmospheric refraction. Titan's transmission spectrum was also reconstructed from *Cassini* solar occultation observations (Robinson et al., 2014). It displayed absorption features from methane, acetylene, ethane, higher-order hydrocarbons, and potentially carbon monoxide. Multiple scattering by the thick haze present in Titan's atmosphere gives the spectrum a distinctive slope increasing toward shorter wavelengths, which is comparable in size to the strong absorption features (Robinson et al., 2014).

## 5.7 DYNAMICS OF EXOPLANETARY SYSTEMS

In order to fully understand the formation and evolution of our Solar system, a variety of computational tools have been developed that allow the dynamical evolution of both the small bodies and the planets themselves to be studied on billion-year timescales. These *n*-body dynamics packages, such as SWIFT (Levison & Duncan, 1994), MERCURY (Chambers, 1999), and REBOUND (Rein & Liu, 2012) can clearly also be used to study the dynamics of exoplanetary systems. They can be used, for example, to study the impact regimes that might be experienced by Earth-like planets in those systems (e.g. Horner & Jones, 2008a, 2009, 2012; Horner, Jones & Chambers, 2010a), or to assess which of the known exoplanetary systems could host potentially habitable planets, based on the gravitational influence of the known planets in the system (e.g. Jones, Underwood & Sleep, 2005; Jones, Sleep & Underwood, 2006; Agnew et al., 2017, 2019; Agnew, Maddison & Horner, 2018a, b; Horner et al., 2020).

A more immediate and important use for such tools in the study of exoplanetary systems is to use them to assess whether the proposed planets within a given system are actually dynamically feasible. As the number of known exoplanets has increased, the rigour with which new discoveries are investigated, prior to being announced, has fallen. For the first few discoveries, researchers and referees worked to the maxim that "*extraordinary claims require extraordinary evidence*", and all possible alternatives to the planetary hypothesis were examined in great detail. In recent years, as the discovery of exoplanets has become routine, an ever increasing population of planets moving on unusual orbits, or orbiting peculiar stars, have been announced (e.g. Wolszczan & Frail, 1992; Lee et al., 2009; Beuermann et al., 2010; Potter et al., 2011; Beuerman, Dreizler & Hessman, 2013; Ramm et al., 2016).

Given the importance of new exoplanet discoveries in helping theorists to better understand planetary formation, it is clearly vital to ensure that each newly proposed planetary system is what it seems to be. Since the great majority of such systems are found by indirect means, there is clearly a risk that some other physical process might be producing a variation that could be misinterpreted as the influence of a planet orbiting a given star. Indeed, the detection of exoplanets is hindered by a plethora of phenomena that can cause stars to vary in a fashion similar to that expected to be caused by planetary companions (such as sunspots, magnetic cycles, periodic variability, companion stars, and many others; e.g. Dumusque et al., 2011; Robertson et al., 2014; Meunier et al., 2015; Fischer et al., 2016; Rajpaul et al., 2016; Nicholson et al., 2019). It is therefore useful to have a "sanity check" for each newly proposed planetary system, in order to make sure that the candidate planets are reasonable.

Dynamical studies of newly discovered planetary systems therefore provide an excellent tool by which the veracity of those systems can be examined. If the candidate planets prove to be dynamically stable on timescales comparable to the lifetime of the system, then they can be considered to be dynamically feasible. On the other hand, if dynamical study of a given system shows that the planets therein interact so strongly that they collide, or eject one another from their host system on timescales far shorter than the system's lifetime, this either suggests that we have fortuitously observed the system during its death throes (a statistically unlikely but certainly plausible possibility) or that the proposed planets do not exist, at least on the orbits claimed.

When such simulations are carried out, the results fall into three broad categories. The first are those systems for which no dynamical instability is observed across the full spread of architectures tested (e.g. Wittenmyer et al., 2014b). These systems prove so dynamically stable that it is perfectly reasonable to assume that the planets truly exist on the orbits that are proposed. Beyond this, however, such results show the systems to be so dynamically stable that there is likely to be plenty of space left for further planets in that system – in other words, they indicate systems where further observations may well prove fruitful in coming years.

The second class of exoplanetary system are those for which such simulations reveal only narrow islands of stability across the phase space studied. Examples of the dynamical maps for two such systems are shown in Figure 22, below. In both cases, the candidate planets move on orbits that are close to mutual mean-motion resonances. In the case of HD 200964 (the left-hand plot; Wittenmyer, Horner & Tinney, 2012), our simulations show that the orbits of the two candidate planets are only stable when they are trapped in mutual 4:3 mean-motion resonance (the narrow island of stability visible at the centre of the dynamical map). Similarly, for HD 204313 (the right-hand plot; Robertson et al., 2012b), the planets are only dynamically stable when trapped in mutual 3:2 mean-motion resonance. In both cases, as soon as the planets are moved away from a mutually resonant solution, they begin to strongly perturb with one another, resulting in one or other of them being flung into the centre star, the two planets colliding with each other, or one being ejected from the system entirely.

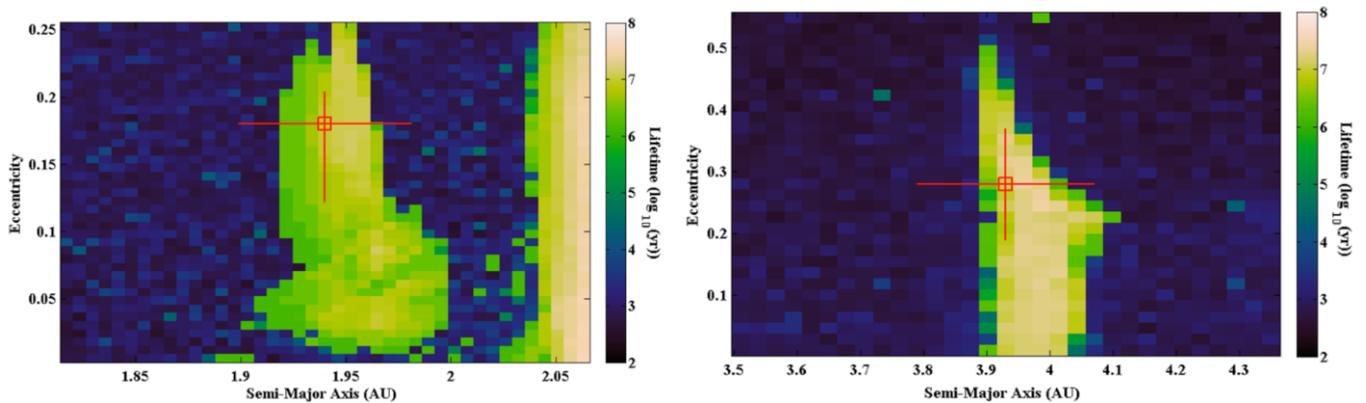

*Figure 22: The dynamical stability of the planetary systems proposed orbiting the stars HD 200964 (Wittenmyer, Horner & Tinney, 2012; left) and HD 204313 (Robertson et al., 2012b; right), as a function of the semi-major axis and eccentricity of the orbit of the outermost planet in that system. The colour at each point denotes the mean lifetime of a total of more than fifty individual simulations that began with the planet at that particular a-e location. For both systems, the dynamically stable solutions make up only a small fraction of the total number of simulations carried out. In both cases, the stability of the system is only ensured when the two planets within are trapped in mutual mean motion resonance - the 4:3 and 3:2 resonances, respectively.*

As a result, we can safely conclude that the planetary systems proposed are dynamically feasible provided that the planets therein are trapped in mutual mean-motion resonance. Such results have an interesting additional outcome – not only do they demonstrate that the systems can be dynamically feasible, such simulations have enabled significant additional constraints to be placed on the orbits of the planets in a number of systems (e.g. Robertson et al., 2012a, b; Wittenmyer et al., 2014a, 2016a), above and beyond those that result purely from the observational data.

The third class of system are those for which the proposed planets in a given system move on orbits that appear highly dynamically unfeasible – sometimes even crossing one another's path (e.g. Horner et al., 2011b, 2012b; 2013, 2014a, 2019; Wittenmyer et al., 2012; Wittenmyer, Horner & Marshall, 2013; Hinse et al., 2014; Marshall et al., 2020). Whilst it is certainly feasible for mutually crossing orbits to be dynamically stable (examples in our own Solar system include the Jovian and Neptunian Trojans, and the Plutinos; e.g. Morbidelli et al., 2005; di Sisto, Brunini & de Elía, 2010; Horner & Lykawka, 2010a; Di Sisto, Ramos & Gallardo, 2019; Pirani et al., 2019), such stability is typically facilitated by mutual mean motion resonance between the orbits of the objects in question. Even for newly discovered objects in our own Solar system, it is critically important to confirm whether their orbits are dynamically stable or dynamically unstable (e.g. Lykawka & Mukai, 2007; Gladman et al., 2008; Horner & Lykawka, 2010a; Horner, Lykawka & Müller, 2012; Horner et al., 2011a; di Sisto, Ramos & Beaugé, 2014; Wu, Zhou & Zhou, 2018; di Sisto, Ramos & Gallardo, 2019). For exoplanetary systems, such dynamical studies are even more important – since strong dynamical instability would infer that the planetary system, as proposed, is simply not feasible, and that either the system is dramatically different to that proposed in the literature, or the observed variations that led to the announcement of planets around a given star must have another explanation.

In recent years, a number of multiple planetary systems have been proposed orbiting very tightly bound eclipsing binary stars, including several around what are known as "post-common envelope binaries" (hereafter PCEBs). Those systems consist of a highly evolved primary star (typically, though not always, a white dwarf) with a secondary stellar companion (often a dim red dwarf star, similar to our Sun's nearest stellar neighbour, Proxima Centauri) orbiting at a remarkably small distance. The two stars, which orbit about their common centre of gravity with periods of a couple of hours, move on orbits that are aligned such that they eclipse one another each time the secondary passes between the primary star and the Earth, along our line of sight. Multiple massive companions have been claimed orbiting several such binaries (e.g. Lee et al., 2009; Beuermann et al., 2010; Potter et al., 2011; Qian et al., 2011) on the basis of variations in the timing of the eclipses between them, with those timing variations proposed to result from the small wobble back and forth of the binary stars under the gravitational influence of distant, unseen companions. Of those systems, however, only two have stood up to dynamical scrutiny - namely, the planets orbiting NN Serpentis (e.g. Beuermann et al., 2010; Horner et al., 2012b; Beuermann, Dreizler & Hessman, 2013; Marsh et al., 2014) and UZ Fornacis (Potter et al., 2011; Horner et al., 2014b). Even in the case of NN Serpentis, however, Mustill et al. (2013) showed that the planets could not have survived the evolution of the binary from its main-sequence to post-main sequence state, leading to the suggestion that, if the planets truly exist, they most likely represent a second-generation of planet formation, which occurred after the ejection of the star's common envelope, at the end of the white dwarf star's evolution away from the main sequence - a finding supported by Völschow, Banerjee & Hessman, 2014.

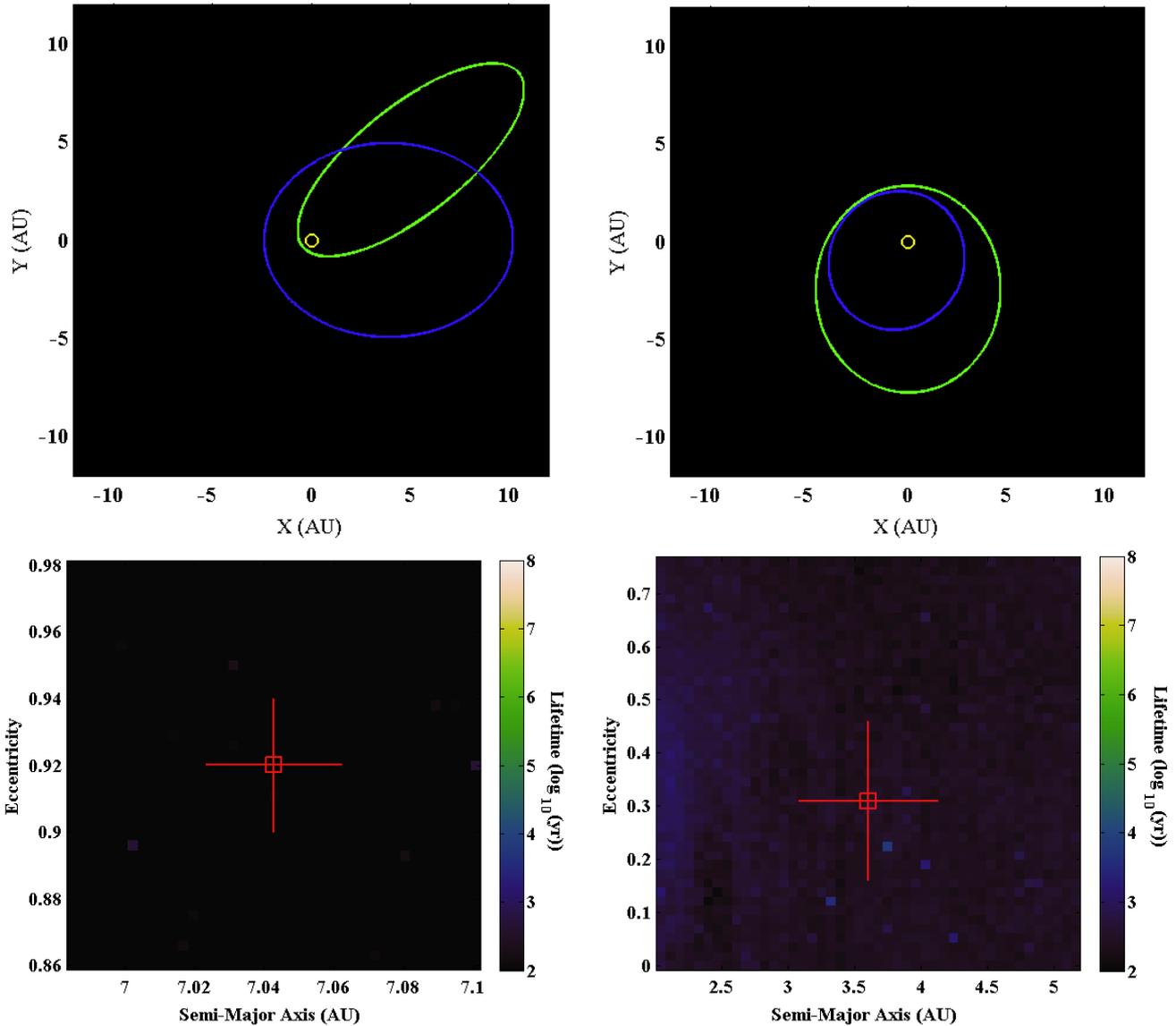

*Figure 23:* *The best-fit orbital solutions (top) and dynamical stability maps (bottom) for the planetary systems proposed to orbit the stars QS Virginis (Lee et al., 2009; Horner et al., 2012c; left) and HW Virginis (Almeida & Jablonski, 2011; Horner et al., 2013; right)*

In Figure 23, we show the best-fit proposed orbital solutions (top) and resulting dynamical stability plots (bottom) for the planets proposed to orbit the eclipsing binary stars QS Virginis (Lee et al., 2009; Horner et al., 2012c; left) and HW Virginis (Almeida & Jablonski, 2011; Horner et al., 2013; right). In both cases, the proposed solutions prove to be catastrophically dynamically unstable across the whole range of possible solutions - with instability occurring on timescales of just a few hundred or few thousand years. Such extreme instability suggests that the origin of the observed variations in the timing of eclipses in these systems can not be planets, at least, not planets moving on the orbits proposed by Lee et al., 2009 and Almeida & Jablonski, 2011. Indeed, given the fact that the planets proposed in the majority of such systems have failed to stand up to dynamical scrutiny, it seems likely that a common non-planetary explanation must be sought to explain the periodic variability in their eclipse timings - a possibility that researchers are continuing to explore as more data becomes available on these fascinating systems (e.g. Goździewski et al., 2015; Völschow et al., 2016; Nasiroglu et al., 2017; Navarrete et al., 2018, 2020).

Regardless of the true origin of the observed eclipse-timing variations for these fascinating objects, the results described above demonstrate conclusively the importance of applying dynamical models, first developed to study the evolution of the Solar system, to the orbital stability of newly discovered exoplanetary systems. In some cases, such simulations will simply reveal that a proposed planetary system is dynamically stable - but in other cases, they might reveal a false positive, or highlight a case where the process of fitting to the observational data has converged on a solution that does not represent the true global minimum of the orbital phase space that the planets could occupy. We therefore contend that such modelling should form a critical component of the planet discovery process, and have incorporated such analysis into the standard procedure of the Anglo-Australian Planet Search. (e.g. Wittenmyer et al., 2014a, 2014b, 2017a).

Whilst the orbital solutions of the transiting and radial velocity exoplanets are well constrained due to the requirement of confirming the cyclical nature of the signal before announcement of detection (along with short orbital periods in the case of transiting planets/hot Jupiters), identification of planets through multi-epoch direct imaging of exoplanet systems (e.g. HR8799; Marois et al., 2008, 2010), looking for co-moving companions to the star, provide only weak constraints on the orbits of the planets as they are, by and large, on wide orbits, with decades-long orbital periods and therefore do not move much in year-to-year imaging. The dynamical stability of HR 8799 has been extensively studied (e.g. Marois et al., 2008; Goździewski & Migaszewski, 2009, 2014, 2018; Marshall, Horner & Carter, 2010; Wang et al., 2018b) leading to a variety of more-or-less stable orbital solutions depending on the assumptions made by the authors of the orbital eccentricities of the planets and the inclination of the system. One consistent result is that three of the planets are trapped in a 4:2:1 mean-motion resonance (a scenario often known as a Laplace resonance, and one that mirrors the orbital architecture of Jupiter's moons Io, Europa and Ganymede). Unstable orbital solutions might then be indicative that the system is undergoing a dynamical rearrangement, analogous to the Late Heavy Bombardment in the Solar system, rather than suggesting the planetary system is somehow spurious - a result that would be in keeping with the bright debris discs observed in the system (e.g. Su et al., 2009; Hughes et al., 2011; Matthews et al., 2014; Contro et al., 2016; Goździewski & Migaszewski, 2018; Wilner et al., 2018; Geiler et al., 2019). In the future, as more multi-planet systems are discovered using techniques such as direct imaging and astrometry, dynamical methods will become even more important in determining which of the many plausible orbital architectures that could fit the observed data are most likely to be a fair representation of the true state of the newly discovered systems.

**5.8 CURRENT AND FUTURE EXOPLANET SURVEYS**
In the first few years of the exoplanet era, the search for alien worlds was dominated by radial velocity surveys that were limited in the number of stars they could observe, and the frequency with which they could carry out their observations (e.g. Latham et al., 1989; Cochran & Hatzes, 1994; Butler et al., 1996; Walker et al., 1995; Baranne et al., 1996; Tinney et al., 2002; Naef et al., 2004). Whilst those surveys were able to build up lengthy temporal baselines, and to begin finding ever more distant planets (e.g. Gregory & Fischer, 2010; Ségransan et al., 2011; Wittenmyer et al., 2017a; Vogt et al., 2017; Kane et al., 2019a; Rickman et al., 2019b), the fact that they could only target a small number of stars limited the number of planets that they could discover.

In the middle of the first decade of the 21st Century, the first exoplanet transit surveys came online.

Unlike the radial velocity surveys, these dedicated programs were able to observe all night, every night. By using wide-field cameras, those transit surveys (such as the Wide-Angle Search for Planets, WASP; e.g. Kane et al., 2004, 2005a, b; Christian et al., 2006; Butters et al., 2010; HAT-Net and HAT-South; e.g. Bakos et al., 2004, 2007, 2013; Hartman et al., 2011; Zhou et al., 2014; and the multi-purpose CoRoT space observatory; e.g. Baglin et al., 2006; Alonso et al., 2008; Barge et al., 2008; Deleuil et al., 2008; Léger et al., 2009) were able to monitor the brightness of thousands of stars at once, continually - playing a numbers game to ensure large numbers of detections. The ultimate expression of that philosophy came with the launch of *Kepler*, in 2009. The *Kepler* spacecraft was a dedicated exoplanet finding machine - with a field of view that allowed it to continually monitor more than 150,000 stars, 24 hours per day, for four years, during which time it achieved a duty cycle in excess of 90% for many of the target stars (e.g. Garcia et al., 2014). In that manner, *Kepler* soon became by far the most successful exoplanet survey to date - finding some 2345 confirmed planets, and a further 2420 candidate planets that still await follow-up[76] (e.g. Borucki et al., 2011; Batalha et al., 2013; Mullally et al., 2015; Coughlin et al., 2016; Morton et al., 2016).

Whilst *Kepler* was a huge success, it also revealed a shortcoming of ground-based exoplanetary science work. Simply put, there were too many candidate planets to follow-up, and too few ground-based facilities to perform that work. Most radial velocity instruments are located on shared-time telescopes, meaning that researchers have to compete for time to make their observations. This, combined with the time taken to perform those observations, and the fact that a given facility can only perform radial velocity observations of a single star at a time, represents a significant bottleneck to the exoplanet detection process - and one that will only get worse as the *TESS* mission continues to deliver new candidate planets for follow-up work (e.g. Ricker et al., 2015; Sullivan et al., 2015; Barclay et al., 2018; Stassun et al., 2018). To address the overwhelming wealth of exoplanet candidates that will be delivered by *TESS* in the coming years, numerous radial velocity facilities are being employed to perform follow-up observations (e.g. Nielsen et al., 2019; Quinn et al., 2019; Vanderburg et al., 2019).

The Anglo-Australian Planet Search (AAPS), operated from 1997 until 2014 using the Anglo-Australia Telescope (AAT), built up a 17-year long database of RV observations for more than two hundred Sunlike stars (e.g. Wittenmyer et al., 2016, 2017a; Kane et al., 2019a). The Keck telescopes have also played a significant role in radial velocity searches, such as the use of HIRES by such teams as the California Planet Search (CPS), including exoplanet occurrence rates (Howard & Fulton 2016). Other facilities, such as the MINiature Exoplanet Radial Velocity Array (MINERVA), and MINERVA Australis (based in Australia), use off-the-shelf telescopes to reduce the cost of a dedicated facility (e.g. Swift et al., 2015; Addison et al., 2019), and have already delivered some key science results (e.g. Rodriguez et al., 2017; Wilson et al., 2019; Addison et al., 2020). Numerous other precision radial velocity instruments operated in conjunction with the European Southern Observatory (ESO) include CORALIE (Queloz et al., 2000), HARPS (Pepe et al., 2000), and ESPRESSO (Pepe et al., 2014). Other spectrographs, such as the NN-explore Exoplanet Investigations with Doppler spectroscopy (NEID) instrument, have been developed to optimize performance in the infra-red in order to provide improved radial velocities for low-mass stars (Halverson et al., 2016). This list of radial velocity instruments is not intended to be exhaustive, and for a more detailed discussion of the future of Radial Velocity facilities, we direct the interested reader to the recent review by Fischer et al. (2016).

Whilst the Exoplanet Era has been dominated, to date, by the radial velocity and transit methods of exoplanet detection, it is likely that, in the future, astrometric and direct imaging observations will begin to yield increasing number of exoplanet discoveries, and play an important role in their characterisation. To date, direct imaging studies have been relatively limited, primarily focussing on young (and therefore hot) planets orbiting nearby stars (e.g. HR8799; Marois et al., 2008, 2010). With recent developments, however, the number of potential targets for such imaging has increased (e.g. Kane, Meshkat & Turbull, 2018). In addition, direct imaging is now finding a growing role in the confirmation of the planetary nature of objects whose existence has been inferred on the basis of observed long-period radial velocity trends (e.g. Kane et al., 2019a). In the future, this complementarity between different observing techniques will become ever more important - and astrometric observations by the *Gaia* spacecraft are likely to contribute large numbers of long-period exoplanets that would otherwise be particularly hard to detect with traditional methods (e.g. Perryman et al., 2014; Gaia Collaboration et al., 2016).

---

[76] Values correct as of 30th July, 2019; taken from the NASA Exoplanet Archive, at
https://exoplanetarchive.ipac.caltech.edu/docs/counts_detail.html

Another method that has long offered the promise of being able to detect significant numbers of exoplanets, and to sample a population of planets that are hard to find using other techniques, is gravitational microlensing (e.g. Gaudi, 2012). Microlensing has numerous advantages over other methods, including the observational requirement of moderate aperture size telescopes and sensitivity to relatively low masses. Disadvantages include the non-repeatability of the observations and degeneracy in modeling the available data (e.g. Dominik, 2009). To date, microlensing observations have discovered almost 100 exoplanets (NASA Exoplanet Archive; Akeson et al., 2013), and it is anticipated that the launch of the *WFIRST* satellite, in 2025, will lead to a new era in microlensing observations, with the potential that the spacecraft could yield hundreds, if not thousands, of new discoveries (e.g. Yee, 2013; Penny et al., 2019).

Planned exoplanet missions for the future span a variety of science goals regarding both detection and characterization of exoplanets. The legacy of space-based transit detection will be continued by such missions as *PLATO* (PLAnetary Transits and Oscillations of stars; Rauer et al., 2014), with smaller space-based telescopes like *CHEOPS* (CHaracterizing ExoPlanet Satellite) performing follow-up observations of known exoplanetary systems (Broeg et al., 2014). Other missions, such as *ARIEL* (Atmospheric Remote-sensing Infrared Exoplanet Large-survey; Encrenaz, Tinetti & Coustenis, 2018; Tinetti et al., 2018) and *JWST* (James Webb Space Telescope; Boccaletti et al., 2005; Seager, Deming & Valenti, 2009), will focus their observational efforts on detecting and characterizing the atmospheres of transiting planets via the method of transmission spectroscopy (e.g. Seager & Sasselov, 2000, Kempton et al., 2018). Beyond these missions lies a pathway towards direct imaging of exoplanets and the potential for direct spectra of exoplanet atmospheres. In the near-term, the *WFIRST* (Wide Field Infrared Survey Telescope) mission will utilize a coronagraph that will enable direct imaging of giant planets at wide separations from their host stars (Kane, Meshkat & Turnbull, 2018; Lacy, Shlivko & Burrows, 2019). Further afield, the design of the Habitable Exoplanet Imaging Mission (*HabEx*) is underway as a means to achieve the dream of finally being able to directly image and characterize the atmospheres of terrestrial planets, particularly those that lie in the Habitable Zone of their host stars (Kopparapu et al., 2018; Wang et al., 2018c; Kawashima & Rugheimer, 2019). As both the scientific and funding landscape is a constantly evolving environment, the precise direction of future mission design and deployment will depend heavily on the discoveries that are made over the intervening years.

# 6. CONCLUSIONS

In the last three decades, we have entered the Exoplanet Era. Where once we knew of only a single planetary system - our own - we now know that planets are ubiquitous. Almost every star has planets, and planet formation has been confirmed as a natural byproduct of star formation. At the time of writing, some 4152 planets have been confirmed orbiting other stars[77] - a number that is rising on an almost daily basis. There is, however, a problem. Whilst we are finding ever more exoplanets, our ability to characterise and study those planets and their planetary systems in depth is severely limited. Whilst we know the planets are there, we are decades from being able to get a truly in depth overview of their host systems.

With the Solar system, we face the opposite problem. The number of known Solar system objects is rapidly approaching a million - and since those objects are in our own celestial backyard, we are able to study them in exquisite detail. For that reason, the Solar system will remain extremely valuable as our only source of in-situ data for a fully characterised planetary system - and it is clearly important to consider our knowledge of exoplanetary systems in the context of our understanding of the Solar system. At the same time, we run into the inverse problem – whilst we can study the Solar system in exquisite detail, it is just one planetary system. To fully understand the formation and history of the Solar system, it is vital to place it in the context of the myriad other systems we are now discovering.

In this review, we have therefore attempted to describe the scope of current Solar system research - with a particular focus on the components that might have detectable analogues in the exoplanetary systems we know today, or will discover in the coming years.

In section two, we presented an overview of the currently known small Solar system objects, from the 205 known planetary satellites and the dwarf planets to the smallest of the small bodies. In section three, we discussed the Solar system's eight planets as physical objects, revealing the wealth of information that we have gleaned as a result of our exploration during the six decades of the space age. In section four, we focussed on what the diverse populations of Solar system bodies have told us about our own planetary system's history, from the giant collisions that sculpted the terrestrial worlds early in the system's youth to the formation and evolution of the known Solar system small body populations. Then, in section five, we turned our gaze outwards, considering the growing body of work studying other planetary systems, with the aim of drawing parallels between that knowledge and studies of the Solar system.

In the coming years, the number of known exoplanets is expected to grow almost exponentially, and it is certain that, in the process, new discoveries will be made that revolutionise our understanding of planet formation, evolution, and our place in the cosmos. As we move from simply finding exoplanets to striving to characterise them, and potentially even search for life upon them, the Solar system will remain the critical touchstone on which that work will be based, with research into our own Solar system shedding new light on those alien worlds, and our studies of exoplanets helping illuminate our own system's past, present and future.

**ACKNOWLEDGEMENTS**

This research has made use of the NASA Exoplanet Archive, which is operated by the California Institute of Technology, under contract with the National Aeronautics and Space Administration under the Exoplanet Exploration Program. This research has made use of NASA's Astrophysics Data System. The authors would like to express their sincere appreciation for the hard work of the anonymous referee, whose comments and suggestions greatly improved the scope and clarity of this paper. Figure 1 is reproduced from Horner & Lykawka, 2011, with permission from Oxford University Press. Figure 8 is reprinted by permission from Springer:Nature Geoscience, and appeared in "Large stationary gravity wave in the atmosphere of Venus", Fukuhara et al., 2017, Nature Geoscience, Volume 10, Issue 2, pp. 85-88, in which it was Figure 1.

---

[77] As of 28th April 2020, according to the NASA Exoplanet Archive, at https://exoplanetarchive.ipac.caltech.edu/

# REFERENCES

Absil, O. et al., 2013, Astronomy and Astrophysics, 555, 104

Acuña, M. H. et al., 1999, Science, 284, 790

Adams, J. B., Smith, M. O. & Johnson, P. E., 1986, Journal of Geophysical Research, 91, 8098

Adams, J. C., 1846, Monthly Notices of the Royal Astronomical Society, 7, 149

Addison, B. C. et al., 2013, The Astrophysical Journal Letters, 774, 9

Addison, B. C. et al., 2019, The Publications of the Astronomical Society of the Pacific, 131, 5003

Addison, B. C. et al., 2020, Monthly Notices of the Royal Astronomical Society, *accepted, arXiv:2001.07345*

Agarwal, J., Jewitt, D., Mutchler, M., Weaver, H. & Larson, S., 2017, Nature, 549, 357

Agnew, M. T., Maddison, S. T., Thilliez, E. & Horner, J., 2017, Monthly Notices of the Royal Astronomical Society, 471, 4494

Agnew, M. T., Maddison, S. T. & Horner, J., 2018a, Monthly Notices of the Royal Astronomical Society, 477, 3646

Agnew, M. T., Maddison, S. T. & Horner, J., 2018b, Monthly Notices of the Royal Astronomical Society, 481, 4680

Agnew, M. T., Maddison, S. T., Horner, J. & Kane, S. R., 2019, Monthly Notices of the Royal Astronomical Society, 485, 4703

Agnor, C. B. & Asphaug, E., 2004, The Astrophysical Journal, 613, 157

Agnor, C. B. & Hamilton, D. P., 2006, Nature, 441, 192

Agnor, C. B. & Lin, D. N. C., 2012, The Astrophysical Journal, 745, 143

Agol, E et al., 2005, Monthly Notices of the Royal Astronomical Society, 359, 567

Airapetian, V. S. et al., 2019, The International Journal of Astrobiology, 1-59. doi:10.1017/S1473550419000132

Akeson, R. L. et al.., 2013, Publications of the Astronomy Society of the Pacific, 125, 989

Alexander, C. M. O'D., Boss, A. P., Keller, L. P., Nuth, J. A. & Weinberger, A., 2007, Protostars and Planets V, 801

Alexander, C. M. O'D., Howard, K. T., Bowden, R., & Fogel, M. L. 2013, Geochimica et Cosmochimica Acta, 123, 244

Alexeev, I. I. et al., 2010, Icarus, 209, 23

Alibert, Y., Mordasini, C., Benz, W. & Winisdoerffer, C., 2005, Astronomy & Astrophysics, 434, 343

Alibert, Y., Mordasini, C. & Benz, W., 2011, Astronomy & Astrophysics, 526, A63

Allen, R. L., Bernstein, G. M. & Malhotra, R., 2001, The Astrophysical Journal, 549, 241

Almeida, L. A. & Jablonski, F., 2011, The Astrophysics of Planetary Systems: Formation, Structure, and Dynamical Evolution, Proceedings of the International Astronomical Union, IAU Symposium, Volume 276, p. 495-496

Alonso, R. et al., 2008, Astronomy & Astrophysics, 482, 21

Alvarez, L. W., Alvarez, W., Asaro, F. & Michel, H. V., 1980, Science, 208, 1095

Alvarez, W. & Muller, R. A., 1984, Nature, 308, 718

Amelin, Y. & Ireland, T., 2013, Elements, 9, 39

Anders, E. & Zinner, E., 1993, Meteoritics, 28, 490

Anderson, M., Li, Z.-Y., Krasnopolsky, R., Blandford, R. D., 2003, The Astrophysical Journal Letters, 590, L107

Anderson, B. J. et al., 2008, Science, 321, 82

Anderson, B. J. et al., 2010, Space Science Reviews, 152, 307

Andrews-Hanna, J. C., Zuber, M. T. & Banerdt, W. B., 2008, Nature, 453, 1212

Arnould, M., Goriely, S. & Meynet, G., 2006, Astronomy & Astrophysics, 453, 653

Arridge, C.S., et al., 2014, Planetary and Space Science, 104, 122

Artymowicz, P. & Clampin, M., 1997, The Astrophysical Journal, 490, 863

Asaduzzaman, A., Muralidharan, K. & Ganguly, J., 2015, Meteoritics & Planetary Science, 50, 578

Asher, D. J., Clube, S. V. M. & Steel, D. I., 1993, Monthly Notices of the Royal Astronomical Society, 264, 93

Asher, D. J. & Clube, S. V. M., 1993, Quarterly Journal of the Royal Astronomical Society, 34, 481

Asher, D. J. & Clube, S. V. M., 1997, Celestial Mechanics and Dynamical Astronomy, 69, 149

Asher, D. J. & Steel, D. I., 1998, Planetary and Space Science, 46, 205

Asher, D. J., 1999, Monthly Notices of the Royal Astronomical Society, 307, 919

Asphaug, E. & Benz, W., 1996, Icarus, 121, 225

Asphaug, E., 1997, Meteoritics and Planetary Science, 32, 965

Asphaug, E. & Reufer, A., 2014, Nature Geoscience, 7, 564

Atreya, S., Crida, A., Guillot, t., Lunine, J. I., Madhusudhan, N. & Mousis, O., 2016, "The Origin and Evolution of Saturn, with Exoplanet Perspective", in "Saturn in the 21st Century", Cambridge University Press, arXiv:1606.04510

Augereau, J. C., Lagrange, A. M., Mouillet, D. & Ménard, F., 1999, Astronomy & Astrophysics, 350, 51

Aumann, H. H. et al., 1984, The Astrophysical Journal, 278, 23

Aumann, H. H. & Good, J. C., 1990, The Astrophysical Journal, 350, 408

Avduevskij, V. S., Avduevsky, V. S., Marov, M. Ya., Rozhdestvenskij, M. K., Rozhdestvensky, M. K., Borodin, N. F. & Kerzhanovich, V. V., 1971, Journal of Atmospheric Science, 28, 263

Babadzhanov, P. B., Obrubov, Yu. V. & Makhmudov, N., 1990, Astronomicheskii Vestnik, 24, 18

Babadzhanov, P. B., 2001, Astronomy & Astrophysics, 373, 329

Babadzhanov, P. B., 2008, Monthly Notices of the Royal Astronomical Society, 386, 1426

Backman, D. E. & Paresce, F., 1993, Protostars and Planets III (A93-42937 17-90), 1253

Backman, D. E., Dasgupta, A. & Stencel, R. E., 1995, The Astrophysical Journal, 450, 35

Baglin, A. et al., 2006, 36th COSPAR Scientific Assembly, 36, 3749

Bailer-Jones, C. A. L., 2018., Astronomy & Astrophysics, 609, 8

Bailey, E. & Batygin, K., 2018, The Astrophysical Journal Letters, 866, 2

Bailey, E., Batygin, K. & Brown, M. E., The Astronomical Journal, 152, 126

Bailey, M. E., 1984, Nature, 311, 602

Bailey, M. E., Chambers, J. E. & Hahn, G., 1992, Astronomy & Astrophysics, 257, 315

Bailey, M. E., Emel'Yanenko, V. V., Hahn, G., Harris, N. W., Hughes, K. W., Muinonen, K. & Scotti, J. V., 1996, Monthly Notices of the Royal Astronomical Society, 281, 916

Baines, K. H. et al., 2007, Science, 318, 226

Bakos, G., Noyes, R. W., Kovács, G., Stanek, K. Z., Sasselov, D. D. & Domsa, I., 2004, The Publications of the Astronomical Society of the Pacific, 116, 266

Bakos, G. Á. et al., 2007, The Astrophysical Journal, 656, 552

Bakos, G. Á. et al., 2013, The Publications of the Astronomical Society of the Pacific, 125, 154

Ballard, S. & Johnson, J. A., 2016, The Astrophysical Journal, 816, 66

Balsiger, H., Altwegg, K. & Geiss, J.: 1995, Journal of Geophysical Research 100, 5827

Balogh, A. et al., 1992, Science, 257, 1515

Balogh, A. et al., 2008, "Mercury", Space Sciences Series of ISSI, editors: Balogh, A., Ksanfomality, L. & von Steiger, R., ISBN 9780387775388, DOI: 10.1007/978-0-387-77539-5

Banerdt, W. et al., 2012, Insight: an integrated exploration of the interior of Mars. Lunar and Planetary Science Conference, 43.

Banerdt, W. B. et al., 2020, Nature Geoscience, 13, 183

Bannister, M. T. et al., 2017, The Astrophysical Journal, 851, 388

Baranne, A. et al., 1996, Astronomy & Astrophysics Supplement, 119, 373

Barath, F. T., Barrett, A. H., Copeland, J., Jones, D. E. & Lilley, A. E., 1964, The Astronomical Journal, 69, 49

Barclay, T. et al., 2013, Nature, 494, 452

Barclay, T., Pepper, J. & Quintana, E. V., 2018, The Astrophysical Journal Supplement Series, 239, 2

Barge, P. et al., 2008, Astronomy & Astrophysics, 482, 17

Barnes, J. W. & O'Brien, D. P., 2002, The Astrophysical Journal, 575, 1087

Barstow, J. K., Aigrain, S., Irwin, P. G. J., Kendrew, S. & Fletcher, L. N., 2016, Monthly Notices of the Royal Astronomical Society, 458, 2657

Bartos, I. & Marka, S., 2019, Nature, 569, 85

Basilevsky, A. T. & Head, J. W., 2012, Planetary and Space Science, 73, 302

Batalha, N. M. et al., 2013, The Astrophysical Journal Supplement Series, 204, 24

Battams, K. & Knight, M. M., 2017, Philosophical Transactions of the Royal Society of London Series A, 375, 20160257

Batygin, K., Adams, F. C., Brown, M. E. & Becker, J. C., 2019, Physics Reports, 805, 1

Batygin, K., Bodenheimer, P. H. & Laughlin, G. P., 2016, The Astrophysical Journal, 829, 114

Batygin, K., Brown, M. E. & Betts, H., 2012, The Astrophysical Journal, 744, 3

Batygin, K. & Brown, M. E., 2016a, The Astronomical Journal, 151, 22

Batygin, K. & Brown, M. E., 2016b, The Astrophysical Journal Letters, 833, 3

Batygin, K. & Morbidelli, A., 2017, The Astronomical Journal, 154, 229

Bayliss, D. R., Winn, J. N., Mardling, R. A. & Sackett, P. D., 2010, The Astrophysical Journal, 722, 224

Beatty, J. K., O'Leary, B. & Chaikin, A., 1981, *The New Solar System*, Cambridge University Press and Sky Publishing Corporation, p. 220

Beaugé, C. & Nesvorný, D., 2012, The Astrophysical Journal, 751, 119

Becker, J. C., Vanderburg, A., Adams, F. C., Rappaport, S. A. & Schwengler, H. M., 2015, The Astrophysical Journal: Letters, 812, 18

Belbruno, E. & Gott, J. R. III., 2005, The Astronomical Journal, 129, 1724

Bellot Rubio, L. R., Ortiz, J. L. & Sada, P. V., 2000, Earth, Moon and Planets, 82, 575

Benedetti, L. R., Nguyen, J. H., Caldwell, W. A., Liu, H., Kruger, M., & Jeanloz, R., 1999, Science, 286, 5437, 100.

Benkhoff, J. et al., 2010, Planetary & Space Science, 58, 2

Bennett, D.P. et al., 2014, The Astrophysical Journal, 785, 2, 155

Benz, W., Slattery, W.L. & Cameron, A.G.W., 1986, Icarus, 66, 515

Benz, W., Slattery, W.L. & Cameron, A.G.W., 1987, Icarus, 71, 30

Benz, W., Cameron, A.G.W. & Slattery, W.L., 1988, Icarus, 74, 516

Benz, W., Cameron, A.G.W. & Melosh, H. J., 1989, Icarus, 81, 113

Benz, W. & Asphaug, E., 1999, Icarus, 142, 58

Benz, W., Anic, A., Horner, J. & Whitby, J.A., 2007, Space Sci. Rev. 132, 189

Bergstralh, J. T., Miner, E. D., & Matthews, M. S., 1991, Uranus. University of Arizona Press

Bessel, F. W., 1844, Monthly Notices of the Royal Astronomical Society, 6, 136

Beuermann, K. et al., 2010, Astronomy & Astrophysics, 521, 60

Beuermann, K., Dreizler, S. & Hessman, F. V., 2013, Astronomy & Astrophysics, 555, 133

Beust, H, 2016, Astronomy and Astrophysics, 590, 2

Biermann, L., Huebner, W. F. & Lust, R., 1983, Proceedings of the National Academy of Science, 80, 5151

Binzel, R. P., Xu, S., Bus, S. J. & Bowell, E., 1992, Science, 257, 779

Birkby, J. L., de Kok, R. J., Brogi, M., de Mooij, E. J. W., Schwarz, H., Albrecht, S. & Snellen, I. A. G., 2013, Monthly Notices of the Royal Astronomical Society, 436, 35

Bizzarro, M., Baker, J. A. & Haack, H., 2004, Nature, 431, 275

Bjerkeli, P., van der Wiel, M. H. D., Harsono, D., Ramsey, J. P. & Jorgensen, J. K., 2016, Nature, 540, 406

Bjork, R. L., 1961, Journal of Geophysical Research, 66, 3379

Bland, P. A., Alard, O., Benedix, G. K., Kearsley, A. T., Menzies, O. N., Watt, L. E. & Rogers, N. W., 2005, Proceedings of the National Academy of Sciences, 102, 755

Bland, P. A. et al., 2012, Australian Journal of Earth Sciences, 59, 177

Blandford, R. D. & Payne, D. G., 1982, Monthly Notices of the Royal Astronomical Society, 199, 883

Blum, J., 2010, Research in Astronomy and Astrophysics, 10, 1199

Boccaletti, A., Baudoz, P., Baudrand, J., Reess, J. M. & Rouan, D., 2005, Advances in Space Research, 36, 1099

Boccaletti, A. et al., 2018, Astronomy & Astrophysics, 614, 52

Bockelée-Morvan, D., Gautier, D., Lis, D. C., et al., 1998, Icarus 133, 147

Bodenheimer, P. & Pollack, J. B., 1986, Icarus, 67, 391

Boehnhardt, H., Schulz, R., Tozzi, G. P., Rauer, H. & Sekanina, Z., 1996, IAUC, 6495, 2

Boley, A. C., Morris, M. A. & Desch, S. J., 2013, The Astrophysical Journal, 776, 101

Bolin, B. T., Delbó, M., Morbidelli, A., & Walsh, K. J., 2017, Icarus, 282, 290

Bolin, B. T., Walsh, K. J., Morbidelli, A., & Delbó, M., 2018., Monthly Notices of the Royal Astronomical Society, 473, 3949

Bolton, S. J. et al., 2017, Space Science Reviews, 213, 5

Bonomo, A. S., Zeng, L., Damasso, M. et al., 2019, Nature Astronomy, 201

Booth, M., Wyatt, M. C., Morbidelli, A., Moro-Martín, A. & Levison, H. F., 2009, Monthly Notices of the Royal Astronomical Society, 399, 385

Borucki, W. J., Koch, D., Basri, G., et al., 2010, Science, 327, 977

Borucki, W. J. et al., 2011, The Astrophysical Journal, 736, 19

Boss, A. P. & Keiser, S. A., 2010, The Astrophysical Journal, Letters, 717, L1

Boss, A. P. & Vanhala, H. A. T., 2000, Space Science Reviews, 92, 13

Bottke, W. F., Jedicke, R., Morbidelli, A., Petit, J.-M. & Gladman, B., 2000, Science, 288, 2190

Bottke, W. F., Vokrouhlický, D., Brož, M., Nesvorný, D., & Morbidelli, A. 2001. Science, 294, 1693–1696.

Bottke, W.F., Morbidelli, A., Jedicke, R., Petit, J., Levison, H.F., Michel, P. & Metcalfe, T.S., 2002, Icarus 156, 399

Bottke, Jr., W. F., Vokrouhlický, D., Rubincam, D. P., & Nesvorný, D. 2006, AREPS, 34, 157

Bottke, W. F., Vokrouhlický, D. & Nesvorný, D., 2007, Nature, 449, 48

Bottke, W. F. et al., 2012, Nature, 485, 7396

Bottke, W. F., Brož, M., O'Brien, D. P., Campo Bagatin, A., Morbidelli, A. & Marchi, S., 2015, Asteroids IV, Patrick Michel, Francesca E. DeMeo, and William F. Bottke (eds.), University of Arizona Press, Tucson, 895 pp. ISBN: 978-0-816-53213-1, 2015., p.701-724

Bottke, W. F. & Norman, M. D., 2017, Annual Review of Earth and Planetary Sciences, 45, 619

Boué, G. & Laskar, J., 2010, The Astrophysical Journal, Letters, 712, L44

Braga-Ribas, F., Sicardy, B., Ortiz, J. L., et al., 2014, Nature, 508, 72

Brandt, J. C. & Hodge, P. W., 1961, Nature, 192, 957

Brasser, R., Schwamb, M. E., Lykawka, P. S. & Gomes, R. S., 2012a, Monthly Notices of the Royal Astronomical Society, 420, 3396

Brasser, R., 2013, Space Science Reviews, 174, 11

Brasser, R., Walsh, K. J. & Nesvorný, D., 2013, Monthly Notices of the Royal Astronomical Society, 433, 3417

Brasser, R. & Lee, M. H., 2015, The Astronomical Journal, 150, 157

Brasser, R., Matsumura, S., Ida, S., Mojzsis, S. J., Werner, S. C., 2016, The Astrophysical Journal, 821, 75

Brasser, R., Mojzsis, S. J., Matsumura, S. & Ida, S., 2017, Earth and Planetary Science Letters, 468, 85

Breuer, D. & Spohn, T., 2003, Journal of Geophysical Research Planets, 108, 5072

Broadfoot, A. L., Kumar, S., Belton, M. J. S. & McElroy, M. B., 1974, Science, 185, 166

Broadfoot, A. L., Shemansky, D. E. & Kumar, S., 1976, Geophysical Research Letters, 3, 577

Broadfoot, A. L. et al., 1979, Science, 204. 979

Broeg, C. et al., 2014, Contributions of the Astronomical Observatory Skalnaté Pleso, 43, 498

Brogi, M., Snellen, I. A. G., de Kok, R. J., Albrecht, S., Birkby, J. & de Mooij, E. J. W., 2012, Nature, 486, 502

Bromley, B. C. & Kenyon, S. J., 2014, The Astrophysical Journal, 796, 141

Bromley, B. C. & Kenyon, S. J., 2016, The Astrophysical Journal, 826, 64

Bromley, B. C. & Kenyon, S. J., 2017, The Astronomical Journal, 153, 216

Bougher, S. W., Hunten, D. M., Phillips, R. J., eds., 1997, "Venus II", Univ. of Arizona Press, Tucson, 1376 p.

Brown, M. E., Trujillo, C. & Rabinowitz, D., 2004, The Astrophysical Journal, 617, 645

Brown, M. E., et al., 2006, The Astrophysical Journal, Letters, 639, L43

Brown, M. E., Barkume, K. M., Ragozzine, D. & Schaller, E. L., 2007, Nature, 446, 294

Brown, M. E., Ragozzine, D., Stansberry, J. & Fraser, W. C., 2010, The Astronomical Journal, 139, 2700

Brown, M. E. & Batygin, K., 2016, The Astrophysical Journal Letters, 824, 23

Brown, M. E., 2017, The Astronomical Journal, 154, 65

Brown, M. E. & Batygin, K., 2019, The Astronomical Journal, 157, 62

Brown, T. M. et al., 2001, The Astronomical Journal, 552, 2

Brož, M. & Vokrouhlický, D., 2008, Monthly Notices of the Royal Astronomical Society, 390, 715

Brunini, A., 1995, Planetary & Space Science, 43, 1019

Bruno, G., 1584, 'De l'infinito, universo e mondi' (On the Infinite, the Universe, & the Worlds)

Bryan, M. L., Knutson, H. A., Lee, E. J., Fulton, B. J., Batygin, K., Bgo, H., & Meshkat, T., 2019, The Astronomical Journal, 157, 52

Bryan, M. L. et al., 2012, The Astrophysical Journal, 750, 84

Bryan, M. L., Knutson, H. A., Lee, E. J., Fulton, B. J., Batygin, K., Ngo, H., & Meshkat, T., 2019, The Astronomical Journal, 157, 52

Budde, G., Burkhardt, C. & Kleine, T., 2019, Nature Astronomy, 326

Buffett, B. A., 2000, Science, 288, 2007

Buffington, A., Bisi, M. M., Clover, J. M., Hick, P. P., Jackson, B. V., Kuchar, T. A. & Price, S. D., 2009, Icarus, 203, 124

Burbine, T. H., McCoy, T. J., Meibom, A., Gladman, B., Keil, K., 2002, in Bottke, Jr., W. F., Cellino, A., Paolicchi, P., Binzel, R.P. (Eds.), Asteroids III. University of Arizona Press, Tucson, Arizona, 653

Burgasser, A. J.,Gelino, C. R., Cushing, M. C. & Kitzpatrick, J. D.., 2012, The Astrophysical Journal, 745, 26

Burns, J. A., Lamy, P. L. & Soter, S., 1979, Icarus, 40, 1

Bus, S. J., A'Hearn, M. F., Schleicher, D. G., et al. 1991, Science, 251, 774

Bus S. J., A'Hearn, M. F., Bowell, E. & Stern, S. A., 2001, Icarus, 150, 94

Butler, R. P., Marcy, G. W., Williams, E., McCarthy, C., Dosanjh, P. & Vogt, S. S., 1996, Publications of the Astronomical Society of the Pacific, 108, 500

Butters, O. W. et al., 2010, Astronomy & Astrophysics, 520, 10

Cabrera, J., & Schneider, J., 2007, Astronomy & Astrophysics, 464, 3

Calame, O. & Mulholland, J. D., 1978, Science, 199, 875

Calvet, N., Hartmann, L. & Strom, S. E., 2000, Protostars and Planets IV, 377

Cameron, A. G. W., 1975, Icarus, 24, 280

Cameron, A. G. W., Fegley, B., Benz, W., & Slattery, W. L., 1988, in Mercury, ed. by F. Vilas, C. Chapman, M.S. Matthews (University of Arizona Press, Tucson, 1988), 692 pp

Cameron, A. G. W., & Benz, W., 1991, Icarus, 92, 204

Cameron, A. G. W., & Truran, J. W., 1977, Icarus, 30, 447

Campante, T. L. et al., 2015, The Astrophysical Journal, 799, 170

Campbell, B., Walker, G. A. H., & Yang, S., 1988, The Astrophysical Journal, 331, 902

Campo Bagatín, A. et al., 1994, Planetary and Space Science, 42, 1079

Canup, R. M. & Asphaug, E., 2001, Nature, 412, 708

Canup, R. M., 2005, Science, 307, 546

Canup, R. M. & Ward, W. R., 2009, Europa, Edited by Robert T. Pappalardo, William B. McKinnon, Krishan K. Khurana; with the assistance of René Dotson with 85 collaborating authors. University of Arizona Press, Tucson, 2009. The University of Arizona space science series ISBN: 9780816528448, p.59

Canup, R. M., 2010, Nature, 468, 943

Canup, R. M., 2011, The Astronomical Journal, 141, 35

Canup, R. M., 2012, Science, 338, 1052

Canup, R. & Salmon, J., 2018, Science Advances, 4, 4

Carlson, R. W. & Judge, D. L., 1974, Journal of Geophysical Research, 74, 3623

Carlson, R. W. et al., 1996, Science, 274, 385

Carr, M. H. & Head, J. W., 2003, Journal of Geophysical Research (Planets) 108, 5042

Cassen, P., 2001, Meteoritics and Planetary Science, 36, 671

Castillo-Rogez, J. C., Matson, D. L., Sotin, C., Johnson, T. V., Lunine, J. I. & Thomas, P. C., 2007, Icarus, 190, 179

Cataldi, G. et al., 2020, The Astrophysical Journal, 892, 99

Cellino, A. et al., 2019, Monthly Notices of the Royal Astronomical Society, 485, 570

Chambers, J. E., 1999, Monthly Notices of the Royal Astronomical Society, 304, 793

Chambers, J. E., 2004, Earth and Planetary Science Letters, 223, 241

Chambers, J. E., 2016, The Astrophysical Journal, 825, 63

Chandler, C. O., Kueny, J., Gustafsson, A., Trujillo, C. A., Robinson, T. D. & Trilling, D. E., 2019, The Astrophysical Journal Letters, 877, 12

Chao, E. C. T., Shoemaker, E. M. & Madsen, B. M., 1960, Science, 132, 220

Chapman, C. R., Pollack, J. B. & Sagan, C., 1968, SAO Special Report #268

Chapman, C.R. & Morrison, D., 1994, Nature 367, 33–40

Charbonneau, D., Brown, T. M., Noyes, R. W. & Gilliland, R. L., 2002, The Astrophysical Journal, 568, 377

Charbonneau, D. et al., 2005, The Astrophysical Journal, 626, 523

Charbonneau, D., Brown, T. M., Noyes, R. W., & Gilliland, R. L., 2002, The Astrophysical Journal, 568, 377

Charnoz, S., Morbidelli, A., Dones, L. & Salmon, J., 2009, Icarus, 199, 413

Chau, A., Reinhardt, C., Helled, R. & Stadel, J., 2018, The Astrophysical Journal, 865, 15

Chauvin, G. et al., 2018, Astronomy and Astrophysics, 617, 76

Chen, C. H. et al., 2006, The Astrophysical Journal Supplement Series, 166, 351

Chen, C. H. et al., 2014, The Astrophysical Journal Supplement, 211, 25

Chiang, E. I. & Brown, M. E., 1999, The Astronomical Journal, 118, 1411

Chiang, E. I. & Jordan, A. B., 2002, The Astronomical Journal, 124, 3430

Chiang, E. I. & Laughlin, G., 2013, Monthly Notices of the Royal Astronomical Society, 431, 3444

Chiang, E. I. et al., 2003, The Astronomical Journal, 126, 430

Choquet, É. et al., 2016, The Astrophysical Journal, 817, 2

Choquet, É. et al., 2018, The Astrophysical Journal, 854, 53

Christian, D. J. et al., 2006, Monthly Notices of the Royal Astronomical Society, 372, 1117

Christner, B. C. et al., 2014, Nature, 512, 310

Christou, A. A., 2010, Monthly Notices of the Royal Astronomical Society, 402, 2759

Christou, A. A., Killen, R. M. & Burger, M. H., 2015, Geophysical Research Letters, 42, 7311

Chulliat, A. et al., 2019, Out-of-Cycle Update of the US/UK World Magnetic Model for 2015-2020.

Chyba, C. F., 1987, Nature, 330, 632

Chyba, C. F. & McDonald, G. D., 1995, Annual Review of Earth and Planetary Sciences, 23, 215

Ciesla, F. J., Lauretta, D. S., Cohen, B. A. & Hood, L.L., 2003, Science 299, 549

Ciesla, F. J., 2005, in Krot, A. N., Scott, E. R. D. & Reipurth, B. eds. Chondrites and the protoplanetary disk. Astronomical Society of the Pacific Conference Series 341, 811

Clark, D. L., Wiegert, P. & Brown, P. G., 2019, Monthly Notices of the Royal Astronomical Society: Letters, 487, 35

Clarke, J. T. et al., 2009, Journal of Geophysical Research, 114, 5210

Clayton, R. N., 2005, in Davis, A. M., Holland, H. D. & Turekian, K. K. eds. Meteorites, comets and planets: treatise on geochemistry, 1, 129

Clement, M. S., Kaib, N. A., Raymond, S. N. & Walsh, K. J., 2018, Icarus, 311, 340

Clement, M. S., Raymond, S. N. & Kaib, N. A., 2019, The Astronomical Journal, 157, 38

Clement, M. S., Kaib, N. A., Raymond, S. N., Chambers, J. E. & Walsh, K. J., 2019b, Icarus, 321, 778

Clement, M. S., Kaib, N. A., Chambers, J. E., 2019c, The Astronomical Journal, 157, 208

Clifford, S. M. & Parker, T. J., 2001, Icarus, 154, 40

Cloutier, R., Tamayo, D. & Valencia, D., 2015, The Astrophysical Journal, 813, 8

Clube, S. V. M. & Napier, W. M., 1984a, Nature, 311, 635

Clube, S. V. M. & Napier, W. M., 1984b, Monthly Notices of the Royal Astronomical Society, 211, 953

Clube, S. V. M., Hoyle, F., Napier, W. M. & Wickramasinghe, N. C., 1996, Astrophysics and Space Science, 245, 43

Cochran, W. D. & Hatzes, A. P., 1994, Astrophysics and Space Science, 212, 281

Coffey, D., Bacciotti, F., Woitas, J., Ray, T.P. & Eislöffel, J., 2004, The Astrophysical Journal, 604, 758

Coffey, D., Bacciotti, F., Ray, T.P., Eislöffel, J. & Woitas, J., 2007, The Astrophysical Journal, 663, 350

Cohen, B.A., Swindle, T.D., Kring, D.A., 2000, Science, 290, 1754

Comstock, G. C., 1902, The Observatory, 25, 62

Concha-Ramírez, F., Wilhelm, M. J. C., Zwart, S. P. & Haworth, T. J., 2019, Monthly Notices of the Royal Astronomical Society, 490, 5678

Connerney, J. E. P. et al., 2017, Science, 356, 826

Connolly, Jr H. C., 2005, in Krot, A. N., Scott, E. R. D. & Reipurth, B. eds. Chondrites and the protoplanetary disk, Astronomical Society of the Pacific Conference Series, 341, 215

Connolly, Jr H. C. & Love, S. G., 1998, Science, 280, 62

Connors, M., Wiegert, P., & Veillet, C., 2011, Nature, 475, 481

Conrath, B. et al., 1973, Journal of Geophysical Research, 78, 4267

Contro, B., Horner, J., Wittenmyer, R. A., Marshall, J. P. & Hinse, T. C., 2016, Monthly Notices of the Royal Astronomical Society, 463, 191

Correia, A. C. M., Laskar, J. & de Surgy, O. N., 2003, Icarus, 163, 1

Correia, A. C. M. & Laskar, J., 2001, Nature, 411, 767

Cossou, C., Raymond, S. N., Hersant, F. & Pierens, A., 2014, Astronomy & Astrophysics, 569, A56

Coughlin, J. L. et al., 2016, The Astrophysical Journal Supplement Series, 224, 12

Craddock, R. A., 2011, Icarus, 211, 1150

Crida, A. & Morbidelli, A., 2007, Monthly Notices of the Royal Astronomical Society, 377, 1324

Crovisier, J., Bockelee-Morvan, D., Gerard, E., Rauer, H., Biver, N., Colom, P. & Jodra, L., 1996, Astronomy & Astrophysics, 310, L17

Cruikshank, D. P., Matthews, M. S., & Schumann, A., 1995, Neptune and Triton. University of Arizona Press

Ćuk, M. & Burns, J. A., 2004, Icarus, 167, 369

Ćuk, M. & Stewart, S. T., 2012, Science, 338, 1047

Ćuk, M., Christou, A. A., & Hamilton, D. P., 2015, Icarus, 252, 339

Cumming, A., Butler, R. P., Marcy, G. W., Vogt, S. S., Wright, J. T. & Fischer, D. A., 2008, Publications of the Astronomical Society of the Pacific, 120, 531

Dalba, P. A., Muirhead, P. S., Fortney, J. J., Hedman, M. M, Nicholson, P. D., & Veyette, M. J., 2015, The Astrophysical Journal, 814, 154

Dalba, P. A. & Muirhead, P. S., 2016, The Astrophysical Journal, 826, 7

Dalba, P. A. & Tamburo, P., 2019, The Astrophysical Journal, 873, 17

Darriba, L. A., de Elia, G. C., Guilera, O. M. & Brunini, A., 2017, Astronomy & Astrophysics, 607 63

Dauber, I. et al., 2018, Space Science Reviews, 214, 132

Dauphas, N., Robert, F. & Marty, B., 2000, 148, 508

Davis, M., Hut, P. & Muller, R. A., 1984, Nature, 308, 715

Davis, A. M., Alexander, C. M. O'D., Ciesla, F. J., Gounelle, M., Krot, A. N., Petaev, M. I. & Stephan, T., 2014, Protostars and Planets VI, 809

Davies, J. H., 2008, Earth and Planetary Science Letters, 268, 376

Dawson, R. I. & Johnson, J. A., 2018, Annual Review of Astronomy and Astrophysics, 56, 175

Debras, F. & Chabrier, G., 2019, The Astrophysical Journal, 872, 100

de la Fuente Marcos, C. & de La Fuente Marcos, R., 2014, Monthly Notices of the Royal Astronomical Society, 443, 59

de la Fuente Marcos, C. & de la Fuente Marcos, R., 2014, Astrophysics & Space Science, 352, 409

de la Fuente Marcos, C., de la Fuente Marcos, R. & Aarseth, S. J., 2018, Monthly Notices of the Royal Astronomical Society, 476, 1

de Winter, N. J. et al., 2020, Paleoceanography and Paleoclimatology, e2019PA003723

Deleuil, M. et al., 2008, Astronomy & Astrophysics, 491, 889

Del Genio, A. D., Brain, D., Noack, L. & Schaefer, L., 2019, accepted review (arXiv:1807.04776)

Dello Russo, N., Vervack, R. J., Jr., Weaver, H. A., Montgomery, M. M., Deshpande, R., Fernández, Y. R. & Martin, E. L., 2008, The Astrophysical Journal, 680, 793

Deming, D., Seager, S., Richardson, L. J. & Harrington, J., 2005, Nature, 434, 740

Dent, W. R. F. et al., 2014, Science, 343, 1490

Deo, S. N., & Kushvah, B. S., 2017, Astronomy and Computing, 20, 97

de Pater, I., Sromovsky, L. A., Fry, P. M., Hammel, H. B., Baranec, C., & Sayanagi, K. M., 2015, Icarus, 252, 121

de Pater, I., & Lissauer, J. J. 2015, Planetary Sciences, Second Edition (Cambridge, UK: Cambridge University Press)

Desch, S. J., Ciesla, F. J., Hood, L. L. & Nakamoto, T., 2005, in Krot, A. N., Scott, E. R. D. & Reipurth, B. eds. Chondrites and the protoplanetary disk, Astronomical Society of the Pacific Conference Series, 341, 849

Desch, S. J., Morris, M. A., Connolly, H. C. & Boss, A. P., 2010, The Astrophysical Journal, 725, 692

Desch, S. J., Kalyaan, A. & Alexander, C.M. O'D., 2018, The Astrophysical Journal Supplement Series, 238, 11

Dessler, A. J., 1983, Physics of the Jovian magnetosphere, Cambridge University Press, Cambridge

Devillepoix, H. A. R. et al., 2018, Meteoritics & Planetary Science, 53, 2212

Devillepoix, H. A. R. et al., 2020, *accepted to appear in Planetary and Space Science*

Di Sisto, R. P. & Brunini, A., 2007, Icarus, 190, 224

Di Sisto, R. P., Brunini, A. & di Elía, G. C., 2010, Astronomy & Astrophysics, 519, 112

Di Sisto, R. P., Ramos, X. S., & Beaugé, C. 2014, Icarus, 243, 287

Di Sisto, R. P., Ramos, X. S. & Gallardo, T., 2019, Icarus, 319, 828

Dodson-Robinson, S. E., Willacy, K., Bodenheimer, P., Turner, N. J. & Beichman, C. A., 2009, Icarus, 200, 672

Delbó, M., Walsh, K., Bolin, B., Avdellidou, C., & Morbidelli, A., 2017, Science, 357, 1026

Delbó, M., Avdellidou, C., & Morbidelli, A., 2019, Astronomy & Astrophysics, 624, 69

Dobrovolskis, A. R. & Ingersoll, A. P., 1980, Icarus, 41, 1

Dohnanyi, J. S., 1969, Journal of Geophysical Research, 74, 2531

Dombard, A. J., Cheng, A. F., McKinnon, W. B. & Kay, J. P., 2010, AGU Fall Meeting Abstracts, 1

Dombard, A. J., Cheng, A. F., McKinnon, W. B. & Kay, J. P., 2012, Journal of Geophysical Research, 117, 3002

Dominik, C., Laureijs, R. J., Jourdain de Muizon, M. & Habing, H. J., 1998, Astronomy & Astrophysics, 329, 53

Dominik, C. & Decin, G., 2003, The Astrophysical Journal, 598, 1, 626

Dominik, M., 2009, Monthly Notices of the Royal Astronomical Society, 393, 816

Dones, L., Weissman, P. R., Levison, H. F. & Duncan, M. J., 2004, "Oort cloud formation and dynamics", in Comets II, M. C. Festou, H. U. Keller, and H. A. Weaver (eds.), University of Arizona Press, Tucson, 745 pp., p.153-174

Dones, L., Brasser, R., Kaib, N. & Rickman, H., 2015, Space Science Reviews, 197, 191

Dormand, J. R. & McCue, J., 1987, Journal of the British Astronomical Association, 98, 23

Dougherty, M. K., Esposito, L. W. & Krimigis, S. M., 2009, Saturn from Cassini-Huygens. Edited by M.K. Dougherty, L.W. Esposito, and S.M. Krimigis. Berlin: Springer, 2009. ISBN 978-1-4020-9216-9

Drake, M. J., 2001, Meteoritics and Planetary Science, 36, 501

Drake, M. J., 2004, Meteoritics and Planetary Science, 39, 5031

Drake, M. J., 2005, Meteoritics and Planetary Science, 40, 519

Draper, Z. H. et al., 2016, The Astrophysical Journal, 826, 147

Dressing, C. D. & Charbonneau, D., 2016, The Astrophysical Journal, 807, 45

Dressler, B. O., Sharpton, V. L., Morgan, J., Buffler, R., Moran, D., Smit, J., Stöffler, D. & Urrutia, J., 2003, EOS Transactions, 84, 125

Drouart, A., Dubrulle, B., Gautier, D. & Robert, F., 1999, Icarus 140, 129–155

Dubinin, E., Fraenz, M., Fedorov, A., Lundin, R., Edberg, N., Duru, F. & Vaisberg, O., 2011, Space Science Reviews, 162, 173

Dumusque, X. et al., 2011, Astronomy & Astrophysics, 525, 140

Duncan, M., Quinn, T. & Tremaine, S., 1988, The Astrophysical Journal, 328, L69–L73.

Duncan M. J., Levison H. F., 1997, Science, 276, 1670

Dunn, T. L., Burbine, T. H., Bottke, W. F. & Clark, J. P., 2013, Icarus, 222, 273

Durda, D. D. & Dermott, S. F., 1997, Icarus, 130, 140

Durda, D. D. & Stern, S. A., 2000, Icarus, 145, 220

Durda, D. D. et al., 2007, Icarus, 186, 498

Ďurech, J., et al., 2018, Astronomy & Astrophysics, 609, A86

Dwarkadas, V. V., Dauphas, N., Meyer, B., Boyajian, P. & Bojazi, M., 2017, The Astrophysical Journal, 851, 147

Eberhardt, P., Meier, R., Krankowsky, D. et al.: 1995, Astronomy and Astrophysics 302, 301

Edgeworth. K. E., 1943, J.B.A.A. 53, 186

Edgeworth, K. E., 1949, Monthly Notices of the Royal Astronomical Society, 109, 600

Ehrenreich, D., et al., 2012, Astronomy & Astrophysics, 537, L2

Eiroa, C., 2010, Astronomy & Astrophysics, 518. 131

Eiroa, C. et al., 2013, Astronomy & Astrophysics, 555, 11

Elliot, J. L., Dunham, E. & Mink, D., 1977, Nature, 267, 328

Elliot, J. L., Kern, S. D., Clancy, K. B. et al., 2005, The Astronomical Journal, 129, 1117

Emel'yanenko, V. V. & Bailey, M. E., 1998, Monthly Notices of the Royal Astronomical Society, 298, 212

Emel'yanenko V. V., Asher D. J. & Bailey M. E., 2005, Monthly Notices of the Royal Astronomical Society, 361, 1345

Emel'yanenko, V. V., Asher, D. J. & Bailey, M. E., 2007, Monthly Notices of the Royal Astronomical Society, 381, 779
134

Encrenaz, T., Tinetti, G & Coustenis, A., 2018, Experimental Astronomy, 46, 31

Endl, M. et al., 2016, The Astrophysical Journal, 818, 34

Ertel, S. et al., 2014, Astronomy & Astrophysics, 570, 128

Ertel, S. et al., 2018, The Astronomical Journal, 155, 194

Ertel, S. et al., 2020, *accepted to appear in the Astronomical Journal, arXiv:2003.03499*

Esposito, M. et al., 2014, Astronomy & Astrophysics, 564, 13

Evans, D. W. et al., 2018, Astronomy & Astrophysics, 616, 4

Evans, J. E. & Maunder, E. W., 1903, Monthly Notices of the Royal Astronomical Society, 63, 488

Faramaz, V. et al., 2019, The Astronomical Journal, 158, 162

Farnocchia, D., Chesley, S. R., Vokrouhlický, D., Milani, A., Spoto, F., & Bottke, W. F., 2013, Icarus, 224, 1

Farinella, P., Froeschlé, Ch.; Froeschlé, Cl., Gonczi, R., Hahn, G., Morbidelli, A. & Valsecchi, G. B., 1994, Nature, 371, 314

Fechtig, H. & Grün, E., 1975, Mitteilungen der Astronomischen Gesellschaft Hamburg, 36, 67

Feng, F. & Bailer-Jones, C. A. L., 2015, Monthly Notices of the Royal Astronomical Society, 454, 3267

Fernández, Y. R., 2002, Earth, Moon & Planets, 89, 3

Fernández, Y. R., 2009, Planetary and Space Science, 57, 1218

Feulner, G., 2012, Reviews of Geophysics, 50, 2006

Filipovic, M. D., Horner, J., Crawford, E. J., Tothill, N. F. H. & White, G. L., 2013, Serbian Astronomical Journal, 187, 43

Fischer, D. A. & Valenti, J., 2005, The Astrophysical Journal, 622, 1102

Fischer, D. A., Marcy, G. W. & Spronck, J. F. P., 2014, The Astrophysical Journal Supplement, 210, 5

Fischer, D. A. et al., 2016, Publications of the Astronomical Society of the Pacific, 128, 6001

Fitzsimmons, A., Snodgrass, C., Rozitis, B., Yang, B., Hyland, M., Seccull, T., Bannister, M. T., Fraser, W. C., Jedicke, R. & Lacerda, P., 2018, Nature Astronomy, 2, 133

Fitzsimmons, A. et al., 2019, The Astrophysical Journal, 855, 9

Fjeldbo, G., Fjeldbo, W. C. & Eschleman, V. R., Journal of Geophysical Research, 71, 2307

Fletcher, L. N. et al., Icarus, 208, 337

Fogg, M. J. & Nelson, R. P., 2007, Astronomy and Astrophysics, 461, 1195

Fogg, M. J. & Nelson, R. P., 2007, Astronomy and Astrophysics, 472, 1003

Fogg, M. J. & Nelson, R. P., 2009, Astronomy and Astrophysics, 498, 575

Folkner, W. M. et al., 2017, Geophysical Research Letters, 44, 4694

Folkner, W. M. et al., 2018, Space Science Reviews, 214, 100

Foreman-Mackey, Daniel, Morton, Timothy D., Hogg, David W., Agol, Eric & Schölkopf, Bernhard, 2016, The Astronomical Journal, 152, 206

Fortney, J. J. & Nettlemann, N., 2010, Space Science Reviews, 152, 423

Fortney, J. J. et al., 2016, The Astrophysical Journal Letters, 824, 25

Fortney, J. J., Helled, R., Nettelmann, N., Stevenson, D. J., Marley, M. S., Hubbard., W. B., & Iess, L., 2018, The Interior of Saturn, in Saturn in the 21st Century (Cambridge University Press: Cambridge) 44-68

Fouchard, M., Froeschlé, Ch., Rickman, H. & Valsecchi, G. B., 2011, Icarus, 214, 334

Fouchet, T., Moses, J. I., & Conrath, B. J., 2009, Saturn: Composition and Chemistry, in Saturn from Cassini-Huygens (Springer Science+Business Media: Dordrecht) 83-112

Frank, A., Ray, T. P., Cabrit, S., Hartigan, P., Arce, H. G., Bacciotti, F., Bally, J., Benisty, M., Eislöffel, J., Güdel, M., Lebedev, S., Nisini, B. & Raga, A., 2014, Protostars and Planets VI, 451

Fraser, W. C. & Brown, M. E., 2012, The Astrophysical Journal, 749, 33

Fraser, W. C., Brown, M. E., Morbidelli, A., Parker, A. & Batygin, K., 2014, The Astrophysical Journal, 782, 100

Fressin, F. et al., 2013, The Astrophysical Journal, 766, 81

Fujiwara, H. et al., 2009, The Astrophysical Journal, 695, 88

Fujiwara, H. et al., 2010, The Astrophysical Journal, 714, 152

Fujiwara, H. et al., 2012, The Astrophysical Journal, 759, 18

Fujiwara, H. et al., 2013, Astronomy and Astrophysics, 550, 45

Fukuhara, T. et al., 2017a, Nature Geoscience, 10, 85

Fukuhara, T. et al., 2017b, Earth Planets Space, 69, 141

Fulton, B. J.. et al., 2017, The Astronomical Journal, 154, 109

Fulton, B. J. & Petigura, E. A., 2018, The Astronomical Journal, 156, 265

Gaia Collaboration et al., 2016, Astronomy & Astrophysics, 595, 1

Gaia Collaboration et al., 2018, Astronomy & Astrophysics, 616, 1

Gaidos, E., Mann, A. W., Kraus, A. L. & Ireland, M., 2016, Monthly Notices of the Royal Astronomical Society, 457, 2877

Gajdoš, P., Vaňko, M. & Parimucha, S., 2019, Research in Astronomy and Astrophysics, 19, 41

Galanti, E., Kaspi, Y., Miguel, Y., Guillot, T., Durante, D., Racioppa, P., & Iess, L., 2019, Journal of Geophysical Research Planets, 46, 616

Galle, J. G., 1846, Monthly Notices of the Royal Astronomical Society, 7, 153

Garcia, R. A. et al., 2014, Astronomy & Astrophysics, 568, 10

García Muñoz, A., Zapatero Osorio, M. R., Barrena, R., Montañés-Rodríguez, P., Martín, E. L. & Pallé, E., 755, 103

Gatewood, G. & Eichhorn, H., 1973, The Astronomical Journal, 78, 769

Gaudi, B. S., 2012, Annual Review of Astronomy and Astrophysics, 50, 411

Gault, D. E., Burns., J. A., Cassen, P. & Strom, R. G., 1977, Annual review of Astronomy and Astrophysics, 15, 97

Gautier, D., Hersant, F., Mousis, O. & Lunine, J. I., 2001, The Astrophysical Journal, 550, 227

Gehrels, T. et al., 1980, Science, 207, 434

Geiler, F., Krivov, A. V., Booth, M. & Löhne, T., 2019, Monthly Notices of the Royal Astronomical Society, 483, 332

Giardini, D. et al., 2020, Nature Geoscience, 13, 205

Gillon, M. et al., 2017, Nature, 542, 456

Gilman, M. & Erenler, H., 2008, International Journal of Astrobiology, 7, 17

Gladman, B., Kavelaars, J. J., Nicholson, P. D., Loredo, T. J. & Burns, J. A., 1998a, The Astronomical Journal, 116, 2042

Gladman, B. J. et al., 1998b, Nature, 392, 897

Gladman, B. J. et al., 2000, Icarus, 147, 320

Gladman, B., Kavelaars, J. J., Petit, J.-M., Morbidelli, A., Holman, M. J. & Loredo, T., 2001, The Astronomical Journal, 122, 1051

Gladman, B., Marsden, B. G., & Vanlaerhoven, C. 2008, in The Solar System Beyond Neptune, ed. Barucci, M. A., Boehnhardt, H., Cruikshank, D. P., & Morbidelli, A. (Tucson: University of Arizona Press), 43

Gladman, B. J., S. M. Lawler, J.-M. Petit, J. J. Kavelaars, R. L. Jones, J. Wm. Parker, C. Van Laerhoven, P. Nicholson, P. Rousselot, A. Bieryla, & M. L. N. Ashby, 2012, The resonant trans-Neptunian populations, The Astronomical Journal, 144, 1

Gladstone, G. R. et al., 2007, Science, 318, 229

Goldblatt, C. & Watson, A. J. J., 2012, Philosophical Transactions of the Royal Society A, 370, 4197

Goldblatt, C., Robinson, T. D., Zahnle, K. J. & Crisp, D., 2013, Nature Geoscience, 6, 661

Goldreich, P. & Tremaine, S., 1982, Annual Review of Astronomy & Astrophysics, 20, 249

Goldreich, P., Lithwick, Y. & Sari, R., 2004, The Astrophysical Journal, 614, 497

Goldreich P, Murray N, Longaretti P-Y, & Banfield D, 1989, Science 245, 500

Goldstein, R. M. & Carpenter, R. L., 1963, Science, 139, 910

Golimowski, D. A. et al., 2006, The Astronomical Journal, 131 3109

Gomes, R. S., 1997, The Astronomical Journal, 114, 396

Gomes, R. S., Morbidelli, A. & Levison, H. F., 2004, Icarus, 170, 492

Gomes, R. S., Fernández, J. A., Gallardo, T. & Brunini, A., 2008, 'The Scattered Disk: Origins, Dynamics, and End States', in The Solar System Beyond Neptune, M. A. Barucci, H. Boehnhardt, D. P. Cruikshank, and A. Morbidelli (eds.), University of Arizona Press, Tucson, 592 pp., p.259-273

Gomes, R., Levison, H. F., Tsiganis, K. & Morbidelli, A., 2005, Nature, 435, 466

Gomes, R. & Nesvorný, D., 2016, Astronomy & Astrophysics, 592, 146

Gomes, R., Nesvorný, D., Morbidelli, A., Deienno, R. & Noguiera, E., 2018, Icarus, 306, 319

Gonczi, R., Rickman, H. & Froeschle, C., 1992, Monthly Notices of the Royal Astronomical Society, 254, 627

Goodricke, J., 1783, Philosophical Transactions of the Royal Society of London, 73, 474

Goździewski, K. & Migaszewski, C., 2009, Monthly Notices of the Royal Astronomical Society: Letters, 397, 16

Goździewski, K. & Migaszewski, C., 2014, Monthly Notices of the Royal Astronomical Society, 440, 3140

Goździewski, K. et al., 2015, Monthly Notices of the Royal Astronomical Society, 448, 1118

Goździewski, K. & Migaszewski, C., 2018, The Astrophysical Journal Supplement Series, 238, 6

Granvik, M., Morbidelli, A., Vokrouhlický, D., Bottke, W. F., Nesvorný, D. & Jedicke, R., 2017, Astronomy & Astrophysics, 598, 52

Gravity Collaboration et al., 2019, Astronomy & Astrophysics, 623, 11

Grazier, K. R., 2016, Astrobiology, 16, 23

Grazier, K. R., Castillo-Rogez, J. C. & Horner, J., 2018, The Astronomical Journal, 156, 232

Grazier, K. R., Horner, J. & Castillo-Rogez, J. C., 2019, Monthly Notices of the Royal Astronomical Society, 490, 4388

Greaves, J. S. et al., 1998, The Astrophysical Journal, 506, 133

Greaves, J. S., 2006, International Journal of Astrobiology 5, 187

Greaves, J. S., 2010, Monthly Notices of the Royal Astronomical Society: Letters, 409, 1, L44

Greaves, J. S. et al., 2016, Monthly Notices of the Royal Astronomical Society, 461, 3910

Green, N. E., 1879, The Observatory, 3, 252

Greenberg, R. & Barnes, R., 2008, Icarus, 194, 847

Gregory, P. C. & Fischer, D. A., 2010, Monthly Notices of the Royal Astronomical Society, 403, 731

Grogan, K.., Dermott, S. F. & Durda, D. D., 2001, Icarus, 152, 251

Grossman, L., 1972, Geochimica et Cosmochimica Acta, 36, 597

Grossman, L., 2010, Meteoritics and Planetary Science, 45, 7

Grün, E. et al., 1985, Icarus, 62, 244

Grün, E. et al., 1993, Nature, 362, 428

Guillot, T., Gautier, D. & Hubbard, W. B., 1997, Icarus, 130, 534

Guillot, T., 1999, Science, 286, 72

Guillot, T. & Gautier, D., 2015, Giant planets, in second edition of the Treatise on Geophysics (Elsevier: Amsterdam) 529-559

Gunnarsson, M., Bockel'ee-Morvan, D., Biver, N., et al., 2008, Astronomy and Astrophysics, 484, 537

Gurwell, M.A., 1995, Nature, 378, 22

Gustafson, B. A. S., 1989, Astronomy & Astrophysics, 225, 533

Gustafson, B. A. S., 1994, Annual Review of Earth and Planetary Sciences, 22, 553

Guzik, P. et al., 2020, Nature Astronomy, 4, 53

Haghighipour, N., 2009, Meteoritics and Planetary Science, 44, 1863

Hahn, J. M. & Malhotra, R., 2005, The Astronomical Journal, 130, 2392

Hallatt, T. & Wiegert, P., 2020, The Astronomical Journal, 159, 147

Halliday, A. N., 2008, Royal Society of London Philosophical Transactions Series A, 366, 4163

Halverson, S., et al., 2016, SPIE, 9908, 99086

Hamann, A. et al., 2019, The Astronomical Journal, 158, 133

Hamano, K., Abe, Y. & Genda, H., 2013, Nature, 497, 607

Hamilton, G. H., 1916, The Observatory, 39, 363

Hamilton, D. P. & Ward, W. R., 2004, The Astronomical Journal, 128, 2510

Hammel, H. B., Beebe , R. F., Ingersoll, A. P., et al., 1995, Science, 267, 1288

Hammond, N. P. & Barr, A. C., 2014, Geology, 42, 931

Han, C., & Han, W., 2002, The Astrophysical Journal, 580, 1

Han, D., Poppe, A. R., Piquette, M., Grün, E. & Horányi, M, 2011, Geophysical Research Letters, 38, 24102

Hanel, R. et al., 1972, Icarus, 17, 423

Hanse, J., Jílková, L., Portegies Zwart, S. F. & Peupessy, F. I., 2018, Monthly Notices of the Royal Astronomical Society, 473, 5432

Hansen, B. M. S., 2009, The Astrophysical Journal, 703, 1131

Hansen, B. M. S. & Murray, N., 2012, The Astrophysical Journal, 751, 158

Hanuš, J. et al., 2019, Astronomy & Astrophysics, 624, 121

Harder, H. & Schubert, G., 2001, Icarus, 151, 118

Harrington, R. S. & van Flandern, T. C., 1979, Icarus, 39, 131

Harris, A. W. & D'Abramo, G., 2015, Icarus, 257, 302

Hart, M. H., 1979, Icarus, 37, 351

Hartman, J. D. et al., 2011, The Astrophysical Journal, 728, 138

Hartmann, W. K., 1975, Icarus, 24, 181

Hartmann, W. K., 2003, Meteoritics & Planetary Science, 38, 579

Hartmann, L., Herczeg, G. & Calvet, N., 2016, Annual Review of Astronomy and Astrophysics, 54, 135

Hartung, J. B., 1976, Meteoritics, 11, 187

Hartung, J. B., 1993, Icarus, 104, 280

Hasegawa, I. & Nakano, S., 2001, Publications of the Astronomical Society of Japan, 53, 931

Hasegawa, I. & Nakano, S., 2003, Monthly Notices of the Royal Astronomical Society, 345, 883

Haus, R. et al., 2016, Icarus, 272, 178

Hayashi, C., 1981, Progress of Theoretical Physics Supplement, 70, 35

Head, J.W., 2014, Geology, 42, 95

Heisler, J., Tremaine, S. & Alcock, C., 1987, Icarus, 70, 269

Helfer, H. L., 1990, Icarus, 87, 228

Helled, R., 2011, The Astrophysical Journal: Letters, 735, L16

Heller, R. 2012, Astronomy & Astrophysics, 545, 1

Heller, R., & Barnes, R., 2013, Astrobiology, 13, 1

Heller, R., & Pudritz, R., 2013, The Astrophysical Journal, 806, 2

Heller, R., 2016, Astronomy & Astrophysics, 588, 1

Heller, R., Rodenbeck, K., Bruno, G. 2019, Astronomy & Astrophysics, 624, 1

Hergenrother, C. W. et al., 2019, Nature Communications, 10, 1291

Hermann, F. & Krivov, A. V., 2007, Astronomy & Astrophysics, 476, 829

Herrick, R.R., 1994, Geology, 22, 703

Herschel, W., 1787, Royal Society of London Philosophical Transactions Series I, 77, 125

Hershey, J. L., 1973, The Astronomical Journal, 78, 421

Herwartz, D., Pack, A., Friedrichs, B. & Bischoff, A., 2014, Science, 344, 1146

Hess, S. L., Henry, R. M., Leovy, C. B., Ryan, J. A. & Tilman, J. E., 1977, Journal of Geophysical Research, 82, 4559

Hezel, D. C. & Palme, H., 2008, Earth and Planetary Science Letters, 265, 716

Higuchi, A. E., Oya, Y. & Yamamoto, S., 2019, The Astrophysical Journal, 885, 39

Hildebrand, A. R., Penfield, G. T., Kring, D. A., Pilkington, M., Camargo, Z. A., Jacobsen, S. B. & Boynton, W. V., 1991, Geology, 19, 861

Hill, M. L. et al., 2018, The Astrophysical Journal, 860, 1

Hills, J. G., 1981, The Astronomical Journal, 86, 1730

Hinkel, N.R., Kane, S.R. 2013, ApJ, 774, 27

Hirabayashi, M., Scheeres, D. J., Sánchez, D. P. & Gabriel, T., 2014, The Astrophysical Journal Letters, 789, L12

Hodges, R. R., 2002, Geophysical Research Letters, 29, 1038

Holland, W. S. et al., 2017, Monthly Notices of the Royal Astronomical Society, 470, 3606

Hollis, W. A., 1908, Nature, 77, 438

Holman M. J., Wisdom J., 1993, The Astronomical Journal, 105, 1987

Holman, M. J. & Murray, N. W., 2005, Science, 307, 1288

Holt, T. R., Brown, A. J., Nesvorný, D., Horner, J. & Carter, B. 2018, The Astrophysical Journal, 859, 2, 97

Holt, T. R., Nesvorný, D., Horner, J., King, R., Brookshaw, L., Carter, B. & Tylor, C., 2020, *submitted*

Hom, J. et al., 2020, The Astronomical Journal, 159, 31

Homma, K., Okuzumi, S., Nakamoto, T. & Ueda, Y., 2019, The Astrophysical Journal, 877, 128

Hood, L. L. & Horanyi, M., 1993, Icarus, 106, 179

Hori, Y., Sano, T., Ikoma, M. & Ida, S., 2008, Exoplanets: Detection, Formation and Dynamics, Proceedings of the International Astronomical Union, IAU Symposium, 249, 163

Horner, J. & Evans, N. W., 2002, Monthly Notices of the Royal Astronomical Society, 335, 641

Horner, J., Evans, N. W., Bailey, M. E. & Asher, D. J., 2003, Monthly Notices of the Royal Astronomical Society, 343, 1057

Horner, J., Evans, N. W. & Bailey, M. E., 2004a, Monthly Notices of the Royal Astronomical Society, 354, 798

Horner, J., Evans, N. W. & Bailey, M. E., 2004b, Monthly Notices of the Royal Astronomical Society, 355, 321

Horner, J. & Evans, N. W., 2006, Monthly Notices of the Royal Astronomical Society, 367, L20

Horner, J., Mousis, O. & Hersant, F., 2007, Earth Moon and Planets 100, 43

Horner, J. & Jones, B. W., 2008a, International Journal of Astrobiology, 7, 251

Horner, J. & Jones, B. W., 2008b, A&G, 49, 1.22

Horner, J., Mousis, O., Alibert, Y., Lunine, J. I. & Blanc, M., 2008, Planetary and Space Science 56, 1585

Horner, J. & Jones, B. W., 2009, International Journal of Astrobiology, 8, 75

Horner, J., Mousis, O., Petit, J. -M. & Jones, B. W., 2009, Planetary and Space Science, 57, 1338

Horner, J., Jones, B. W. & Chambers, J., 2010a, International Journal of Astrobiology 9, 1

Horner, J. & Jones, B. W., 2010b, Astronomy & Geophysics, 51, 6.16

Horner, J. & Jones, B. W., 2010c, International Journal of Astrobiology, 9, 273

Horner, J. & Lykawka, P. S., 2010a, Monthly Notices of the Royal Astronomical Society, 402, 13

Horner, J. & Lykawka, P. S., 2010c, Monthly Notices of the Royal Astronomical Society, 405, 49

Horner, J. & Lykawka, P. S., 2010b, Monthly Notices of the Royal Astronomical Society, 405, 1375

Horner, J. & Lykawka, P. S., 2010d, International Journal of Astrobiology, 9, 227

Horner, J, & Jones, B. W., 2011, Astronomy and Geophysics, 52, 16

Horner, J. & Lykawka, P. S., 2011, Astronomy and Geophysics, 52, 24

Horner, J., Marshall, J. P., Wittenmyer, R. A. & Tinney, C. G., 2011b, Monthly Notices of the Royal Astronomical Society, 416, 11

Horner, J. & Jones, B. W., 2012, International Journal of Astrobiology, 11, 147

Horner, J., Lykawka, P. S., Bannister, M. T. & Francis, P., 2012a, Monthly Notices of the Royal Astronomical Society, 422, 4125

Horner, J., Wittenymer, R. A., Hinse, T. C. & Tinney, C. G., 2012b, Monthly Notices of the Royal Astronomical Society, 425, 749

Horner, J., Hinse, T. C., Wittenmyer, R. A., Marshall, J. P. & Tinney, C. G., 2012c, Monthly Notices of the Royal Astronomical Society, 427, 2812

Horner, J., Lykawka, P. S. & Müller, T., 2012, Monthly Notices of the Royal Astronomical Society, 423, 2587

Horner, J. & Lykawka, P. S., 2012, *in the proceedings of the 11th Annual Australian Space Science Conference, arXiv: 1202.5729*

Horner, J., Wittenmyer, R. A., Hinse, T. C., Marshall, J. P., Mustill, A. J. & Tinney, C. G., 2013, Monthly Notices of the Royal Astronomical Society, 435, 2033

Horner, J., Wittenmyer, R. A., Hinse, T. C. & Marshall, J. P., 2014a, Monthly Notices of the Royal Astronomical Society, 439, 1176

Horner, J., Wittenmyer, R., Hinse, T., Marshall, J. & Mustill, A., 2014, *in the proceedings of the 13th Annual Australian Space Science Conference; arXiv:1401.6742*

Horner, J. et al., 2019, The Astronomical Journal, 158, 100

Horner, J. et al., 2020, The Astronomical Journal, 159, 10

Hoppe, P., Strebel, R., Eberhardt, P., Amari, S. & Lewis, R. S., 1996, Science, 272, 1314

Hosono, N., Karato, S. -I., Makino, J. & Saitoh, T. R., 2019, Nature Geoscience, 12, 418

Howard, A. W. et al., 2010, Science, 330, 653

Howard, A. W. et al., 2012, The Astrophysical Journal: Supplement Series, 201, 15

Howard, A.W., Fulton, B.J. 2016, Publications of the Astronomical Society of the Pacific, 128, 114401

Howell, S. B. et al., 2014, Publications of the Astronomical Society of the Pacific, 126, 398

Howie, R. M., Paxman, J., Bland, P. A., Towner, M. C., Cupak, M., Sansom, E. K. & Devillepoix, H. A. R., 2017, Experimental Astronomy, 43, 237

Hsieh, H. H., Jewitt, D. & Fernández, Y. R., 2004, The Astronomical Journal, 127, 2997

Hsieh, H. H. & Jewitt, D., 2006, Science, 312, 561

Hsieh, H. H., Jewitt, D., Lacerda, P., Lory, S. C. & Snodgrass, C., 2010, Monthly Notices of the Royal Astronomical Society, 403, 363

Hsieh, H. H. et al., 2015, Icarus, 248, 289

Hsu, D. C., Ford, E. B., Ragozzine, D. & Ashby, K., 2019, The Astronomical Journal, 158, 109

Hubbard, W. B. * MacFarlane, J. J., 1980, Journal of Geophysical Research: Solid Earth, 85, 225

Hubickyj, O., Bodenheimer, P. & Lissauer, J. J., 2005, Icarus, 179, 415

Hueso, R. et al., 2018, Astronomy & Astrophysics, 617, 68

Hughes, A. M., Wilner, D. J., Andrews, S. M., Williams, J. P., Su, K. Y. L., Murray-Clay, R. A. & Qi, C., 2011, The Astrophysical Journal, 740, 38

Hughes, A. M., Duchêne, G. & Matthews, B. C., 2018, Annual Review of Astronomy and Astrophysics, 56, 541

Hung, L.-W. et al., 2015, The Astrophysical Journal, 815, 14

Hut, P., 1984, Nature, 311, 638

Huygens, C., 1659, "Systema Saturnium sive de causis mirandorum Saturni phaenomenon et comite ejus planeta novo"

Ida, S., Larwood, J. & Burkert, A., 2000, The Astrophysical Journal, 528, 351

Ida, S., Bryden, G., Lin, D. N. C. & Tanaka, H., 2000, The Astrophysical Journal, 534, 428

Ida, S. & Lin, D. N. C., 2004, The Astrophysical Journal, 604, 388

Ida, S. & Lin, D. N. C., 2008, The Astrophysical Journal, 673, 487

Ida, S. & Lin, D. N. C., 2010, The Astrophysical Journal, 719, 810

Iess, L. et al., 2019, Science, 364. 2965

Inaba, S., Wetherill, G. W. & Ikoma, M., 2003, Icarus, 166, 46

Ingersoll, A.P., 1969, The Journal of Atmospheric Science, 26, 1191

Iorio, L., 2014, Monthly Notices of the Royal Astronomical Society, 444, L78

Ip, W.-H., 1988, Astronomy and Astrophysics, 199, 340

Ip, W.-H., 2006, Geophysics Research Letters, 33, 16203

Irwin, P. G. J., et al., 1998, Journal of Geophysical Research Planets, 103, 23001

Irwin, P. G. J., Barstow, J. K., Bowles, N. E., Fletcher, L. N., Aigrain, S., & Lee, J. -M., 2014, Icarus, 242, 172

Ishiguru, M. et al., 2013, The Astrophysical Journal, 767, 75

Ivanov, M. A. & Head, J. W., 2011, Planetary & Space Science, 59, 1559

Ivezić, Ž. et al., 2019, The Astrophysical Journal, 873, 111

Izidoro, A., de Souza Torres, K., Winter, O. C. & Haghighipour, N., 2013, The Astrophysical Journal, 767, 54

Izidoro, A., Morbidelli, A. & Raymond, S. N., 2014, The Astrophysical Journal, 794, 11

Izidoro, A., Haghighipour, N., Winter, O. C. & Tsuchida, M., 2014, The Astrophysical Journal, 782, 31

Izidoro, A., Raymond, S. N., Morbidelli, A., Hersant, F. & Pierens, A., 2015a, The Astrophysical Journal Letters, 800, L22

Izidoro, A., Raymond, S. N., Morbidelli, A. & Winter, O. C., 2015b, Monthly Notices of the Royal Astronomical Society, 453, 3619

Izidoro, A. & Raymond, S. N., 2018, 'Formation of Terrestrial Planets', chapter in the Handbook of Exoplanets, ISBN 978-3-319-55332-0. Springer International Publishing AG, part of Springer Nature, 2018, id.142

Jackson, A. P. & Wyatt, M. C., 2012, Monthly Notices of the Royal Astronomical Society, 425, 657

Jackson, A. P., Wyatt, M. C., Bonsor, A. & Veras, 2014, Monthly Notices of the Royal Astronomical Society, 440, 3757

Jackson, A. P., Gabriel, T. S. J. & Asphaug, E. I., 2018, Monthly Notices of the Royal Astronomical Society, 474, 2924

Jacobsen, R. A., 1998, The Astronomical Journal, 115, 1195

Jacobsen, S. B. & Ranen, M. C., 2006, Geochimica et Cosmochimica Acta Supplement, 70, 286

Jakosky, B. M. & Phillips, R. J., 2001, Nature, 412, 327

Janes, D. M. & Melosh, H. J., 1988, Journal of Geophysics Research, 93, 3127

Jenniskens, P. & Lyytinen, E., 2005, The Astronomical Journal, 130, 1286

Jenniskens, P., 2006, Meteor Showers and their Parent Comets, by Peter Jenniskens, pp. . ISBN 0521853494. Cambridge, UK: Cambridge University Press, 2006.

Jenniskens, P. & Vaubaillon, J., 2007, The Astronomical Journal, 134, 1037

Jenniskens, P., Duckworth, H. & Grigsby, B., 2012, WGN, Journal of the International Meteor Organization, 40, 98

Jetsu, L., Porceddu, S., Lyytinen, J., Kajatkari, P., Lehtinen, J., Markkanen, T. & Toivari-Viitala, J., 2013, The Astrophysical Journal, 733, 1

Jewitt, D. 2009, The Astronomical Journal, 137, 4296

Jewitt, D., Luu, J. & Trujillo, C., 1998, The Astronomical Journal, 115, 2125

Jewitt, D. C., Trujillo, C. A. & Luu, J. X., 2000, The Astronomical Journal, 120, 1140

Jewitt, D. & Sheppard, S., 2005, Space Science Reviews, 116, 441

Jewitt, D. & Haghighipour, N., 2007, Annual Reviews of Astronomy and Astrophysics, 45, 261

Jewitt, D., 2009, The Astronomical Journal, 137, 4296

Jewitt, D., Yang, B. & Haghighipour, N., 2009, The Astronomical Journal, 137, 4313

Jewitt, D. & Li, J., 2010, The Astronomical Journal, 140, 1519

Jewitt, D., Weaver, H., Mutchler, M., Larson, S. & Agarwal, J., 2011, The Astrophysical Journal Letters, 733, L4

Jewitt, D., 2012, The Astronomical Journal, 143, 66

Jewitt, D., Agarwal, J., Weaver, H., Mutchler, M., & Larson, S., 2013, The Astrophysical Journal Letters, 778, L21

Jewitt, D., Luu, J., Rajagopal, J., Kotulla, R., Ridgway, S., Liu, W. & Augusteijn, T., 2017, The Astrophysical Journal, 850, 36

Jewitt, D. & Luu, J., 2019, The Astrophysical Journal, 886, 29

Johns, D., Marti, C., Huff, M., McCann, J., Wittenmyer, R. A., Horner, J. & Wright, D. J., 2018, The Astrophysical Journal Supplement Series, 239, 14

Johnson, B. C. et al., 2012, The Astrophysical Journal, 761, 45

Johnson, J. A., et al., 2006, The Astrophysical Journal, 652, 1724

Johnson, J. A., Aller, K. M., Howard, A. W. & Crepp, J. R., 2010, Publications of the Astronomical Society of the Pacific, 122, 905

Johnson, J. A. et al., 2011, The Astronomical Journal, 141, 16

Johnson, C. L. et al., 2020, Nature Geoscience, 13, 199

Jolliff, B. L., Shearer, C. K., Cohen, B. A., 2012, Annual Meeting of the Lunar Exploration Analysis Group, 3047

Jones, R. H., Lee, T., Connolly, Jr. H. C., Love, S. G. & Shang, H., 2000, Protostars and Planets IV, 927

Jones, B. W., Underwood, D. R. & Sleep, P. N., 2005, The Astrophysical Journal, 622, 1091

Jones, B. W., Sleep, P. N. & Underwood, D. R., 2006, International Journal of Astrobiology, 5, 251

Jones, G. H. et al., 2018, Space Science Reviews, 214, 20

Jourdain de Muizon, M. et al., 1999, Astronomy & Astrophysics, 350, 875

Kaib, N. A. & Chambers, J. E., 2016, Monthly Notices of the Royal Astronomical Society, 455, 3561

Kains, N., Wyatt, M. C. & Greaves, J. S., 2011, Monthly Notices of the Royal Astronomical Society, 414, 2486

Kanani, S. J. et al., 2010, Journal of Geophysical Research A, 115, 6207

Kane, S. R. et al., 2004, Monthly Notices of the Royal Astronomical Society, 353, 689

Kane, S. R. et al., 2005a, Monthly Notices of the Royal Astronomical Society, 362, 117

Kane, S. R. et al., 2005b, Monthly Notices of the Royal Astronomical Society, 364, 1049

Kane, S. R. & von Braun, K., 2008, The Astrophysical Journal, 689, 492

Kane, S. R., 2011, Icarus, 214, 327

Kane, S. R., & Gelino, D.M., 2012, Publications of the Astronomical Society of the Pacific, 124, 323

Kane, S. R., Barclay, T. & Gelino, D. M., 2013, The Astrophysical Journal, 770, L20

Kane, S. R., Kopparapu, R. K., Domagal-Goldman, S. D., 2014, The Astrophysical Journal, 794, L5

Kane, S. R. et al., 2016, The Astrophysical Journal, 830, 1

Kane, S. R., Ceja, A. Y., Way, M. J. & Quintana, E. V., 2018, The Astrophysical Journal, 869, 46

Kane, S. R., Meshkat, T. & Turnbull, M. C., 2018, The Astronomical Journal, 156, 267

Kane, S. R. et al., 2019a, The Astronomical Journal, 157, 252

Kane, S. R. et al., 2019b, Journal of Geophysical Research: Planets, 124, 2015

Karkoschka, E., 2001, Icarus, 151, 69

Kasting, J. F., Whitmire, D. P. & Reynolds, R. T., 1993, Icarus, 101, 108

Kavelaars, J. J., Mousis, O., Petit, J.-M. & Weaver, H. A., 2011, The Astrophysical Journal Letters, 734, 30

Kawashima, Y. & Rugheimer, S., 2019, The Astronomical Journal, 157, 213

Kegerreis, J. A. et al., 2018, The Astrophysical Journal, 861, 52

Keldysh, M. V., 1977, Icarus, 30, 605

Kempton, E. M.-R., Perna, R. & Heng, K., 2014, The Astrophysical Journal, 795, 24

Kempton, E. M.-R et al., 2018, Publications of the Astronomical Society of the Pacific, 130, 114401

Kennedy, G. M. & Wyatt, M. C., 2013, Monthly Notices of the Royal Astronomical Society, 433, 2334

Kennedy, G. M. et al., 2015, The Astrophysical Journal Supplement Series, 216, 23

Kenyon, S. J. & Bromley, B. C., 2001, The Astronomical Journal, 121, 538

Kenyon, S. J. & Bromley, B. C., 2004, Nature, 432, 598

Kenyon, S. J. & Bromley, B. C., 2008, The Astrophysical Journal Supplement Series, 179, 451

Kenyon, S. J. & Bromley, B. C., 2016, The Astrophysical Journal, 817, 51

Khalafinejad, S. et al., 2017, Astronomy & Astrophysics, 598, 131

Kiefer, F. et al., 2014, Nature, 514, 462

Killen, R. M. & Hahn, J. M., 2015, Icarus, 250, 230

Kim, M. et al., 2018, Astronomy & Astrophysics, 618, 38

Kipping, D. M., Fossey, S. J., & Campanella, G. 2009, MNRAS, 400, 398

Kipping, D. M. et al., 2012, The Astrophysical Journal, 750, 115

Kipping, D. M. et al., 2013, The Astrophysical Journal, 770, 101

Kipping, D. M. et al., 2014, The Astrophysical Journal, 795, 25

Kipping, D. M. et al., 2016, The Astrophysical Journal, 820, 112

Kita, N. T., Huss, G. R., Tachibana, S., Amelin, Y., Nyquist, L. E. & Hutcheon, I. D., 2005, in Krot, A. N., Scott, E. R. D. & Reipurth, B. eds. Chondrites and the protoplanetary disk. Astronomical Society of the Pacific Conference Series 341, 558

Klein, H. P. et al., 1976, Science, 194, 99

Klein, H. P., 1977, Journal of Geophysical Research, 82, 4677

Klein, H. P., 1978, Icarus, 34, 666

Kleine, T., Münker, C., Mezger, K. & Palme, H., 2002, Nature, 418, 952

Kleine, T., Mezger, K., Palme, H., Scherer, E. & Münker, C, 2005, Geochimica et Cosmochimica Acta, 69, 5805

Knutson, H. A. et al., 2007, Nature, 477, 183

Knutson, H. A., Benneke, B., Deming, D. & Homeier, D., 2014, Nature, 505, 66

Kobayashi, H. & Ida, S., 2001, Icarus, 153, 416

Kobayashi, H., Ida, S. & Tanaka, H., 2005, Icarus, 177, 246

Kobayashi, H. et al., 2011, Earth, Planets, and Space, 63, 1067

Kobayashi, H. & Dauphas, N., 2013, Icarus, 225, 122

Koch, F. E. & Hansen, B. M. S., 2011, Monthly Notices of the Royal Astronomical Society, 416, 1274

Komabayashi, M.,1967, Journal of the Meteorological Society of Japan, 45, 137

Kong, D., Zhang, K. Schubert, G., & Anderson, J. D., 2018, Proceedings of the National Academy of Sciences, 115, 8499

Königl, A. & Salmeron, R., 2011, in Physical processes in circumstellar disks around young stars, 283

Kopparapu, R. K. et al., 2013, The Astrophysical Journal, 765, 131

Kopparapu, R. K., Ramirez, R. M., SchottelKotte, J., Kasting, J. F., Domagal-Goldman, S. & Eymet, V., 2014, The Astrophysical Journal, 787, 29

Kopparapu, R. K. et al., 2016, The Astrophysical Journal, 819, 84

Kopparapu, R. K. et al., 2018, The Astrophysical Journal, 856, 122

Koschny, D. et al., 2019, Space Science Reviews, 215, 34

Koskinen, T. T. & Guerlet, S., 2018, Icarus, 307, 161

Kral, Q., Matrà, L., Wyatt, M. C. & Kennedy, G. M., 2017, Monthly Notices of the Royal Astronomical Society, 469, 521

Krasnopolsky, V.A., Mumma, M.J., Randall Gladstone, G., 1998, Science, 280, 1576

Krauss, O. & Wurm, G., 2005, The Astrophysical Journal, 630, 1088

Krauss, O., Wurm, G. Mousis, O., Petit, J.-M., Horner, J. & Alibert, Y., 2007, Astronomy & Astrophysics, 462, 3, 977

Kreidberg, L., Luger, R., Bedell, M. 2019, ApJ, 877, 2

Kresák, L., 1976, Bulletin of the Astronomical Institute of Czechoslovakia, 27, 35

Kresák, L., 1978, Bulletin of the Astronomical Institute of Czechoslovakia, 29, 129

Kreutz, H. C. F., 1888, "Untersuchungen uber das comentesystem 1843 I, 1880 I und 1882 II.", Kiel, Druck von C. Schaidt, C. F. Mohr nachfl., 1888.

Kring, D. A., & Cohen, B. A., 2002, Journal of Geophysical Research, 107, E2, 5009

Krivov, A. V. et al., 2006, Astronomy & Astrophysics, 455, 509

Krivov, A. V., 2010, Research in Astronomy & Astrophysics, 10, 383

Krivov, A. V. & Booth, M., 2018, Monthly Notices of the Royal Astronomical Society, 479, 3300

Kronk, G. W., 1999, Cometography, by Gary W. Kronk, pp. 579. ISBN 052158504X. Cambridge, UK: Cambridge University Press, September 1999

Krot, A. N., Keil, K., Scott, E. R. D., Goodrich, C. A. & Weisberg, M. K., 2007, In Treatise on Geochemistry, edited by Holland H. D. and Turekian K. K. Oxford: Pergamon. pp. 1–52.

Krot, A. N. et al., 2009, Geochimica et Cosmochimica Acta, 73, 4963

Krot, A. N. et al., 2015, in Sources of Water and Aqueous Activity on the Chondrite Parent Asteroids, ed. P. Michel, F. E. DeMeo, & W. F. Bottke, 635

Kuang, W., & Bloxham, J., 1997, Nature, 389, 371

Kuchner, M. J. & Lecar, M., 2002, The Astrophysical Journal, 574, 87

Kuchner, M. J. & Stark, C. C., 2010, The Astronomical Journal, 140, 1007

Kuchner, M., 2012, Astrophysics Source Code Library, record ascl: 1202.002

Kuiper, G. P., 1951a, Proceedings of the National Academy of Science, 37, 717

Kuiper, G. P., 1951b, in Astrophysics: A Topical Symposium, ed. J. A. Hynek, New York: McGraw-Hill.

Kulikov, Y. N. et al., 2006, Planetary and Space Science, 54, 1425

Kurosawa, K., 2015, Earth and Planetary Science Letters, 429, 181

Laakso, T., Rantala, J. & Kaasalainen, M., 2006, Astronomy and Astrophysics, 456, 373

Lacy, B., Shlivko, D. & Burrows, A., 2019, The Astronomical Journal, 157, 132

Lagrange, A.–M., Vidal-Madjar, A., Deluil, M., Emerich, C., Beust, H. & Ferlet, R., 1995, Astronomy & Astrophysics 296, 499

Lam, K. W. F. et al., 2020, The Astronomical Journal, 159, 120

Lammer, H., Lichtenegger, H. I. M., Kolb, C., Ribas, I., Guinan, E. F., Abart, R. & Bauer, S. J., 2003, Icarus, 165, 9

Lambrechts, M. & Johansen, A., 2012. Astronomy and Astrophysics, 544, A32

Langlais, B., Purucker, M. E. & Mandea, M., 2004, Journal of Geophysical Research, 109, 2008

Laplace, P. S., 1796, *Exposition du Système du Monde*. Engl. Transl. by H. H. Harte 1830, *The System of the World*. Dublin: Dublin Univ. Press

Latham, D. W., Stefanik, R. P., Mazeh, T., Mayor, M. & Burki, G., 1989, Nature, 339, 38

Lattimer, J. M., Schramm, D. N. & Grossman, L., 1978, The Astrophysical Journal, 219, 230

Lauretta, D. S. et al., 2017, Space Science Reviews, 212, 925

Le Sergeant d'Hendecourt, L. B. & Lamy, Ph. L., 1980, Icarus, 43, 350

Le Verrier, U. J., 1846, Astronomische Nachrichten, 25, 65

Lécuyer, C., Simon, L., Guy, F., 2000, Earth and Planetary Science Letters 181, 33

Lee, J. W., Kim, S.-L., Kim, C.-H., Koch, R. H., Lee, C.-U., Kim, H.-I. & Park, J.-H., 2009, The Astronomical Journal, 137, 3181

Léger, A et al., 2009, Astronomy & Astrophysics, 506, 287

Lehmann, I., 1936, P'[P-prime], 1936. Bureau Central Seismologique International, Series A, Travaux Scientifiques, 14, 87

Leinert, C., 1975, Space Science Reviews, 18, 281

Leinert, C., Richter, I., Pitz, E. & Planck, B., 1981, Astronomy & Astrophysics, 103, 177

Leinhardt, Z. M., Marcus, R. A. & Stewart, S. T., 2010, The Astrophysical Journal, 714, 1789

Leinhardt, Z. M. & Stewart, S. T., 2012, The Astrophysical Journal, 745, 79

Leitch, E. M. & Vasisht, G., 1998, New Astronomy, 3, 51

Leleu, A., Jutzi, M. & Rubin, M., 2018, Nature Astronomy, 2, 555

Levin, G. V. & Straat, P. A., 1976, Science, 194, 1322

Levison, H. F. & Duncan, M. J., 1994, Icarus, 108, 18

Levison, H. F. & Duncan, M. J., 1997, Icarus, 127, 13

Levison, H. F., Shoemaker, E. M. & Shoemaker, C. S., 1997, Nature, 385, 42

Levison, H. F. & Stern, S. A., 2001, The Astronomical Journal, 121, 1730

Levison, H. F., Dones, L. & Duncan, M. J., 2001, The Astronomical Journal, 121, 2253

Levison, H. F., Morbidelli, A., 2003, Nature, 426, 419

Levison, H. F., Morbidelli, A. & Dones, L., 2004, The Astronomical Journal, 128, 2553

Levison, H. F., Duncan, M. J., Dones, L. & Gladman, 2006, Icarus, 184, 619

Levison, H. F., Morbidelli, A., Gomes, R., & Backman, D., 2007, Planetary migration in planetesimal disks, in Protostars and Planets V Compendium, edited by Reipurth, B., Jewitt, D., & Keil, K., 669, University of Arizona Press, Tucson.

Levison, H. F., Morbidelli, A., Vanlaerhoven, C., Gomes, $. & Tsiganis, K., 2008, Icarus, 196, 258

Levison, H. F., Bottke, W. F., Gounelle, M., Morbidelli, A., Nesvorný, D. & Tsiganis, K., 2009, Nature, 460, 364

Levison, H. F., Duncan, M. J., Brasser, R. & Kaufmann, D. E., 2010, Science, 329, 187

Levison, H., 2012, AAS/Division of Dynamical Astronomy Meeting, 43, #02.01

Levison, H. F., Kretke, K. A., Walsh, K. J. & Bottke, W. F., 2015, Proceedings of the National Academy of Sciences, 112, 14180

Levison, H. F., Kretke, K. A., & Duncan, M. J., 2015, Nature, 524, 322

Lewis, J. S., 1972, Earth and Planetary Science Letters, 15, 286

Lewis, K. M., Sackett, P. D., & Mardling, R.A., 2008, The Astrophysical Journal, 685, 2

Lewis, A. R., Quinn, T. & Kaib, N. A., 2013, The Astronomical Journal, 146, 16

Li, J. & Jewitt, D., 2013, The Astronomical Journal, 145, 154

Li, M. & Xiao, L., 2016, The Astrophysical Journal, 820, 36

Lichtenberg, T., Golabek, G. J., Dullemond, C. P., Schönbächler, M., Gerya, T. V. & Meyer, M. R., 2018, Icarus 302, 27

Liseau, R. et al., 2008, Astronomy & Astrophysics, 480, 47

Liseau, R. et al., 2010, Astronomy & Astrophysics, 518, 132

Lissauer, J. J., 1987, Icarus, 69, 249

Lissauer, J. J., 1993, Annual Review of Astronomy and Astrophysics, 31, 129

Lissauer, J. J., et al., 2011, Nature, 470, 53

Lisse, C. M., Chen, C. H., Wyatt, M. C., Morlok, A., Song, I., Bryden, G. & Sheehan, P., 2009, The Astrophysical Journal, 701, 2019

Lisse, C. M. et al., 2013, Icarus, 222, 752

Lithwick, Y., Xie, J. & Wu, Y., 2012, The Astrophysical Journal, 761, 122

Liu, H.-S. & O'Keefe, J. A., 1965, Science, 150, 1717

Liu, M. C., Deacon, N. R., Magnier, E. A., et al., 2011, The Astrophysical Journal, 740, 32

Liu, S.-F., Hori, Y., Müller, S., Zheng, X., Helled, R., Lin, D. & Isella, A., 2019, Nature, 572, 355

Lodders, K., Palme, H. & Gail, H.-P., 2009, Landolt-Börnstein, 4B, 712

Lognonné, P. et al., 2019, Space Science Reviews, 215, 12

Löhne, T. et al., 2012, Astronomy & Astrophysics, 537, 110

Lopez, E. D. & Fortney, J. J., 2014, The Astrophysical Journal, 792, 1

López-Puertas, M., et al., 2018, Astronomical Journal, 156, 169

Lowell, P., 1895, Popular Astronomy, 2, 255

Lowell, P., 1908, Mars as the Abode of Life, by Percival Lowell; 8vo, illus., xix + 287 pp.; New York: The Macmillan Company (1908)

Lowry, S., Fitzsimmons, A., Lamy, P. & Weissman, P., 2008, The Solar System Beyond Neptune, editors: M. A. Marucci, H. Boehnhardt, D. P. Cruikshank & A. Morbidelli, University of Arizona Press, Tucson, 592, 397

Lozovsky, M., Helled, R., Rosenberg, E. D. & Bodenheimer, P., 2017, The Astrophysical Journal, 836, 227

LSST Science Collaboration et al., 2009, "LSST Science Book, Version 2.0", arXiv:0912.0201, available at full resolution at http://www.lsst.org/lsst/scibook

Luhman, K. L., Burgasser, A. J. & Bochanski, J. J., 2011, The Astrophysical Journal, Letters, 730, L9

Luhman, K. L., 2014, The Astrophysical Journal, 781, 4

Luu, J. X. & Jewitt, D. C., 1990, The Astronomical Journal, 100, 913

Lykawka, P. S. & Horner, J. 2010, Monthly Notices of the Royal Astronomical Society, 405, 1375

Lykawka, P. S. & Mukai, T., 2007, Icarus, 189, 213

Lykawka, P. S. & Mukai, T., 2008, The Astronomical Journal, 135, 1161

Lykawka, P. S., Horner, J., Jones, B. W. & Mukai, T., 2009, Monthly Notices of the Royal Astronomical Society, 398, 1715

Lykawka, P. S., Horner, J., Jones, B. W. & Mukai, T., 2010, Monthly Notices of the Royal Astronomical Society, 404, 1272

Lykawka, P. S., Horner, J., Nakamura, A. M. & Mukai, T., 2012, Monthly Notices of the Royal Astronomical Society, 421, 1331

Lykawka, P. S., 2012, MEEP, Vol. 1, No. 3, 121

Lykawka, P. S. & Ito, T., 2019, The Astrophysical Journal, 883, 130

Lyons, T. W. & Reinhard, C.T., 2009, Nature, 461, 179

Lyttleton, R. A., 1936, Monthly Notices of the Royal Astronomical Society, 96, 559

Lyttleton, R. A., 1936, Monthly Notices of the Royal Astronomical Society, 97, 108

MacDonald, M. G. et al., 2016, The Astronomical Journal, 152, 105

MacGregor, M. A. et al., 2017, The Astrophysical Journal, 852, 8

MacGregor, M. A. et al., 2018, The Astrophysical Journal, 869, 75

MacGregor, M. A. et al., 2019, The Astrophysical Journal, 877, 32

Madiedo, J. M., Ortiz, J. L., Morales, N. & Cabrera-Caño, J., 2014, Monthly Notices of the Royal Astronomical Society, 439, 2364

Madiedo, J. M., Ortiz, J. L., Morales, N. & Santos-Sanz, P., 2019, Monthly Notices of the Royal Astronomical Society, 486, 3380

Malhotra, R., 1993, Nature, 365, 819

Malhotra, R., 1995, The Astronomical Journal, 110, 420

Mann, C. R., Boley, A. C. & Morris, M. A., 2016, The Astrophysical Journal, 818, 103

Marcialis, R. & Greenberg, R., 1987, Nature, 328, 227

Marchi, S. et al., 2012, Science, 336, 690

Marcus, R. A., Ragozzine, D. R., Murray-Clay A., Holman, M. J., 2011, The Astrophysical Journal, 733, 40

Marcy, G., Isaacson, H., Howard, A. W., et al., 2014, The Astrophysical Journal Supplement, 210, 20

Marino, S. et al., 2016, Monthly Notices of the Royal Astronomical Society, 460, 2933

Marino, S. et al., 2017, Monthly Notices of the Royal Astronomical Society, 465, 2595

Marino, S. et al., 2018, Monthly Notices of the Royal Astronomical Society, 479, 5423

Marino, S. et al., 2019, Monthly Notices of the Royal Astronomical Society, 484, 1257

Marinova, M. M., Aharonson, O. & Asphaug, E., 2008, Nature, 453, 1216

Margot, J.-L., 2015, The Astrophysical Journal, 150, 185

Marois, C. et al., 2008, Science, 322, 1348

Marois, C., Zuckerman, B, Konopacky, Q. M., Macintosh, B. & Barman, T., 2010, Nature, 468, 1080

Marsden, B. G., 1967, The Astronomical Journal, 72, 1170

Marsden, B. G., 1989, The Astronomical Journal, 98, 2306

Marshall, J. P., Horner, J. & Carter, A., 2010, International Journal of Astrobiology, 9, 259

Marshall, J. P. et al., 2011, Astronomy & Astrophysics, 529, 117

Marshall, J. P. et al., 2014a, Astronomy & Astrophysics, 565, 15

Marshall, J. P. et al., 2014b, Astronomy & Astrophysics, 570, 114

Marshall, J. P. et al., 2018, The Astrophysical Journal, 869, 10

Marshall, J. P. et al., 2020, Monthly Notices of the Royal Astronomical Society, *available through advance access,* doi:10.1093/mnras/staa847

Marsh, T. R. et al., 2014, Monthly Notices of the Royal Astronomical Society, 437, 475

Martin, R. & Livio, M., 2016, The Astrophysical Journal, 822, 90

Marov, M. Y., 1978, Annual Review of Astronomy & Astrophysics, 16, 141

Marty, B., 2012, Earth and Planetary Science Letters, 313, 56

Marzari, F., Davis, D. & Vanzani, V., 1995, Icarus, 113, 168

Marzari, F., Farinella, P. & Vanzani, V., 1995, Astronomy & Astrophysics, 299, 267

Marzari, F., Dotto, E., Davis, D. R., Weidenschilling, S. J. & Vanzani, V., 1998, Astronomy & Astrophysics, 333, 1082

Masset, F. & Snellgrove, M., 2001, Monthly Notices of the Royal Astronomical Society, 320, L55

Masset, F. S. & Paploizou, J. C. B., 2003, The Astrophysical Journal, 588, 494

Masuda, K., 2014, The Astrophysical Journal, 783, 53

Matese, J. J., Whitman, P. G. & Whitmire, D. P., 1999, Icarus, 141, 354

Matese, J. J. & Lissauer, J. J., 2004, Icarus, 170, 508

Matese, J. J. & Whitmire, D. P., 2011, Icarus, 211, 926

Matousek, S., 2007, Acta Astronautica, 61, 932

Matrà, L. et al., 2017, Monthly Notices of the Royal Astronomical Society, 464, 1415

Matrà, L., Marino, S., Kennedy, G. M., Wyatt, M. C., Öberg, K. I. & Wilner, D. J., 2018, The Astrophysical Journal, 859, 72

Matrà, L. et al., 2019, The Astronomical Journal, 157, 135

Matthews, B. C. et al., 2010, Astronomy & Astrophysics, 518, 135

Matthews, B. C. et al., 2014, The Astrophysical Journal, 780, 97

Maunder, E. W., 1888, The Observatory, 11, 345

Mayer, C.H., McCullough, T.P. & Sloanaker, R.M.,1958, The Astrophysical Journal, 127, 1

Mayor, M. & Queloz, D. 1995, Nature, 378, 355

Mayor, M., Udry, S., Lovis, C., Pepe, F., Queloz, D., Benz, W., Bertaux, J.-L., Bouchy, F., Mordasini, C., Segransan, D., 2009, Astronomy & Astrophysics, 493, 639

Mayorga, L.C., et al., 2016, Astronomical Journal, 152, 209

McCauley, J. F. et al., 1972, Icarus, 17, 289

McCue, J. & Dormand, J. R., 1993, Earth Moon and Planets, 63, 209

McGill, G. E. & Squyres, S. W., 1991, Icarus, 93, 386

McGovern, W. E., 1965, Nature, 208, 375

McIntosh, B. A., 1990, Icarus, 86, 299

McKay, A. J., Cochran, A. L., Dello Russo, N. & DiSanti, M. A., 2020, The Astrophysical Journal, 889, 10

McKay, C. P., 1991, Icarus, 91, 93

McKay, C. P., 2014, Proceedings of the National Academy of Sciences, 111, 12628

McKinnon, W. B., 1984, Nature, 311, 355

McKinnon, W. B., 1988, Nature, 333, 701

McKinnon, W. B., 1989, The Astrophysical Journal, Letters, 344, L41

McKinnon, W. B., Zahnle, K. J., Ivanov, B. A. & Melosh, H. J., 1997, in Bouger, S.W., et al., eds., Venus II: Tucson, Arizona, University of Arizona Press, p. 969-1014

McKinnon, W. B., 2008, American Astronomical Society, DPS meeting #40, id.38.03; Bulletin of the American Astronomical Society, Vol. 40, p.464

McKinnon, W. B. et al., 2017, Icarus, 287, 2

McSween, H., Mittlefehldt, D., Beck, A., Mayne, R., McCoy, T., 2011, Space Science Reviews, 163, 141

Meech, K. J. & Belton, M. J. S., 1990, The Astronomical Journal, 100, 1323

Meech, K. J., Weryk, R., Micheli, M. et al., 2017, Nature, 552, 378

Meier, R., Owen, T. C., & Matthews, H. E. et al., 1998, Science 279, 842

Meisner, A. M., Bromley, B. C., Kenyon, S. J. & Anderson, T. E., 2018, The Astronomical Journal, 155, 166

Melis, C. et al., 2010, The Astrophysical Journal, 717, 57

Melis, C., Zuckerman, B., Rhee, J. H., Song, I., Murphy, S. J. & Bessell, M. S., 2012, Nature, 487, 74

Melita, M. D., Larwood, J., Collander-Brown, S., Fitzsimmons, A., Williams, I. P. & Brunini, A., 2002, In: Proceedings of Asteroids, Comets, Meteors - ACM 2002. International Conference, 29 July - 2 August 2002, Berlin, Germany. Ed. Barbara Warmbein. ESA SP-500. Noordwijk, Netherlands: ESA Publications Division, ISBN 92-9092-810-7, 2002, p. 305 - 308

Melita, M. D., Larwood, J. D. & Williams, I. P., 2005, Icarus, 173, 559

Melosh, H. J., 1977, Icarus, 31, 221

Meng, H. Y. A. et al., 2012, The Astrophysical Journal, 751, 17

Meng, H. Y. A. et al., 2014, Science, 345, 1032

Meng, H. Y. A. et al., 2015, The Astrophysical Journal, 805, 77

Mennesson, B. et al., 2014, The Astrophysical Journal, 797, 119

Meshkat, T. et al., 2017, The Astronomical Journal, 154, 245

Meunier, N. et al., 2015, Astronomy & Astrophysics, 583, 118

Michels, D. J., Sheeley, N. R., Howard, R. A. & Koomen, M. J., 1982, Science, 215, 1097

Milani, A., Cellino, A., Knežević, Z., Novaković, B., Spoto, F. & Paolicchi, P., 2014, Icarus, 239, 46

Militzer, B., Soubiran, F., Wahl, S. M. & Hubbard, W., 2016, Journal of Geophysical Research: Planets, 121, 1552

Millar-Blanchaer, M. A. et al., 2016, The Astronomical Journal, 152, 128

Miner, E. D., Wessen, R. R., & Cuzzi, J. N., 2007, The discovery of the Jupiter ring system, in Planetary Ring Systems (Springer Praxis Books: Chichester) 49-59

Minton, D. A. & Malhotra, R., 2009, Nature, 457, 1109

Minton, D. A. & Malhotra, R., 2011, The Astrophysical Journal, 732, 53

Mitchell, D. G., Carbary, J. F., Cowley, S. W. H., Hill, T. W., & Zarka, P., 2009, The Dynamics of Saturns' Magnetosphere, in Saturn from Cassini-Huygens (Springer Science+Business Media: Dordrecht), 257-280

Mizuno, H., 1980, Progress of Theoretical Physics, 64, 544

Mojzsis, S. J., Brasser, R., Kelly, N. M., Abramov, O. & Werner, S. C., 2019, The Astrophysical Journal, 881, 44

Molter, E. et al., 2019, Icarus, 321, 324

Montalto, M., Riffeser, A., Hopp, U., Wilke, S. & Carraro, G., 2008, Astronomy & Astrophysics, 479, 45

Montañés-Rodríguez, P., González-Merino, B., Pallé, E., López-Puertas, M., & García-Melendo, E., 2015, The Astrophysical Journal Letters, 801, L8

Montesinos, B. et al., 2016, Astronomy & Astrophysics, 593, 51

Moór, A. et al., 2015, The Astrophysical Journal, 814, 42

Moór, A. et al., 2017, The Astrophysical Journal, 849, 123

Moór, A. et al., 2017, The Astrophysical Journal, 884, 108

Morbidelli, A., 1999, Celestial Mechanics and Dynamical Astronomy, 73, 39

Morbidelli, A., Chambers, J., Lunine, J. I., Petit, J. M., Robert, F., Valsecchi, G. B. & Cyr, K. E., 2000, Meteoritics and Planetary Science, 35, 1309

Morbidelli, A., Bottke, W. F., Jr., Froeschlé, C. & Michel, P., 2002, Asteroids III, 409

Morbidelli, A., Jedicke, R., Bottke, W. F., Michel, P. & Tedesco, E. F. 2002, Icarus, 158, 329

Morbidelli, A. & Vokrouhlický, D., 2003, Icarus, 163, 120

Morbidelli, A. & Levison, H. F., 2004, The Astronomical Journal, 128, 2564

Morbidelli, A., Levison, H. F., Tsiganis, K., & Gomes, R. 2005, Nature, 435, 462

Morbidelli, A. & Crida, A., 2007, Icarus, 191, 158

Morbidelli, A., Bottke, W. F., Nesvorný, D. & Levison, H. F., 2009, Icarus, 204, 558

Morbidelli, A., Brasser, R., Gomes, R., Levison, H. F. & Tsiganis, K., 2010, The Astronomical Journal, 140, 1391

Morbidelli, A., Lunine, J. I., O'Brien, D. P., Raymond, S. N., & Walsh, K. J., 2012, AREPS, 40, 251

Mordasini, C., Alibert, Y., Benz, W. & Naef, D. 2009, Astronomy & Astrophysics, 501, 1161

Morgan, J., Warner, M., the Chicxulub Working Group, et al., 1997, Nature, 390, 472

Moro-Martín, A. & Malhotra, R., 2003, The Astrophysical Journal, 125, 2255

Moro-Martín, A., 2013, *Dusty Planetary Systems,* in Planets, Stars and Stellar Systems, by Oswalt, Terry D.; French, Linda M.; Kalas, Paul, ISBN 978-94-007-5605-2. Springer Science+Business Media Dordrecht, 2013, 431

Moro-Martín, A. et al., 2015, The Astrophysical Journal, 801, 143

Morota, T. et al., 2009, Meteoritics and Planetary Science, 44, 1115

Morris, M. A. & Desch, S. J., 2010, The Astrophysical Journal, 722, 1474

Morton, T. D., Bryson, S. T., Coughlin, J. L., Rowe, J. F., Ravichandran, G., Petigura, E. A., Haas, M. R. & Batalha, N. M., 2016, The Astrophysical Journal, 822, 86

Mostefaoui, S., Lugmair, G. W. & Hoppe, P., 2005, The Astrophysical Journal, 625, 271

Mosqueira, I. & Estrada, P. R., 2003a, Icarus, 163, 198

Mosqueira, I. & Estrada, P. R., 2003b, Icarus, 163, 232

Mosqueira, I., Estrada, P. & Turrini, D., 2010, Space Science Reviews, 153, 431

Moudens, A., Mousis, O., Petit, J.-M., Wurm, G., Cordier, D. & Charnoz, S., 2011, Astronomy & Astrophysics, 531A, 106

Mouginis-Mark, 2018, Chemie der Erde, 78, 397

Mousis, O., 2004, Astronomy and Astrophysics, 413, 373

Mousis, O., Petit, J.-M., Wurm. G., Krauss, O., Alibert, Y. & Horner, J., 2007, Astronomy & Astrophysics, 466, 9

Mousis, O., Petit, J.-M., Wurm, G., Krauss, O., Alibert, Y. & Horner, J., 2007, Proceedings of the Annual meeting of the French Society of Astronomy and Astrophysics, held in Grenoble, France, July 2-6 2007, Eds. J. Bouvier, A. Chalabaev & C. Charbonnel, p416

Movshovitz, N., Nimmo, F., Korycansky, D. G., Asphaug, E. & Owen, J. M., 2015, Geophysical Research Letters, 42, 256

Movshovitz, N. et al., 2020, The Astrophysical Journal, 891, 109

Muirhead, P. S. et al., 2015, The Astrophysical Journal, 801, 18

Mullally, F. et al., 2015, The Astrophysical Journal Supplement Series, 217, 31

Mura, A. et al., 2019, Journal of Geophysical Research Planets, 44, 5308

Murray, B. C. et al., 1974, Science, 185, 169

Murphy, S. J., Mamajek, E. E. & Bell, C. P. M., 2018, Monthly Notices of the Royal Astronomical Society, 476, 3290

Murray, J. B., 1999, Monthly Notices of the Royal Astronomical Society, 309, 31

Murray-Clay, R. A., & Chiang, E. I., 2005, The Astrophysical Journal, 619, 623

Mustill, A. J. & Wyatt, M. C., 2009, Monthly Notices of the Royal Astronomical Society, 399, 1403

Mustill, A. J., Marshall, J. P., Villaver, E., Veras, D., Davis, P. J., Horner, J. & Wittenmyer, R. A., 2013, Monthly Notices of the Royal Astronomical Society, 436, 2515

Muzerolle, J., Calvet, N., Hartmann, L. & D'Alessio, P., 2003, The Astrophysical Journal Letters 597, L149

Naef, D. et al., 2001, Astronomy & Astrophysics, 375, L27

Naef, D., Mayor, M., Beuzit, J. L., Perrier, C., Queloz, D., Sivan, J. P. & Udry, S., 2004, Astronomy & Astrophysics, 414, 351

Nagasawa, M., Tanaka, H. & Ida, S., 2000, The Astronomical Journal, 119, 1480

Nakajima, S., Hayashi, Y. Y. & Abe. Y., 1992, The Journal of Atmospheric Science, 49, 2256

Nakamura T. et al., 2011, Science, 333, 1113

Napier, W. M., 2010, Monthly Notices of the Royal Astronomical Society, 405, 1901

Napier, W. M., 2015, Monthly Notices of the Royal Astronomical Society, 448, 27

Nasiroglu, I. et al., 2017, The Astronomical Journal, 153, 137

Navarrete, F. H., Schleicher, D. R. G., Zamponi Fuentealba, J. & Völschow, M., 2018, Astronomy & Astrophysics, 615, 81

Navarrete, F. H., Schleicher, D. R. G., Käpylä, P. J., Schober, J., Völschow, M., Mennickent, R. E., 2020, Monthly Notices of the Royal Astronomical Society, 491, 1043

Ness, N. F., Behannon, K. W., Lepping, R. P., Whang, Y. C. & Schatten, K. H., 1974, Science, 185, 151

Ness, N. F., Behannon, K. W., Lepping, R. P. & Whang, Y. C., 1975, Journal of Geophysical Research, 60, 2708

Nesvorný, D., Bottke, W. F., Jr., Dones, L. & Levison, H. F., 2002, Nature, 417, 720

Nesvorný, D., Alvarellos, J. L. A., Dones, L. & Levison, H. F., 2003, The Astronomical Journal, 126, 398

Nesvorný, D., Beaugé, C. & Dones, L., 2004, The Astronomical Journal, 127, 1768

Nesvorný, D., & Bottke, W. F., 2004, Icarus, 170, 324

Nesvorný, D., Vokrouhlický, D. & Morbidelli, A., 2007, The Astronomical Journal, 133, 1962

Nesvorný, D., & Vokrouhlicky, D., 2009, The Astronomical Journal, 137, 5003

Nesvorný, D., Vokrouhlický, D. & Morbidelli, A., 2013, The Astrophysical Journal, 768, 45

Nesvorný, D., Jenniskens, P., Levison, H. F., Bottke, W. F., Vokrouhlický, D. & Gounelle, M., 2010, The Astrophysical Journal, 713, 2, 816

Nesvorný, D., Morbidelli, A., 2012, The Astronomical Journal, 144, 117

Nesvorný, D., Broz, M., & Carruba, V. 2015. In P. Michel, F. E. DeMeo, & W. F. Bottke (Eds.), *Asteroids IV*. Tucson, AZ: University of Arizona Press.

Nesvorný, D., 2015, The Astronomical Journal, 150, 73

Nesvorný, D., & Vokrouhlický, D., 2016, The Astrophysical Journal, 825:94 (18pp)

Nesvorný, D. et al., 2018. Nature Astronomy. 2:11. 878-882

Nesvorný, D., 2018, Annual Review of Astronomy and Astrophysics, 56, 137

Nesvorný, D., Vokrouhlický, D., Dones, L., Levison, H. F., Kaib, N. & Morbidelli, A., 2017, The Astrophysical Journal, 845, 25

Nettelmann, N., 2011, Astrophysics and Space Science, 336, 47

Neukum, G., Ivanov, B.A., Hartmann, W.K., 2001, *Space Sci. Rev.,* 96, 55

Neves, V., Bonfils, X., Santos, N. C., Delfosse, X., Forveille, T., Allard, F. & Udry, S., 2013, Astronomy & Astrophysics, 551, 36

Nicholson et al., 2018, Monthly Notices of the Royal Astronomical Society, 480, 1754

Nielsen, L. D. et al., 2019, Astronomy & Astrophysics, 623, 100

Niemann, H. B. et al., 1998, Journal of Geophysical Research, 10322831

Nimmo, F. & McKenzie, D., 1998, Annual Review of Earth and Planetary Sciences, 26, 23

Nimmo, F., 2002, Geology, 30, 987

Nittler, L. R., 2003, Earth and Planetary Science Letters, 209, 259

Noll, K. S. et al., 1995, Science, 267, 1307

Norman, M.D., Duncan, R.A., Huard, J.J., 2010, *Geochim. Cosmochim. Acta*, 74, 763

Noyola, J. P., Satyal, S., Musielak, Z.E., 2014, The Astrophysical Journal, 791, 1

Null, G. W., Lau, E. L., Biller, E. D. & Anderson, J. D., 1981, The Astronomical Journal, 86, 456

O'Brien, D. P. & Greenberg, R., 2003, Icarus, 164, 334

O'Brien, D. P., Morbidelli, A. & Bottke, W. F., 2007, Icarus, 191, 434

O'Brien, D. P., Walsh, K. J., Morbidelli, A., Raymond, S. N. & Mandell, A. M., 2014, Icarus, 239, 74

O'Brien, D. P., Izidoro, A., Jacobson, S. A., Raymond, S. N. & Rubie, D. C., 2018, Space Science Reviews, 214, 47

O'Donoghue, J., Moore, L., Connerney, J., Melin, H., Stallard, T., Miller, S. & Baines, K. H., 2019, Icarus, 322, 251

Oberbeck, V.R. & Fogleman, G., 1989, Nature, 339, 434

O'Neill, C., Jellinek, A. M. & Lenardic, A., 2007, Earth and Planetary Science Letters, 261, 20

O'Neill, C., Marchi, S., Zhang, S. & Bottke, W., 2017, Nature Geoscience, 10, 793

O'Neill, C., Marchi, S., Bottke, W. & Fu, R., 2019, Geology, doi: https://doi.org/10.1130/G46533.1

Ohtsuka, K., Nakano, S. & Yoshikawa, M., 2003, Publications of the Astronomical Society of Japan, 55, 321

Oldham, R. D., 1906, Quarterly Journal of the Geological Society, 62, 456

Olech, A. et al., 2017, Monthly Notices of the Royal Astronomical Society, 469, 2077

Olofsson, J. et al., 2018, Astronomy and Astrophysics, 617, 109

Oort, J. H., 1950, Bulletin of the Astronomical Institutes of the Netherlands, 11, 91

Öpik, E., 1932, Proceedings of the American Academy of Arts and Sciences, 67, 169

Opitom, C. et al., 2019, Astronomy & Astrophysics, 631, 8

Oppenheimer, B. R. et al., 2013, The Astrophysical Journal, 768, 24

Orosei, R. et al., 2018, Science, 361, 490

Ortiz, J. L., Duffard, R., Pinilla-Alonso, N., et al., 2015, Astronomy and Astrophysics, 576, A18

Ortiz, J. L., Quesada, J. A., Aceituno, J., Aceituno, F. J. & Bellot Rubio, L. R., 2002, The Astrophysical Journal, 576, 567

Ortiz, J. L., Santos-Sanz, P., Sicardy, B., et al., 2017, Nature, 550, 219

Orton, G. S., Fletcher L. N., Lisse, C. M., et al., 2011, Icarus, 211, 587

Osborn, H. P. et al., 2016, Monthly Notices of the Royal Astronomical Society, 457, 2273

Owen, T. & Bar-Nun, A., 1995, Icarus 116, 215

Pál, A. et al., 2015, Astronomy & Astrophysics, 583, 93

Pallé, Enric, Zapatero Osorio, María Rosa, Barrena, Rafael, Montañés-Rodríguez, Pilar & Martín, Eduardo L., Nature, 459, 814

Pan, Margaret, & Wu, Yanqin 2016, The Astronomical Journal, 821, article id. 18

Panić, O. et al., 2013, Monthly Notices of the Royal Astronomical Society, 435, 1037

Paolicchi, P., Spoto, F., Knežević, Z., & Milani, A., 2019, Monthly Notices of the Royal Astronomical Society, 484, 1815

Parisi, M. G. & Brunini, A., 1997, Planetary & Space Science, 45, 181

Parker, A. H., 2015, Icarus, 247, 112

Parmentier, E. M. & Hess, P. C., 1992, Geophysical Research Letters, 19, 2015

Partridge, E. A. & Whitaker, H. C., 1896, Popular Astronomy, 3, 408

Pappalardo, R. T., Reynolds, S. J. & Greeley, R., 1997, Journal of Geophysics Research, 102, 13369

Pasquini, L., Döllinger, M. P., Weiss, A., Girardi, L., Chavero, C., Hatzes, A. P., da Silva, L. & Setiawan, J., 2007, Astronomy and Astrophysics, 473, 979

Pawellek, N., Moór, A., Pascucci, I., Krivov, A. V., 2019, Monthly Notices of the Royal Astronomical Society, 487, 5874

Pearl, J. & Conrath, B., 1991, Journal of Geophysical Research: Space Physics, 96, 18921

Peixinho, N., Thirouin, A., Tegler, S. C., Di Sisto, R. P., Delsanti, A., Guilbert-Lepoutre, A. & Bauer, J. G., 2020, *"From Centaurs to Comets – 40 years", in "The Transneptunian Solar System", Editors: Dina Prialnik, Maria Antonietta Barucci, and Leslie Young, Publisher: Elsevier, p 307-329*

Penny, M. T. et al., 2019, The Astrophysical Journal Supplement Series, 241, 3

Pepe, F. et al., 2000, Proceedings of SPIE, 4008, 582

Pepe, F., et al., 2014, Astronomische Nachrichten, 335, 8

Perrin, M. D. et al., 2015, The Astrophysical Journal, 799, 182

Perryman, M., Hartman, J., Bakos, G. Á. & Lindegren, L., 2014, The Astrophysical Journal, 797, 14

Perryman, M., 2018, *The Exoplanet Handbook, Second Edition, Cambridge University Press*

Petaev, M. I. & Wood, J. A., 2005, in Krot, A. N., Scott, E. R. D. & Reipurth, B. eds. Chondrites and the protoplanetary disk, Astronomical Society of the Pacific Conference Series 341, 373

Peterson, S. V., Dutton, A. & Lohmann, K. C., 2016, Nature Communications, 7, 12079

Petit, J. -M., Morbidelli, A. & Chambers, J., 2001, Icarus 153, 338

Petit, J. -M., Kavelaars, J. J., Gladman, B. J. et al., 2011, The Astronomical Journal, 143, 131

Pettengill, G. H. & Dyce, R. B., 1965, Nature, 206, 1240

Pettengill, G. H, Eliason, E., Ford, P. G., Loriot, G. B. & Masursky, H. G. E., 1980, Journal of Geophysical Research Space Physics 85, A13

Pfalzner, S. et al., 2015, Physica Scripta, 90, 068001

Pfalzner, S., Bhandare, A., Vincke, K. & Lacerda, P., 2018, The Astrophysical Journal, 863, 45

Pirani, S., Johansen, A., Bitsch, B., Mustill, A. J. & Turrini, D., 2019, Astronomy & Astrophysics, 623, 169

Plavchan, P., et al., 2015, ExoPAG SAG 8 final report, eprint arXiv:1503.01770

Pokorný, P., Sarantos, M. & Janches, D., 2017, The Astrophysical Journal Letters, 842, 17

Pollack, J. B., Hubickyj, O., Bodenheimer, P., Lissauer, J. J., Podolak, M. & Greenzweig, Y., 1996, Icarus, 124, 62

Porco, C. C. et al., 2005, Science, 307, 1226

Porubčan, V., Kornoš, L. & Williams, I. P., 2006, Contributions of the Astronomical Observatory Skalnaté Pleso, 36, 103

Potter, S. B., et al., 2011, Monthly Notices of the Royal Astronomical Society, 416, 2202

Poynting, J. H., 1904, Philosophical Transactions of the Royal Society of London, 202, 525

Prialnik, D. & Rosenberg, E. D., 2009, Monthly Notices of the Royal Astronomical Society: Letters, 399, L79

Proudfoot, B. C. N. & Ragozzine, D., 2019, The Astronomical Journal, 157, 230

Pudritz, R. E., Ouyed, R., Fendt, C. & Brandenburg, A., 2007, Protostars and Planets V, 277

Qian, S.-B. et al., 2011, Monthly Notices of the Royal Astronomical Society, 414, 16

Quarles, B. L. & Lissauer, J. J., 2015, Icarus, 248, 318

Quarles, B. & Kaib, N., 2019, The Astronomical Journal, 157, 67

Queloz, D. et al., 2000, Astronomy & Astrophysics, 354, 99

Quinn, S. N. et al., 2019, The Astronomical Journal, 158, 177

Ragent, B., Colburn, D. S., Rages, K. A., Knight, T. C. D., Avrin, P., Orton, G. S., Yanamandra-Fisher, P. A., & Grams, G. W., 1998, Journal of Geophysical Research Planets, 103, 22891

Ragozzine, D. & Brown, M. E., 2007, The Astronomical Journal, 134, 2160

Rajpaul, V. et al., 2016, Monthly Notices of the Royal Astronomical Society, 456, 6

Ramirez, R. M., 2017, Icarus, 297. 71

Ramm, D. J. et al., 2016, Monthly Notices of the Royal Astronomical Society, 460, 3706

Rampino, M. R. & Caldeira, K., 1994, Annual Review of Astronomy & Astrophysics, 32, 83

Randall, L. & Reece, M., 2014, Physical Review Letters, 112, 1301

Rauer, H. et al., 2014, Experimental Astronomy, 38, 249

Raup, D. M. & Sepkoski, J. J., 1984, Proceedings of the National Academy of Science, 81, 801

Raup, D. M. & Sepkoski, J. J., 1986, Science, 231, 833

Raymond, S. N. et al., 2011, Astronomy & Astrophysics, 530, 62

Raymond, S. N., Kokubo, E., Morbidelli, A., Morishima, R., Walsh, K. J., 2014, in Protostars and Planets VI, ed. H. Beuther et al. ( Tucson, AZ: Univ. Arizona Press ), 595

Raymond, S. N. & Izidoro, A., 2017, Science Advances, 3, e1701138

Reach, W. T., 1992, The Astrophysical Journal, 392, 289

Reach, W. T., Morris, P., Boulanger, F. & Okumura, K., 2003, Icarus, 164, 384

Reddy, V., Emery, J. P., Gaffey, M. J., Bottke, W. F., Cramer, A. & Kelley, M. S., 2009, Meteoritics and Planetary Science, 44, 1917

Rein, H. & Liu, S.-F., 2012, Astronomy & Astrophysics, 537, 128

Ren, B. et al., 2019, The Astrophysical Journal, 882, 64

Renne, P. R., Sprain, C. J., Richards, M. A., Self, S., Vanderkluysen, L. & Pande, K., 2015, Science, 350, 76

Reufer, A., Meier, M. M. M., Benz, W. & Wieler, R., 2012, Icarus, 221, 296

Reuter, D. C. et al., 2007, Science 318, 223

Ribas, Á, Merín, B., Bouy, H. & Maud, L. T., 2014, Astronomy & Astrophysics, 561, 54

Richert, A. J. W. et al., 2018, Monthly Notices of the Royal Astronomical Society, 477, 5191

Richter, F. M., Davis, A. M., Ebel, D. S. & Hashimoto, A., 2002, Geochimica et Cosmochimica Acta, 66, 521

Ricker, G. R. et al., 2015, Journal of Astronomical Telescopes, Instruments, and Systems, 1, 4003

Rickman, E. L. et al., 2019b, Astronomy & Astrophysics, 625, 71

Rickman, H., Fouchard, M., Froeschlé, Ch. & Valsecchi, G. B., 2012, Planetary and Space Science, 73, 124

Rickman, H., Błęcka, M. I., Gurgurewicz, J., Jørgensen, U. G., Słaby, E., Szutowicz, S., & Zalewska, N., 2019a, Planetary and Space Science, 166, 70

Rieke, G. H. et al., 2005, The Astrophysical Journal, 620, 1010

Roberge, A. et al., 2012, Publications of the Astronomical Society of the Pacific, 124, 799

Roberts, J. H. & Zhong, S., 2006, Journal of Geophysical Research, 111, 6013

Robertson, H. P., 1937, Monthly Notices of the Royal Astronomical Society, 97, 423

Robertson, P. et al., 2012a, The Astrophysical Journal, 749, 39

Robertson, P. et al., 2012b, The Astrophysical Journal, 754, 50

Robertson, P. et al., 2014, Science, 345, 440

Robinson, T. D., Maltaglitati, L., Marley, M. S., & Fortney, J. J., 2014, Publications of the National Academies of Sciences, 111, 9042

Rodrigeuz, J. A. P. et al., 2015, Scientific Reports, 5, 13404

Rodrigeuz, J. E. et al., 2017, The Astrophysical Journal, 826, 209

Roelfsema, P. R. et al., 2018, Publications of the Astronomical Society of Australia, 35, 30

Rogers, Leslie A., Bodenheimer, Peter, Lissauer, Jack J. & Seager, Sara, 2011, The Astrophysical Journal, 738, 59

Rogers, L. A., 2015, The Astrophysical Journal, 801, 41

Rogers, L. A. & Seager, S., 2010, The Astrophysical Journal, 712, 974

Roig, F., Nesvorný, D. & DeSouza, S. R., 2016, The Astrophysical Journal Letters, 820, 30

Ronca, L. B. & Green, R. R., 1970, Astrophysics and Space Science, 8, 69

Roosen, R. G., 1970, Icarus, 13, 184

Rothery, D. A., 2015, Planet Mercury: From Pale Pink Dot to Dynamic World, Springer Praxis Books. ISBN 978-3-319-12116-1. Berlin: Springer-Verlag

Rousselot, P., 2008, Astronomy and Astrophysics, 480, 543

Rowan, D. et al., 2016, The Astrophysical Journal, 817, 104

Rowe, J. F. et al., 2014, The Astrophysical Journal, 784, 45

Rubin, A. E., 2000, Earth Science Reviews, 50, 3

Rufu, R. & Canup, R. M., 2017, The Astronomical Journal, 154, 208

Runcorn, S., 1956, Geological Society of America Bulletin, 67, 301

Ruprecht, J. D., Bosh, A. S., Person, M. J., et al., 2013, AAS/Division, Planet. Sci. Meet. Abstr., 45, 414.07

Russell, S. S., Krot, A. N., Huss, G. R., Keil, K., Itoh, S., Yurimoto, H. & MacPherson, G. J., 2005, in Krot, A. N., Scott, E. R. D. & Reipurth, B. eds. Chondrites and the protoplanetary disk. Astronomical Society of the Pacific Conference Series 341, 317

Ryder, G., 1990, *EOS Trans. AGU*, 71, 322

Ryder, G., 2002, *J. Geophys. Res.,* 107, E4, 1-14

Safronov, V. S., 1966, Soviet Astronomy, 9, 987

Safronov, V. S., 1969, *Evoliutsiia doplanetnogo oblaka*, book LCCN: 78-447534 (PREM); CALL NUMBER: QB981 .S26

Safronov, V. S., 1972, Evolution of the protoplanetary cloud and formation of the earth and planets, Translated from Russian. Jerusalem (Israel): Israel Program for Scientific Translations, Keter Publishing House

Sagan, C., 1962, Icarus, 1, 151

Sagan, C. et al., 1973, Journal of Geophysical Research, 78, 4163

Sagan, C., 1975, Icarus, 25, 602

Salmeron, R. & Ireland, T. R., 2012a, Earth and Planetary Science Letters, 327-328, 61

Salmeron, R. & Ireland, T. R., 2012b, Meteoritics and Planetary Science, 47, 1922

Sánchez-Lavega, A., Wesley, A., Orton, G., et al., 2010, The Astrophysical Journal Letters, 715, L155

Sandwell, D. & Schubert, G., 2010, Icarus, 210, 817

Sansom, E. K. et al., 2019, Icarus, 321, 388

Santerne, A. et al., 2016, Astronomy and Astrophysics, 587, 64

Santos, N. C., Israelian, G., Mayor, M., Rebolo, R. & Udry, S., 2003, Astronomy and Astrophysics, 398, 363

Sasaki, T., Stewart, G. R. & Ida, S., 2010, The Astrophysical Journal, 714, 1052

Sato, T., Okuzumi, S. & Ida, S., 2006, Astronomy & Astrophysics, 589, 15

Saumon, G. & Guillot, T., 2004, The Astrophysical Journal, 609, 1170

Schaber, G. G. et al., 1992, Journal of Geophysical Research, 97, 13257

Schaefer, L. & Fegley, B., 2011, The Astrophysical Journal, 729, 6

Scharf, C.A. 2006, The Astrophysical Journal, 648, 1196

Schlaufman, K. C., Lin, D. N. C. & Ida, S., 2009, The Astrophysical Journal, 691, 1322

Schlichting, H. E. & Sari, R., 2009, The Astrophysical Journal, 700, 1242

Schneider, G. et al., 1999, The Astrophysical Journal, 513, 127

Schneider, G. et al., 2014, The Astronomical Journal, 148, 59

Schubert, G. et al., 1980, Journal of Geophysical Research, 85, 8007

Schulte, P. et al., 2010, Science, 327, 5970

Schüppler, Ch. et al., 2015, Astronomy & Astrophysics, 581, 97

Schwartz, R. D. & James, P. B. 1984, Nature, 308, 712

Scott, E. R. D. & Krot, A. N., 2005a, in Krot, A. N., Scott, E. R. D. & Reipurth, B. eds. Chondrites and the protoplanetary disk. Astronomical Society of the Pacific Conference Series 341, 15

Scott, E. R. D. & Krot, A. N., 2005b, in Davis A. M., Holland H. D. & Turekian K. K. eds. Meteorites, comets and planets: treatise on geochemistry, 1, 143

Scott, E. R. D. & Krot, A. N., 2014, in Chondrites and their Components, ed. A. M. Davis, 65

Seager, S. & Sasselov, D. D., 2000, The Astrophysical Journal, 537, 916

Seager, S., Deming, D. & Valenti, J. A., 2009, Astrophysics in the Next Decade, Astrophysics and Space Science Proceedings, ISBN 978-1-4020-9456-9, Springer Netherlands, 123

Ségransan, D. et al., 2011, Astronomy & Astrophysics, 535, 54

Sekanina, Z., Larson, S. M., Hainaut, O., Smette, A. & West, R. M., 1992, Astronomy & Astrophysics, 263, 367

Sekanina, Z., 2000, The Astrophysical Journal, 542, 147

Sekanina, Z., 2001, Publications of the Astronomical Institute of the Academy of Sciences of the Czech Republic, 89, 78

Sekanina, Z. & Chodas, P. W., 2004, The Astrophysical Journal, 607, 620

Sekanina, Z. & Chodas, P. W., 2005, The Astrophysical Journal Supplement Series, 161, 551

Sekanina, Z. & Chodas, P. W., 2007, The Astrophysical Journal, 663, 657

Sekanina, Z. & Chodas, P. W., 2012, The Astrophysical Journal, 757, 127

Sekanina, Z. & Kracht, R., 2013. The Astrophysical Journal, 778, 24

Sekine, Y., Genda, H., Kamata, S. & Funatsu, T., 2017, Nature Astronomy, 1, 31

Shankman, C. et al., 2017, The Astronomical Journal, 154, 50

Sheppard, S. S. & Jewitt, D. C., 2003, Nature, 423, 261

Sheppard, S. S., Jewitt, D. & Kleyna, J., 2005, The Astronomical Journal, 129, 518

Sheppard, S. S. & Trujillo, C. A., 2006, Science, 313, 511

Sheppard, S. S. & Trujillo, C., 2016, The Astronomical Journal, 152, 221

Shoemaker, E. M., 1960, The Moon Meteorites and Comets, Edited by Gerard P. Kuiper, and Barbarra Middlehurts. Chicago: The University of Chicago Press, 1963, p.301

Shoemaker, E. M. & Chao, E. C. T., 1961, Journal of Geophysical Research, 66, 3371

Showalter, M. R. & Lissauer, J. J., 2003, IAU Circular 8209

Showalter, M. R. & Lissauer, J. J., 2006, Science, 311, 973

Showalter, M. R. & Hamilton, D. P., 2015. Nature, 522, 45

Shu, F. H., Shang, H. & Lee, T., 1996, Science, 271, 1545

Sibthorpe, B., Kennedy, G. M., Wyatt, M. C., Lestrade, J.-F., Greaves, J. S., Matthews, B. C. & Duchêne, G., 2018, Monthly Notices of the Royal Astronomical Society, 475, 3046

Siegfried, R. W. II & Solomon, S. C., 1974, Icarus, 23, 192

Sierchio, J. M., Rieke, G. H., Su, K. Y. L. & Gáspár, A., 2014, The Astrophysical Journal, 785, 33

Singer, S. F., 1975, Icarus, 25, 484

Singer, K. N. et al., 2019, Science, 363, 955

Siscoe, G. L., Davis, L. Jr., Coleman, P. J. Jr., Smith, E. J. & Jones, D. E., 1968, Journal of Geophysical Research, 73, 61

Sissa, E., 2018, Astronomy and Astrophysics, 613, 6

Siverd, R. J. et al., 2012, The Astrophysical Journal, 761, 123

Slattery, W. L., Benz, W. & Cameron, A. G. W., 1992, Icarus, 99, 167

Slipher, V. M., 1903, Astronomische Nachrichten, 163, 35

Smit, J., 1999, Annual Review of Earth and Planetary Sciences, 27, 75

Smith, B. A. et al., 1979, Science, 204, 951

Smith, B. A., et al., 1981, Science, 212, 163

Smith, B. A., et al., 1982, Science, 215, 504

Smith, B. A. & Terrile, R. J., 1984, Science, 226, 1421

Smith, B. A., et al., 1986, Science, 233, 43

Smith, B. A. et al., 1989, Science, 246, 1422

Smith, D. E. et al., 2012, Science, 336, 214

Smith, E. J., Davis, L. Jr., Coleman, P. J. Jr. & Jones, D. E., 1965, Science, 149, 1241

Smith, E. J. et al., 1974, Journal of Geophysical Research, 79, 3501

Snellen, Ignas A. G., de Kok, Remco J., de Mooij, Ernst J. W. & Albrecht, Simon, Nature, 2010, 465, 1049

Solomon, S. C., McNutt, R. L., Gold, R. E. & Domingue, D. L., 2007, Space Science Reviews, 131, 3

Solomon, S. C., 1977, Physics of the Earth and Planetary Interiors, 15, 135

Solomon, S. C. et al., 2008, Science, 321, 59

Song, I., Zuckerman, B., Weinberger, A. J. & Becklin, E. E., 2005, Nature, 436, 363

Spahn, F., Sachse, M., Seiß, M, Hsu, H.-W., Kempf, S. & Horányi, M., 2019, Space Science Reviews, 215, 11

Spalding, C., Batygin, K. & Adams, F. C., 2016, The Astrophysical Journal, 817, 18

Spencer, J. R. et al., 2007, Science, 318, 240

Spohn, T. et al., 2018, Space Science Reviews, 214, 96

Spoto, F., Milani, A., & Knežević, Z., 2015, Icarus, 257, 275

Spudis, P., Wilhelms, D.E., Robinson, M.S., 2011, 42nd Lunar and Planetary Science Conference, 1365

Spurný, P., Borovička, J., Mucke, H. & Svoreň, J., 2017, Astronomy & Astrophysics, 605, 68

Stanley, S. & Bloxham, J., 2004, Nature, 428, 151

Stansberry, J. A., Van Cleve, J., Reach, W. T., et al., 2004, The Astrophysical Journal Supplement Series, 154, 463

Stassun K. G. et al., 2018, The Astronomical Journal, 156, 102

Steel, D. I., Asher, D. J. & Clube, S. V. M., 1991, Monthly Notices of the Royal Astronomical Society, 251, 632

Stern, S. A., 1996, The Astronomical Journal, 112, 1203

Stern, S. A. & Colwell, J. E., 1997, The Astrophysical Journal, 490, 879

Stern, S. A. et al., 2006, Nature. 439, 946

Stern, S. A. et al., 2019, Science, 364, 9771

Stevenson, D. J. & Salpeter, E. E. (1977a) The Astrophysical Journal Supplement Series, 35, 221

Stevenson, D. J. & Salpeter, E. E. (1977b) The Astrophysical Journal Supplement Series, 35, 239

Stevenson, D. J., Spohn, T. & Schubert, G., 1983, Icarus, 54, 466

Stevenson, D. J., 1987, Annual Review of Earth and Planetary Sciences, 15, 271

Stevenson, D. J., 2001, Nature, 412, 214

Stevenson, D. J., 2003, Earth and Planetary Science Letters, 208, 1

Stickle, A. M. & Roberts, J. H., 2018, Icarus, 307, 197

Stoffler, D. & Ryder, G., 2001, Space Sci. Rev., 96, 9

Stone, R. G. et al., 1992, Science, 257, 1524

Stooke, P. J., 1994, Earth Moon and Planets, 65, 31

Strand, K. Aa., 1943, Publications of the Astronomical Society of the Pacific, 55, 29

Strubbe, L. E. & Chiang, E. I., 2006, The Astrophysical Journal, 648, 652

Su, K. Y. L. et al., 2009, The Astrophysical Journal, 705, 314

Su, K. Y. L. et al., 2019, The Astronomical Journal, 157, 202

Sullivan, P. W. et al., 2015, The Astrophysical Journal, 809, 77

Swedenborg, E., 1734, *Opera Philosophica et Mineralia, (Principia rerum naturalium. Regnum subterraneum sive minerale de ferro, de cupro et orichalco).*, sumpt. F. Hekelii

Swift, J. J. et al., 2015, Journal of Astronomical Telescopes, Instruments, and Systems, 1, 7002

Sykes, M. V. & Greenberg, R., 1986, Icarus, 65, 51

Sykes, M. V., 1990, Icarus, 85, 267

Tackley, P. J., Stevenson, D. J., Glatzmaier, G. A., & Schubert, G., 1994, Journal of Geophysical Research: Solid Earth, 99, 15877

Tamuz, O. et al., 2008, Astronomy & Astrophysics, 480, L33

Tatischeff, V., Duprat, J. & de Séréville, N., 2010, The Astrophysical Journal Letters, 714, L26

Teachey, A., Kipping, D.M., & Schmitt, A.R., 2017, The Astrophysical Journal, 155, 1

Teachey, A., & Kipping, D.M., 2018, Science Advances, 4, 10

Teanby, N. A., 2015, Icarus, 256, 49

Tedesco E. F. & Desert F. -X., 2002, The Astronomical Journal, 123, 2070

Tera, F., Papanastassiou, D.A., Wasserburg, G.J., 1974, Earth Planet Sci. Lett., 22 , 1

Testi, L., Birnstiel, T., Ricci, L., et al., 2014, Protostars and Planets VI, 339

Thébault, P., Augereau, J. C. & Beust, H., 2003, Astronomy & Astrophysics, 408, 775

Thébault, P. & Augereau, J. -C., 2007, Astronomy & Astrophysics, 472, 169

Thébault, P. & Wu, Y., 2008, Astronomy & Astrophysics, 481, 713

Thébault, P. & Kral, Q., 2019, Astronomy & Astrophysics, 626, 24

Thomas, P. C., 2010, Icarus, 208, 395

Thomas, R. J., Rothery, D. A., Conway, S. J. & Anand, M., 2014, Geophysical Research Letters, 41, 6084

Thommes, E. W., Duncan, M. J. & Levison, H. F., 1999, Nature, 402, 635

Thommes, E. W., Duncan, M. J. & Levison, H. F., 2002, The Astronomical Journal, 123, 2862

Thompson, S. E. et al., 2018, The Astrophysical Journal Supplement Series, 235, 38

Thorsett, S. E., Arzoumanian, Z. & Taylor, J. H., 1993, The Astrophysical Journal, Letters, 412, L33

Thureau, N. D. et al., 2014, Monthly Notices of the Royal Astronomical Society, 445, 2558

Tinetti, G. et al., 2018, Experimental Astronomy, 46, 135

Tinney, C. G., Butler, R. P., Marcy, G. W., Jones, H. R. A., Penny, A. J., McCarthy, C. & Carter, B. D., 2002, The Astrophysical Journal, 571, 528

Tiscareno, M. S. & Malhotra, R., 2003, The Astronomical Journal, 126, 3122

Tomkinson, T., Lee, M. R., Mark. D. F. & Smith, C. L., 2013, Nature Communications, 4, 2662

Torbett, M. V. & Smoluchowski, R., 1984, Nature, 311, 641

Toulmin, P. III. Et al., 1977, Journal of Geophysical Research, 82, 4625

Trail, D., Mojzsis, S.J., Harrison, M.T., 2007, Geochim Cosmochim. Acta, 71, 4044

Trilling, D. E. et al., 2008, The Astrophysical Journal, 674, 1086

Trujillo, C. A. & Brown, M. E., 2001, The Astrophysical Journal, 554, 95

Trujillo, C. A. & Sheppard, S. S., 2014, Nature, 507, 471

Tsiganis, K., Gomes, R., Morbidelli, A. & Levison, H. F., 2005, Nature, 435, 459

Tsuda, Y., Yoshikawa, M., Abe, M., Minamino, H. & Nakazawa, S., 2013, Acta Astronautica, 91, 356

Turner, G., Knott, S.F., Ash, R.D., & Gilmour, J.D., 1997, *Geochim. Cosmochim. Acta*, 61, 3835

Twicken, J. D. et al., 2016, The Astronomical Journal, 152, 158

Valeille, A., Brougher, S. W., Tenishev, V., Combi, M. R. & Nagy, A. F., 2010, Icarus, 206, 28

Valencia, D., Sasselov, D. D. & O'Connell, R. J., 2007, The Astrophysical Journal, 665, 1413

Van Borstel, I. & Blum, J., 2012, Astronomy & Astrophysics, 548, A96, 13

van de Kamp, P. & Lippincott, S. L., 1951, The Astronomical Journal, 56, 49

van de Kamp, P., 1963, The Astronomical Journal, 68, 515

van de Kamp, P., 1969, The Astronomical Journal, 74, 238

van de Kamp, P., 1969, The Astronomical Journal, 74, 757

Vanderburg, A. et al., 2019, The Astrophysical Journal: Letters, 881, 19

Vasavada, A. R. et al., 1999, Journal of Geophysical Research, 104, 27133

Vasavada, A. R. & Showman, A. P., 2005, Reports on Progress in Physics, 68, 1935

Vican, L., Schneider, A., Bryden, G., Melis, C., Zuckerman, B., Rhee, J. & Song, I., 2016, The Astrophysical Journal, 833, 263

Vilenius, E. et al., 2018, Astronomy & Astrophysics, 618, 136

Villanueva, G. L., Mumma, M. J., Bonev, B. P., Di Santi, M. A., Gibb, E. L., Böhnhardt, H. & Lippi, M., 2009, The Astrophysical Journal, Letters, 690, L5

Villanueva, S., Dragomir, D., Gaudi, B.S., 2019, The Astronomical Journal, 157, 84

Vitense, Ch., Krivov, A. V., Kobayashi, H. & Löhne, T., 2012, Astronomy & Astrophysics, 540, 30

Vogt, S. et al., 2010, The Astrophysical Journal, 708, 1366

Vogt, S. S. et al., 2017, The Astronomical Journal, 154, 181

Vokrouhlický, D., Brož, M., Bottke, W. F., Nesvorný, D. & Morbidelli, A., 2006, Icarus, 182, 118

Vokrouhlický, D., Nesvorný, D. & Levison, H. F., 2008, The Astronomical Journal, 136, 1463

Vokrouhlicky, D., Bottke, W. F., Chesley, S. R., Scheeres, D. J., & Statler, T. S., 2015, In P. Michel, F. E. DeMeo, & W. F. Bottke (Eds.), *Asteroids IV*. Tucson, AZ: University of Arizona Press.

Vokrouhlický, D., Nesvorný, D. & Dones, L., 2019, The Astronomical Journal, 157, 181

Volk, K. & Malhotra, R., 2008, The Astrophysical Journal, 687, 714

Volk, K. & Malhotra, R., 2012, Icarus, 221, 106

Volk, K. & Malhotra, R., 2013, Icarus, 224, 66

Volk, K. & Malhotra, R., 2019, The Astronomical Journal, 158, 64

Völschow, M., Banerjee, R. & Hessman, F. V., 2014, Astronomy & Astrophysics, 562, 19

Völschow, M., Schleicher, D. R. G., Perdelwitz, V. & Banerjee, R., 2016, Astronomy & Astrophysics, 587, 34

von Braun, K., Kane, S.R. & Ciardi, D.R., 2009, The Astrophysical Journal, 702, 779

Wada, K., Kokubo, E. & Makino, J., 2006, The Astrophysical Journal, 638, 1180

Wadhwa, M. & Russell, S. S., 2000, Protostars and Planets IV, 995

Wahl, S. M. et al., 2017, Geophysical Research Letters, 44, 4649

Walker. J. C. G., 1975, Journal of Atmospheric Science, 32, 1248

Walker, G. A. H., Walker, A. R., Irwin, A. W., Larson, A. M., Yang, S. L. S. & Richardson, D. C., 1995, Icarus, 116, 359

Wallace, J. M. & Hobbs, P. V., 2006, Atmospheric Science – an Introductory Survey (2nd edn.), Academic Press

Walsh, K. J., Morbidelli, A., Raymond, S. N., O'Brien, D. P., & Mandell, A. M., 2011, Nature, 475, 206

Walsh, K. J., Morbidelli, A., Raymond, S. N., O'Brien, D. P. & Mandell, A. M., 2012, Meteoritics & Planetary Science, 47, 1941

Wang, J., Fischer, D. A., Horch, E. P. & Huang, X., 2015, The Astrophysical Journal, 799, 229

Wang, J., Mawet, D., Fortney, J. J., Hood, C., Morley, C. V. & Benneke, B., 2018a, The Astronomical Journal, 156, 272

Wang, J., Mawet, D., Hu, R., Ruane, G., Delorme, J.-R. & Kilmovitch, N., 2018c, Journal of Astronomical Telescopes, Instruments and Systems, 4, 035001

Wang, J.-H. & Brasser, R., 2014, Astronomy & Astrophysics, 563, 122

Wang, J. J. et al., 2018b, The Astronomical Journal, 156, 192

Wang, X., & Hou, X., 2017, Monthly Notices of the Royal Astronomical Society, 471, 243

Ward, P.D. & Brownlee, D., 2000, Rare Earth: Why Complex Life is Uncommon in the Universe, New York

Ward, W. R. & Hamilton, D. P., 2004, The Astronomical Journal, 128, 2501

Watters, T. R., Robinson, M. S. & Cook, A. C., 1998, Geology, 26, 991

Way, M.J., et al., 2016, Geophysical Research Letters, 43, 8376

Webb, W. A., 1955, Publications of the Astronomical Society of the Pacific, 67, 283

Weisberg, M. K., McCoy, T. J. & Krot, A. N., 2006, in Meteorites and the Early Solar System II, ed. D. S., Lauretta & H. Y., McSween, 19

Weiss, B. P. et al., 2002, Earth and Planetary Science Letters, 201, 449

Weiss, L. M. & Marcy, G. W., 2014, The Astrophysical Journal Letters, 783, L6

Weidenschilling, S. J., 1977, Astrophysics and Space Science, 51, 153

Wells, L. E., Armstrong, J. C. & Gonzalez, G., 2003, Icarus 162, 38

Werner, S. C., 2009, Icarus, 201, 44

West, R. A., Baines, K. H., Karkoschka, E., & Sánchez-Lavega, A., 2009, Saturn: Composition and Chemistry, in Saturn from Cassini-Huygens (Springer Science+Business Media: Dordrecht), 161-180

Wetherill, G. W., 1988, Icarus, 76, 1

Wetherill, G. W., 1994, Astrophysics and Space Science, 212, 23

Whipple. F. L., 1951, The Astrophysical Journal, 113, 464

Whitby, J., Burgess, R., Turner, G., Gilmour, J. & Bridges, J., 2000, Science, 288, 1819

Whitmire, D. P. & Jackson, A. A., 1984, Nature, 308, 713

Wiechert, U., Halliday, A. N., Lee, D. -C., Snyder, G. A., Taylor, L. A. & Rumble, D., 2001, Science, 294, 345

Wieczorek, M. A., Joliff, B. L., Khan, A., et al., 2006, Reviews in Mineralogy & Geochemistry, 60, 221

Wiegert, P. & Tremaine, S., 1999, Icarus, 137, 84

Wiegert, P. A., Brown, P. G. Vaubaillon, J. & Schijns, H., 2005, Monthly Notices of the Royal Astronomical Society, 361, 638

Wierzchos, K., Womack, M., & Sarid, G. 2017, The Astronomical Journal, 153, 230

Wilhelms, D. E. & Squyres, S. W., 1984, Nature, 309, 138

Wilhelms, D.E., 1987, *The Geologic History of the Moon*, U.S. Geological Survey Professional Paper, 1348

Williams, I. P. & Wu, Z., 1993, Monthly Notices of the Royal Astronomical Society, 262, 231

Williams, D.M., Kasting, J.F., & Wade, R.A., 1997, Nature, 385, 234

Williams, J. -P. & Nimmo, F., 2004, Geology, 32, 97

Williams J. P. & Cieza L. A., 2011, Annual Reviews of Astronomy and Astrophysics, 49, 67

Wilner, D. J., MacGregor, M. A., Andrews, S. M., Hughes, A. M., Matthews, B. & Su, K., 2018, The Astrophysical Journal, 855, 56

Wilson, H. F. & Militzer, B., 2012, The Astrophysical Journal, 745, 54

Wilson, M. L. et al., 2019, The Publications of the Astronomical Society of the Pacific, 131, 5001

Windmark, F., Birnstiel, T., Güttler, C., Blum, J., Dullemond, C. P. & Henning, Th., 2012a, Astronomy & Astrophysics, 540, 73

Windmark, F., Birnstiel, T., Ormel, C. W. & Dullemond, C. P., 2012b, Astronomy & Astrophysics, 2012b, 544, 16

Winn, J. N., Fabrycky, D., Albrecht, S. & Johnson, J. A., 2010, The Astrophysical Journal, 718, 145

Winn, J. N. & Fabrycky, D. C., 2015, Annual Review of Astronomy and Astrophysics, 53, 409

Winter, O. C., Borderes-Motta, G. & Ribeiro, T., 2019, Monthly Notices of the Royal Astronomical Society, 484, 3765

Wisdom, J. & Tian, Z., 2015, Icarus, 256, 138

Withers, P., Vogt, M.F., 2017, Astrophysical Journal, 836, 114

Wittenmyer, R. A., Endl, M., Cochran, W. D., Hatzes, A. P., Walker, G. A. H., Yang, S. L. S. & Paulson, D. B., 2006, The Astronomical Journal, 132, 177

Wittenmyer, Robert A., Tinney, C. G., Butler, R. P., O'Toole, Simon J., Jones, H. R. A., Carter, B. D., Bailey, J. & Horner, J., 2011a, The Astrophysical Journal, 738, 81

Wittenmyer, R. A., Tinney, C. G., O'Toole, S. J., Jones, H. R. A., Butler, R. P., Carter, B. D. & Bailey, J., 2011b, The Astrophysical Journal, 727, 102

Wittenmyer, R. A., Horner, J. & Tinney C. G., 2012, The Astrophysical Journal, 761, 165

Wittenmyer, R. A., Horner, J., Marshall, J. P., Buters, O. W. & Tinney, C. G., 2012, Monthly Notices of the Royal Astronomical Society, 419, 3258

Wittenmyer, R. A., Horner, J. & Marshall, J. P., 2013, Monthly Notices of the Royal Astronomical Society, 431, 2150

Wittenmyer, R. A. et al., 2014a, The Astrophysical Journal, 780, 140

Wittenmyer, R. A. et al., 2014b, The Astrophysical Journal, 783, 103

Wittenmyer, R. A. & Marshall, J. P., 2015, The Astronomical Journal, 149, 86

Wittenmyer, R. A. et al., 2016a, The Astrophysical Journal, 818, 35

Wittenmyer, R. A. et al., 2016b, The Astrophysical Journal, 819, 28

Wittenmyer, R. A. et al., 2017a, The Astronomical Journal, 153, 167

Wittenmyer, R. A. et al., 2017b, The Astronomical Journal, 154, 274

Witze, A., 2019, Nature, 576, 348

Wolfgang, A., Rogers, L. A. & Ford, E. B., 2016, The Astrophysical Journal, 825, 19

Wolstencroft, R. D., 1967, Planetary & Space Science, 15, 1081

Wolszczan, A. & Frail, D. A., 1992, Nature, 355, 145

Wolszczan, A., 1994, Science, 264, 538

Womack, M., Sarid, G., & Wierzchos, K. 2017, Publications of the Astronomical Society of the Pacific, 129, 31001

Womack, M., & Stern, S. A. 1999, Solar System Research, 33, 187

Wood, J. A., 1962, Geochimica et Cosmochimica Acta, 26, 739

Wood, J., Horner, J., Hinse, T. C., et al., 2017, The Astronomical Journal, 153, 245

Wood, J., Horner, J., Hinse, T. C., et al., 2018, The Astronomical Journal, 155, 2

Woolfson, M. M., 1964, Royal Society of London Proceedings Series A, 282, 485

Woolfson, M. M., 1999, Monthly Notices of the Royal Astronomical Society, 304, 195

Woolfson, M. M., 2013, Earth, Moon, and Planets, 111, 1

Wright, E. L. 2007. Wide-field Infrared Survey Explorer: (http://www.astro.ucla.edu/wright/WISE/index.html).

Wu, R. J., Zhou, L. & Zhou, L. Y., 2018, Acta Astronomica Sinica, 59, 36

Wyatt, M. C. & Dent, W. R. F., 2002, Monthly Notices of the Royal Astronomical Society, 334, 589

Wyatt, M. C., 2005, Astronomy & Astrophysics, 433, 1007

Wyatt, M. C., Smith, R., Su, K. Y. L., Rieke, G. H., Greaves, J. S., Beichman, C. A. & Bryden, G., 2007, The Astrophysical Journal, 663, 365

Wyatt, M. C., Booth, M., Payne, M. J. & Churcher, L. J., 2010, Monthly Notices of the Royal Astronomical Society, 402, 657

Wyatt, M. C., Clarke, C. J. and Booth, M., 2011, Celestial Mechanics and Dynamical Astronomy, 111, 1-2, 1

Wyatt, M. C., et al., 2012, Monthly Notices of the Royal Astronomical Society, 424, 1206

Wyatt, M. C. & Jackson, A. P., 2016, Space Science Reviews, 205, 231

Wyatt, M. C., Bonsor, A., Jackson, A. P., Marino, S. & Shannon, A., 2017, Monthly Notices of the Royal Astronomical Society, 464, 3385

Wyatt, S. P. & Whipple, F. L., 1950, The Astrophysical Journal, 111, 134

Yang, L., Ciesla, F. J. & Alexander, C. M. O. 'D., 2013, Icarus, 226, 256

Yano, H. et al., 2006, Science, 312, 1350

Yao, X., et al., 2019, The Astronomical Journal, 157, 37

Ye, Q. & Clark, D. L., 2019, The Astrophysical Journal Letters, 878, 34

Yee, J. C., 2013, The Astrophysical Journal, 770, 31

Yin, Q., Jacobsen, S. B. & Yamashita, K., 2002, Nature, 415, 881

Yoshida, F. & Nakamura, T., 2005, The Astronomical Journal, 130, 2900

Young, E. D., 2016, The Astrophysical Journal, 826, 129

Young, E. D., Kohl, I. E., Warren, P. H., Rubie, D. C., Jacobson, S. A. & Morbidelli, A., 2016, Science, 351, 493

Yu, Q. & Tremaine, S., 1999, The Astronomical Journal, 118, 1873

Yurimoto, H. et al, 2011, Science, 333, 1116

Zahnle, K. & Mac Low, M.-M., 1994, Icarus, 108, 1

Zahnle, K. et al., 2007, Space Science Reviews, 129, 35

Zanda, B., Hewins, R. H., Bourot-Denise, M., Bland, P. A. & Albare de F., 2006, Earth and Planetary Science Letters, 248, 650

Zechmeister, M. et al., 2013, Astronomy & Astrophysics, 552, 78

Zharkov, V.N., 1983, The Moon and the Planets, 29, 139

Zhou, G. et al., 2014, The Astronomical Journal, 147, 144

Zhou, G. et al., 2019, The Astronomical Journal, 158, 141

Zhu, W., Petrovich, C., Wu, Y., Dong, S. & Xie, J., 2018, The Astrophysical Journal, 860, 101

Zhu, W. & Wu, Y., 2018, The Astronomical Journal, 156, 92

Zollinger, R.R., Armstrong, J.C., & Heller, R. 2017, Monthly Notices of the Royal Astronomical Society, 472, 1

Zolotov, M. Y., Fegley, B. & Lodders, K., 1997, Icarus, 130, 475

Zotkin, I. T. & Tsikulin, M. A., 1966, Soviet Physics Doklady, 11, 183

Zotkin, I. T. & Chigorin, A. N., 1991, Astronomicheskii Vestnik, 25, 613

Zsom, A., Seager, S., de Wit, J. & Stamenković, V., 2013, The Astrophysical Journal, 778, 109

Zuckerman, B., Fekel, F. C., Williamson, M. H., Henry, G. W. & Muno, M. P., 2008, The Astrophysical Journal, 688, 1345

Zurbuchen, T. H. et al., 2008, Science, 321, 5885

**Appendix I: Plots of individual Solar system small body populations (greyscale)**

In Figures 2 - 6, we presented the distribution of small bodies throughout the Solar system, with a variety of colours employed to help the reader distinguish between the various populations of small bodies shown in those plots. In this appendix, we present grayscale plots that show the distribution of objects in those individual population. For each population, we provide four plots - two that show the spatial distribution of objects at epoch 2000-01-01 00:00:00 UT (in the X-Y and X-Z planes), and two that show the orbital element distribution of the objects (in the semi-major axis -- inclination and semi-major axis -- eccentricity planes). As with the data plotted in Figures 2 - 6, these data were taken from the JPL HORIZONS database, which can be found at https://ssd.jpl.nasa.gov/horizons.cgi, on 13th February 2019.

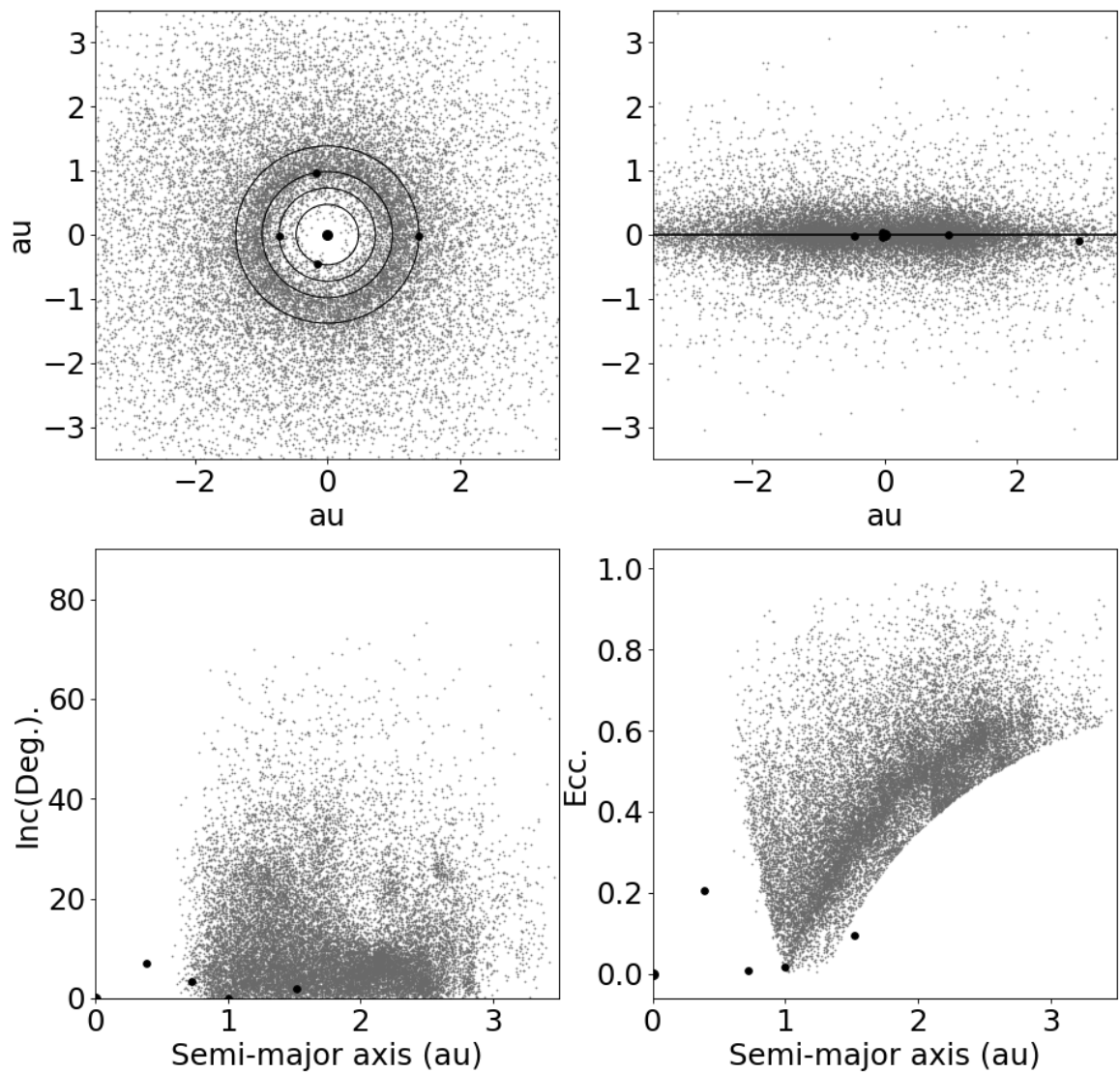

**Figure A1:** The distribution of all known near-Earth objects. The impact of observational biases on the distribution of known objects is readily apparent - particularly in the lower right panel, where an excess of known objects with perihelion distances of ~1 au is clearly seen, curving upwards and to the right from the dark dot that denotes the orbit of the Earth.

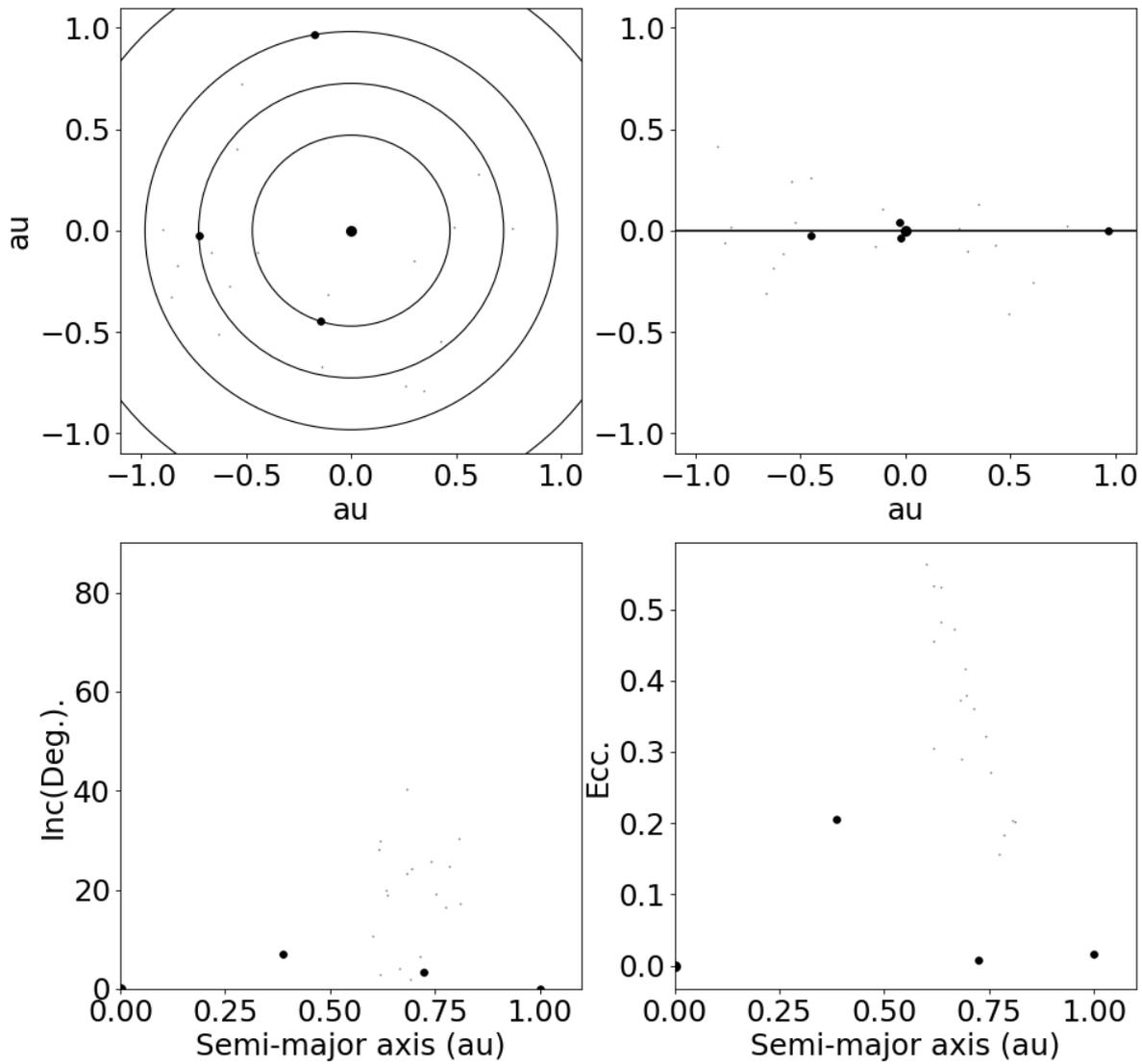

**Figure A2:** The distribution of all known Atira asteroids - asteroids whose orbits are entirely inside that of the Earth (i.e. whose aphelia lie closer to the Sun than the Earth's perihelion distance of 0.983 au). Very few such objects are currently known - a combination of their intrinsic scarcity (it is estimated that the Atiras make up only 2% of the total NEA population) and the challenges inherent in their discovery.

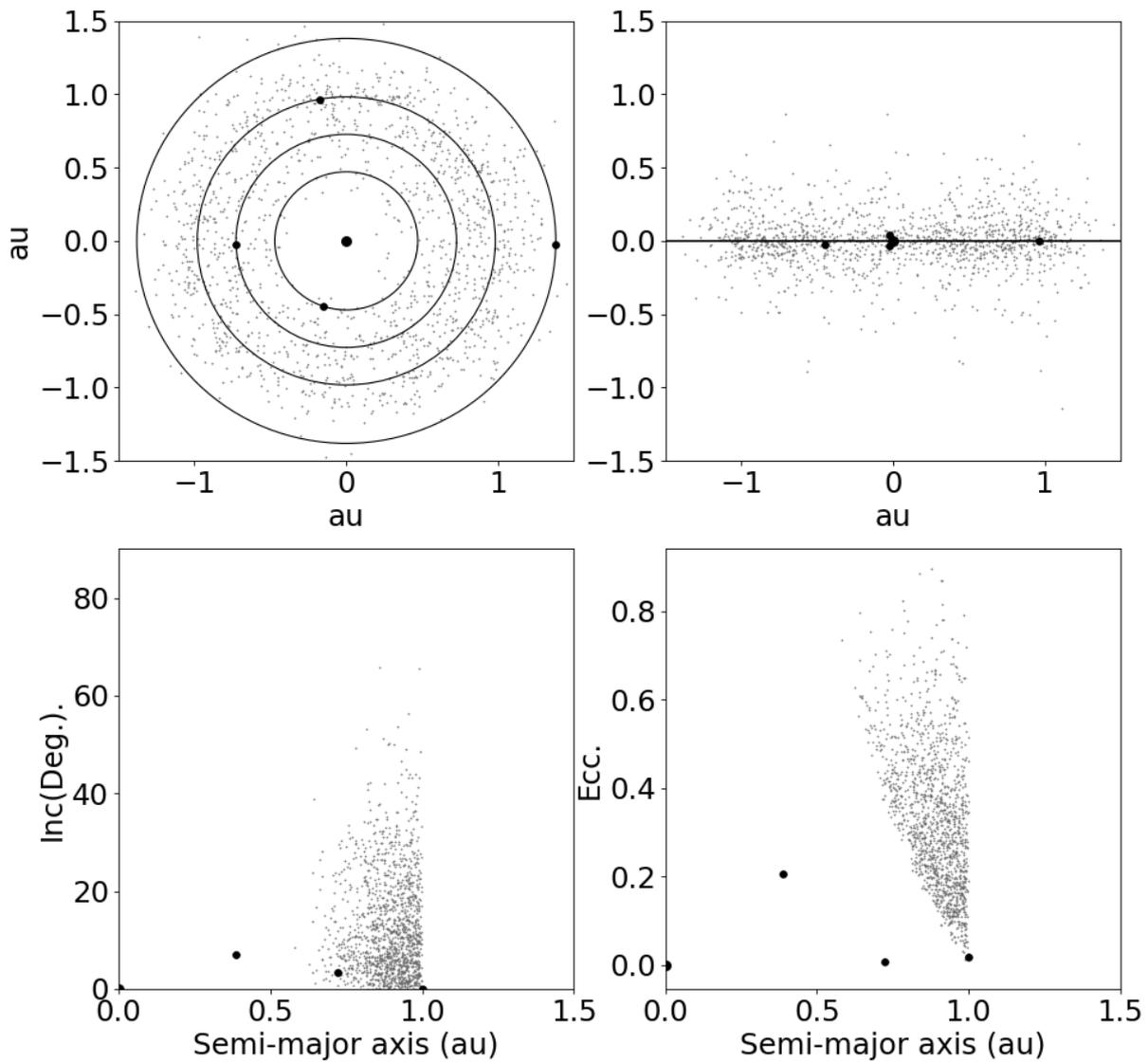

**Figure A3:** The distribution of all known Aten asteroids - those near-Earth asteroids whose orbital semi-major axes are less than that of the Earth (1 au). As they move on orbits with semi-major smaller than that of the Earth, their orbital periods are less than one year. The Aten asteroid population includes the Atira objects, which are generally viewed as a subset of the Aten population, rather than a distinct grouping in their own right - although this may change as more objects are discovered in coming years.

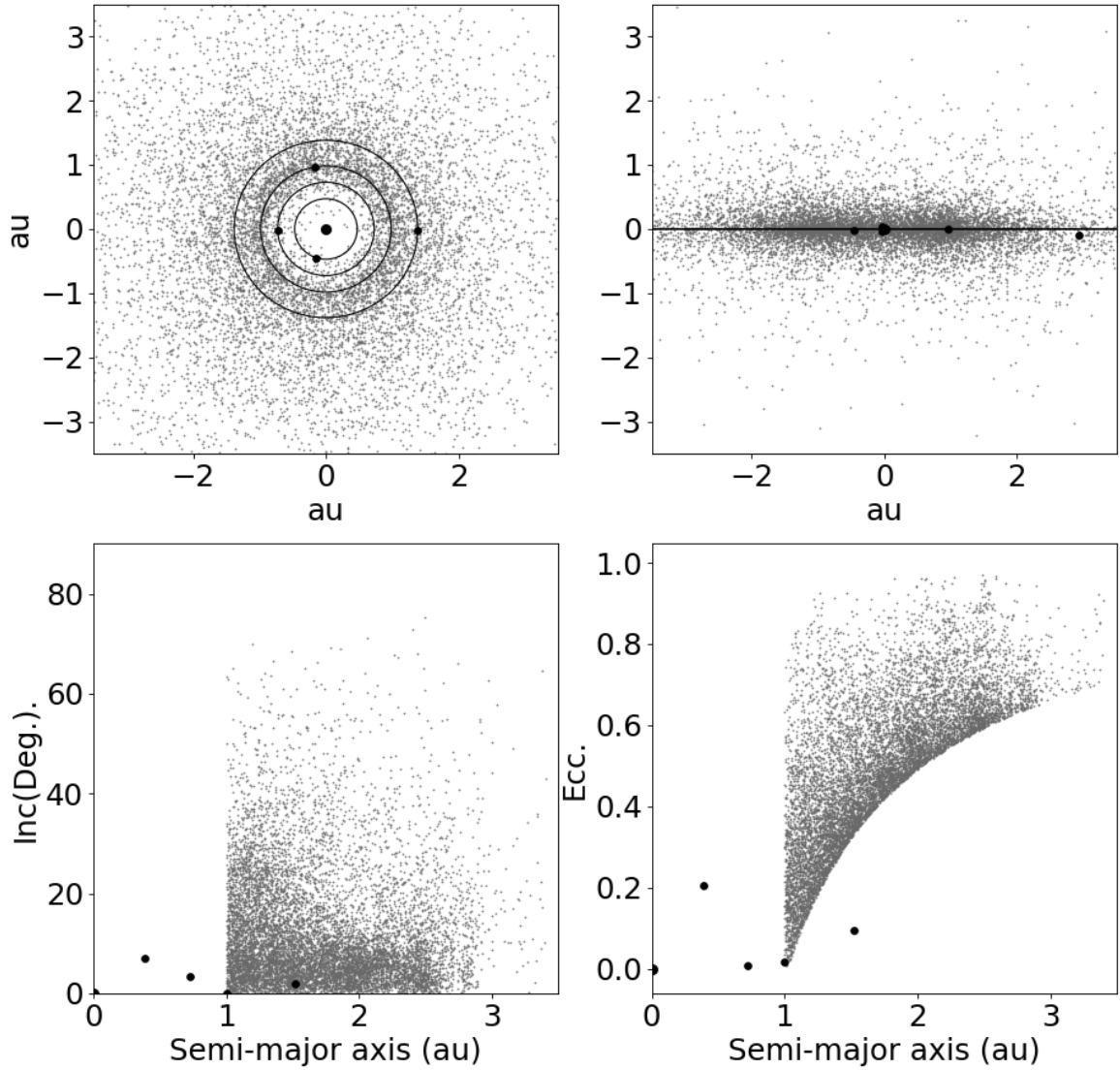

**Figure A4:** The distribution of the Apollo asteroids - near-Earth asteroids whose orbital semi-major axes are greater than that of the Earth (i.e. a > 1 au), but whose orbits cross that of our planet. The two boundaries of the population are clearly seen in the lower right-hand plot, with the curved outer edge of the population in phase space tracing a line of constant perihelion distance (where the perihelion distance for objects on that line is equal to Earth's aphelion distance, approximately 1.0167 au).

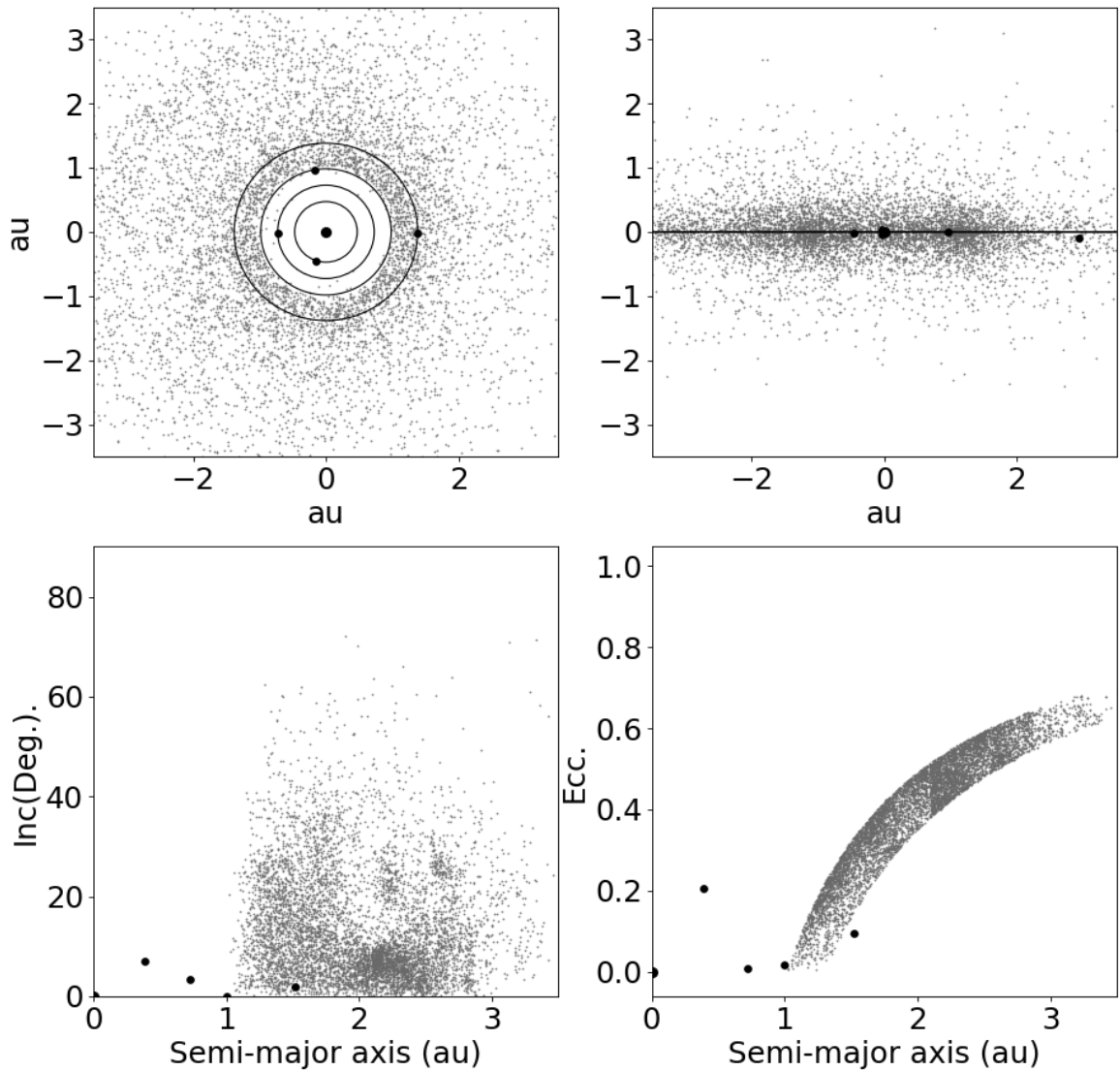

**Figure A5:** The orbital distribution of the Amor asteroids - near-Earth asteroids whose perihelia fall beyond the distance of Earth's aphelion. In other words, whilst Amor asteroids can be Earth-approaching, they are not Earth-crossing. The outer edge of the Amor population is typically set at a perihelion distance of around 1.3 au - asteroids that pass perihelion at greater distances from the Sun are not considered to be near-Earth asteroids, but instead are typically included within the population of the Asteroid belt.

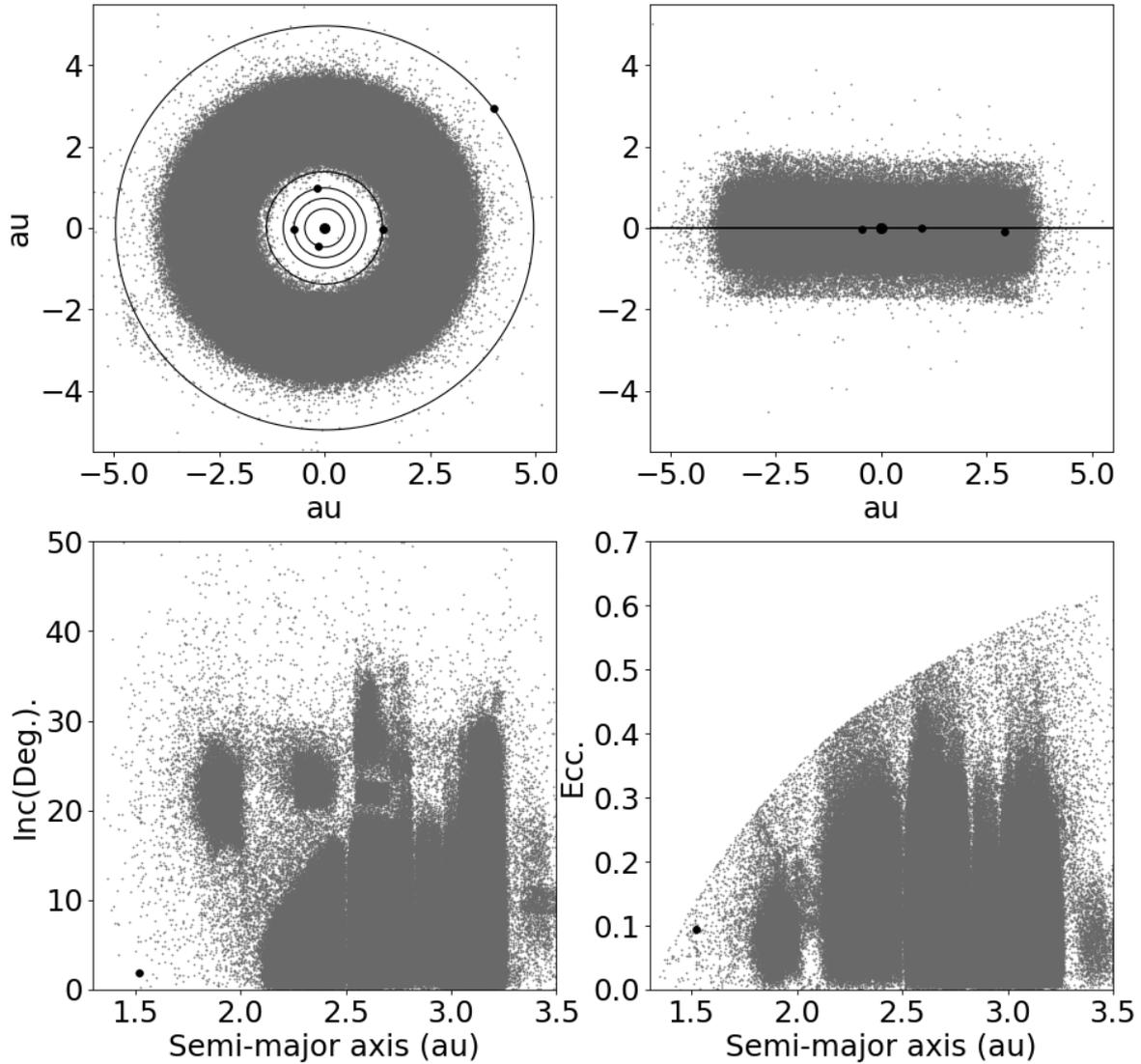

**Figure A6:** The distribution of the main belt asteroids. The Kirkwood gaps are clearly visible - specific regions of the belt in semi-major axis space that relatively free of asteroids. Those regions are the locations of mean-motion resonances with the giant planet Jupiter, and objects moving on such orbits would be destabilised and removed from the asteroid belt relatively quickly as a result of the giant planet's influence. In addition, the influence of secular resonances in sculpting the belt can be seen, particularly in the lower-left hand panel (showing the asteroids in semi-major axis vs inclination space). The curved boundary between the dense and lightly population regions extending upwards from ~2.1 au is the location of one such resonance - the $\nu_6$ secular resonance. Objects in the region just Sunward of that sharp boundary (or at higher inclinations) will gradually undergo eccentricity excitation as a result of secular perturbations from Saturn, with the eventual result that they will become Mars crossing, and be ejected from the belt entirely.

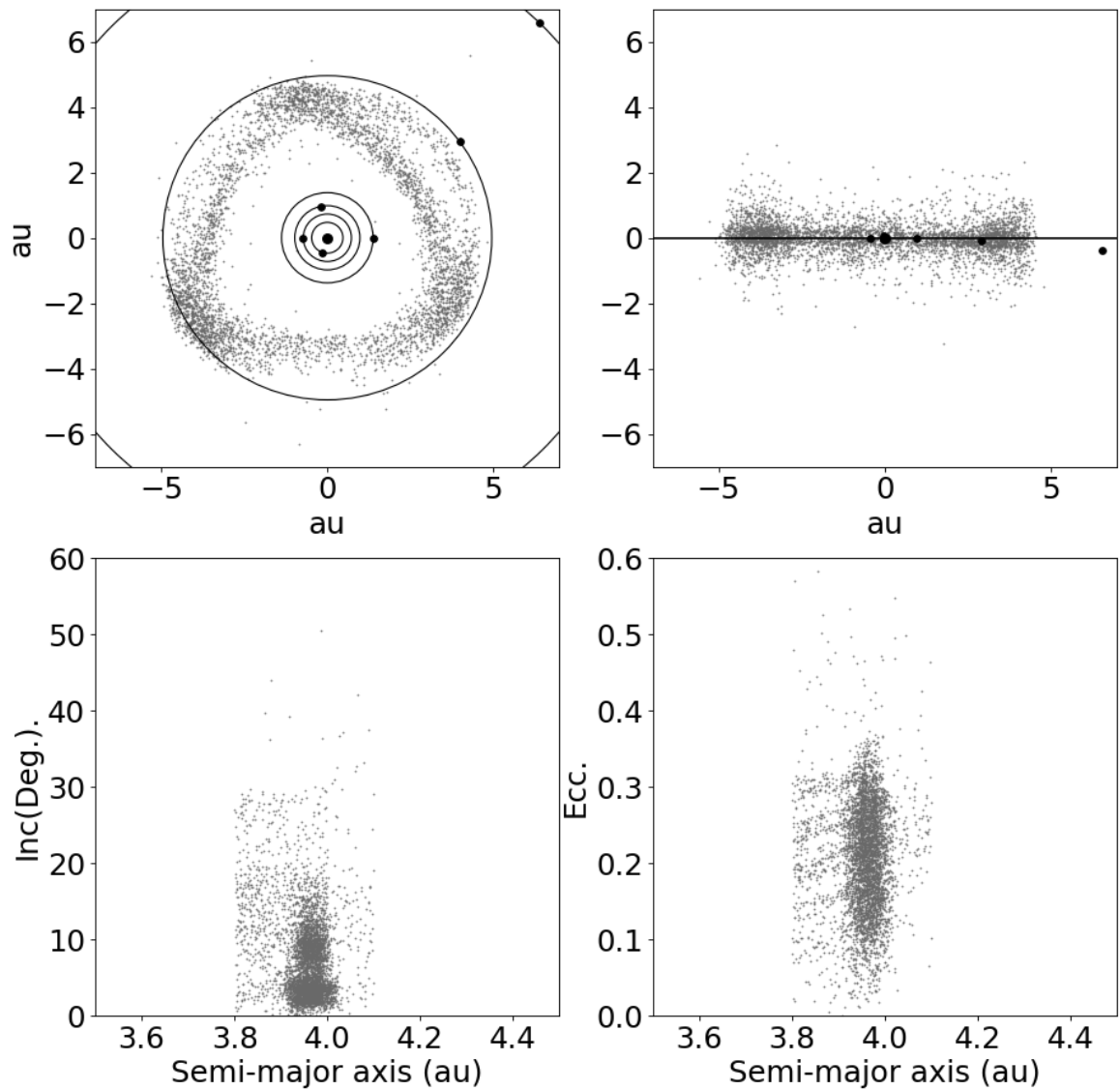

**Figure A7:** The distribution of the Hilda asteroids, trapped in 3:2 mean-motion resonance with Jupiter. When plotted in Cartesian coordinates, the Hildas trace an approximately triangular shape, with the vertices located at the locations of the Jovian $L_4$, $L_3$ and $L_5$ Lagrange points (working around in an anti-clockwise direction from the location of Jupiter). As the objects orbit the Sun, their eccentric paths are such that, whenever they are located between Jupiter and the Sun, they are close to perihelion, and hence far from the giant planet. As Jupiter orbits the Sun, the locations of the vertices of the triangle rotate with the planet, keeping station with the Lagrange points. A given Hilda will, through the course of three complete orbits around the Sun (two orbits of Jupiter) pass through each of the three vertices of the triangle. Given the number of objects located away from the main concentration of the Hildas (including those in the top left panel that seem relatively close to Jupiter, and those in the lower panels between ~3.8 and 3.9 au), it seems likely that a reasonable number of objects classified here as Hildas are actually interlopers, moving on unstable orbits whose 'Hilda-like' behaviour is a transient phenomenon (e.g. Brož & Vokrouhlický, 2008).

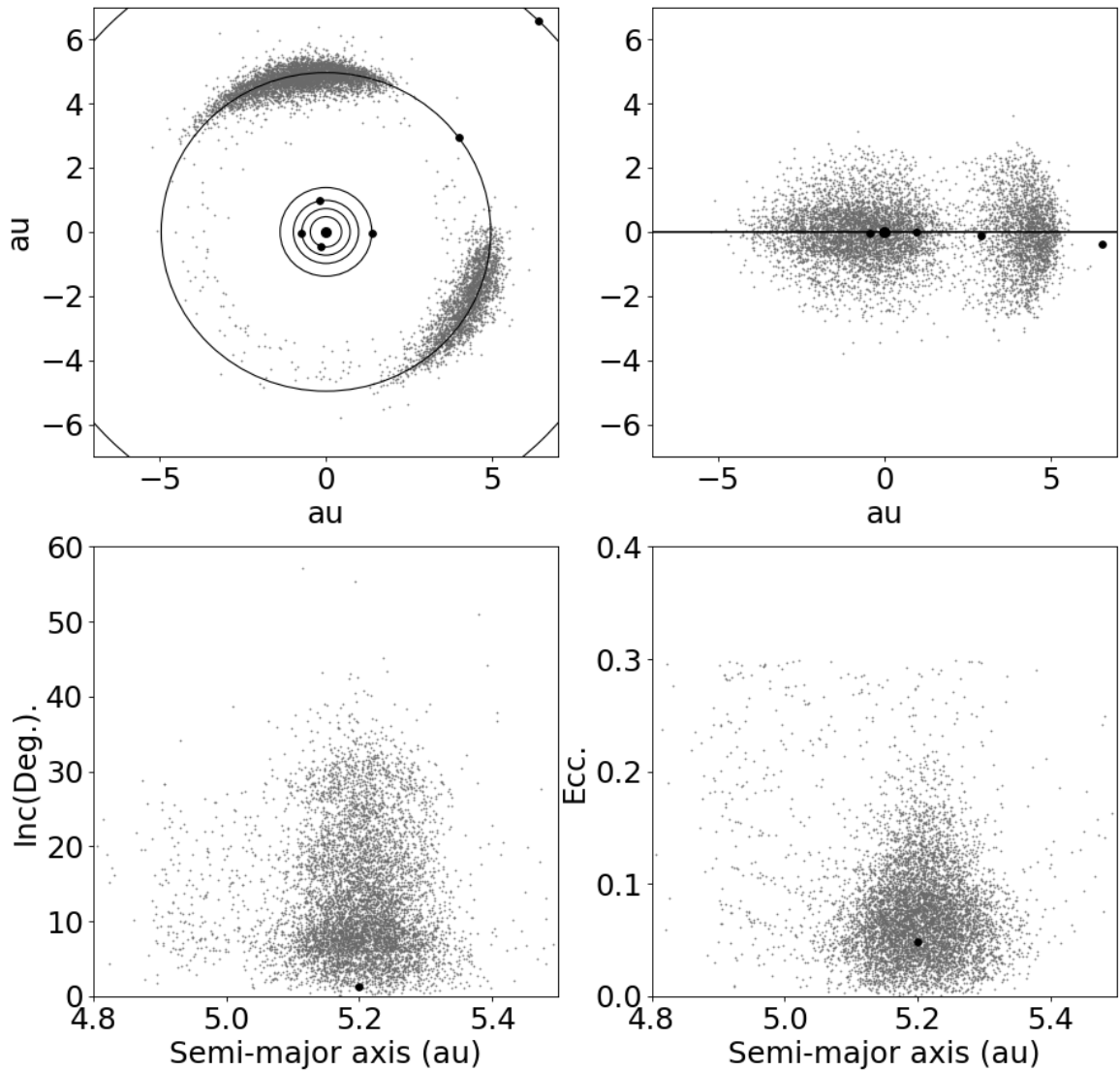

**Figure A8:** The distribution of the Jovian Trojans, trapped in 1:1 mean-motion resonance with Jupiter. The great majority of these objects move on 'tadpole' orbits, librating around the Jovian $L_4$ and $L_5$ Lagrange points. As a result, in the top-down view (at top left), the great bulk of the Trojan population are concentrated in two lengthy clouds, centred on those two Lagrange points. Rarer, and typically less dynamically stable, are the 'horseshoe' Trojans - objects whose libration within the 1:1 resonance takes them around from the $L_4$ to the $L_5$ points, and back again, passing through the $L_3$ Lagrange point, opposite Jupiter in its orbit. It should be noted that not all of the objects plotted here are true Trojans - there may well be a number of interlopers - objects temporarily captured into 1:1 resonance with Jupiter, and those experiencing short-term satellite capture by the giant planet (such as those closest to Jupiter in the top-left hand plot).

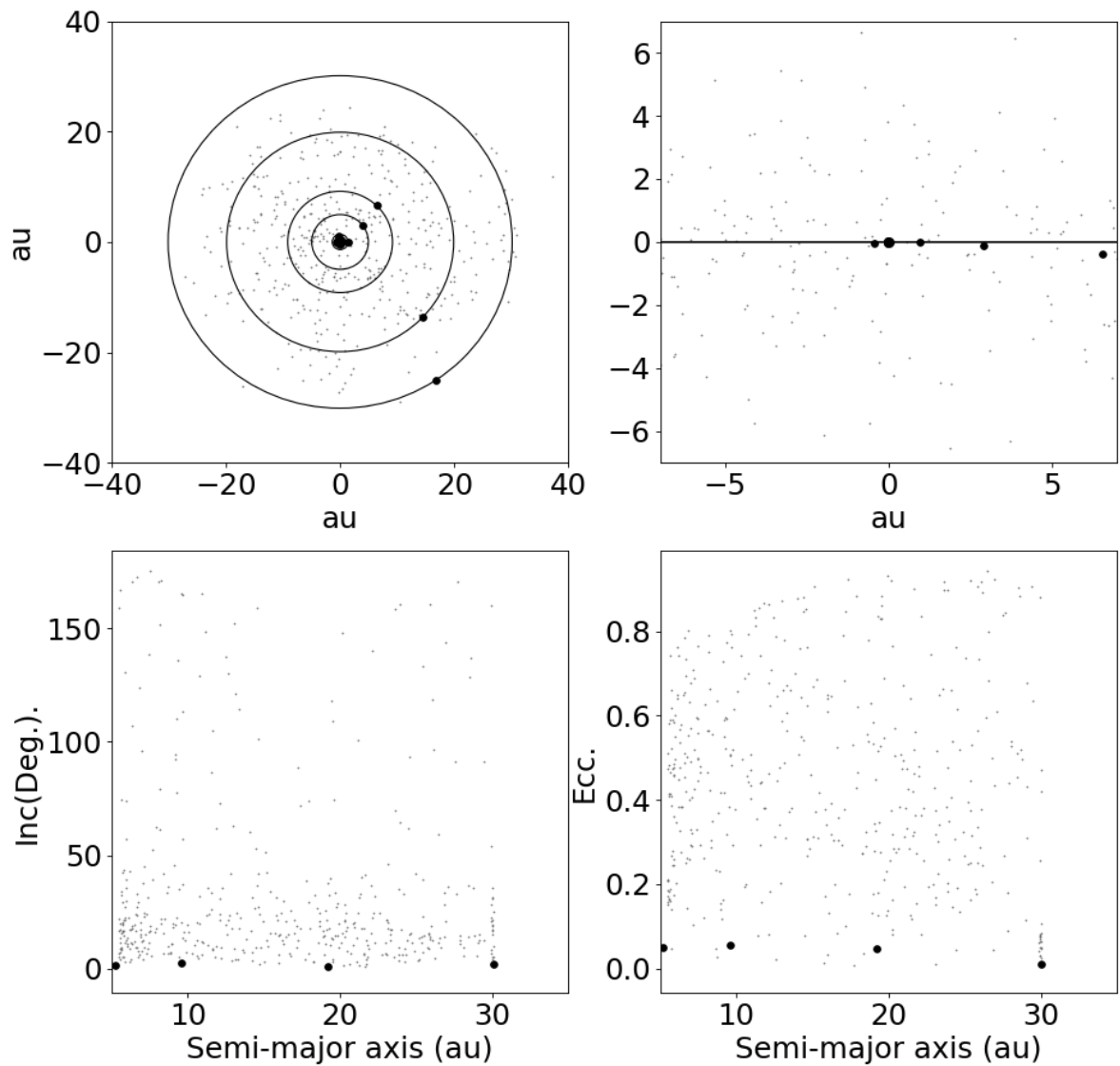

**Figure A9:** The Solar system's known Centaur population - dynamically unstable objects moving on orbits between those of the giant planets. The majority of these objects move on orbits with relatively low inclinations - less than ~30 degrees - a fact that reflects their expected origins in the Scattered Disc, Neptunian and Jovian Trojans, and the Edgeworth-Kuiper belt. The more inclined Centaurs are thought to be objects that originated in the Oort cloud, captured by the giant planets.

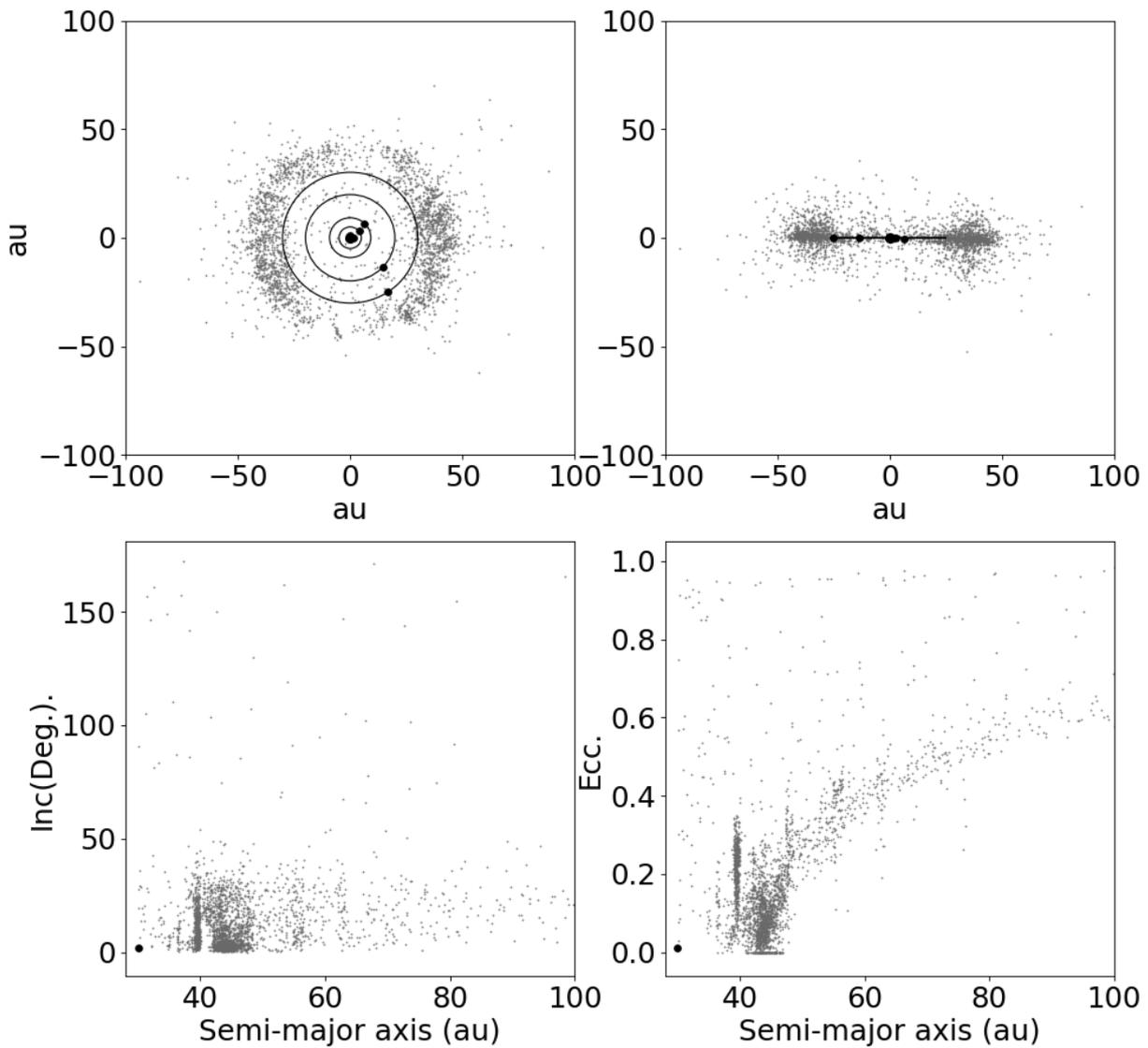

**Figure A10:** All known trans-Neptunian objects - including members of the Scattered Disc, the Edgeworth-Kuiper belt, the detached population, and the resonant TNOs. The objects clustered together in vertical 'spikes' in the lower two figures are the resonant TNOs - the most prominent population of which are the Plutinos, located at around 39.5 au. The Scattered disc objects, with perihelia between ~30 and ~39.5 au from the Sun can be clearly seen as the population that curves upwards to the right, whilst the large population of Edgeworth-Kuiper belt objects are found between ~40 and ~48 au from the Sun.

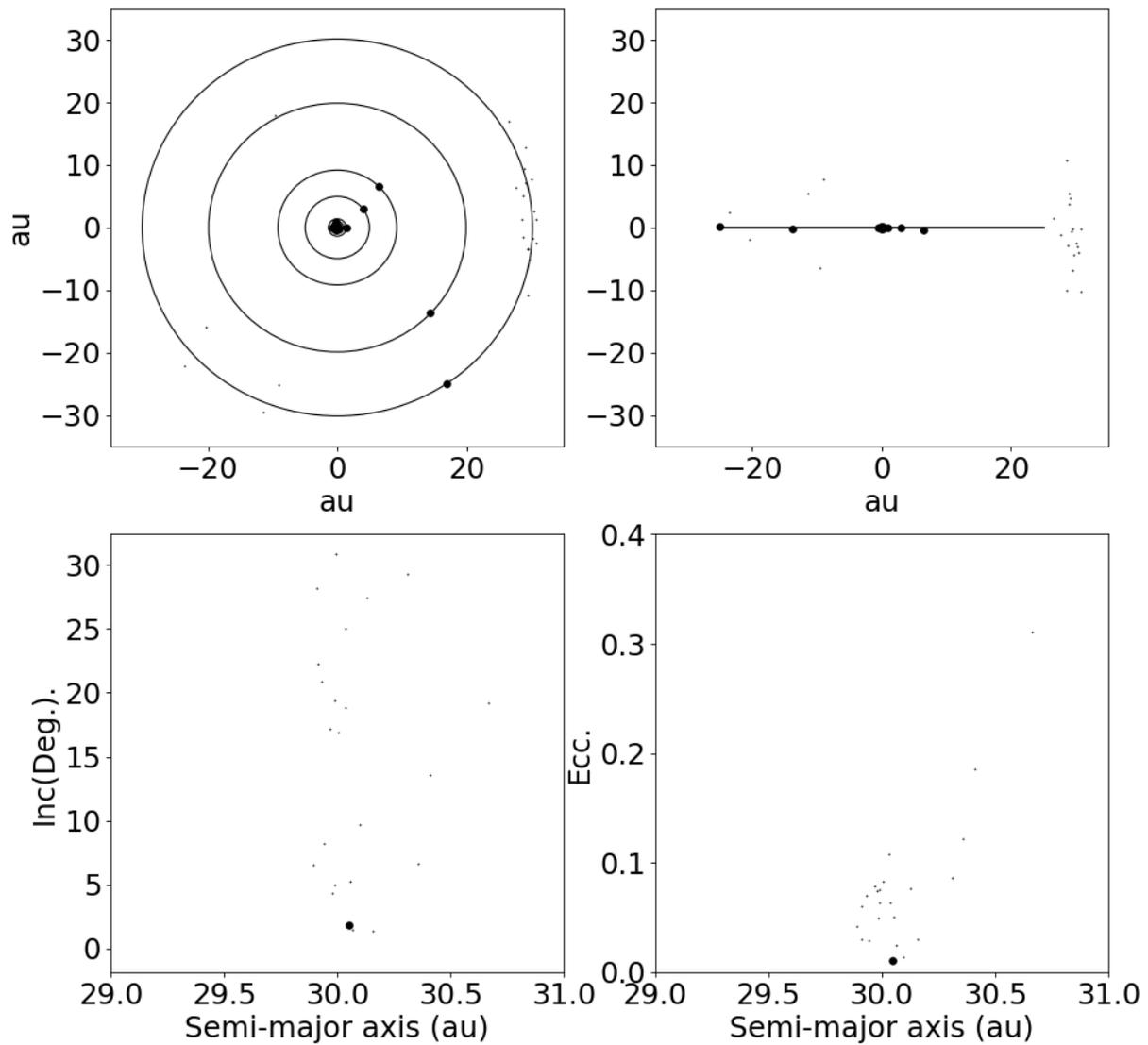

**Figure A11:** The distribution of those objects considered to be 'Neptune Trojans' - trapped in 1:1 mean motion resonance with Neptune. Whilst this population is still relatively small, and includes some temporarily captured objects, similarities can be seen to the Jovian Trojan population (shown in Figure A8). The Neptune Trojans are found in two clouds, around the $L_4$ and $L_5$ Lagrange points in Neptune's orbit.

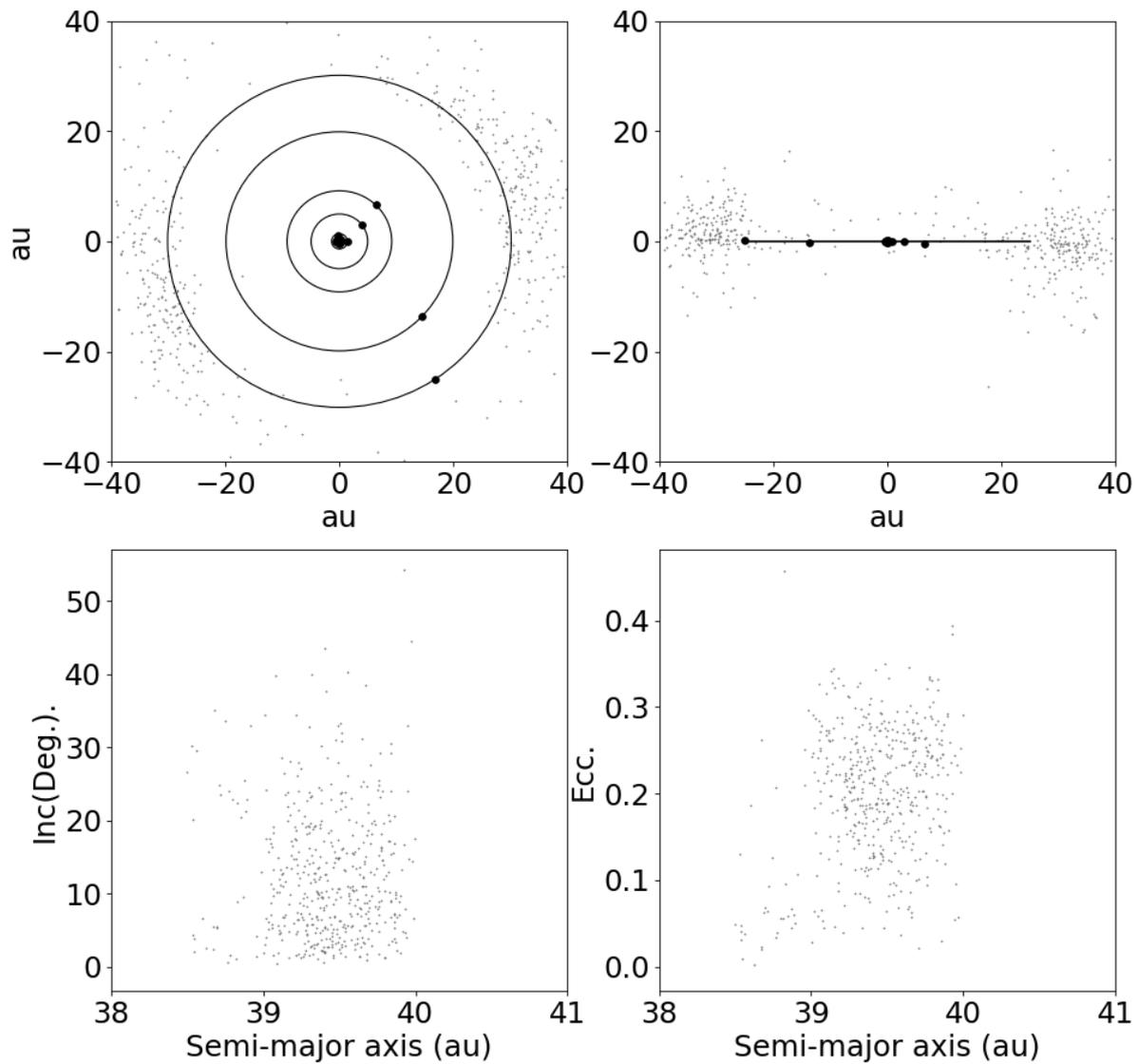

**Figure A12:** The distribution of the known Plutinos - objects trapped in 2:3 mean-motion resonance with Neptune. These objects are thought to have been captured by the giant planet into that resonance during its outward migration through the early Solar system - a process that resulted in the gradual orbital excitation of captured members (which explains the broad range of orbital inclinations and eccentricities exhibited by the population). Those Plutinos with orbital eccentricities greater than ~0.24 move on Neptune-crossing orbits - but their resonant motion prevents them from experiencing close encounters with the giant planet. This is the reason that, in the top-down view (top left), all those objects currently in the vicinity of Neptune's orbit are located either well ahead, or well behind, the planet in its orbit.

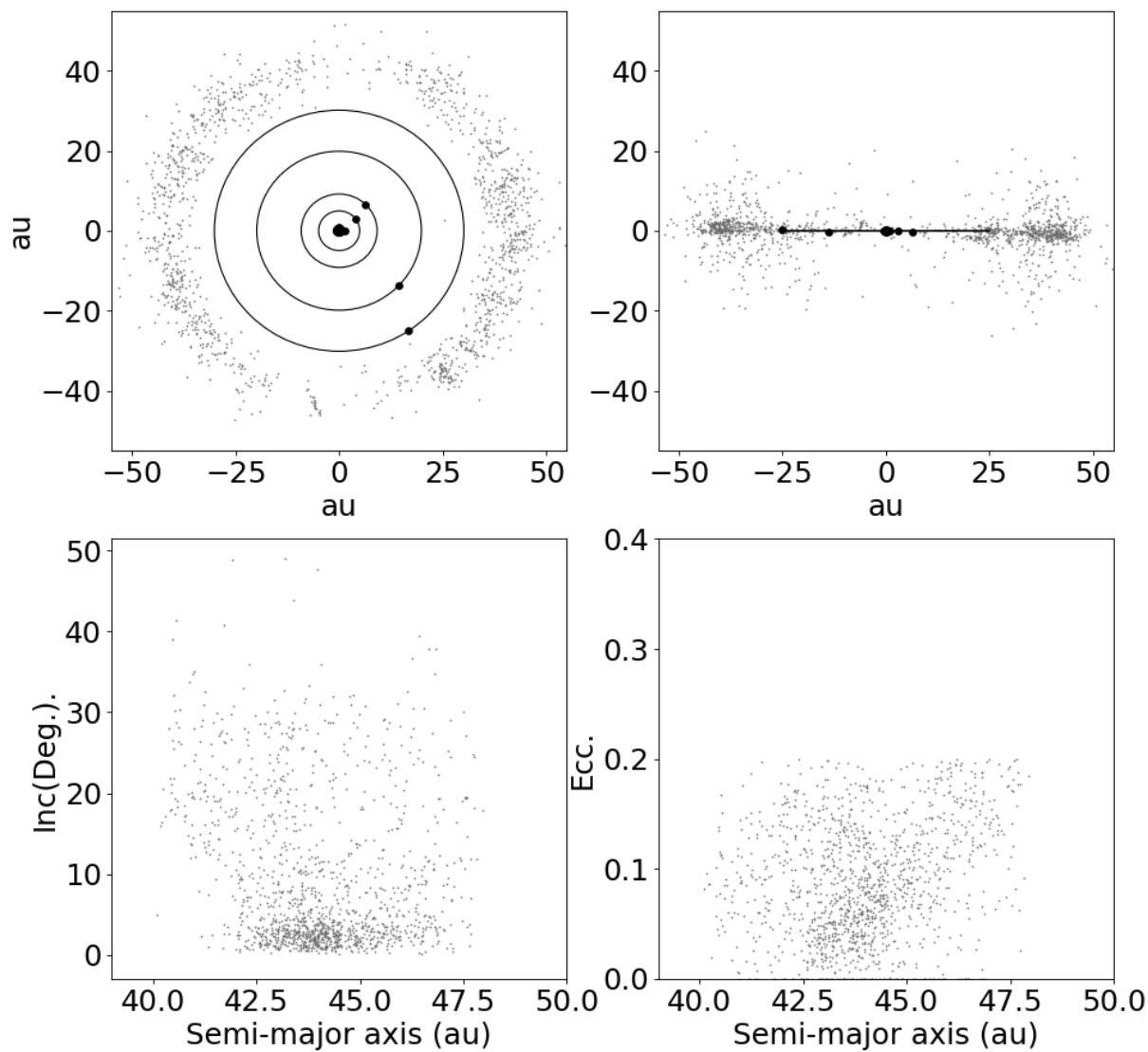

**Figure A13:** The distribution of objects in the classical Edgeworth-Kuiper belt. The top-down view here (top left) reflects the observational biases implicit in the discovery of these distant objects. The broad gap towards the bottom of the figure, where few TNOs can be found, corresponds to the direction of the centre of the Galaxy (the Milky Way), in Scorpius and Sagittarius, and represents the most challenging part of the sky in which to detect faint, star-like points of light. A similar, but less significant, rarefaction can be seen towards the top of that plot - which corresponds to the other location in which the plane of the ecliptic crosses the Galactic plane - between Taurus and Gemini. Thanks to the lower stellar density in that region, compared with that towards the Galactic centre, more objects can readily be found. The dynamically 'cold' Edgeworth-Kuiper belt objects can be seen as an excess of objects with inclinations less than ~5 degrees. The dynamically 'hot' population are more widely dispersed, with inclinations ranging up to, and sometimes exceeding, 30 degrees.

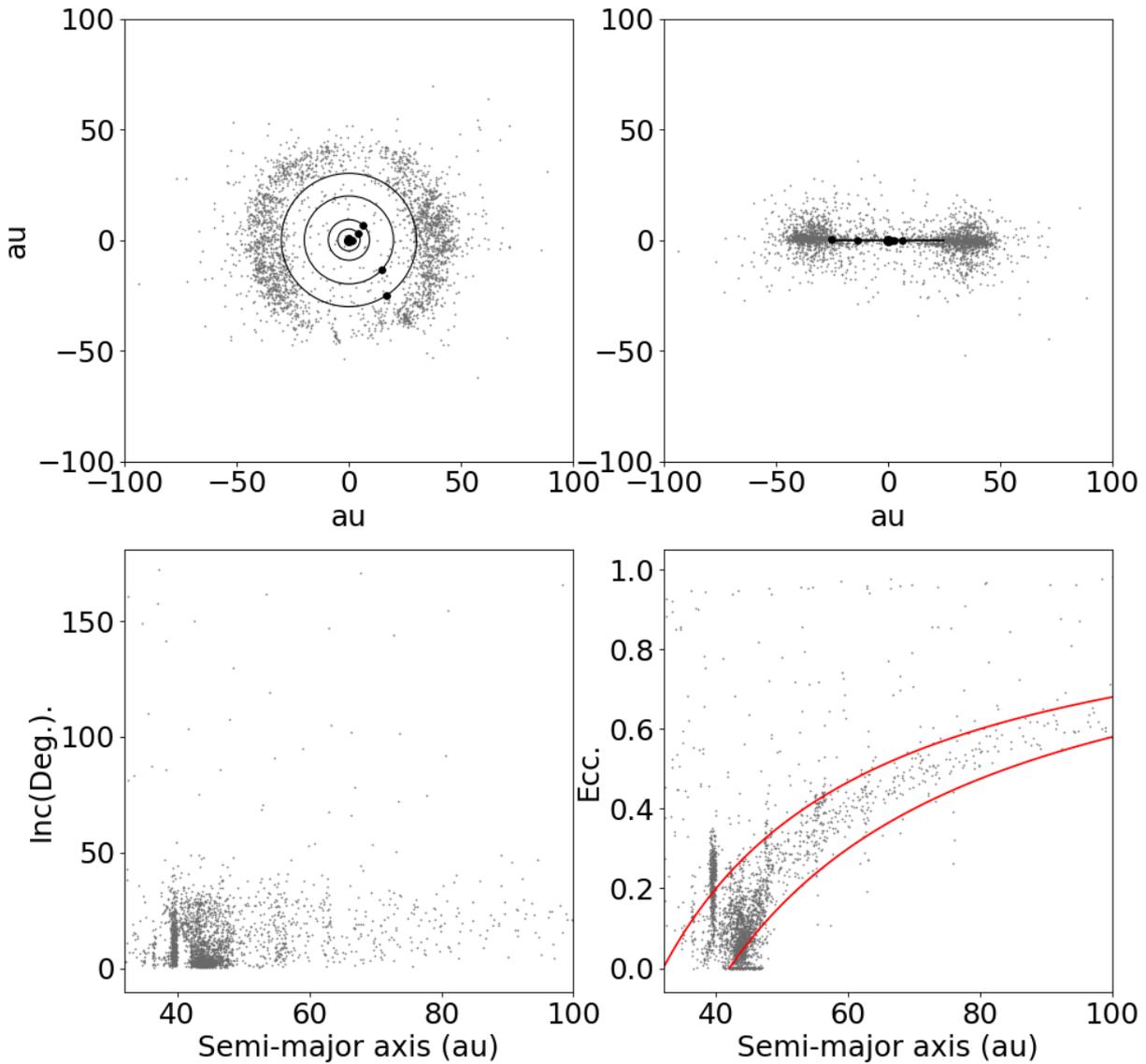

**Figure A14:** The distribution of the Scattered Disc and detached objects. The two red lines in the lower right-hand plot show the boundaries for the Scattered Disc (as per Table 2), defined as perihelia of ~32 au and ~40 au. Objects that fall below the lower of these two lines are considered to be 'detached'. The effect of resonant sticking can be clearly seen, with clusters of objects at specific semi-major axes. Whilst Scattered Disc objects would be expected to random walk in semi-major axis, as a result of perturbations by the giant planets (principally Neptune), it is possible for them to experience temporary (but lengthy) captures in various distant Neptunian mean-motion resonances, during which time their orbital eccentricity can very dramatically, sometimes even transferring those objects to the detached population, lifting their perihelia well beyond Neptune's immediate control. Objects above the upper of the two red lines are considered to be Centaurs (again, as per Table 2), though they appear in many lists of Scattered Disc objects as a result of the blurred nature of the divide between the two linked populations. The concentration of objects at $39 < a < 40$ au are the Plutinos, some of which fulfil the criteria of being either Centaurs or Scattered Disc objects. Here, the fact those objects are trapped in Neptune's 2:3 mean motion resonance supersedes their other orbital parameters in determining their classification, but they are included here for completeness.

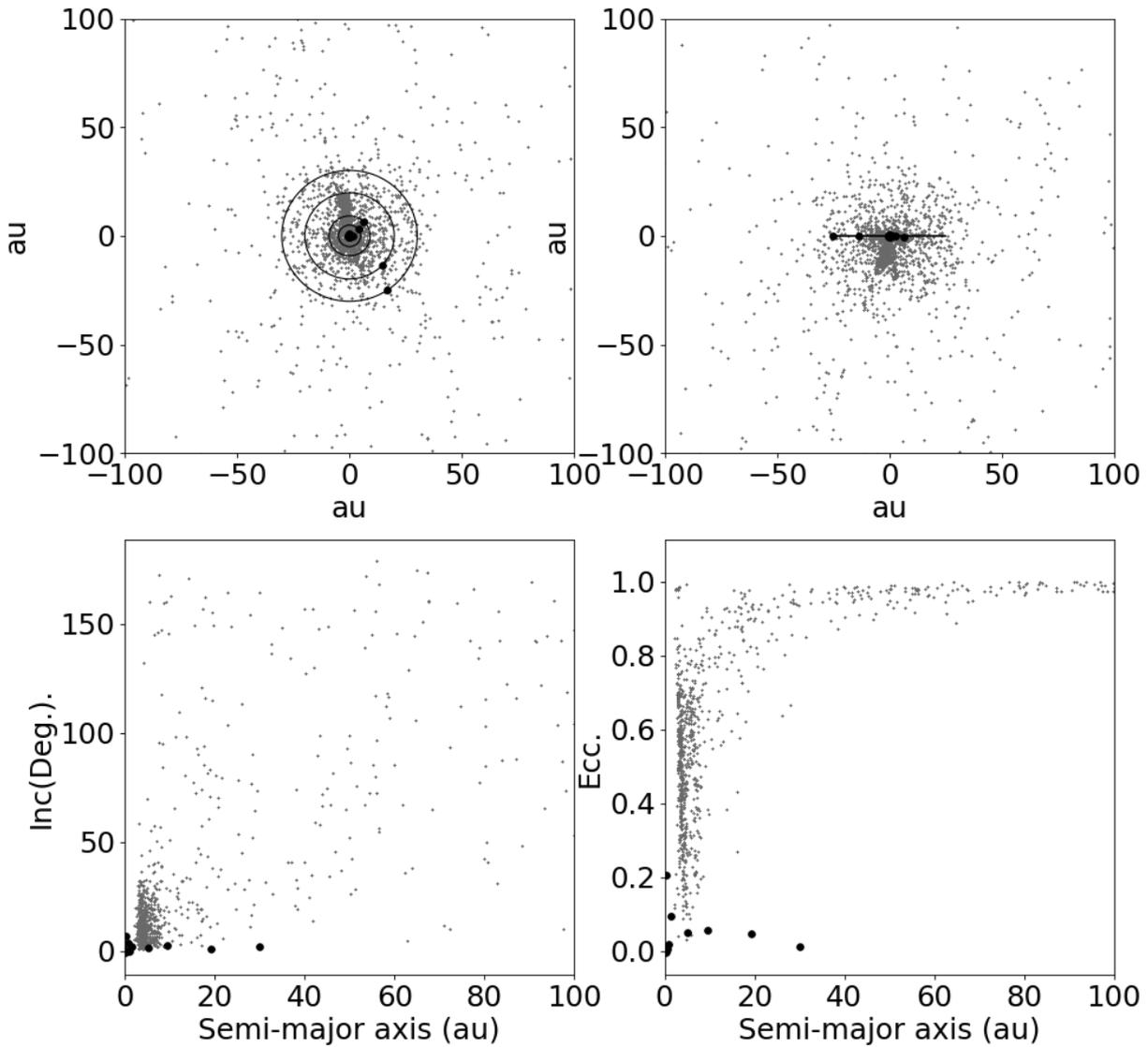

**Figure A15:** The distribution of all known comets. The concentration of objects in the top-left hand plot stretching outward slightly to the left of the vertical is the result of vast number of known members of the Kreutz Sungrazing comet population, the majority of which were discovered by the SOHO spacecraft. When plotted in semi-major axis space, the Jupiter-family comets stand out clearly, with semi-major axes smaller than the orbit of Saturn. Because of the truncation of the semi-major axis plots at 100 au, many long period comets do not appear on the plot - the tail of objects to very high (~1) eccentricities continues out to semi-major axes of over 100,000 au.

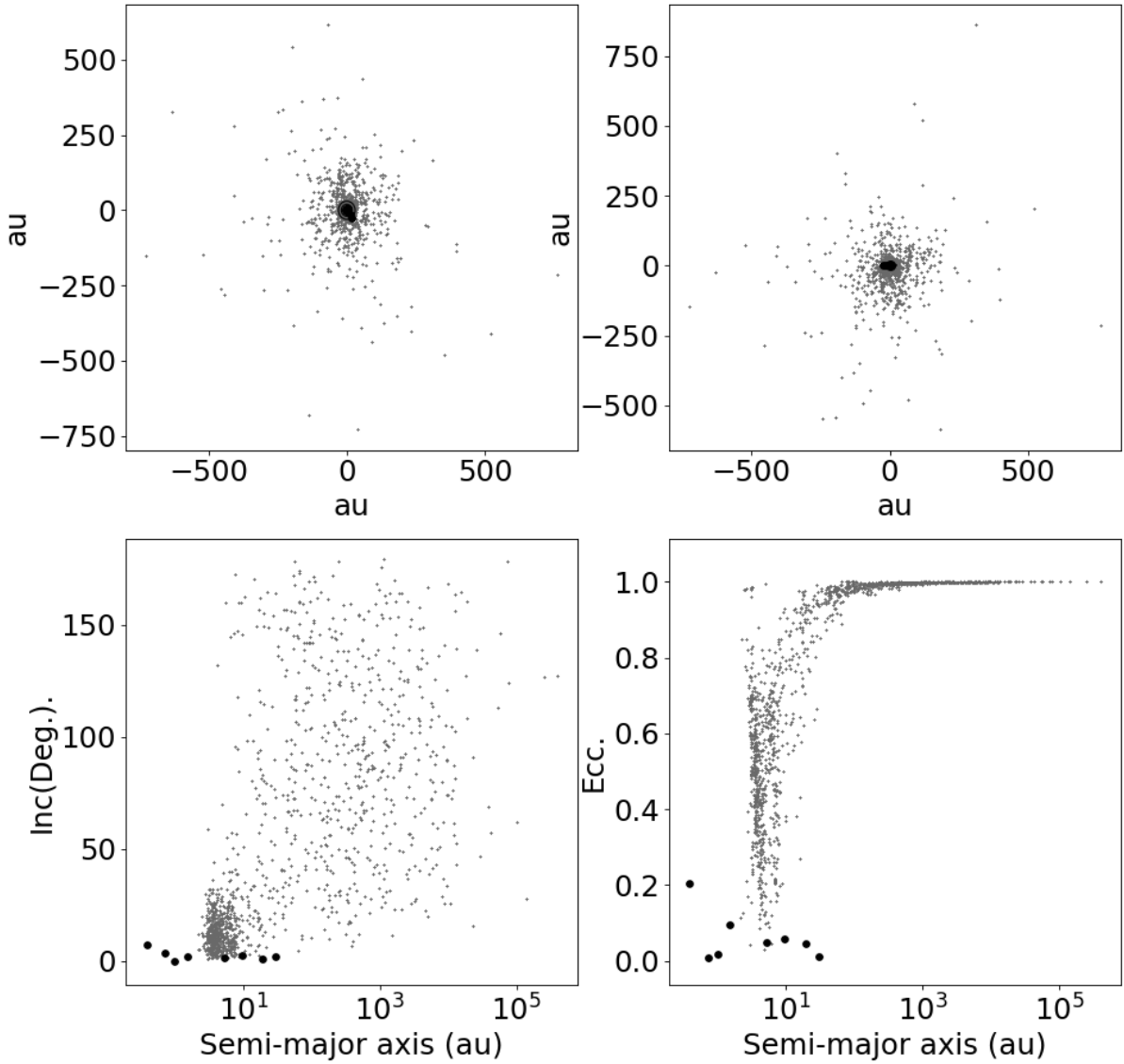

**Figure A16:** As with Figure A15, we show here the orbital distribution of all known comets. The upper panels now show all comets within a cubic space spanning approximately 1500 au on a side (i.e. roughly ±750 au in X, Y and Z), whilst the lower panels show the distribution in orbital element space, with a logarithmic semi-major axis scale, to show the extent of the long-period comet population. The isotropic nature of the long-period/Oort cloud comet flux is clearly seen in the lower left-hand panel, with those comets evenly distributed in orbital inclination - a stark contrast to the Jupiter-family/short period comets, whose orbital inclinations are typically less than ~30 degrees. It is this division that highlights the different origins of these two populations, with the long-period comets sourced from the Oort cloud, which is to first order isotropically distributed around the Sun.

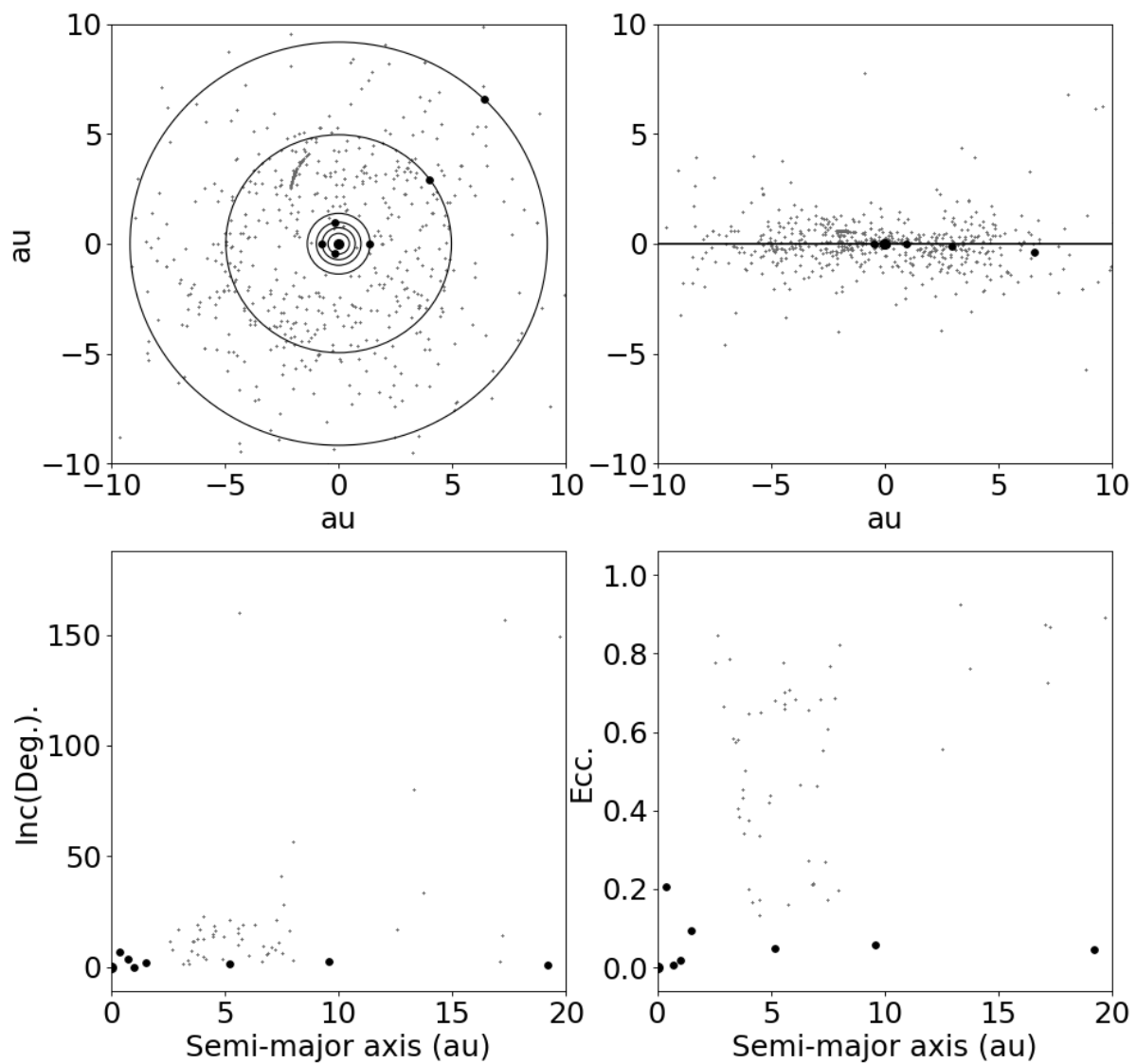

**Figure A17:** The known Jupiter-Family comets. The great majority of Jupiter-Family comets move on orbits with relatively small inclinations (< 30 degrees), reflecting their direct association with the Centaur and trans-Neptunian populations. Those at higher inclinations (particularly those with retrograde orbits - i.e. inclinations > 90 degrees) are likely more accurately considered to be Halley-type comets - but are included here based on their classifications in the data harvested from the JPL Horizons database. The 'arc' feature in the top-down plot (top-left), which lies interior to Jupiter's orbit, just above and to the left of centre, is a cluster of comet fragments, created when comet 73P/Schwassmann-Wachmann 3 broke apart in 1995. To date, almost 70 separate fragments of that comet have been identified, each moving on a slightly different orbit. In the ~25 years since the fragmentation of the comet, each fragment has completed almost five full orbits of the Sun, and as a result, the fragments have dispersed along a lengthy arc of the orbit of their parent comet.

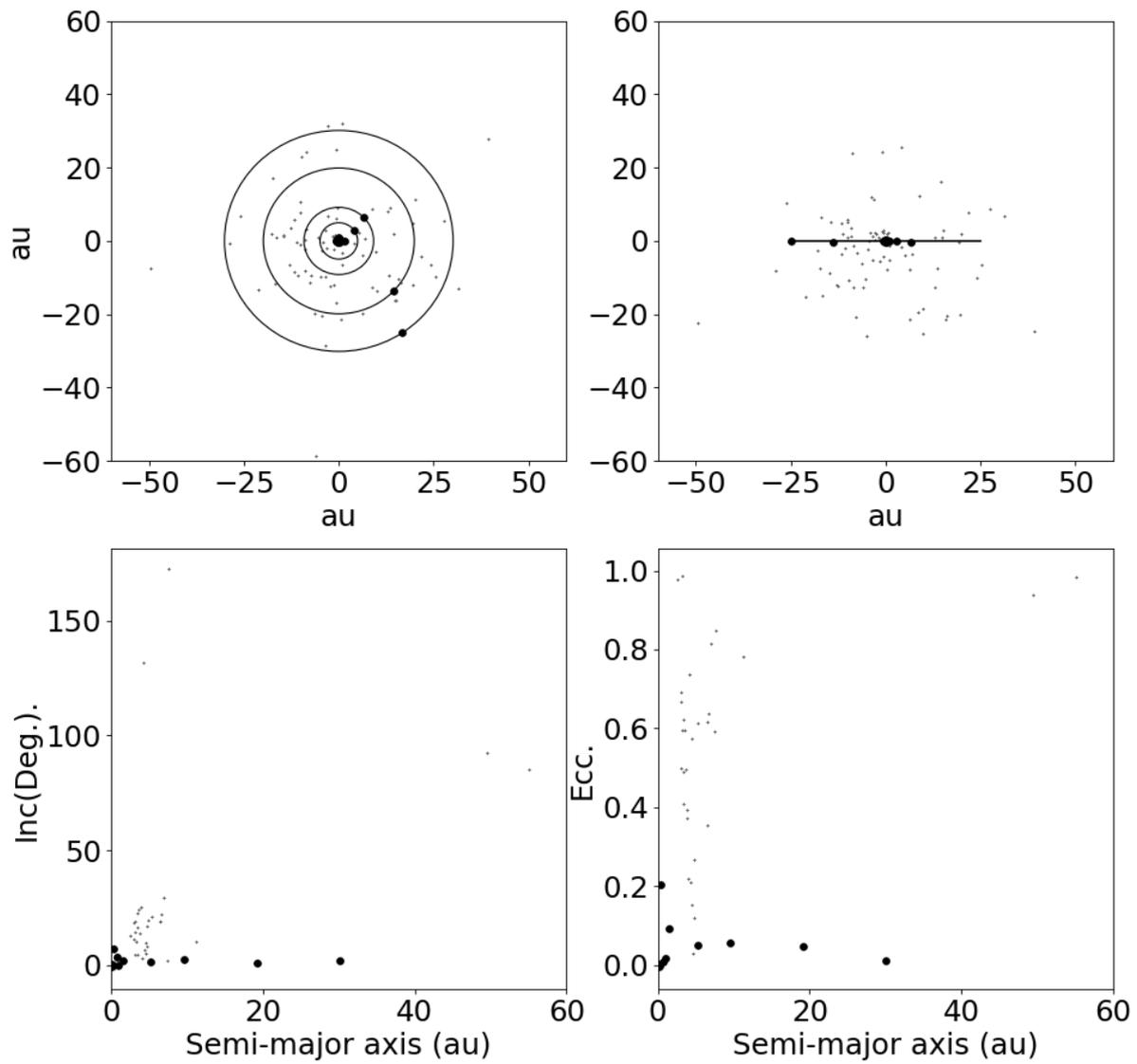

**Figure A18:** The known Halley-type comets. Where the bulk of the Jupiter-Family comets moved on orbits with low inclinations, the Halley Type comets feature a wide dispersal in orbital inclination, with several notable members moving on retrograde orbits (such as comet 1P/Halley itself, with an orbital inclination of ~162 degrees).